\documentclass[longauth]{aa}
\pdfoutput=1
\usepackage{graphicx}
\usepackage{txfonts}
\usepackage[utf8]{inputenc}
\usepackage{geometry}
\usepackage{longtable}
\usepackage{epstopdf}
\usepackage{adjustbox}
\usepackage{multirow}
\usepackage{lscape}

\titlerunning{XXL-1000-AGN}
\authorrunning{S. Fotopoulou et al.}

\begin{document}

    \title{The XXL Survey: VI. The 1000 brightest X-ray point sources\thanks{Based on observations obtained with XMM-Newton, an ESA science mission with instruments and contributions directly funded by ESA Member States and NASA. Based on observations made with ESO
Telescopes at the La Silla and Paranal Observatories under programme ID 089.A-0666 and LP191.A-0268} }

    \author{S. Fotopoulou\inst{\ref{Geneva}}
            \and F. Pacaud\inst{\ref{Bonn}}
            \and S. Paltani\inst{\ref{Geneva}}
            \and P. Ranalli\inst{\ref{NOA},\ref{Lund}}
            \and M. E. Ramos-Ceja\inst{\ref{Bonn}}
            \and L. Faccioli\inst{\ref{CEA}}
            \and M. Plionis\inst{\ref{Thessaloniki},\ref{Mexico},\ref{NOA}}
            \and C. Adami\inst{\ref{LAM}}
            \and A. Bongiorno\inst{\ref{Rome}}
            \and M. Brusa\inst{\ref{INAF_Bologna},\ref{Bologna}}
            \and L. Chiappetti\inst{\ref{INAF_Milan}}
            \and S. Desai\inst{\ref{LMU},\ref{ExcellenceCluster}}
            \and A. Elyiv\inst{\ref{Bologna},\ref{Ukraine_a}}
            \and C. Lidman\inst{\ref{AAO}}
            \and O. Melnyk\inst{\ref{Ukraine_b},\ref{Croatia}}
            \and M. Pierre\inst{\ref{CEA}}
            \and E. Piconcelli\inst{\ref{Rome}}
            \and C. Vignali\inst{\ref{Bologna},\ref{INAF_Bologna}}
            \and S. Alis\inst{\ref{Instabul}} 
            \and F. Ardila\inst{\ref{Florida}}
            \and S. Arnouts\inst{\ref{LAM2}}
            \and I. Baldry\inst{\ref{Liverpool}}
            \and M. Bremer\inst{\ref{Bristol}}
            \and D. Eckert\inst{\ref{Geneva}}            
            \and L. Guennou\inst{\ref{KwaZulu_Natal}} 
            \and C. Horellou\inst{\ref{Onsala}}
            \and A. Iovino\inst{\ref{INAF_Brera}}
            \and E. Koulouridis\inst{\ref{CEA},\ref{NOA}}
            \and J. Liske\inst{\ref{Uni_Hamburg}}
            \and S. Maurogordato\inst{\ref{Nice}} 
            \and F. Menanteau\inst{\ref{Illinois_a},\ref{Illinois_b}}
            \and J. J. Mohr\inst{\ref{LMU},\ref{ExcellenceCluster},\ref{MPE}}
            \and M. Owers\inst{\ref{UMac}}
            \and B. Poggianti\inst{\ref{INAF_Padova}}           
            \and E. Pompei\inst{\ref{ESO_chile}}
            \and T. Sadibekova\inst{\ref{CEA}} 
            \and A. Stanford\inst{\ref{UDavis}} 
            \and R. Tuffs\inst{\ref{Heidelberg}}
            \and J. Willis\inst{\ref{Victoria}}           
        }

    \institute{Department of Astronomy, University of Geneva, ch. d'Ecogia 16, 1290 Versoix, Switzerland\label{Geneva}\\ \email{sotiria.fotopoulou@unige.ch} 
    \and Argelander Institut f\"ur Astronomie, Universit\"at Bonn, Auf dem Huegel 71, 53121 Bonn, Germany\label{Bonn} 
    \and IAASARS, National Observatory of Athens, GR-15236 Penteli, Greece\label{NOA} 
    \and Lund Observatory, PO Box 43, 22100 Lund, Sweden \label{Lund} 
    \and Service d’Astrophysique AIM, CEA Saclay, F-91191 Gif sur Yvette\label{CEA} 
    \and Aristotle University of Thessaloniki, Physics Department, Thessaloniki, 54124, Greece\label{Thessaloniki} 
    \and Instituto Nacional de Astrofísica Óptica y Electrónica, AP 51 y 216, 72000 Puebla, Mexico\label{Mexico} 
    \and Université Aix Marseille, CNRS, LAM (Laboratoire d’Astrophysique de Marseille) UMR 7326, 13388, Marseille, France\label{LAM} 
    \and Osservatorio Astronomico di Roma (INAF), Via Frascati 33, 00078 Monte Porzio Catone (Rome), Italy\label{Rome} 
    \and INAF - Osservatorio Astronomico di Bologna, via Ranzani 1, 40127, Bologna, Italy\label{INAF_Bologna} 
    \and Dipartimento di Fisica e Astronomia, Universit\`a di Bologna, viale Berti Pichat 6/2, 40127 Bologna, Italy\label{Bologna} 
    \and INAF, IASF Milano, via Bassini 15, I-20133 Milano, Italy\label{INAF_Milan} 
    \and Faculty of Physics, Ludwig-Maximilians-Universit\"{a}t, Scheinerstr.\ 1, 81679 M\"{u}nchen, Germany\label{LMU}
        \and Excellence Cluster Universe, Boltzmannstr.\ 2, 85748 Garching, Germany\label{ExcellenceCluster}  
        \and Max Planck Institute for Extraterrestrial Physics, Giessenbachstrasse 1, 85748 Garching Germany \label{MPE} 
    \and Main Astronomical Observatory, Academy of Sciences of Ukraine, 27 Akademika Zabolotnoho St., 03680 Kyiv, Ukraine\label{Ukraine_a} 
    \and Australian Astronomical Observatory, PO Box 915,North Ryde NSW 1670, Australia\label{AAO} 
    \and Department of Physics, University of Zagreb, Bijenicka cesta 32, HR-10000 Zagreb, Croatia\label{Croatia} 
    \and Astronomical Observatory, National Taras Schevchenko University of Kyiv, 3 Observatorna St., 04053 Kyiv, Ukraine\label{Ukraine_b} 
    \and Department of Astronomy and Space Sciences, Faculty of Science, Istanbul University, 34119 Istanbul, Turkey\label{Instabul} 
    \and Department of Astronomy, University of Florida, Gainesville, FL 32611, USA\label{Florida} 
    \and Laboratoire d'Astrophysique de Marseille Pôle de l’Étoile Site de Château-Gombert 38, rue Frédéric Joliot-Curie 13388 Marseille cedex 13 France \label{LAM2} 
    \and Astrophysics Research Institute, Liverpool John Moores University, IC2, Liverpool Science Park, 146 Brownlow Hill, Liverpool L3 5RF, UK\label{Liverpool} 
    \and H.H. Wills Physics Laboratory, University of Bristol, Tyndall Avenue, Bristol, BS8 1TL, UK\label{Bristol} 
    \and Astrophysics and Cosmology Research Unit, University of KwaZulu-Natal, Durban, 4041, South Africa\label{KwaZulu_Natal} 
    \and Chalmers University of Technology Onsala Space Observatory SE-439 92 Onsala, Sweden \label{Onsala} 
    \and INAF - Osservatorio Astronomico di Brera, via Brera, 28, 20159 Milano, Italy\label{INAF_Brera} 
    \and Universit{\"a}t Hamburg, Hamburger Sternwarte, Gojenbergsweg 112, 21029 Hamburg, Germany \label{Uni_Hamburg} 
    \and Laboratoire Lagrange, UMR 7293, Université de Nice Sophia Antipolis, CNRS, Observatoire de la Côte d’Azur, 06304 Nice, France\label{Nice} 
    \and National Center for Supercomputing Applications, University of Illinois at Urbana-Champaign, 1205 W. Clark St., Urbana, IL 61801, USA\label{Illinois_a} 
    \and Department of Astronomy, University of Illinois at Urbana-Champaign, W. Green Street, Urbana, IL 61801, USA\label{Illinois_b} 
    \and Department of Physics and Astronomy, Macquarie University, NSW 2109, Australia and Australian Astronomical Observatory PO Box 915, North Ryde NSW 1670, Australia\label{UMac}  
    \and INAF-Astronomical Observatory of Padova, vicolo dell'Osservatorio 5, 35122 Padova, Italy\label{INAF_Padova} 
    \and European Southern Observatory, Alonso de Cordova 3107, Vitacura, Santiago de Chile, Chile\label{ESO_chile} 
    \and University of California, Davis, US\label{UDavis} 
    \and Astrophysics Department Max Planck Institut f\"ur Kernphysik Saupfercheckweg 1 D-69117 Heidelberg Germany\label{Heidelberg} 
    \and Department of Physics and Astronomy, University of Victoria, 3800 Finnerty Road, Victoria, BC, Canada\label{Victoria} 
}


    \abstract
    {X-ray extragalactic surveys are ideal laboratories for the  study of  the evolution and clustering of active galactic nuclei (AGN). Usually, a combination of deep and wide surveys is necessary to create a complete picture of the population. Deep X-ray surveys provide the faint population at high redshift, while wide surveys provide the rare bright sources. Nevertheless, very wide area surveys often lack the ancillary information available for modern deep surveys. The XXL survey spans two fields of a combined $\rm{50\,deg^2}$ observed for more than 6Ms with XMM-Newton, occupying the parameter space that lies between deep surveys and very wide area surveys; at the same time it benefits from a wealth of ancillary data.}
     {This paper marks the first release of the XXL point source catalogue including four optical photometry bands and redshift estimates. Our sample is selected in the $\rm{2-10\,keV}$ energy band with the goal of providing a sizable sample useful for AGN studies. The limiting flux is $\rm{F_{2-10\,keV}=4.8\,10^{-14}\rm{erg\,s^{-1}\,cm^{-2}}}$.}
     {We use both public and proprietary data sets to identify the counterparts of the X-ray point-like sources by means of a likelihood ratio test. We improve upon the photometric redshift determination for AGN by applying a Random Forest classification trained to identify for each object the optimal photometric redshift category (passive, star forming, starburst, AGN, QSO). Additionally, we assign a probability to each source that indicates whether it might be a star or an outlier. We apply Bayesian analysis to model the X-ray spectra assuming a power-law model with the presence of an absorbing medium.}
     {We find that the average unabsorbed photon index is $\rm{<\Gamma>=1.85\pm0.40}$ while the average hydrogen column density is $\rm{\log<N_{\rm{H}}>=21.07\pm1.2\,cm^{-2}}$. We find no trend of $\rm{\Gamma}$ or $\rm{N_{\rm{H}}}$ with redshift and a fraction of 26\% absorbed sources ($\rm{\log N_{\rm{H}}>22}$) consistent with the literature on bright sources ($\rm{\log L_x>44}$). The counterpart identification rate reaches 96.7\% for sources in the northern field, 97.7\% for the southern field, and 97.2\% in total. The photometric redshift accuracy is 0.095 for the full XMM-XXL with 28\% catastrophic outliers estimated on a sample of 339 sources.}
      {We show that the XXL-1000-AGN sample number counts extended the number counts of the COSMOS survey to higher fluxes and are fully consistent with the Euclidean expectation. We constrain the intrinsic luminosity function of AGN in the $\rm{2-10\,keV}$ energy band where the unabsorbed X-ray flux is estimated from the X-ray spectral fit up to $\rm{z=3}$. Finally, we demonstrate the presence of a supercluster size structure at redshift 0.14, identified by means of percolation analysis of the XXL-1000-AGN sample. The XXL survey, reaching a medium flux limit and covering a wide area, is a stepping stone between current deep fields and planned wide area surveys.}


    \maketitle

\section{Introduction}

Supermassive black holes in the centres  of galaxies participate in the evolution of their hosts. This is demonstrated by the relationships between their masses and the host galaxy properties, such as bulge luminosity or bulge stellar velocity dispersion \citep{Magorrian1998,Gebhardt2000,Ferrarese2000}. These relations show that the phenomena that form the galactic stellar mass and make the black holes grow are connected, although the precise physical mechanism that shapes them is still unclear.

Black hole growth is observable throughout the universe in the objects collectively called active galactic nuclei (AGN). These objects accrete galactic matter in an accretion disk which produces intense radiation before penetrating inside the central black hole. The connection between stellar mass and black hole mass ultimately results from the shared history of star formation and accretion, which follow the same pattern of steady increase until a peak redshift at $z\sim 1-2$, followed by a steep decline until present times \citep[e.g.][]{HopkinsSomerville2006,Silverman2008,Watson2009}. In addition, star formation and accretion seemingly share the same anti-hierarchical scenario, moving from massive galaxies to less massive galaxies \citep{Hirschmann2012}. Different versions of negative feedback affecting both star formation and accretion and thereby regulating the growth of stellar and black hole masses have been proposed.  Observations of activity in galaxies in the green valley, transiting from the blue cloud to the red sequence, put strong constraints on these mechanisms \citep{Schawinski2009}.

The role of black holes in galaxy evolution calls for a systematic census of AGN  in order to study the distribution of their parameters (e.g. luminosity, accretion rate or black hole mass) and their evolution, in order to relate them to the stellar properties. Because of the diversity of the AGN phenomenon, largely due to the presence of obscuring torus on parsec-scales in the immediate vicinity of the AGN that strongly affects  their observational properties, there is no unique way to build AGN samples, and each method is subject to some level of incompleteness.  Numerous studies of these distributions have therefore been performed using AGN samples selected with different means, such as optical spectroscopy \citep[e.g.][]{Bongiorno2007,Masters2012,Ross2013}, X-rays \citep[e.g.][]{Aird2010,Ueda2014,Miyaji2015}, infrared \citep[e.g.][]{Han2012}.

Among the different tools to select AGN, X-ray surveys are among the most efficient. With the exception of extended sources, AGN are often X-ray bright, at least above 2\,keV, and the vast majority of high-latitude X-ray sources turn out to be AGN down to extremely faint fluxes around $10^{-17}$ erg s$^{-1}$ cm$^{-2}$ in the 0.5--2\,keV band where a significant population of normal galaxies appears \citep{Lehmer2012}. Several important X-ray surveys have been conducted, from the widest all-sky surveys performed by ROSAT \citep{Voges1999}, ASCA \citep{Ueda2001,Ueda2005, Nandra2003}, BeppoSAX \citep{Fiore2001, Verrecchia2007}, and MAXI \citep{Ueda2011} to the deepest small-field surveys like the COSMOS field with XMM-Newton \citep{Hasinger2007} and Chandra \citep{Elvis2009} or the Chandra Deep Fields \citep{Giacconi2001,Alexander2003,Xue2011}. Thanks to its large collecting area, XMM-Newton is able to efficiently cover large sky areas reaching at the same time medium flux depth. For example, the Hard Bright Serendipitous Survey (HBSS) covers $\rm{25\,deg^2}$ reaching a flux limit of $\rm{F_{4.5-10\,keV}=7\times10^{-14}erg\,s^{-1},cm^2}$ \citep{DellaCeca2004}. The XMM-LSS survey \citep{Pierre2006}, a survey with XMM-Newton covering contiguously 11\,deg$^2$, is a good compromise between sky area and depth which has provided several results  on the environmental properties of AGN \citep[][Koulouridis et al., in press, Paper XII]{Elyiv2012,Melnyk2013,Koulouridis2014}, showing in particular different behaviors between objects having soft and hard X-ray spectra.

AGN surveys also efficiently probe the large-scale structure of the Universe. The clustering pattern of AGN and its evolution can provide important constraints on the mass of their dark matter halo hosts,  leading to somewhat conflicting results depending on the AGN selection 
\citep{Coil2007,Coil2009,Koutoulidis}, and shed light on the influence of the environment on the nuclear activity.  Furthermore, the dependence of AGN clustering on luminosity can also provide important
 constraints on the AGN triggering mechanism and on studies of AGN-galaxy coevolution, since different AGN fueling modes make distinct predictions for the environment of galaxies that host AGN 
\citep{Shankar09,Fanidakis13a}. The AGN triggering mechanism  adopted by models of galaxy formation are either major mergers for the most luminous AGN \citep{DiMatteo05,Hopkins06,Marulli09}, for which only weak 
luminosity dependence on clustering is expected \citep{Hopkins05, Bonoli09}, or secular disk instabilities for the lowest luminosity AGN \citep{HopkinsHernquist06,Bournaud11}, for which  luminosity 
dependence on clustering should be strong. Observational indications of varying strength for such dependencies have been reported in the literature 
\citep{Plionis08,Krumpe10,Cappelluti10,Koutoulidis,Fanidakis13}.

The Ultimate XMM-Newton Extragalactic X-ray survey, or XXL (Pierre et al., in press, Paper I), is an extension of the XMM-LSS survey at the same depth, but covering 50\,deg$^2$ in two 25\,deg$^2$ fields XXL-N (RA=02h20, DEC=-5d00) and XXL-S (RA=23h30, DEC=-55d00). While the XXL Survey has been primarily designed to build a consistent sample of galaxy clusters for cosmology (Pacaud et al. in press, Paper II), an immediate by-product of the survey is the identification of numerous point sources, most  of which will turn out to be AGN. In this paper, number VI of the first XXL release, we introduce the XXL point source catalogue and present the basic properties of the 1000 brightest objects.

The paper outline is as follows. In Section \S \ref{sec:Xobs} we describe the X-ray observations for XXL, namely the source detection, the extraction, and the analysis of the X-ray spectra. While X-rays are a powerful tool used to uncover nuclear activity in galaxies, additional information about the AGN and their galaxy hosts are needed.
In Section \S \ref{sec:multiwaveobs} we give an overview of the photometric data sets available in the XXL field assembled from both publicly available and proprietary surveys, and we describe the creation of the photometric catalogue and the counterpart assignment for the X-ray detections. We also give a summary of the photometric redshift estimation (photo-z hereafter), and the use of the Random Forest method to classify our sources.

In Section \S \ref{sec:multiwave_properties} we present the results of the X-ray spectral analysis also as a function of redshift and the median SEDs for our sample. In Section \S \ref{sec:web}, we put our sample within the context of the cosmic web, and we present the observed number counts, the determination of the X-ray AGN luminosity function up to redshift $z=3$, and the percolation analysis in 10Mpc and 25Mpc. In Section \S \ref{sec:release} we describe the released catalogue that accompanies this paper. Finally, in Section \S \ref{sec:conclusions} we close with our results and the future prospects of the XMM-XXL survey.

\section{X-ray observations}\label{sec:Xobs}
In this section we describe the X-ray source detection, the definition of the XXL-1000-AGN sample, and the X-ray spectral analysis.  We model the source X-ray spectra as a power-law distribution with the presence of an absorbing medium, determining both the photon index ($\rm{\Gamma}$) and the hydrogen column density ($\rm{N_{\rm{H}}}$).

\subsection{Source detection and sample selection}\label{sec:det}

The primary goal of XXL is the accurate detection of clusters of galaxies, which appear as extended emission in the $\rm{0.5-2.0\,keV}$ energy band. To this end, we  use the dedicated pipeline described in \citet{pacaud06} (\textsc{Xamin}).  As a by-product, the accurate detection of point-like sources is also possible. \textsc{Xamin} was   used previously in the XMM-LSS survey, the pilot of XXL, and was described in detail in \citet{Chiappetti2013}. In this work we are using \textsc{Xamin 3.3}, which uses the latest calibration files for the XMM-Newton observations.

\textsc{Xamin} proceeds in three stages: 1) images from the three EPIC detectors are combined and a smoothed image is obtained using a multiresolution wavelet algorithm tuned to the low-count Poisson regime \citep{starck98}, 2) source detection is then performed on this smoothed image via Sextractor \citep{bertin96} and a list of candidate sources is produced, and 3)  a maximum likelihood (ML) fit based on C$-$statistic \citep{cash79} is performed for each candidate source; only sources with a detection likelihood from the ML fit $>$ 15 are considered significant, which corresponds to 0.5 spurious sources within $\rm{10'}$ of the field of view (FOV) \citep{pacaud06}. This process is performed independently for the soft ($\rm{0.5-2\,keV}$) and hard ($\rm{2-10\,keV}$) bands.

The final stage of the catalogue creation is the ingestion of the source list in the database. At this stage the association of the soft and hard bands is performed. The complete procedure is described in \citet{Chiappetti2013}; here we briefly describe the main points of the process for completeness. We use a search radius of $\rm{10''}$ to match sources between the two detection bands. We also allow for a source to be below the ML fit $<$ 15, at most in one band. If a source is present in more than one pointing, we favour sources present in good pointings ($\rm{>7\,ks}$ exposure time and background level $\rm{<1.5\times10^{-5}\,cts\;s^{-1}\,pixel^{-1}}$, quality flag 0), over sources originating from pointings with low exposure time ($\rm{>3\,ks}$ exposure time, quality flag 1) and/or{\footnote{When both conditions of low exposure time and high background level are met, quality level 3 is assigned.}} high background level (background level $\rm{<4.5\times10^{-5}\;cts\,s^{-1}\,pixel^{-1}}$, quality flag 2). If a source is found in more than one good pointing, we favour the detection that has the smaller off-axis angle.

Finally, in order to define the count-to-flux conversion factors we assume a single power-law spectrum with fixed photon index $\rm{\Gamma=1.7}$ and hydrogen column density $\rm{N_{\rm{H}}=2.6\times10^{20}cm^{-2}}$. The following conversion factors, corresponding to 1 $\rm{ct\;s^{-1}}$, are used: for the PN camera $\rm{1.5\times10^{-12}erg\,s^{-1}cm^{-2}}$ and $\rm{7.9\times10^{-12}erg\,s^{-1}cm^{-2}}$ for the $\rm{0.5-2\,keV}$ and $\rm{2-10\,keV}$ energy bands, respectively, while for the MOS cameras the corresponding factors are $\rm{5\times10^{-12}erg\,s^{-1}cm^{-2}}$ and $\rm{23\times10^{-12}erg\,s^{-1}cm^{-2}}$ \citep[see also Table 3 in][]{Chiappetti2013}. Each source is assigned a unique flux in the soft ($\rm{0.5-2\,keV}$) and hard ($\rm{2-10\,keV}$) bands by taking the average flux from all respective PN and MOS observations.

Since the XXL survey was designed for cosmological studies using X-ray selected galaxy clusters as probes, the identification of extended X-ray sources is of central importance in our collaboration. The definition of extended sources is based on the $\rm{0.5-2\,keV}$ energy band and it is detailed in Paper II. Briefly, a detection enters the extended candidate list when it has an extent greater than 5'' and extension likelihood greater than 15. This list is then split into two categories: C1 for objects with extension likelihood greater than 33 and detection likelihood greater than 32 and C2 for the remaining candidates. The C1 class is mostly free from contamination, while the C2 class is 50\% contaminated by spurious detections or blended point-like sources (see Paper II for more details).

In the first release of our catalogue we present the brightest 1000 X-ray point sources, consisting of detections belonging to neither C1 nor C2.  By merging the source lists of the two fields, we selected the 1000 brightest X-ray point-like sources based on the $\rm{2-10\,keV}$ flux estimated by the \textsc{Xamin 3.3} and assuming a fixed photon index of $\rm{\Gamma=1.7}$. We chose to select our sources in the $\rm{2-10\,keV}$ band since we are mainly interested in AGN. We  refer to this sample throughout the paper as XXL-1000-AGN. Even though the sample contains a small number of unidentified objects (2.8\%), stars (2.3\%), and normal galaxies (3\%), the vast majority of the sources have an estimated luminosity of $\rm{L_{2-10\,keV}>10^{42}erg\,s^{-1}}$ (>90\%), the typical threshold used to separate AGN from starburst galaxies. 

In Table \ref{tab:Xray} we give an overview of the XXL-1000-AGN sample. The minimum flux, as estimated by the pipeline, is $\rm{F_{0.5-2\,keV}=1.35\times10^{-15}\,erg\,s^{-1}cm^{-2}}$ and $\rm{F_{2-10\,keV}=4.85\times10^{-14}\,erg\,s^{-1}cm^{-2}}$ for the soft and hard bands, respectively. Column $\rm{N_{Xspec}}$ gives the number of sources with good-quality X-ray spectra (signal-to-noise ratio S/N >3; see \S\ref{sec:Xspec_analysis}). Column $\rm{N_{ctp}}$  gives the number of identified counterparts using either optical or near infrared images (see \S \ref{sec:counterpart}). Columns $\rm{N_{spec-z}}$ and $\rm{N_{photo-z}}$ give the number of sources with spectroscopic redshift (spec-z) and photometric redshift (photo-z) determination, respectively. Finally, from $\rm{N_{median}}$ we see that the median redshift of this flux limited sample is approximately $\rm{0.63}$.

    \begin{table*}
    \centering
    \caption{Summary of 1000 brightest XXL AGN detections, fluxes refer to \textsc{Xamin 3.3} determined values. Column 1: sample name, column 2: area of each XXL field, columns 3-4: minimum flux detected in the $\rm{0.5-2\,keV}$ and $\rm{2-10\,keV}$ energy band, respectively, column 5: number of detected sources, column 6: number of sources with X-ray spectra, column 7: number of sources with identified optical counterparts, column 8: number of sources with spectroscopic redshift, column 9: number of sources with photometric redshifts when spectroscopic redshift is not available, column 10: median redshift of sample using spec-z when available and photo-z otherwise.}
    \label{tab:Xray}
    \begin{tabular}{lrrrrrrrrr}\hline
     sample     & area & $\rm{F_{0.5-2\,keV, min}}$  & $\rm{F_{2-10\,keV, min}}$  & $\rm{N_{det}}$ & $\rm{N_{Xspec}}$ & $\rm{N_{ctp}}$ & $\rm{N_{spec-z}}$ & $\rm{N_{photo-z}}$ & $\rm{z_{median}}$\\
         &  \small$\rm{(deg^2)}$ & \multicolumn{1}{c}{\small$\rm{(erg\,s^{-1}\,cm^{-2})}$ } & \multicolumn{1}{c}{\small$\rm{(erg\,s^{-1}\,cm^{-2})}$ }  &  &  & &  & &\\
    \hline
     XXL-N & 26.9 &  $1.35\times10^{-15}$ & $4.86\times10^{-14}$ &  558 & 481 & 540 & 339 & 201 & 0.621\\
     XXL-S  &  23.6 & $1.83\times10^{-15}$ & $4.86\times10^{-14}$ & 442 & 424 & 432 & 250 & 182 & 0.637\\     \hline
     XXL-1000-AGN  & 50.5 & $1.35\times10^{-15}$ & $4.86\times10^{-14}$ & 1000 & 902 & 972 & 589 & 383 & 0.632\\
    \hline
    \end{tabular}    
    \end{table*}

\subsection{Spectral extraction}\label{sec:sample}

For each source in the XXL-1000-AGN sample, a circular extraction region and an annulus-shaped background region were computed. The radii were chosen in order to maximise the expected S/N of the spectrum, as described below. 

Since XXL is a mosaic of many overlapping XMM-Newton pointings, any
source may be present on more than one of them and at different
off-axis angles. For the spectral analysis, we make use of all
pointings. However, our pipeline analyses each pointing individually,
which does not guarantee that a detection is available on all
pointings for each source. Therefore, the expected total counts $C$
available for spectral analysis from all exposures were estimated as
\begin{equation}
C = A \times ( C_\mathrm{B, ref} + C_\mathrm{CD, ref} )
,\end{equation}
where $C_\mathrm{B, ref}$ and $C_\mathrm{CD, ref}$ are the counts in
the reference exposure for the 0.5--2 and 2--10 keV bands,
respectively, and $A$ is a factor which corrects for the presence of
other exposures in addition to the reference one. The expected total
background surface brightness was also defined in the same way.

The $A$ factor depends on the number of overlapping exposures, on
their relative length, and on their relative overlap (through the
effective area of the EPIC CCDs, which depends on the off-axis
angle\footnote{See the XMM-Newton Users' Handbook, Sect.~3.2.2.2 and
  Fig.~13;
  http://xmm.esa.int/external/xmm\_user\_support/documentation/uhb/\\effareaoffaxis.html}). For
simplicity, we used $A=1.5$ for all sources, which is justified in the
following cases which we assumed as representative of sources in the
XXL mosaic:
\begin{itemize}
\item a source is detected in the reference pointing within $3\arcmin$
  of the boresight position, and in a second pointing, with the same length, at
  $9\arcmin$--$10\arcmin$ off-axis;
\item an exposure has been repeated, e.g.\ because of high
  background. Therefore, a source is detected in the reference
  pointing, and then in another one at the same off-axis angle, but
  the clean time on the non-reference exposure will be shorter.
\end{itemize}
Had we computed $A$ for each source individually, the expected
variation on $A$ would only have had a minor effect on the radii of
the extraction regions. A non-optimal radius would either have included more
background (if the radius had been larger than optimal) or cause some flux
loss (if the radius had been smaller than optimal). However, for the bright
sources discussed in this paper, such an effect is likely negligible.

The spectral extraction regions were computed using the {\tt autoregions} software \citep{Ranalli2015} which for every source $i$ 
\begin{itemize}
\item defines the source regions as circles of radius $r_i$ and the
  background regions as annuli of inner and outer radii $1.5r_i$ and
  $2r_i$, respectively;
\item given the expected source counts and background surface
  brightness, finds the $r_i$ which maximises the expected
  S/N of the spectrum;
\item checks for neighbour sources. If there are any within
  $12\arcsec$, source $i$ is considered confused and dropped from the
  sample. Else, for any neighbour $j$ within $30\arcsec$ of source
  $i$, it excises a circular area of radius $r_j$ from both the source
  and background regions of $i$.
\end{itemize}

The spectra were finally extracted using the {\tt cdfs-extract} software
\citep{Ranalli2015}, which for any pair of source $i$ and pointing $k$
(treating PN, MOS1, MOS2 exposures individually)
\begin{itemize}
\item checks if source $i$ is present in the FOV of $k$;
\item checks that no more than 40\% of the flux of $i$ is lost in chip
  gaps, missing MOS1 CCDs, and borders of the FOV, otherwise the
  ($i$,$k$) pair is dropped;
\item extracts source and background spectra, and computes responses
  using the standard XMM SAS tools ({\tt evselect}, {\tt rmfgen}, {\tt
    arfgen}).
\end{itemize}

\subsection{Spectral analysis setup}\label{sec:Xspec_analysis}

For each detected source we produced their spectral products (PHA and response files) per camera and per exposure as described in the previous section. In addition, a background region was  associated to each source and its spectral products were extracted. Thus, we obtain a varying number of source spectra and corresponding background spectra for each source. Because of the relatively shallow depth of the XXL survey, spectra  generally have few counts, even for the brightest sources. This is especially true for background spectra, meaning that they are subject to large relative uncertainties. We therefore do not subtract the background spectra from the source+background spectra, but we use them as source-free fields to constrain the background parameters. 

Following the approach of \citet{2008A&A...486..359L}, we model the background in all observations (source and source-free)  using two components: X-ray background (XB) and non-X-ray background (NXB). The XB itself consists of three components: the cosmic X-ray background (CXB), the Local Bubble emission, and the Galactic Halo \citep[see Fig. B.1 in][]{2008A&A...486..359L}. We model the first  with a power law and the last two with a thermal plasma emission (\texttt{apec} model in \texttt{xspec}). The slope of the CXB component is fixed to the value of 1.46 \citep{2004A&A...419..837D}, but its normalisation is allowed to vary for any given sky position (source and corresponding source-free regions are fixed to the same normalisation). The shape of the Local-Bubble and Galactic Halo emissions are fixed, but their normalisations are determined using ROSAT data. We extracted from HEASARC's X-Ray Background Tool\footnote{{http://heasarc.gsfc.nasa.gov/cgi-bin/Tools/xraybg/xraybg.pl}} the ROSAT spectra from regions covering the XXL-N and XXL-S areas \citep{Snowden1997}. The normalisations of the two components were fixed by fitting the emission models (including CXB) over these two areas. Depending on the instrument,  the NXB consists of a combination of several emission lines typically resulting from internal fluorescence and continuum components phenomenologically represented  with a combination of power laws and/or a blackbody. The CXB and NXB models and parameters were derived from detailed studies of the XMM-Newton/EPIC background  \citep{2004A&A...419..837D,2008A&A...486..359L,2008A&A...478..575K,2014A&A...570A.119E}.

Considering the very complex data sets, background models, and the large number of sources that prevent manual analysis of each source separately, we adopt here a Bayesian approach. We  avoid fitting the models in the traditional sense with the goal of providing point estimates for all model parameters, which would include the parameters of the CXB and NXB. Instead, we  build credible intervals for the parameters of interest and marginalise over the parameters of the CXB and NXB, which are considered as nuisance parameters for this work. Non-informative priors are selected for all parameters (using the Jeffreys prior in the case of norm parameters). We determine the Bayesian evidence using the nested sampling algorithm \citep{Skilling2004}. For this purpose, we use the BXA \texttt{xspec} interface \citep{Buchner2014} of the MultiNest implementation \citep{2008MNRAS.384..449F,2009MNRAS.398.1601F,2013arXiv1306.2144F} of this algorithm. MultiNest directly builds a posterior sample from the full distribution, which we marginalise over the nuisance parameters to recover the joint posterior distribution of the parameters of interest (the source's parameters).

    \begin{figure}
    \begin{tabular}{c}
    \includegraphics[width=\columnwidth]{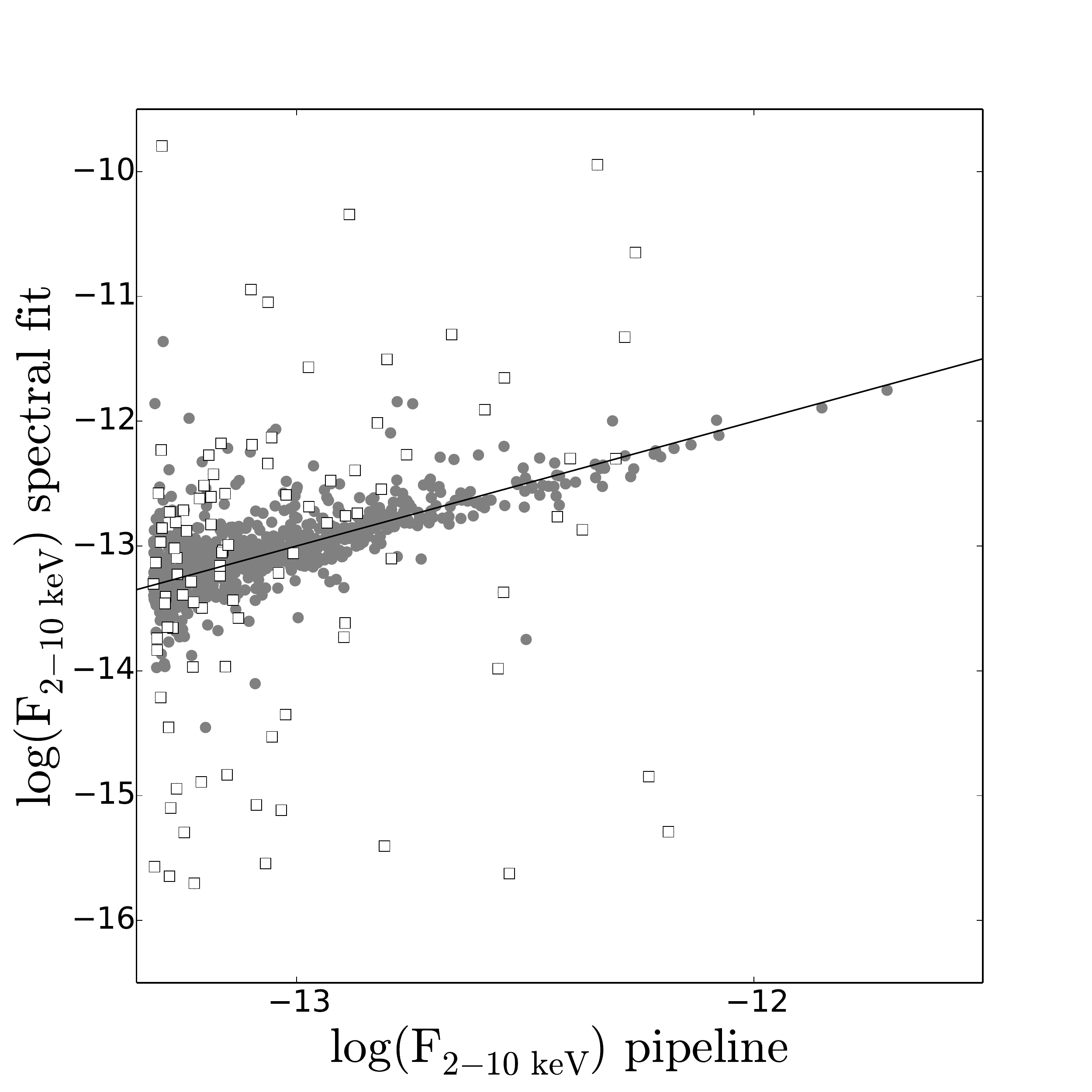} \\
    \end{tabular}
    \caption{Comparison between the $\rm{2-10\,keV}$ flux estimated by \textsc{Xamin 3.3} assuming a power-law spectrum with $\rm{\Gamma=1.7}$, $\rm{N_{\rm{H}}=2.6\times10^{20}\,cm^{-2}}$, and the spectral fit value (grey points). The black line shows the one-to-one relation. The open squares denote the sources that have low-quality X-ray spectra, (flag=4, see Table \ref{tab:RatioF}). 
    \label{fig:fluxfit}}
    \end{figure}

    \begin{table*}
    \centering
    \caption{X-ray spectral quality classes defined based on the XMM pointing quality and the signal-to-noise ratio (S/N) of the spectrum. The XMM pointing quality classes are: 0 = good quality ($\rm{>7\,ks}$ exposure time and $\rm{<1.5\times10^{-5}cts\;s^{-1}cm^{-2}}$ background level, 1 = low exposure time ($\rm{>3\,ks}$), 2 = high background $\rm{<4.5\times10^{-5}cts\;s^{-1}cm^{-2}}$). Pointings with quality class 3 corresponding to both low exposure time and high background were not used for spectrum extraction. }
    \label{tab:RatioF}
    \begin{tabular}{ccrcccc|rrrrrrrrr}\hline
     \multicolumn{2}{c}{Quality class} & N & \multicolumn{3}{c}{$F_{\textsc{Xamin}}/F_{fit}$} &  Flag & \multicolumn{3}{c}{PN Counts} & \multicolumn{3}{c}{MOS1 Counts} & \multicolumn{3}{c}{MOS2 Counts} \\ \cline{1-2} \cline{4-6} \cline{8-10} \cline{11-13} \cline{14-16}
      Pointing & S/N & & med & mean & $\rm{\sigma}$ & & min & max & med & min & max & med& min & max & med \\
    \hline
     0     & >6 & 547 & 1.00 & 1.0 & 0.3 & 1 & 25 & 11194 & 268 & 13 & 2626 & 105 & 6 & 2651 & 107 \\
     0/1/2 & >4 & 281 & 1.01 & 1.1 & 0.5 & 2 & 8 & 1879 & 103 & 3 & 244 & 43 & 6 & 415 & 40  \\
     0/1/2 & 3-4 & 75 & 1.07  & 2   & 3 & 3 & 6 & 246 & 43 & 2 & 49 & 23 & 4 & 81 & 20  \\
     0/1/2 & <3 & 96  & 0.78  & 52  & 191 & 4  & 3 & 177 & 16 & 1 & 34 & 6 & 1 & 55 & 5 \\
     NoData & -- & 1  & --    & --  & -- & 5 & -- & -- & -- & 2 & 2 & -- & 1 & 1 & --  \\
    \hline
    \end{tabular}    
    \end{table*}

We perform a fit appropriate for AGN, assuming a power law absorbed by our Galaxy, allowing the presence of absorption in the host galaxy at the redshift of the source (\texttt{phabs*zphabs*powerlaw} in \texttt{xspec}); therefore, the redshift of the host galaxy absorber needs  to be determined (see Sect.\ 4). It is either fixed to the spectroscopic redshift when available, or let free in the 68\% confidence interval of the photometric redshift. The full 0.5--12\,keV energy range is used, and the allowed range for the intrinsic photon index is $\rm{\Gamma=[1, 3]}$, while for the intrinsic absorption $\rm{\log{N_{\rm{H}}}=[19, 24.5]}$. We include the 10--12\,keV range, as it efficiently constrains the continuum part of the NXB. We rebin all spectra individually so that they have at least three counts in each bin to avoid bins with zero counts. The Galactic hydrogen column density was determined from the hydrogen maps of \citet{1990ARA&A..28..215D}. The resulting fixed Galactic absorption, modelled  with the \texttt{phabs} \texttt{xspec} model, is applied to all components except the Local Bubble emission and the NXB. All components except the source are scaled to the area of the extraction region. As the NXB does not have an astrophysical origin, we did not apply the effective area correction (ARF) on the NXB models;  however, we applied the redistribution matrix (RMF) to take into account the instrument resolution and redistribution.

In the released catalogue we provide a flag (see Tab. \ref{tab:RatioF}) showing the combined quality of the X-ray pointing and the S/N of the X-ray spectrum. From Tab. \ref{tab:RatioF}, we see that sources with S/N above four have the highest quality spectra. Fig. \ref{fig:fluxfit} shows the comparison between the $\rm{2-10\,keV}$ flux determined by \textsc{Xamin 3.3} and the spectral fit. The black line shows the one-to-one relation. The scatter of the points around the one-to-one relation is due to 1) the combined quality of the XMM pointing and the extracted spectrum and 2) the real spectrum of the source. Both are expected to cause a scatter of the estimated flux around the one-to-one relation since lower quality pointings introduce noise in the spectra, while the real spectrum of the observed sources will deviate from the power law of constant photon index assumed in the pipeline. In Table \ref{tab:RatioF}, we summarise the median, mean, and standard deviation of the ratio $\rm{F_{2-10\,keV,pipe}/F_{2-10\,keV,fit}}$. The scatter between $F_{2-10\,keV,pipe}$ and $F_{2-10\,keV,fit}$ actually seems to increase as we move to fainter sources, but no systematic offset is easily observed. The open squares show sources with S/N<3 (flag=4). The challenging nature of flag=4 sources is the result of a combination of factors including the low quality of the X-ray spectrum, the difficulty in constraining the background, and the uncertainty on the redshift of the source. We suggest  avoiding the use of these sources in a source-by-source scientific analysis based on the X-ray spectral fits. These sources are not used in \S \ref{sec:Xspec_results} where the results of the X-ray spectral analysis are discussed in detail.

In what follows, unless otherwise stated, we also consider  sources with flag 3 since estimated flux shows  a  slightly larger scatter ($\rm{\sigma=3}$) than  the pipeline, but not a significant offset (flux ratio median = 1.07). In Tab. \ref{tab:RatioF} we also present the observed counts for PN, MOS1, and MOS2 split according to the quality of the X-ray spectra. For the highest quality spectra (flag=1) the PN counts range from 25 to $\rm{10^4}$ counts, while the median observed counts are 268, 105, and 107 for PN, MOS1, and MOS2, respectively. Only one source belongs to the category with flag=5, for which it was not possible to extract spectral data on any of the pointings. The sample considered in the rest of the paper (flag<4) has median counts $\rm{C_{PN}=175}$, $\rm{C_{MOS1}=70}$, and $\rm{C_{MOS2}=65}$. Further results of the X-ray spectral analysis are discussed in section \S \ref{sec:Xspec_results}.

\section{Multiwavelength observations}\label{sec:multiwaveobs}

The XXL fields benefit from ancillary photometric observations ranging from the ultraviolet to infrared wavelengths consisting of both private and public surveys. A summary of all observations available across the electromagnetic spectrum targeting the XXL field is given in Paper I, Table 2. Here, we describe briefly  the data sets used for this work. In Tables \ref{tab:photN} and \ref{tab:photS} we gather the survey information we used to construct the spectral energy distributions (SEDs) of sources in the XXL-N and XXL-S, respectively. The limiting magnitudes refer to the third quantile of the aperture magnitude distribution. 
    \begin{table*}
    \begin{center}
    \caption{XXL-N: Ancillary photometric data set used in this work and measured limiting magnitudes\tablefootmark{a}}
    \label{tab:photN}
    \begin{tabular}{lcccccccc} \hline 
        {\bf Telescope}                 & {\bf Survey} & {\bf Version}  & \multicolumn{6}{c}{{\bf Limiting magnitude} }  \\ \hline \hline
                                        &         &            & {\bf FUV}       & {\bf NUV}     \\ 
        \multirow{3}{*}{\bf GALEX}      & AIS     &            & 22.18     & 22.47   \\
                                        & DIS     & GR6/7      & 24.57     & 24.18   \\
                                        & GI      &            & 23.51     & 23.69   \\ \hline
                                        &         &            & {\bf u} & {\bf g} & {\bf r} & {\bf i} & {\bf y}\tablefootmark{b} & {\bf z}           \\ 
        \multirow{5}{*}{\bf CFHT}       & W1      & \multirow{2}{*}{T0007}      & 24.96 & 25.30 & 24.74 & 24.39 & 24.50 & 23.40    \\
                                        & D1      &       & 26.26 & 26.62 & 26.34 & 26.08 & 25.61 & 25.21    \\
                                        & WA      &            & --    & 24.82 & 24.49 & --    & --    & 22.86    \\
                                        & WB      & PI         & --    & 24.98 & 24.52 & --    & --    & 23.55    \\
                                        & WC      &            & --    & 24.88 & --    & --    & --    & --       \\
        {\bf SDSS}                      &         & DR10       & 20.73 & 19.77 & 19.92 & 19.96 & --    & 20.13    \\ \hline
                                        &         &            & {\bf z} & {\bf Y} & {\bf J} & {\bf H} & {\bf K}    \\ 
        \multirow{3}{*}{\bf VISTA}      & VHS     & DR2        & --    & 20.85 & 20.98 & 20.70 & 20.27    \\
                                        & VIKING  & DR1        & 22.81 & 22.10 & 21.50 & 21.36 & 21.25    \\
                                        & VIDEO   & DR3        & 25.45 & 24.65 & 24.65 & 24.25 & 23.90    \\
        {\bf WIRcam}                    &         & PI         & --    & --    & --    & --    & 22.20    \\
        \multirow{2}{*}{\bf UKIDSS}     & DXS     & \multirow{2}{*}{DR10}       & --    & --    & 22.37 & --    & 22.17    \\
                                        & UDS     &            & --    & --    & 24.95 & 24.37 & 24.64    \\ \hline
                                        &         &            & {\bf 3.6$\mu$ m}  & {\bf 4.5$\mu$ m} &     \\ 
       {\bf IRAC}                       &         & PI         & 21.50  & 21.34    \\ \hline
                                        &         &            & {\bf W1}    & {\bf W2}    & {\bf W3}    & {\bf W4}    \\ 
       {\bf WISE}                       & ATLAS   & ALLWISE    & 20.38 & 20.35 & 18.85 & 17.50  \\
    \hline
    \end{tabular}
    \tablefoot{
    \tablefoottext{a}{Magnitudes given in AB. As limiting magnitude we quote the third quantile of the respective magnitude distribution.
    }\\
    \tablefoottext{b}{CFHT replacement i-filter.}
    }
    \end{center}

    \begin{center}
    \caption{XXL-S: Ancillary photometric data set used in this work and measured limiting magnitudes\tablefootmark{a}}
    \label{tab:photS}
    \begin{tabular}{lcccccc} \hline 
        {\bf Telescope} & {\bf Survey} & {\bf Version}  & \multicolumn{4}{c}{{\bf Limiting magnitude}} \\ \hline \hline
                        &         &            & {\bf FUV}       & {\bf NUV}     \\ 
                        & AIS     &            & 22.42 & 22.82    \\
        {\bf GALEX}     & MIS     & GR6/7      & --    & 23.61    \\
                        & GII     &            & 23.92 & 24.03    \\ \hline
                        &         &            & {\bf g} & {\bf r} & {\bf i} & {\bf z}           \\ 
        {\bf BCS}       &         &            & 24.14 & 24.06  & 23.23 & 21.68  \\
        {\bf DECam}     &         & PI         & 25.73 & 25.78  & 25.6  & 24.87    \\  \hline
                        &         &            & {\bf J} & {\bf H} & {\bf K}    \\         
        {\bf VISTA}     & VHS     & DR2        & 21.1  & 20.77 & 20.34    \\ \hline
                        &         &            & {\bf 3.6$\mu$ m}  & {\bf 4.5$\mu$ m}     \\ 
        {\bf IRAC}      &   SSDF  &            & 21.5  & 21.45    \\ \hline
                        &         &            & {\bf W1} & {\bf W2}  & {\bf W3}  & {\bf W4}    \\ 
        {\bf WISE}      & ATLAS   & ALLWISE    & 20.32 & 20.35 & 18.68 & 17.26    \\
    \hline
    \end{tabular}
    \tablefoot{
    \tablefoottext{a}{Magnitudes given in AB. As limiting magnitude we quote the third quantile of the respective magnitude distribution.
    }
    }
    \end{center}
    \end{table*}
    
\subsection{XXL-N and XXL-S}

Observations that are in common for the two XXL fields:
\begin{itemize}
\item{{\bf GALEX}} {\it{Galaxy Evolution Explorer\footnote{http://www.galex.caltech.edu/}}} (GALEX) released the final mission catalogue (GR6/7) with full sky coverage in the far ultraviolet ($\lambda_{FUV}=1516\AA$) and near ultraviolet ($\lambda_{NUV}=2267\AA$). We retrieved the relevant images through the Mikulski Archive for Space Telescopes\footnote{http://archive.stsci.edu/} (MAST). The full XXL-N field is covered by one or more of the following surveys: the All-Sky Imaging Survey (AIS), the Deep Imaging Survey (DIS),  and guest investigator programs (GI). The DIS and GI surveys mainly focus on the XMM-LSS area. The XXL-S field is fully covered by GALEX GR6/7 with AIS, Medium Imaging Survey (MIS), and/or GI programs.

\item{{\bf VISTA}} Three public European Southern Observatory (ESO) large programme surveys have observed the XXL field in the near infrared (z, Y, J, H, K filters: $\rm{0.8 - 2.1\mu m}$) with the {\it{Visible and Infrared Survey Telescope for Astronomy}\footnote{http://www.eso.org/public/teles-instr/surveytelescopes/vista/}} (VISTA) in various depths. XXL-N is covered by all three surveys: {\it{ VISTA Hemisphere Survey}} (VHS, PI: R., McMahon), {\it{VISTA Kilo-degree Infrared Galaxy Survey }} (VIKING, PI: W. Sutherland), and {\it{VISTA Deep Extragalactic Observations Survey}} (VIDEO, PI: M. Jarvis), while the XXL-S is covered only by the VHS survey.
\item{}{\bf IRAC} Both XXL fields have been targeted with the {\it{Infrared Array Camera}} (IRAC) on board the Spitzer Space Telescope\footnote{http://ssc.spitzer.caltech.edu/} providing imaging at $\rm{3.6\mu m}$ and $\rm{4.5\mu m}$ over the whole area (XXL-N; PI: M. Bremer, XXL-S; SPT-Spitzer Deep Field, SSDF, \citet{Ashby2014}).

\item{}{\bf WISE} The {\it{ Wide-field Infrared Survey Explorer}\footnote{http://wise.ssl.berkeley.edu/}} (WISE) is an all-sky mission observing the sky from $\rm{3.4\mu m}$ to $\rm{22\mu m}$. In this work we  use the ALLWISE data release\footnote{http://wise2.ipac.caltech.edu/docs/release/allwise/}, which provides imaging for the whole XXL field in all four WISE filters.
\end{itemize}

\subsection{XXL-N}

Observations covering exclusively the XXL-N area:
\begin{itemize}
\item{{\bf CFHT}} The {\it{Canada-France-Hawaii Telescope}\footnote{http://www.cfht.hawaii.edu/}} (CFHT) has observed the parts of the XXL-N area as part of the CHFT-Legacy Survey\footnote{http://www.cfht.hawaii.edu/Science/CFHTLS/} (CFHTLS) in two configurations a) wide area (W1 field) and b) deep field (D1 field) observed with MEGACam, providing imaging in the filters: $u$, $g$, $r$, $i/y$, $z$ ($\rm{3800-8800\AA}$). In this work we  use the T0007 data release downloaded from Terapix\footnote{http://terapix.iap.fr/rubrique.php?id\_rubrique=268}.
We complemented the coverage of CFHT with three additional $\rm{1\,deg^2}$ fields to the north of W1. The WA and WB fields were observed in the $g$, $r$, $z$ filters while the WC field was observed in the $g$ filter (PI: M. Pierre). 
Additionally, XXL-N has been observed with the WIRcam camera on CFHT in the $K_s$ band ($\rm{2.2\mu m}$, MIRACLES survey). Details on the observations and data reduction are provided in Moutard et al. (submitted).

\item{{\bf SDSS}} The {\it{Sloan Digital Sky Survey}\footnote{http://www.sdss.org/}} (SDSS) provides observations in the whole XXL-N area in the filters: $u$, $g$, $r$, $i$, $z$ ($\rm{3800-8800\AA}$). In this work we use  the DR10 data release \citep{Ahn2014}.

\item{{\bf UKIDSS}} The {\it{UKIRT Infrared Deep Sky Survey}\footnote{The UKIDSS project is defined in \citet{Lawrence2007}. UKIDSS uses the UKIRT Wide Field Camera \citep[WFCAM][]{Casali2007}. The photometric system is described in \citet{Hewett2006}, and the calibration is described in \citet{Hodgkin2009}. The pipeline processing and science archive are described in \citet{Hambly2008}.}} \citep[UKIDSS,][]{Dye2006} has two fields overlapping with XXL-N: the Ultra Deep Survey (UDS),  and the Deep Extragalactic Survey (DXS) targeting the XMM-LSS area in $J$, $H$, $K_s$ and $J$, $K_s$ bands, respectively. Here we use  the DR10 data release downloaded from the WFCAM Science Archive\footnote{http://surveys.roe.ac.uk/wsa/}.
\end{itemize}

\subsection{XXL-S}

Observations covering exclusively the XXL-S area:
\begin{itemize}
\item{{\bf BCS}} The Blanco Cosmology Survey\footnote{http://www.usm.uni-muenchen.de/BCS/}\citep[BCS;][]{Desai2012} has targeted an area of $\rm{80\,deg^{2}}$ which overlaps with the XXL-S field using the MOSAIC II imager at the {\it{Cerro Tololo Inter-American Observatory}\footnote{http://www.ctio.noao.edu/noao/}} (CTIO). The observations cover the $g$, $r$, $i$, $z$ bands ($\rm{4850-9000\AA}$). The images used in this work were analysed as described in \citet{Menanteau2009}.

\item{{\bf DECam}} The {\it{Dark Energy Camera}\footnote{http://www.ctio.noao.edu/noao/content/dark-energy-camera-decam}} (DECam) is the successor of the MOSAIC II camera at CTIO. We  observed the XXL-S field (PI: C. Lidman) in the $g$, $r$, $i$, $z$ bands ($\rm{4850-9000\AA}$), but at deeper depth compared to BCS (see Table \ref{tab:photS}). The details of the observations and data reduction are  in Gardiner et al. (in preparation). The stacked images used in this work were performed as described in \citet{Desai2012}.
\end{itemize}

We  retrieved all the images available for these surveys. All the optical and near-infrared images were  rescaled to zero-point 30 for consistency and ease during the photometry extraction. GALEX, IRAC, and WISE data  already have homogeneous zero-points (per survey) so there was no need for rescaling.

\section{Multiwavelength catalogue}\label{sec:Optcatalog}

The photometry of publicly available catalogues is usually a generic extraction which suffers from false detections, non-detected real objects, blended photometry, etc. To make the most out of the existing observations, we performed our own photometric extraction. Since AGN are rare objects (we have about $10^4$ X-ray detections in the $\rm{50\,deg^2}$ of XXL compared to $\sim10^6$ optically detected galaxies in the same area), it is crucial that the photometry is 1) accurate, in terms of photometric calibration, and 2) complete, in terms of detected sources in each filter. At the same time, the computation of photometric redshift for AGN requires special treatment, particularly true in the case of bright X-ray sources \citep{Salvato2011}. The detailed procedure of the photometry pipeline and the photometric redshift estimation is outside the scope of this work and it will be presented in detail in a companion paper (Fotopoulou et al., in prep.). 

\subsection{Photometry}\label{sec:multiwavecatalog}

Similarly to the XMM-Newton pointings, the ancillary multiwavelength observations are organised in tiles corresponding to the FOV of each telescope. Since the photometry is not obtained simultaneously, different observing conditions will cause variations to the PSF and the zero-point of each image.
We extracted the photometry per filter and per tile for each survey. Each catalogue is then corrected for Galactic extinction according to the \citet{Schlegel1998} maps. The aperture magnitudes are corrected to total from the difference between the $\rm{3''}$ aperture magnitude and the mag$_{auto}$ estimated with Sextractor (Bertin 1996). For each survey and filter (see Tables \ref{tab:photN} and \ref{tab:photS}) we built a source catalogue keeping track of the PSF size, aperture correction, and zero-point calibration for each tile separately.

In order to merge the detections with a given survey, it is necessary to remove duplicate detections due to overlapping tiles. For each tile, we identify the distance of the nearest neighbour for each source. We then concatenate the individual tile catalogues and search for sources that lie within that radius. If there are positive matches, they are considered as one source and the best is kept in the catalogue. We consider as `best' the source with the smallest photometric uncertainty. With this procedure we create the catalogue of primary detections for each filter.

In order to create the multiwavelength catalogue, we impose the condition that a source must be detected in at least two filters. Since our optical and near-IR images have a typical seeing of ${\rm 0.8''}$ it is sufficient to use positional matching to associate the sources detected on our images. We adopted a matching radius $\rm{0.7''}$.
From this step we identify matched sources and single-band detections. As a second step, we match GALEX, IRAC, and WISE catalogues within $\rm{ 1''}$  of the multiwavelength catalogue from the previous step and to the individual single-detection catalogues.

\subsection{X-ray counterpart association}\label{sec:counterpart}
    Source association is trivial when the image PSF is small (<$\rm{1''}$) for which  a positional match usually suffices. This is the case for our  optical and near-IR. However, in order to ensure a correct association between X-ray detections and sources detected in the optical we must employ a statistical approach. We chose to use the method of \citet{Sutherland1992} which has been used frequently in the literature. The basis of the method lies in the fact that X-ray sources are rare events; bright optical sources are also rare events, so the observation of an X-ray source and a bright optical source in the same region of the sky is considered a non-random event.

    Since the available optical observations are highly inhomogeneous we have to search for counterparts in multiple catalogues. We start from the brightest catalogues (i.e. SDSS), which usually cover a larger area on the sky, and we use successively surveys with deeper photometric observations available (i.e. CFHTLS-D1). Following the procedure used in previous XMM surveys \citep{Brusa2005,Brusa2007,Rovilos2011} we use the optical $i$ band as a starting point and successively we use the $K$ band, and $\rm{3.6\,\mu m}$ in cases where a counterpart is not found. With this method, we are able to identify 540 counterparts out of 558 X-ray detected sources for the XXL-N and 432 out of the 442 X-ray detected sources for the XXL-S. In total we reach a 97.2\% identification rate for the XXL-1000-AGN.

\subsection{Redshifts and classification}\label{sec:photoz}

    \begin{figure*}
    \begin{tabular}{cc}
    \includegraphics[width=\columnwidth]{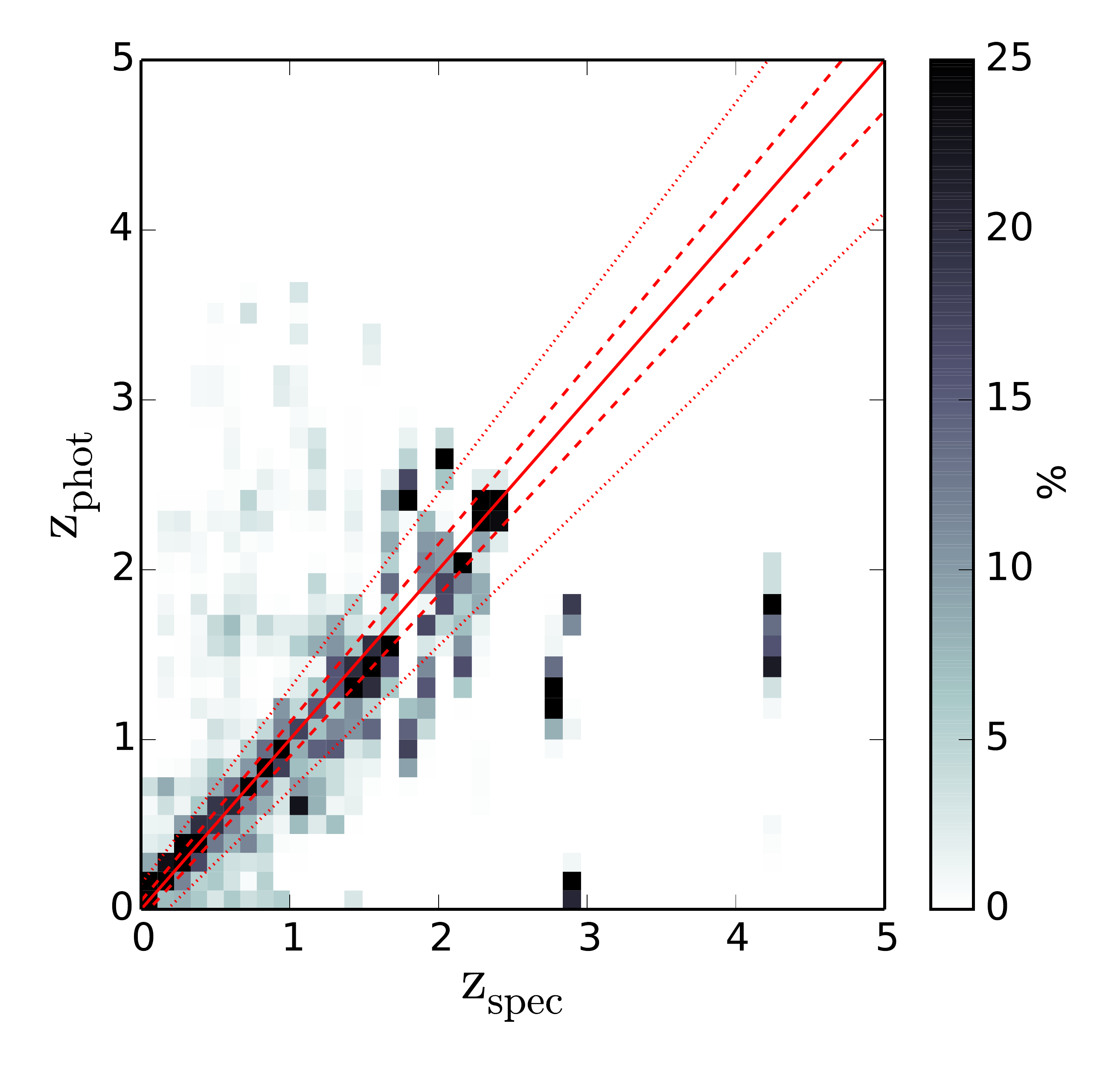} &
    \includegraphics[width=\columnwidth]{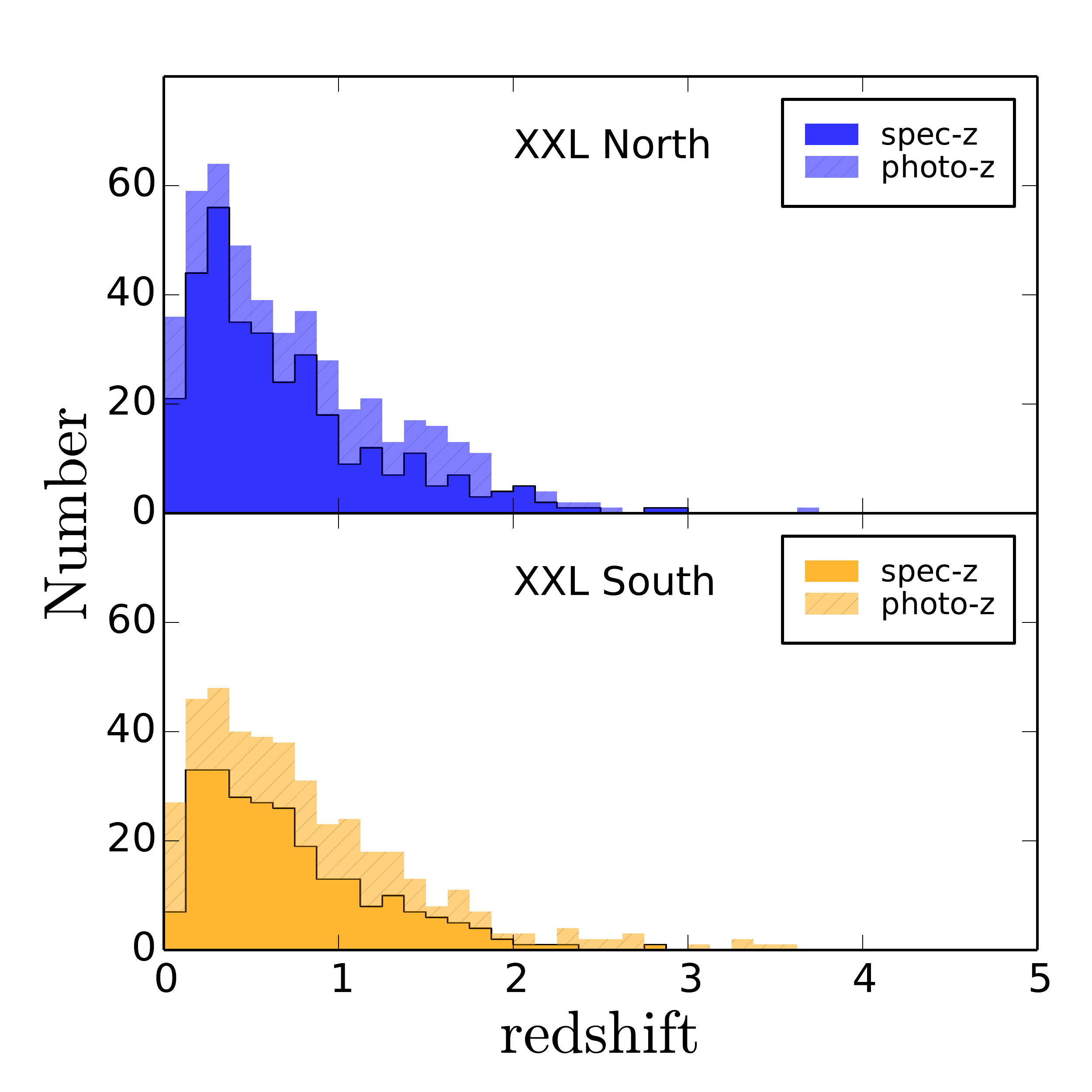} \\
    (a) & (b) \\
    \end{tabular}
    \caption{Panel (a) Spectroscopic versus photometric redshift for the 339 XXL-1000-AGN sources not included in the Random Forest training. The plot is a 2D histogram of the stacked photo-z probability distribution functions and the colourbar shows the probability enclosed in each cell for a given spectroscopic redshift slice ($\rm{\Delta z=0.1}$). The solid red line is the diagonal, the dashed and dotted lines are $\rm{0.10\cdot(1+z_{spec})}$ and $\rm{0.15\cdot(1+z_{spec})}$ respectively. Sources outside the dotted line are considered outliers. Panel (b) Redshift distribution of XXL-N (blue) and XXL-S (orange) fields. The solid histograms show the distribution of spectroscopic redshift, the stacked hatched bars show the distribution of additional photometric redshift when spectroscopic redshifts are not available. The combined height shows the total redshift distribution in each field. 
    \label{fig:zphotzHist}}
    \end{figure*}
    \begin{figure*}
    \begin{tabular}{cc}
    \includegraphics[width=0.5\textwidth]{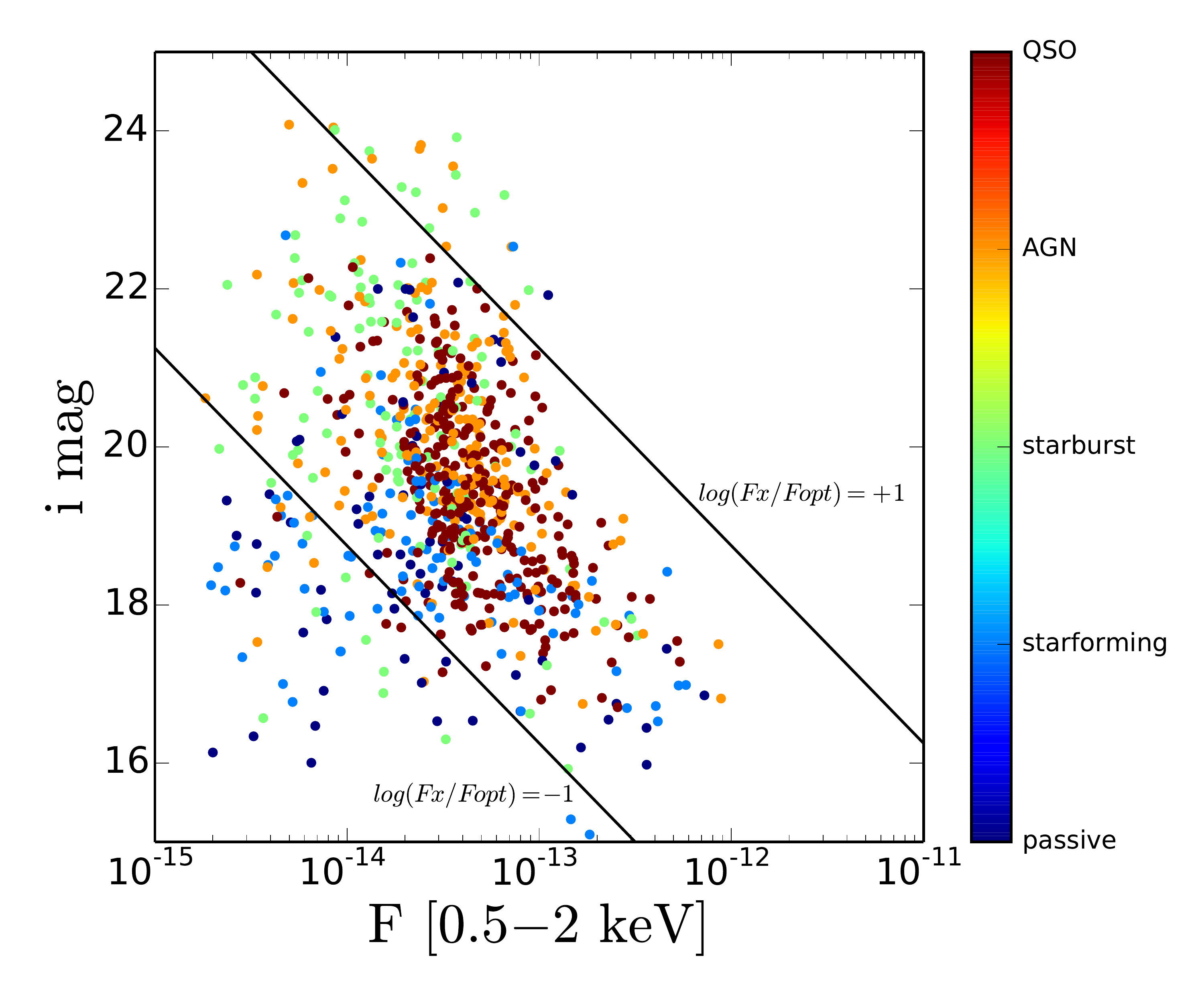} &
    \includegraphics[width=0.4\textwidth]{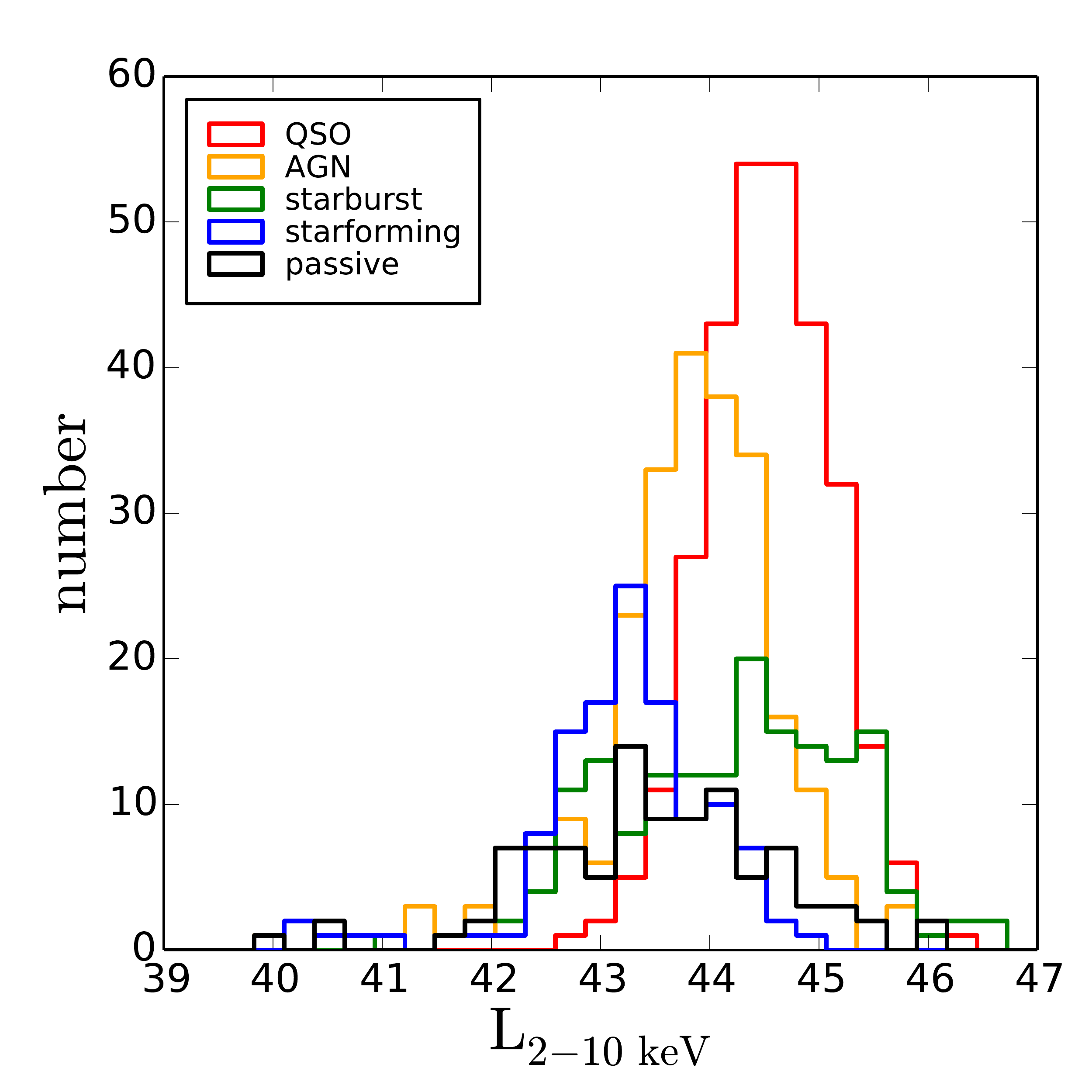} \\
    (a) & (b) \\
    \end{tabular}
    \caption{(a) $\rm{F_{\rm{X}}/F_{\rm{opt}}}$ diagram for the XXL-1000-AGN. The majority of the objects classified as QSO/AGN (red and orange dots) fall between the $\rm{\log(F_{\rm{X}}/F_{\rm{opt}})=\pm1}$ lines, while passive objects (dark blue) and stars are gathered in the bottom left corner of the plot. (b) X-ray luminosity per class. QSO classified objects have the highest $\rm{2-10\,keV}$ luminosity. The colour-coding indicates the photo-z class assigned by the Random Forest in both plots: QSO - red, AGN - orange, starburst - green, star forming - blue, passive - black.
    \label{fig:RFclass}}
    \end{figure*}

A significant effort has been made to gather a large number of spectroscopic redshifts for the XXL sources. We  use a combination of publicly available data (e.g. SDSS\footnote{https://www.sdss3.org/dr10/}) and data obtained through collaborations with other consortia, namely with the VIMOS Public Extragalactic Redshift Survey\footnote{http://vipers.inaf.it/} (VIPERS) \citep{Guzzo2014} and Galaxy and Mass Assembly\footnote{http://www.gama-survey.org/} (GAMA) \citep{Driver2009,Driver2011,Hopkins2013,Liske2015}. Additionally, through the efforts of our collaboration, the XXL benefits from more than 20 000 high-quality spectra. As is the case for the photometric observations, the XXL-N field  also has a better spectroscopic coverage (Adami et al., in prep). For the XXL-N there is a large compilation of spectra for normal galaxies, while in the XXL-S through targeted observations pursued by our collaboration, more than 3000 spectra were obtained to study X-ray selected AGN \citet[][Paper XIV]{Lidman2016}.

The XXL-1000-AGN sample has 540 high-quality spectroscopic redshifts. For the remaining 44.4\% of the sample, we computed photometric redshifts adopting the following procedure.
Building on the legacy of the COSMOS survey on photometric for AGN described in detail in \citet{Salvato2009,Salvato2011}; instead of three classes -- namely normal galaxies, AGN, QSO, -- we increased the number to five classes: passive, star forming, starburst, AGN, QSO. We used the COSMOS templates to estimate a photometric redshift solution for a given class. More specifically, we used templates 1-7 from \citet{Ilbert2009}  to describe the passive class, templates 8-18 for the star forming class, and templates 19-31 for the starburst class. For the active galaxies from \citet{Salvato2009} we used the pure quasar templates with their extension to the UV to describe the QSO class and the hybrid templates to describe the AGN class. The same set-up of extinction laws and redshift step was used as stated in these papers. Namely, we searched for the best fit model within each class in the redshift range z=(0, 6) with zstep=0.01. The extinction laws of \citet{Calzetti2000} was applied to the star forming and starburst classes, while the extinction law of \citet{Prevot1984} was applied to the QSO and AGN classes.
Therefore,  for each object we have five photo-z estimates with the corresponding probability distribution function (PDF). According to the COSMOS schema, X-ray flux, morphology (point-likeness), and variability are used to distinguish between the three categories: normal, AGN, QSO. Here instead we  use an extended attribute set consisting of 84 attributes: 

\begin{itemize}
\item All colour combinations using the bands, $g$, $r$, $i$, $z$, $J$, $H$, $K$, $\rm{3.6\,\mu m;}$
\item FWHM scaled to the corresponding PSF;
\item Half-light radius scaled to the corresponding PSF;
\item X-ray flux;
\item Hardness ratio;
\item $\chi^2$ values of the five model catagories fitted to the data.
\end{itemize}

For the classification we  used a Random Forest classifier \citep{Breiman2001} through the \texttt{sci-kit.learn} package in Python.
We used first the spectroscopic sample from the whole XXL field keeping only sources with high-quality spec-z determination. We split the sample into three equal parts: training, test, validation. Each sample consists of about 8000 galaxies (including AGN/QSO). The training sample is used by the classifier to create the decision trees and the test sample is used again by the classifier to estimate the quality of the classification. In order to label our sources as belonging in one of the five categories, we use the proximity of the photo-z solution to the true spec-z value. The validation sample is never seen by the classifier, and it is used to assess the performance once the Random Forest is created. The forest can be easily saved and applied to all sources in our field.

We used two more Random Forest classifiers to assign probability for a source to be 1) a star or 2) an outlier. Going through the same procedure as described above of class assignment-training-testing, we used the spectroscopically identified stars for the star classifier (1600 objects) and the sources that had photo-z solution in all five categories $\rm{|z_{spec}-z_{phot}|>0.15\cdot(1+z_{spec})}$ as outliers (3000 objects). These classifiers are able to predict the correct class for about 98\% and 96\% of the respective test sample of the star and outlier class.

Figure \ref{fig:zphotzHist} (a) shows the photo-z performance of the XXL-1000-AGN sample, using the validation sample, i.e. the sample that was never seen by our classifier during training. In order to visualise all the information included in the redshift PDF, we create spec-z slices, $\rm{\Delta z_{spec}=0.1}$. We then take the PDFs of all sources that fall in a specific bin. Using the same binning $\rm{\Delta z_{phot}=0.1}$ on the photo-z axis, we integrate the PDF in a given bin. The integral gives the probability of each source  falling in that particular bin cell. Since our sample consists of independent sources, we add the calculated probabilities of each source within a given bin. We then divide the cell value by the number of sources present in the spec-z slice. The grey scale colourbar shows the percentage of this probability. In the ideal case of perfect photometric redshift estimates, our cells would have 100\% probability along the diagonal and 0\% elsewhere. This is not achievable using only broadband photometry. Instead, as seen in Fig. \ref{fig:zphotzHist} (a), there are areas above and below the diagonal with low probability values up to $\rm{z=2.5}$.

To compare these data  with previous results we use the normalised median absolute deviation (NMAD) as an estimator of the accuracy defined as $\rm{\sigma_{NMAD}=1.48\times|z_{phot}-z_{spec}|/(1+z_{spec})}$. The accuracy in the validation sample of 339 sources is $\rm{\sigma_{N_{MAD}}}=0.095$ with $\eta=28.3\%$ catastrophic outliers (i.e. percentage of sources with $\rm{|z_{phot}-z_{spec}|/(1+z_{spec})>0.15}$). As expected, very luminous X-ray sources with photometry observed not simultaneously in all bands is affected by intrinsic variability, which in turn makes the photometric redshift estimation challenging. Once we exclude sources that have a probability of being outliers of more than 0.2\% estimated using our classifier, the accuracy is $\rm{\sigma_{N_{MAD}}}=0.071$ with $\eta=16.3\%$, but the sample has been reduced by half. 

Figure \ref{fig:zphotzHist} (b) shows the redshift distribution of the XXL-1000-AGN in the XXL-N (top) and the XXL-S (bottom). The solid histograms represent the spec-z sources, while the stacked hatched bars show the additional photo-z. We see that the photo-z distribution follows loosely the spec-z distribution.

\subsection{Classification results}

Our classifier returns the optimal photometric class per source for 70\% of the sources in the test sample. We note that the scope of the current set-up is not to identify correctly the exact nature of the source, but rather to identify the class that will give us the optimal photo-z solution. Therefore, misclassification between normal galaxy classes would not necessarily cause a dramatic effect on the photo-z estimates, if a) enough photometric bands are available and b) characteristic features such as the Balmer break are present in the SED. Figure \ref{fig:RFclass} (a) shows the fx/fopt diagram \citep{Maccacaro1988, Lehmann2001, Hornschemeier2003, Brusa2005, Fotopoulou2012}. The AGN and QSO are expected to lie largely between $\log(fx/fopt)=\pm1$, calculated as $\rm{\log(f_{\rm{X}}/f_{\rm{opt}})=\log f_x + I/2.5 + 5.5}$ (black lines). The colour-coding denotes the class assigned by the Random Forest. We see that the AGN (orange) and QSO (red) lie between the two lines, while at the same time the galaxies labelled as passive objects (dark blue) lie below the $\log(fx/fopt)=-1$ line and star forming (cyan) and starburst (green) objects appear to have lower X-ray fluxes than  AGN and QSO in agreement with the observation of \citet{Salvato2011} in the COSMOS survey. This diagram gives us additional confidence on the classifier's results.

In Fig. \ref{fig:RFclass} (b) we plot the histogram of the intrinsic $\rm{L_{2-10\,keV}}$, splitting the sample according to the Random Forest class assignment. To calculate the $\rm{L_{2-10\,keV}}$ we used  the absorption corrected flux estimated by the X-ray spectrum with the corresponding photon index using spectroscopic redshift when available or photometric redshift otherwise. We observe that the two dominant classes are indeed QSO (red) and AGN (orange). Even by using information such as broadband colours and morphology estimates, our classifier correctly identified QSOs, which a posteriori we find to be high-luminosity objects ($\rm{\log Lx_{peak}\sim44.5}$). Similarly, AGNs show $\rm{\log Lx_{peak}\sim43.9}$, while the star forming and passive galaxies, show $\rm{\log Lx_{peak}\sim43.0}$. Interestingly, the starburst population shows a broad distribution in $\rm{\log Lx}$ extending up to high luminosities. This population includes objects that show strong absorption in the blue part of the SED, while at the same time enhanced infrared emission. Therefore, they are indeed AGN as we  concluded by the X-ray luminosity, but the photometric redshift is best constrained by a starburst galaxy template.

In Appendix \ref{cutouts} we present a subsample of the XXL-1000-AGN sources. For each of the Random Forest categories -- QSO, AGN, starburst, star forming, and passive -- we show the X-ray spectrum, the multiwavelength SED along with the best fit model for the photo-z solution (red line) and the star model (grey line). Additionally, we show a single filter image and a false colour image. The green dashed circle is centred at the X-ray position, while the chosen counterpart is marked with a red circle.

\section{XXL-1000-AGN multiwavelength properties}\label{sec:multiwave_properties}

In this section we discuss  further  the XXL-1000-AGN X-ray spectral analysis results and SED fitting. We show that on average, the XXL-1000-AGN sample comprises  unabsorbed sources both in the X-rays ($\rm{\log N_{\rm{H}}}\sim10^{21}cm^{-2}$) and the optical ($\rm{E(B-V)<0.1}$), with photon index $\rm{<\Gamma>=1.85}$, and average intrinsic luminosity $\rm{\log L_x}\sim44$. We also provide a recipe for creating an expected $\rm{N_{\rm{H}}}$ distribution, given a hardness ratio value.
    \begin{figure}
    \begin{tabular}{c}
    \includegraphics[width=\columnwidth]{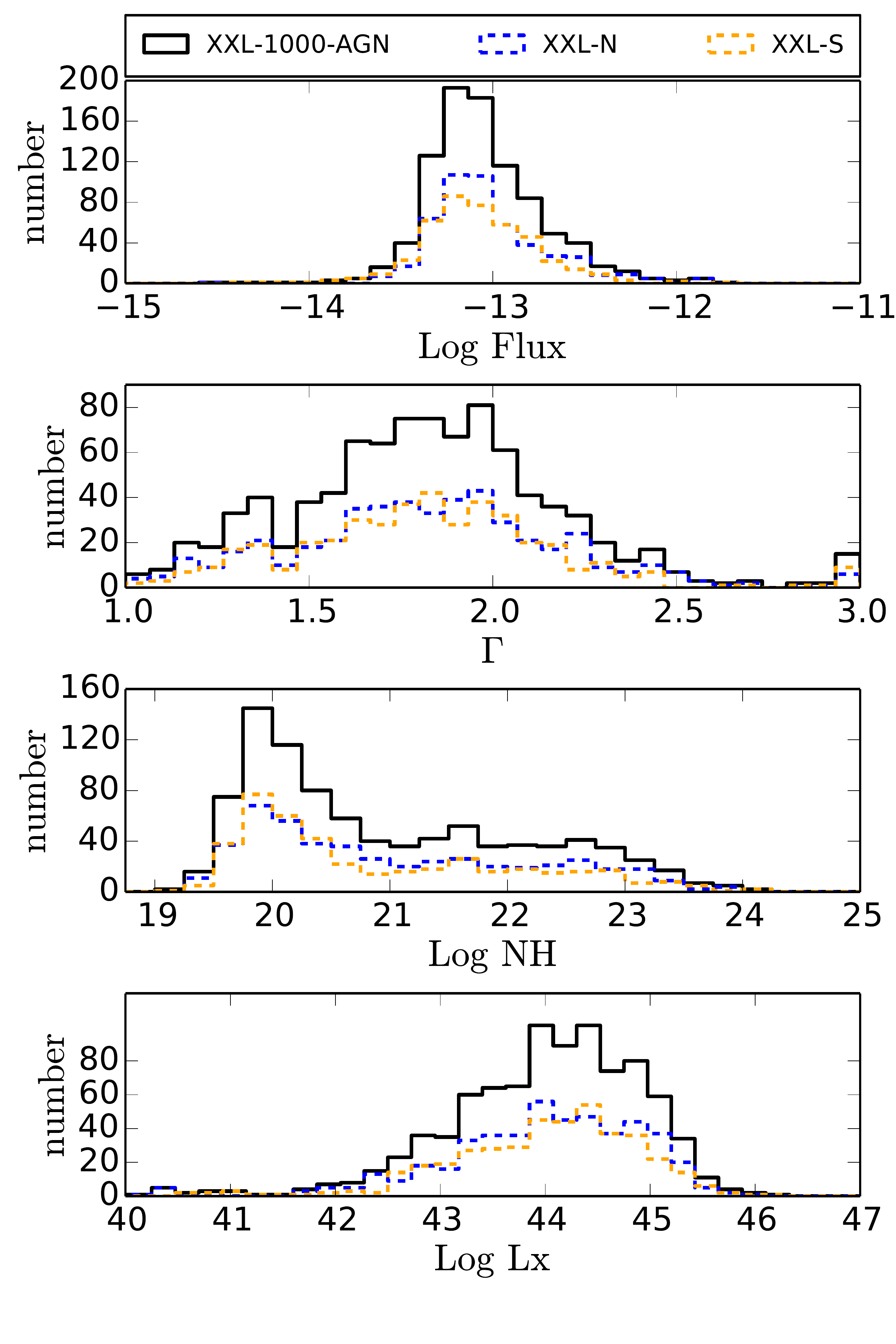}
    \end{tabular}
    \caption{Panels from top to bottom: Distribution of logarithms of the intrinsic flux determined by the absorbed power law,
    the photon index $\Gamma$, the logarithm of the hydrogen column density, the logarithm of the intrinsic luminosity in the $\rm{2-10\,keV}$ energy band. In all panels we show the distribution for XXL-N (blue lines) and XXL-S (orange lines) separately. The black line shows the XXL-1000-AGN sample.
    \label{fig:fluxdistr}}
    \end{figure}

\subsection{X-ray spectral properties}\label{sec:Xspec_results}
The selection of the XXL-1000-AGN sample was based on the $\rm{2-10\,keV}$ estimated by our pipeline, where the spectrum of a source is considered to be a power law with a universal slope of $\rm{\Gamma=1.7}$. With the brightest sample we are in a position to more accurately determine the shape of the spectrum as described in \S \ref{sec:Xspec_analysis}. For this work we adopt a power law with absorbing
medium as a sufficient description of the X-ray spectrum.  We find that there are ten sources that would clearly require an extra modelling component for the soft X-ray emission, but seven out of these ten sources are associated with stars based on visual inspection of the SEDs and optical images.

In Fig. \ref{fig:fluxdistr} we show the distribution of the X-ray spectral parameters determined by the fitting (top to bottom), $\rm{\log\,Flux}$, $\rm{\Gamma}$, $\rm{\log\,N_{\rm{H}}}$, and the distribution of the intrinsic luminosity $\rm{L_x}$. The photon index $\rm{\Gamma}$ is in the range  [1.0, 3.0] during the fitting. We find an average value of $<\Gamma>=1.85\pm0.4$, in agreement with previous observations \citep{Mainieri2002, Mainieri2007, Tozzi2006, Buchner2014, Corral2015}. The relative uncertainties associated with our fitting results vary as a function of X-ray spectral quality flag (as defined in \S \ref{sec:Xspec_analysis}). We find that the $\rm{2-10\,keV}$ flux shows relative uncertainty of 10\%, 20\%, and  35\% for classes 1-3. Similarly, the photon index shows relative uncertainty of 5\%, 10\%, and 20\%, for classes 1-3. The hydrogen column density is the least well constrained parameter with relative uncertainty of about 80\% for class 1. Only $\sim$10\% of our sources have $\rm{N_{\rm{H}}}$ relative uncertainty below 30\%. This is a combination of low count rate X-ray spectra and the fact that we are incorporating the uncertainty interval of the photometric redshift estimation in the fitting process. However, using the full probability distribution function in our scientific analysis we are able to propagate correctly the knowledge (or lack thereof) of the intrinsic absorption.

\begin{figure*}
    \begin{tabular}{cc}
    \includegraphics[width=\columnwidth]{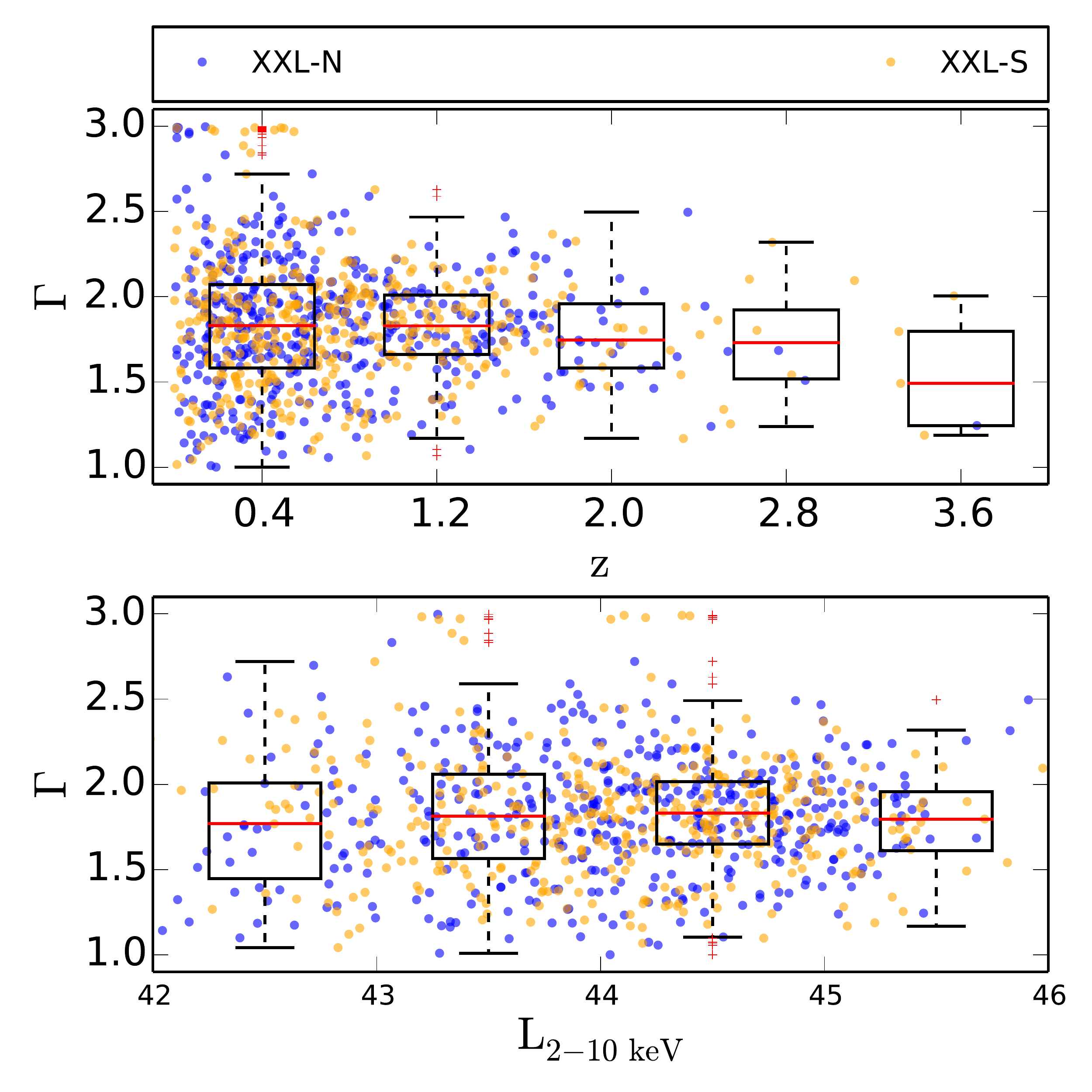} &
    \includegraphics[width=\columnwidth]{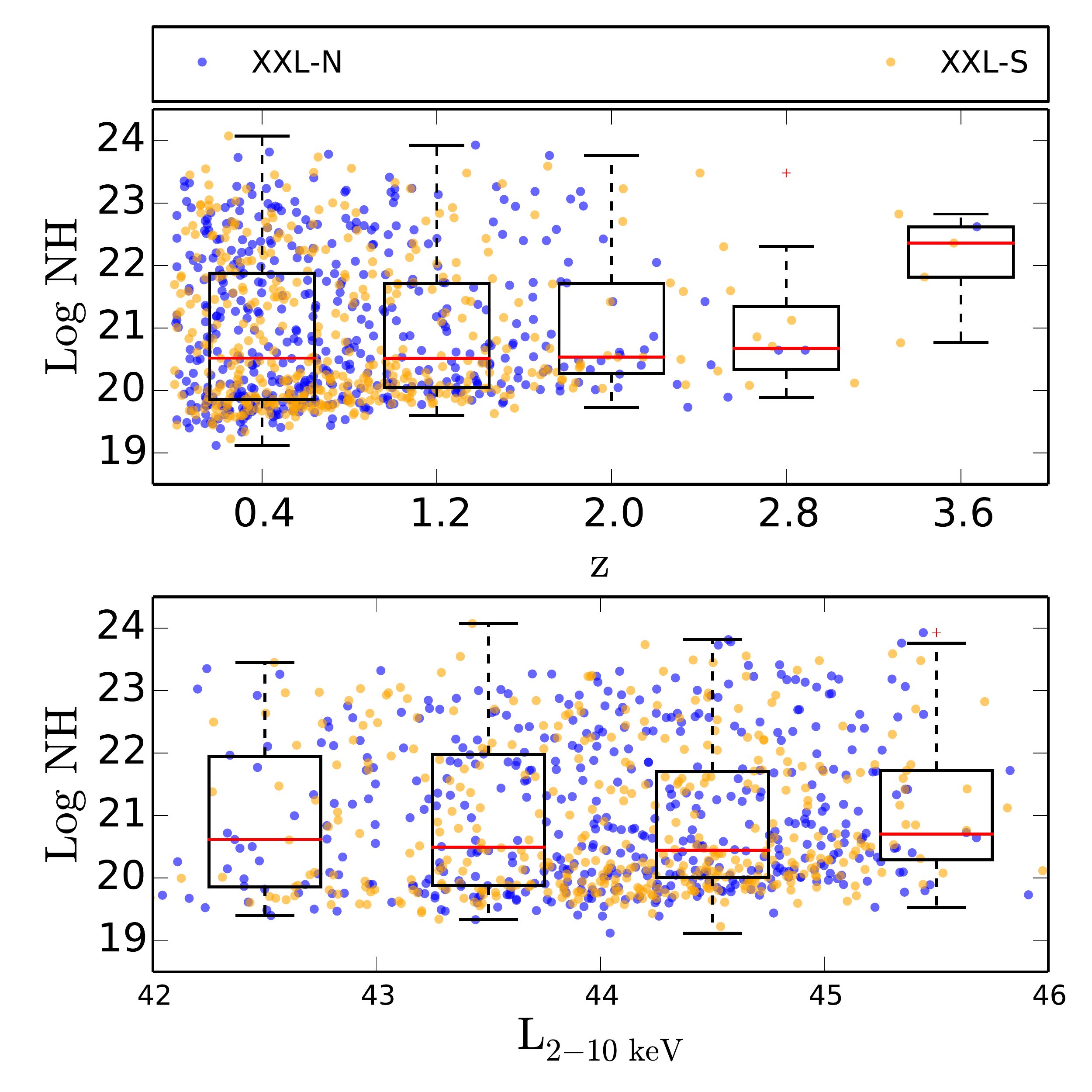} \\
    (a) & (b)
    \end{tabular}
    \caption{{Photon index $\rm{\Gamma}$ (a) and hydrogen column density $\rm{\log\,N_{\rm{H}}}$ (b) as a function of redshift (top) and luminosity (bottom). Splitting the sample into bins, we show with boxplots the median (red line) and 50\% of the distribution within each bin.}
    \label{fig:NHfraction}}

    \begin{tabular}{cc}
    \includegraphics[width=\columnwidth]{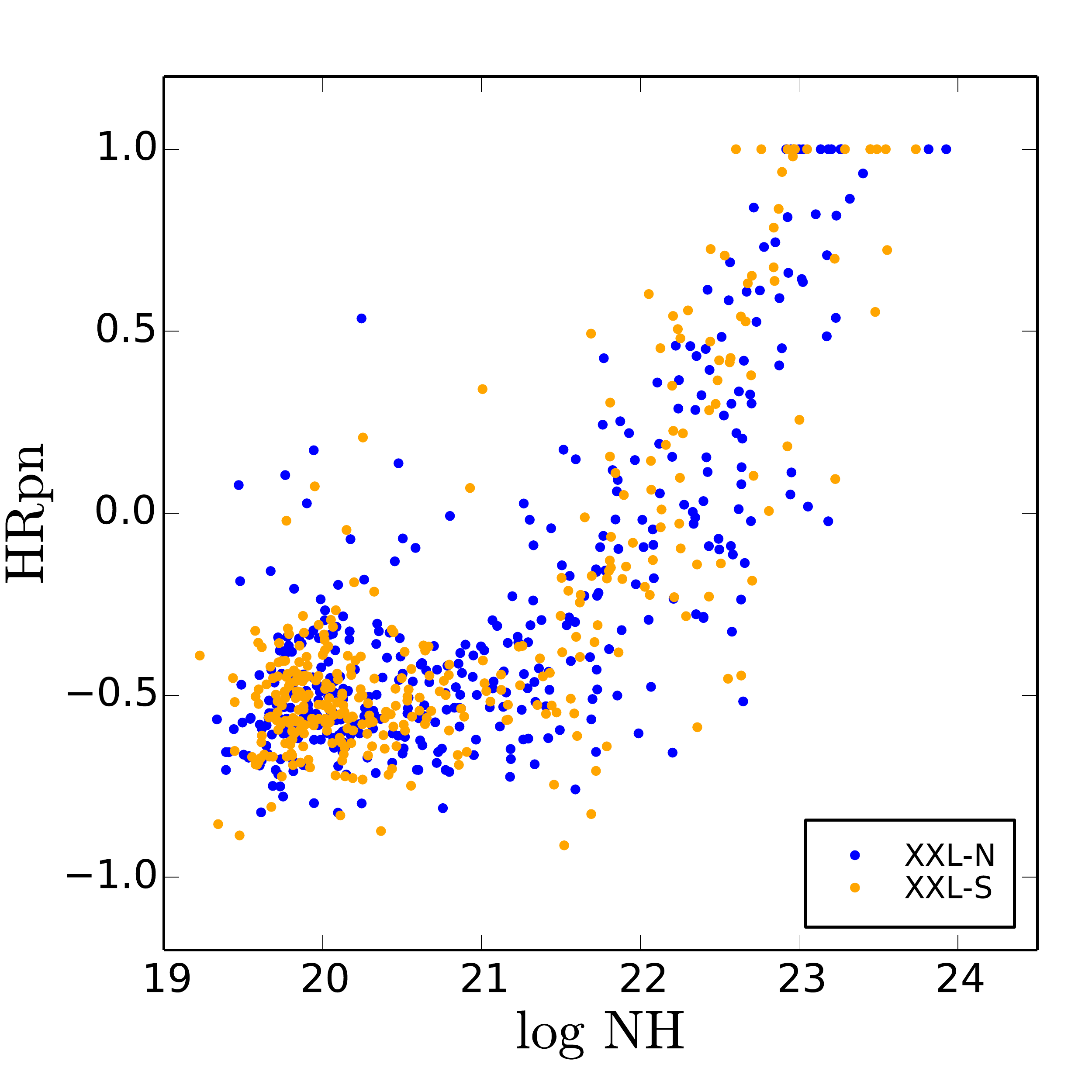} &
    \includegraphics[width=\columnwidth]{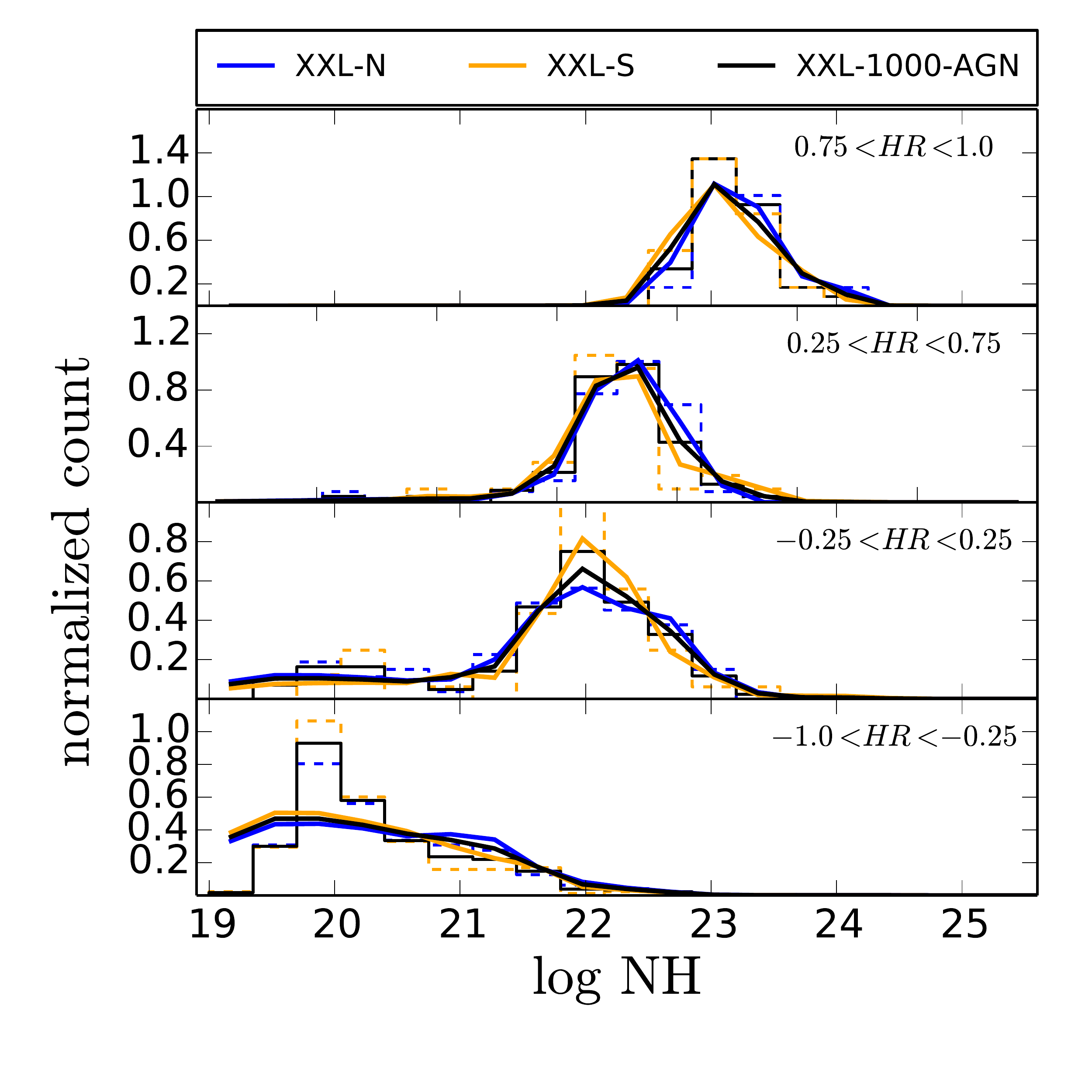} \\
    (a) & (b) \\
    \end{tabular}
    \caption{(a) Hardness ratio versus $\rm{\log{\rm{N_{\rm{H}}}}}$ for the XXL-N (blue points) and XXL-S (orange points); the hardness ratio is in good agreement with the estimated $\rm{\log{N_{\rm{H}}}}$ from the X-ray spectra. (b) $\rm{\log{N_{\rm{H}}}}$ for four hardness ratio bins. The normalised dashed histograms represent the observed values in the XXL-N (blue) and XXL-S (orange) and the combined XXL-1000-AGN sample (solid black histogram). The solid lines show the combined $\rm{\log{N_{\rm{H}}}}$  PDF of the sources in each bin (see Appendix \ref{NH_PDF} for the tabulated values of these curves).\label{fig:NH}}
    \end{figure*}

Fig. \ref{fig:NHfraction} (a) shows the estimated photon index $\rm{\Gamma}$, as a function of redshift (upper panel) and intrinsic luminosity (lower panel). The points show the values obtained for each source in our sample (XXL-N: blue points, XXL-S: orange points). In order to investigate if there is any trend of photon index with redshift or luminosity, we also plot boxplots. We use five bins in redshift and luminosity and show the median of the photon index per bin (red line). The boxes enclose 50\% of the distribution, while the dashed lines extend to the minimum and maximum values. The red dots are considered outliers\footnote{For a boxplot diagram we consider as outliers the points that deviate from the best fit Gaussian distribution within each bin.}. Both panels show that there is no significant change on the median $\rm{\Gamma}$ value. Similar findings are also reported in \citet{Piconcelli2005, Shemmer2005}. 

Equivalently, in Fig. \ref{fig:NHfraction} (b) we investigate the presence of any possible trends of $\rm{\log N_{\rm{H}}}$ with redshift and intrinsic luminosity. We do not find any trends of change in the median value of $\log N_{\rm{H}}$ within the XXL-1000-AGN sample. There is an ongoing debate  in the literature regarding the fraction of obscured AGN and its dependence on redshift and/or luminosity. Studies in the literature\footnote{Taking into account the available cosmological volume in order to avoid biasing the fraction in favour of the unabsorbed population.} find that the fraction of obscured objects  evolves with both redshift and luminosity \citep{LaFranca2005,Akylas2006, Gilli2007, DellaCeca2008, Hasinger2008, Aird2015, Buchner2015}. Nevertheless, \cite{Merloni2014} have shown the impact of the method used to estimate the obscuration (X-rays versus optical classification). With the current XXL-1000-AGN we do not detect a strong decline in the fraction of obscured AGN with luminosity. We obtained an observed fraction of approximately 26\% obscured objects ($\rm{\log\,N_{\rm{H}}>10^{22}cm^{-2}}$) in XXL-1000-AGN, which is consistent with the fraction of obscured high-luminosity objects ($\rm{\log\,L_x}>44$) reported in the deeper X-ray surveys noted previously, and in rough agreement with earlier wide area XMM-surveys \citep{Piconcelli2003,Perola2004} (see also \citet{Gilli2007}, Fig. 16). This number is not surprising if we consider that the fraction of absorbed sources is the highest (up to 60-80\%) for low-luminosity objects at higher redshifts. The flux limit of the XXL-1000-AGN simply does not allow this parameter space  to be covered.

\paragraph{\bf{Hardness ratio distribution}}
Hardness ratios (HR) are used commonly in the literature as a rough estimator of the absorption in the absence of good quality X-ray spectra. For example, in \citet{Ueda2003} the authors use hardness ratio estimates to take into account absorption effects in their estimation of the X-ray luminosity function. We wish to derive a simple relationship between the observed hardness ratio and the true $\log{N_{\rm{H}}}$.  In Fig. \ref{fig:NH} (a), we show the comparison between the hardness ratio computed from the PN count rate and the $\log{N_{\rm{H}}}$ estimated from the spectra with quality flag=1-2. We use only PN count rate here since the sensitivity of PN is higher than  the MOS instrument. We have verified, however, that the plot is similar in the case of HR estimated from the MOS count rate. The hardness ratio is calculated from the count rate (CR) as
\begin{equation}
\rm{HR = \frac{CR_{2-10,keV}-CR_{0.5-2\,keV}}{CR_{0.5-2\,keV}+CR_{2-10\,keV}}}
.\end{equation}
Typically, $\rm{HR\sim-0.5}$ corresponds to unabsorbed or moderately absorbed sources ($\log{N_{\rm{H}}}<22$) over a broad range of redshifts. Higher values of HR correspond to higher absorption systems, while HR=1.0 denotes systems that are only detected in the $\rm{2-10\,keV}$ energy band. We observe a good agreement between the HR and the $\rm{\log{N_{\rm{H}}}}$, albeit with a large scatter and some outliers owing to the degeneracy between the determination of photon index $\rm{\Gamma}$, absorption $\rm{N_{\rm{H}}}$, and redshift effects. Panel (b) of Fig. \ref{fig:NH}, shows the $\log{N_{\rm{H}}}$ histogram for four bins of HR. The dashed histograms show the $\rm{\log{N_{\rm{H}}}}$ distribution when the median of the PDF is used as a point estimate of the hydrogen column density. In order to incorporate the full uncertainty estimation on the $\rm{N_{\rm{H}}}$ parameter as determined by our X-ray spectral fitting, we summed the individual PDFs. The result is normalised to an area of one (Fig. \ref{fig:NH} solid lines). In Appendix \ref{NH_PDF} we provide the combined $\rm{\log{N_{\rm{H}}}}$ PDF for the XXL-1000-AGN sample (black lines) for the four hardness ratio bins. These curves can be used to draw random values of $\rm{\log{N_{\rm{H}}}}$ when only the HR is available.

\subsection{Median SEDs far-UV to mid-IR}\label{sec:sed}

The galaxy - AGN coevolution has been a very active research field in the past decades. One of the main components still under debate in the literature involving observations, simulations, and theory is the interplay between the AGN and the host galaxy. The universal presence of supermassive black holes in the centres of galaxies with bulges and the scaling relation between black hole mass and bulge luminosity, point in the direction of a shared history.

\begin{figure*}
    \begin{tabular}{cc}
    \includegraphics[width=\columnwidth]{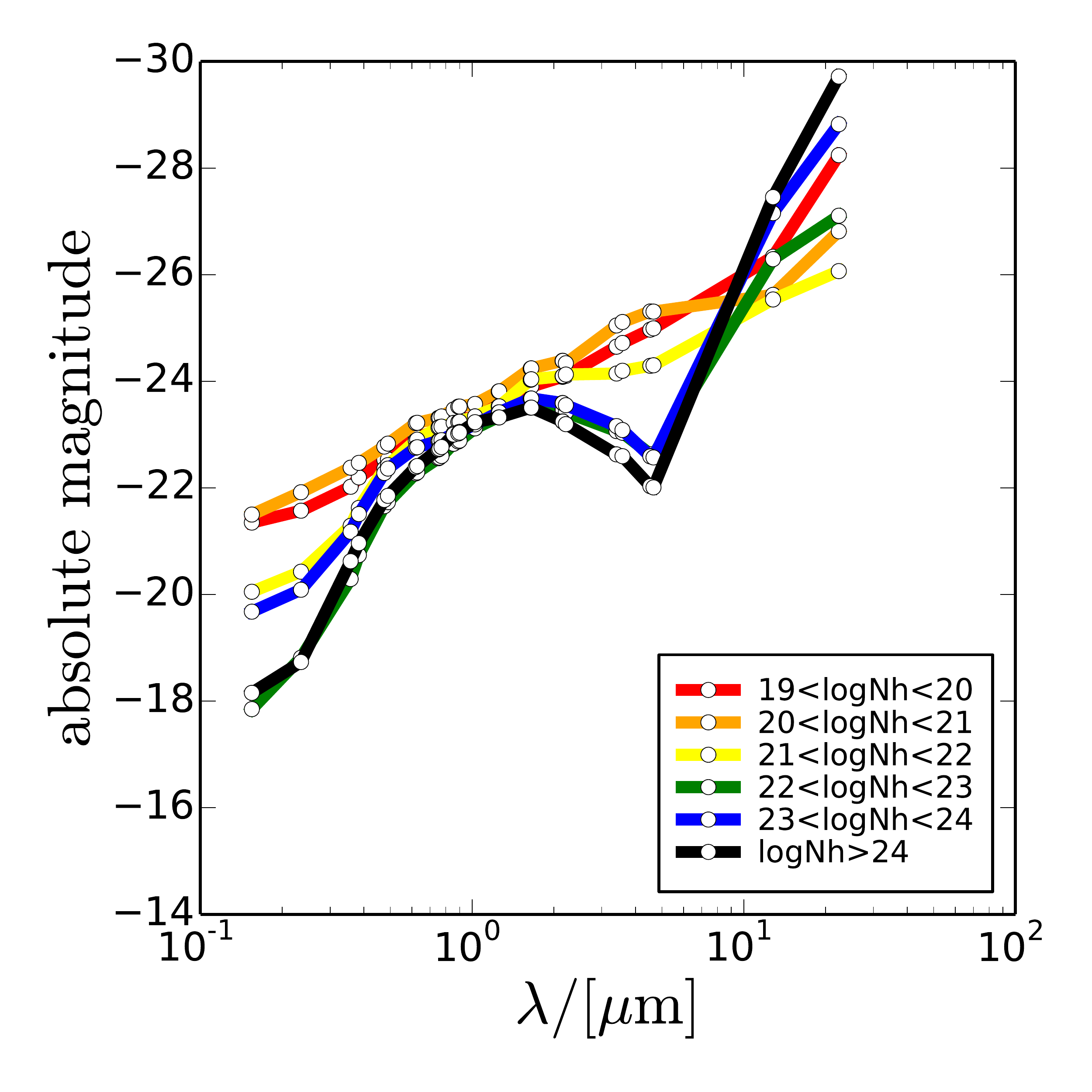} &
    \includegraphics[width=\columnwidth]{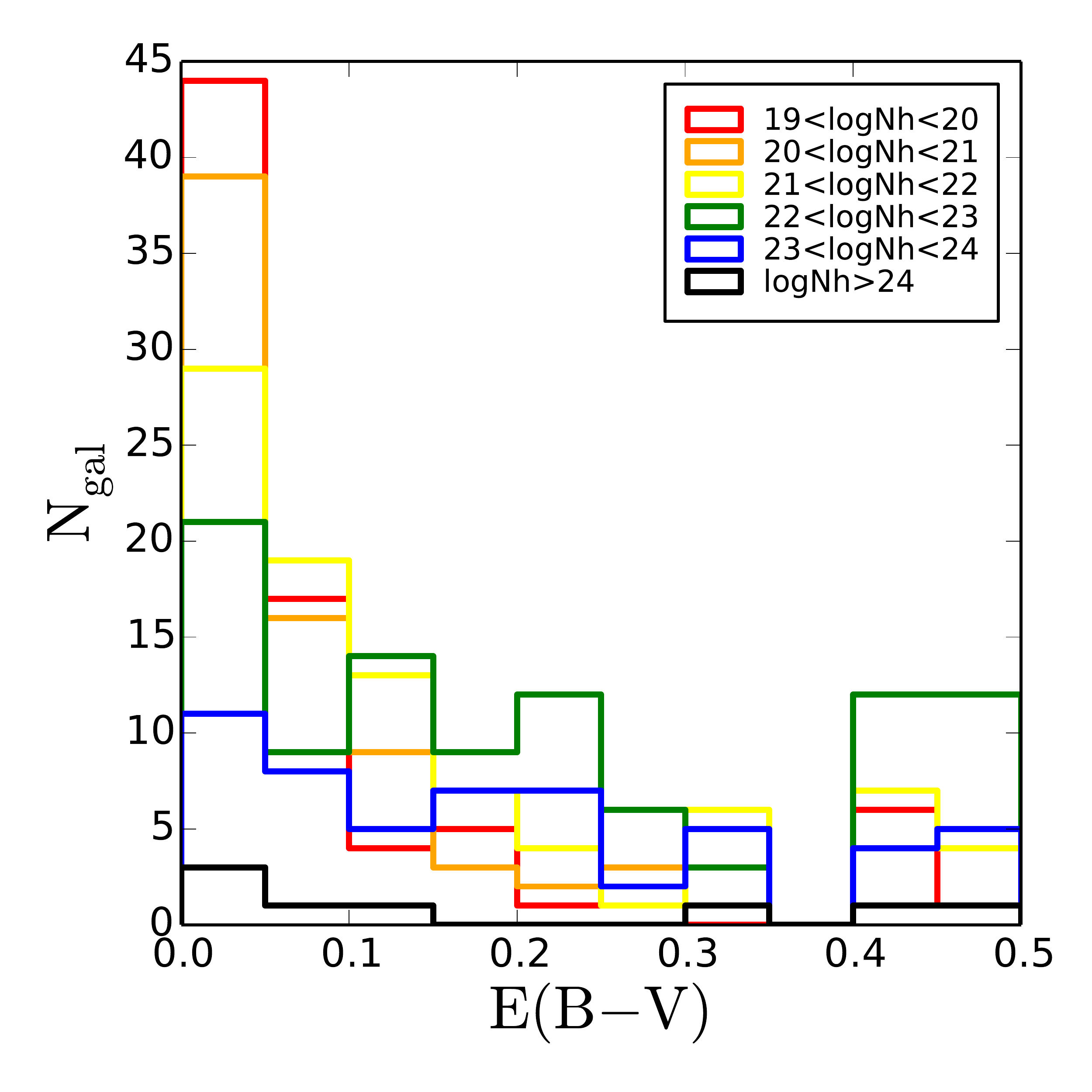} \\
    (a) & (b)\\
    \end{tabular}
    \caption{(a) Median rest-frame SEDs for X-ray counterparts from UV to mid-infrared for XXL-N sources and (b) Values of optical absorption E(B-V) estimated from the broadband SEDs and split into $\rm{N_{\rm{H}}}$ bins determined from the X-ray spectrum, ranging from unabsorbed (red) to Compton-thick sources (black).
    \label{fig:medianSEDNH}}
    \end{figure*}

In Fig. \ref{fig:medianSEDNH} (a) we show the median SEDs from the ultraviolet to the mid-infrared for XXL-N. We split the SED according to the intrinsic absorption $\log{N_{\rm{H}}}$ estimated from the X-ray spectrum. We see that the SEDs of objects with low intrinsic absorption $\log{N_{\rm{H}}}<21$ appear to have QSO-like SEDs.  At higher amounts of absorption ($21<\log{N_{\rm{H}}}<22)$   the central engine does not outshine the host galaxy and the SED starts to have features that are expected in normal galaxies, namely the stellar bump at about $\rm{1\mu m}$. At higher levels of absorption, the host galaxy resembles a passive galaxy, but with a very enhanced mid-infrared emission. According to the unified model of AGN, this is the radiation that is absorbed and re-emitted by the torus close to the AGN.

For comparison in Fig. \ref{fig:medianSEDNH} (b) we show the estimated absorption $\rm{E(B-V)}$ from the SED model fitting, using the same split according to the  $\log{N_{\rm{H}}}$ estimated from the X-ray spectrum. In general, we observe that systems that are less absorbed  in the X-rays also appear  less absorbed in the optical, showing small  $\rm{E(B-V)}$  values (less than 0.1). For moderate and higher hydrogen column densities, where the central engine is not powerful enough to outshine the galaxy, we see no correlation between the absorption close to the black hole and the absorption in the galaxy. This picture is consistent with observing an AGN seeing the host galaxy face-on and the torus aligned with the host galaxy. In this case, both optical and X-ray radiation escape the system with minimal losses due to absorption.

\begin{figure*}
    \begin{tabular}{c}
    \includegraphics[width=\linewidth]{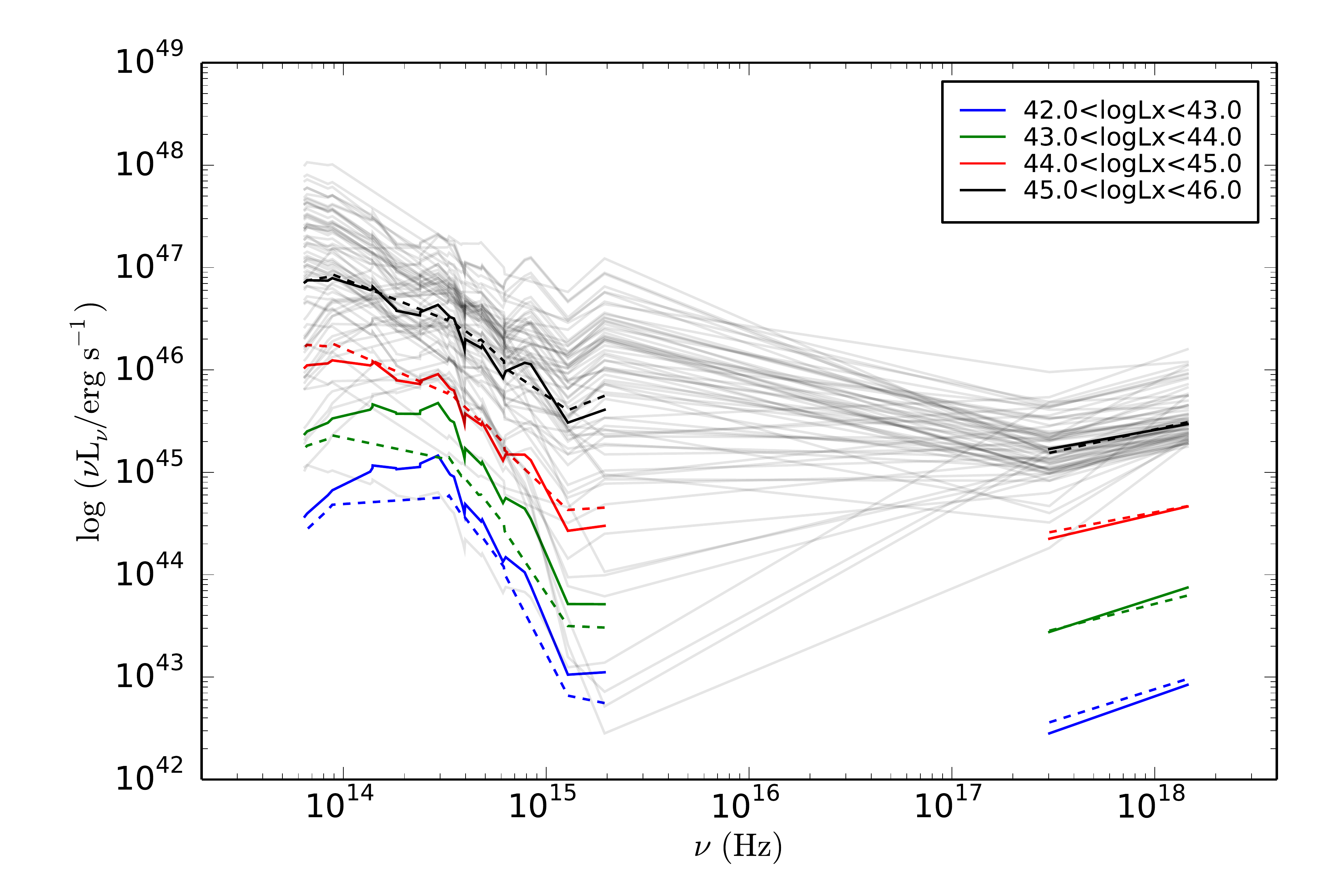} 
    \end{tabular}
    \caption{Median rest-frame SEDs for X-ray counterparts from X-rays to mid-infrared, split into X-ray luminosity bins. Solid lines refer to XXL-N sources, dashed lines to XXL-S sources. The grey lines show the SEDs of all sources in the $\rm{45<logLx<46}$ bin as an example of the large diversity of AGN hosts.
    \label{fig:medianSEDLx}}
    \end{figure*}

In Fig. \ref{fig:medianSEDLx} we show the median rest-frame host galaxy SED ranging from the mid-IR to the X-rays, split into four X-ray luminosity bins. The solid lines correspond to XXL-N sources, while the dashed lines to XXL-S sources. In order to appreciate the large scatter in host galaxy emission, we show with  the individual sources that comprise the most luminous sample ($\rm{45<log\,L_{2-10\,keV}<46}$, grey lines). This plot clearly shows the impact of the presence of a luminous central engine on the broadband SED. Comparing the two extreme cases of the most luminous AGN ($\rm{45<\log L_x<46}$, black lines) with the least luminous ($\rm{42<\log L_x<43}$, blue lines), we see  the enhancement of the emission both in the blue part of the SED ($\rm{v\sim10^{15}Hz}$) originating from the accretion disk close to the black hole at the centre of the galaxy and in the near infrared $\rm{v\sim10^{14}Hz}$) originating from the dusty torus \citep[see also][]{Lusso2011}.

\section{XXL-1000-AGN in the cosmic web}\label{sec:web}

In this section we explore the XXL-1000-AGN sample within the context of the cosmic web by means of number counts, evolution of the X-ray luminosity function, and large-scale structure analysis. In this section we include the flag=4 sources in the analysis, since the sensitivity of the survey is defined according to the \textsc{XAmin 3.3} detections independently of the X-ray spectra.

\subsection{Number counts}\label{sec:counts}

\begin{figure*}
    \begin{tabular}{cc}
    \includegraphics[width=0.99\columnwidth]{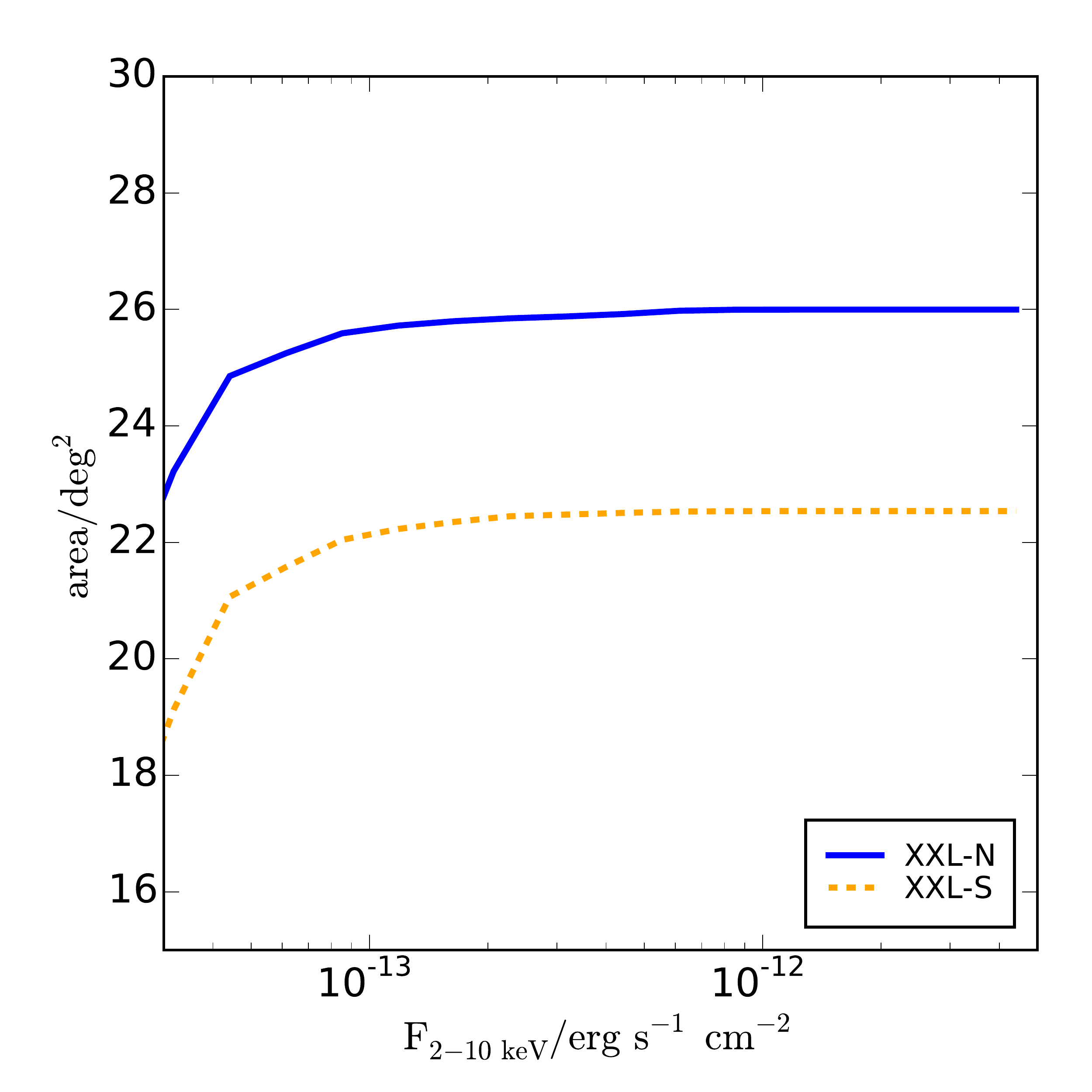} &
    \includegraphics[width=\columnwidth]{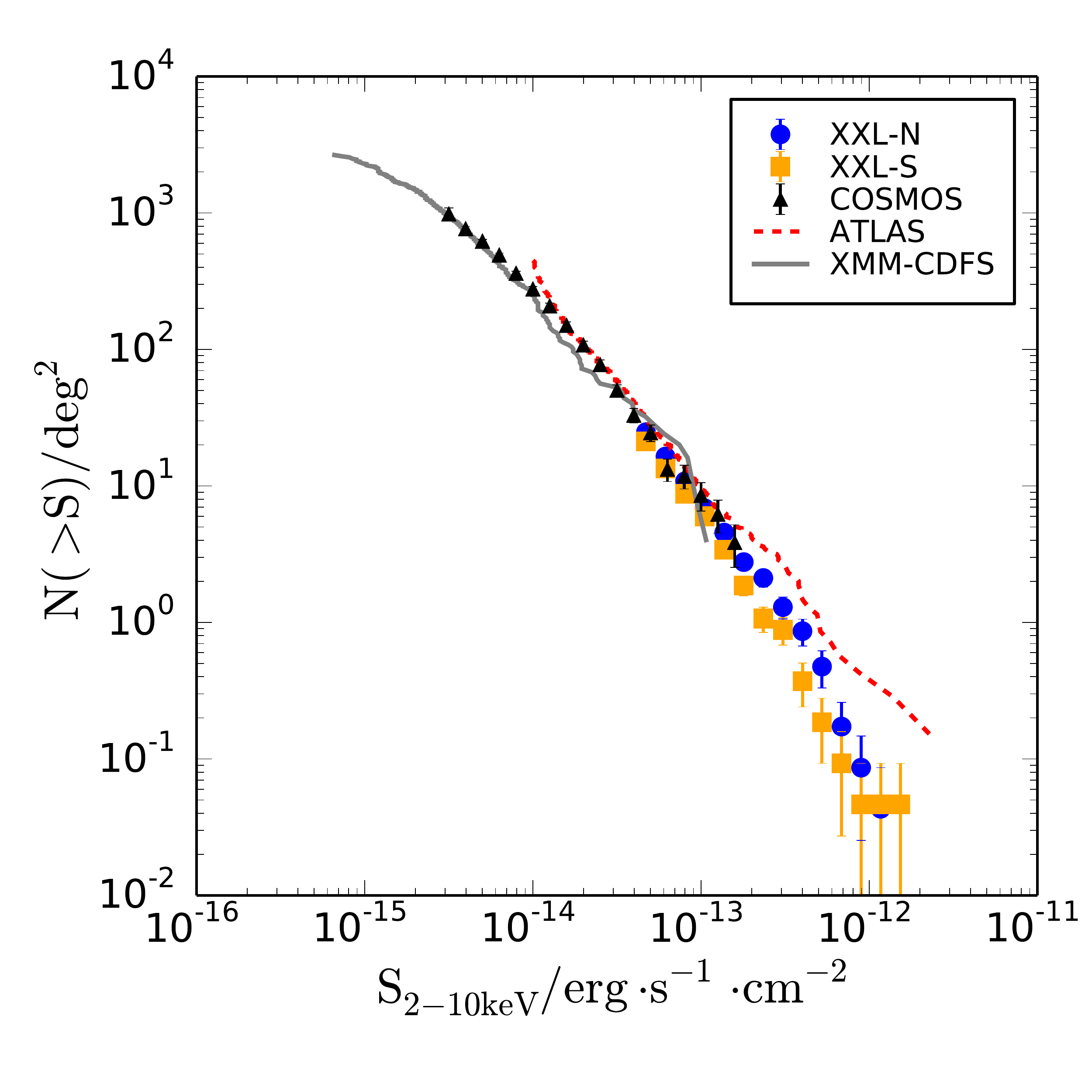} \\
    (a) & (b)\\
    \end{tabular}
    \caption{(a) Area detection efficiency as a function of $\rm{2-10\,keV}$ X-ray flux for the XXL-1000-AGN sample. These curves were used for the calculation of the number counts and the AGN luminosity function. (b) logN-logS for the XXL-1000-AGN in the XXL-N field (blue circles) and in the XXL-S field (orange squares). The black points are the number counts of XMM-COSMOS, the red line shows the cumber counts from the H-ATLAS, and the grey line the number counts of XMM-CDFS. \label{fig:Ncounts}}
    \end{figure*}

The XXL survey with the large contiguous areas and medium X-ray flux limit fills the sparsely explored parameter space between deep and all-sky surveys \citep[see Paper I and][for a review]{Brandt2015}. Using the $\rm{2-10\,keV}$ flux determined by \textsc{Xamin 3.3} we estimated the cumulative number counts for the XXL survey using the equation
\begin{equation}
N(>S) = \sum_{i=1}^N\frac{1}{\Omega_i}
\end{equation}
and the associated uncertainty is estimated by
\begin{equation}
\sigma = \sqrt{\sum_{i=1}^N\left(\frac{1}{\Omega_i}\right)^2}
,\end{equation}
where $\rm{\Omega}$ the corresponding area that is sensitive to the flux limit $S$. The area curve for the XXL is determined following the same approach as for XMM-LSS described in \citet{Elyiv2012} estimated for the source detection of \textsc{Xamin 3.3}(Fig. \ref{fig:Ncounts} (a) ).
In Fig. \ref{fig:Ncounts} (b) we show the number counts estimated using the XXL-1000-AGN sample (XXL-N: blue circles, XXL-S: orange squares). The XXL-1000-AGN number counts are in good agreement with the observed number counts from deeper and narrower XMM-Newton surveys such as COSMOS \citep[$\rm{2\,deg^2}$][]{Cappelluti2007}, H-ATLAS \citep[$\rm{16\,deg^2}$][Fig. 8]{Ranalli2015}, XMM-CDFS \citep[$\rm{0.2\,deg^2}$][]{Ranalli2013}.

The XXL-1000-AGN sample has a brighter flux limit ($\rm{S_{2-10\,keV}=4.8\times10^{-14}erg\,s\,cm^{-2}}$) than the characteristic flux break observed in deeper surveys \citep[e.g.  at $\rm{S_{2-10\,keV}=1.5\times10^{-15}erg\,s\,cm^{-2}}$ in ][]{Cappelluti2007}. Therefore, a single power law is sufficient to describe the observed distribution. The weighted linear least squares fit on both XXL-N and XXL-S points gives $\rm{\log N = (-1.6\pm0.12) \log S -(20\pm2)}$, consistent with the expectation of Euclidean number counts (i.e. $\rm{N(>S)\propto S^{3/2}}$).

The same slope has been reported in \citet{Ueda1999} using a sample of 44 sources detected in the $\rm{2-10,keV}$ energy band in the ASCA Large Sky Survey (ALSS). Similar findings have been reported in the \citet{Fiore2001} using 147 BeppoSAX detections over $\rm{85\,deg^2}$ in the $\rm{4.5-10\,keV}$ energy band. Contrary, \citet{Baldi2002}, one of the earliest attempts to constrain the bright end of the $\rm{2-10\,keV}$ number count distribution with XMM-Newton observations, using 495 detections in the HELLAS2XMM survey reported a sub-Euclidean slope. Similarly, \citet{Mateos2008} found a discrepancy with the ASCA counts at the bright end ($\rm{F_{\rm{2-10\,keV}}>1\times10^{-13}erg\,s^{-1}cm^{-2}}$) of the $\rm{\log N - \log S}$ distribution in 2XMMi sample ($\sim 9000$ sources). They concluded that the observed deviation cannot be explained by  cross-calibration uncertainties alone.

\subsection{2-10\,keV luminosity function}\label{sec:xlf}

The AGN  X-ray luminosity function (XLF) traces the growth of supermassive black holes throughout the history of the Universe.
Usually, because of  the low number density of AGN, independent observations are combined coherently to create a comprehensive survey \citep{Avni1980}. This approach also ensures that the result is not affected by cosmic variance. Numerous deep pencil beam X-ray surveys both with XMM and Chandra have explored the faintest and furthest AGN reaching depths, for example of  $\rm{F_{2-10\,keV}\sim10^{-16}erg\,s^{-1}cm^{-2}}$. Nevertheless, authors still rely on the ASCA catalogues of \citet{Akiyama2003} and \citet{Ueda2005} to introduce the rarest and brightest objects in their samples.

In Fig. \ref{fig:Lxz} we show the coverage of the luminosity - redshift plane of the XXL-1000-AGN sample (black points) compared to a few current X-ray surveys -- red: XMM-COSMOS \citep{Cappelluti2007, Brusa2010}, dark blue: AEGIS \citep{Nandra2015}, cyan: Lockman Hole \citep[LH,][]{Brunner2008, Fotopoulou2012}, green: XMM-CDFS \citep{Ranalli2015, Hsu2014}, pink: CDFN \citep{Alexander2003}, and orange: XMM medium sensitivity survey \citep[XMS][]{Barcons2007}). The orange shaded area shows the anticipated coverage of the full XXL catalogue using a provisional flux limit of $\rm{F_{2-10\,keV}=3\times10^{-15}\,erg\,s^{-1}\,cm^{-2}}$. The yellow area shows the coverage by the eRosita all-sky survey, which is expected to reach a flux limit of $\rm{F_{2-10\,keV}=2\times10^{-13}\,erg\,s^{-1}\,cm^{-2}}$ after four years of observations \citep{Merloni2012}. The XXL-1000-AGN sample with its wide area coverage can help to set constraints on the faint end of the AGN XLF at redshifts  less than 0.5, around the characteristic break luminosity $\rm{\log{L_0}\sim44}$ at redshifts $\rm{0.5<z<1.0,}$ and at the bright end at redshifts $\rm{z>1.0}$.

Early works on the $\rm{2-10\,keV}$ XLF, using samples of $\rm{\sim10^2}$ sources, have shown that there is a strong evolution with redshift \citep[e.g.][]{LaFranca2002,Ueda2003,LaFranca2005,Ebrero2009,Aird2010}, while the exact behaviour of the XLF, particularly at high redshift (z>2) has been heavily debated.
More recent works in the literature have used numerous samples of AGN (1000-4000 objects) to constrain the AGN XLF and its evolution up to redshift $\rm{z=5}$ \citep{Ueda2014, Aird2015, Miyaji2015, Buchner2015}, utilising several approaches in treating the intrinsic absorption $\rm{N_{\rm{H}}}$, which -- depending on the redshift -- can affect the observed $\rm{2-10\,keV}$ flux significantly. They confirm previous studies of the lower redshift Universe showing that the XLF is adequately described by a broken power law which evolves with redshift, showing a decline in the number density of AGN high redshift and low luminosities observed both in the soft ($\rm{0.5-2.0\,keV}$) and hard ($\rm{2.0-10.0\,keV}$) X-ray bands. \citep{Miyaji2000,Ueda2003,Hasinger2005,LaFranca2005,Ebrero2009,Aird2010}.

In this work, we present the $\rm{2-10\,keV}$ intrinsic XLF where the unabsorbed flux is estimated by the power-law continuum emission corrected for absorption (see \S \ref{sec:Xspec_analysis}). In \S \ref{sec:Xspec_results} we  show that the XXL-1000-AGN sample includes mostly unabsorbed sources. However, the determination of $\rm{\log\,N_{\rm{H}}}$ remains challenging especially for the spectra with lower counts (flag=4, 9.6\% of the sample). In the following analysis we include all the sources in our sample bearing in mind that the determination of the XLF is subject to these uncertainties.

 We estimate the shape and evolution of the XLF adopting the {\it{luminosity dependent density evolution}} (LDDE)  model \citep{Miyaji2000} using the \cite{Ueda2003} parametrisation. \citet{Fotopoulou2016} have shown that the parametrisation of \citet{Ueda2003}, thanks to its simplicity, is more favourable for samples at least up to $\rm{z=4}$ than the more complex \citet{Ueda2014} model, which incorporates enough flexibility to account for an exponential number density cut-off at redshift $\rm{z>3}$.
The LDDE model we used is given by
        \begin{equation}
        \frac{d\phi(L,z)}{d\log L} = \frac{d\phi(L,z=0)}{d\log L}\times e(L,z)
        .\end{equation}
The local XLF is modelled as a broken power-law distribution
    \begin{equation}\label{eq:LF0}
        \frac{d\phi(L,z=0)}{d\log{L}}=\frac{N}{\left(\frac{L}{L_0}\right)^{\gamma_1}+\left(\frac{L}{L_0}\right)^{\gamma_2}}
    ,\end{equation}
    where $L_0$, is the luminosity at which the `break' occurs and $\gamma_1$ and $\gamma_2$ are the slopes of the power-law distributions below and above $L_0$.
The evolution factor with $z$ and $L$ is given by
        \begin{equation}
        e(z, L) = \frac{(1+z_c)^{p_1}+(1+z_c)^{p_2}}{\left(\frac{1+z}{1+z_c}\right)^{-p_1}+\left(\frac{1+z}{1+z_c}\right)^{-p_2}}
        \end{equation}
with
        \begin{equation}\label{eq:zc}
        z_c(L) = 
            \begin{cases}
                z_c^{*} & L\ge L_a\\
                z_c^{*}\times\left(\frac{L}{L_a}\right)^a & L< L_a\\
            \end{cases}
        .\end{equation}
Similarly to \citet{Aird2010} and \citet{Fotopoulou2016} we used  spectroscopic redshifts when available and the probability distribution function of the photometric redshift estimates for sources without spec-z. For the parameter estimation we used  PyMultiNest\footnote{https://github.com/JohannesBuchner/PyMultiNest}\citep{Buchner2014}, the python interface of  MultiNest \citep{2008MNRAS.384..449F,2009MNRAS.398.1601F,2013arXiv1306.2144F}. MultiNest performs Nested Sampling introduced by \citet{Skilling2004} and it is able to explore the posterior even in the case of multimodal distributions.

    \begin{center}
    \begin{figure*}
    \centering
    \begin{tabular}{c}
    \includegraphics[width=0.8\textwidth]{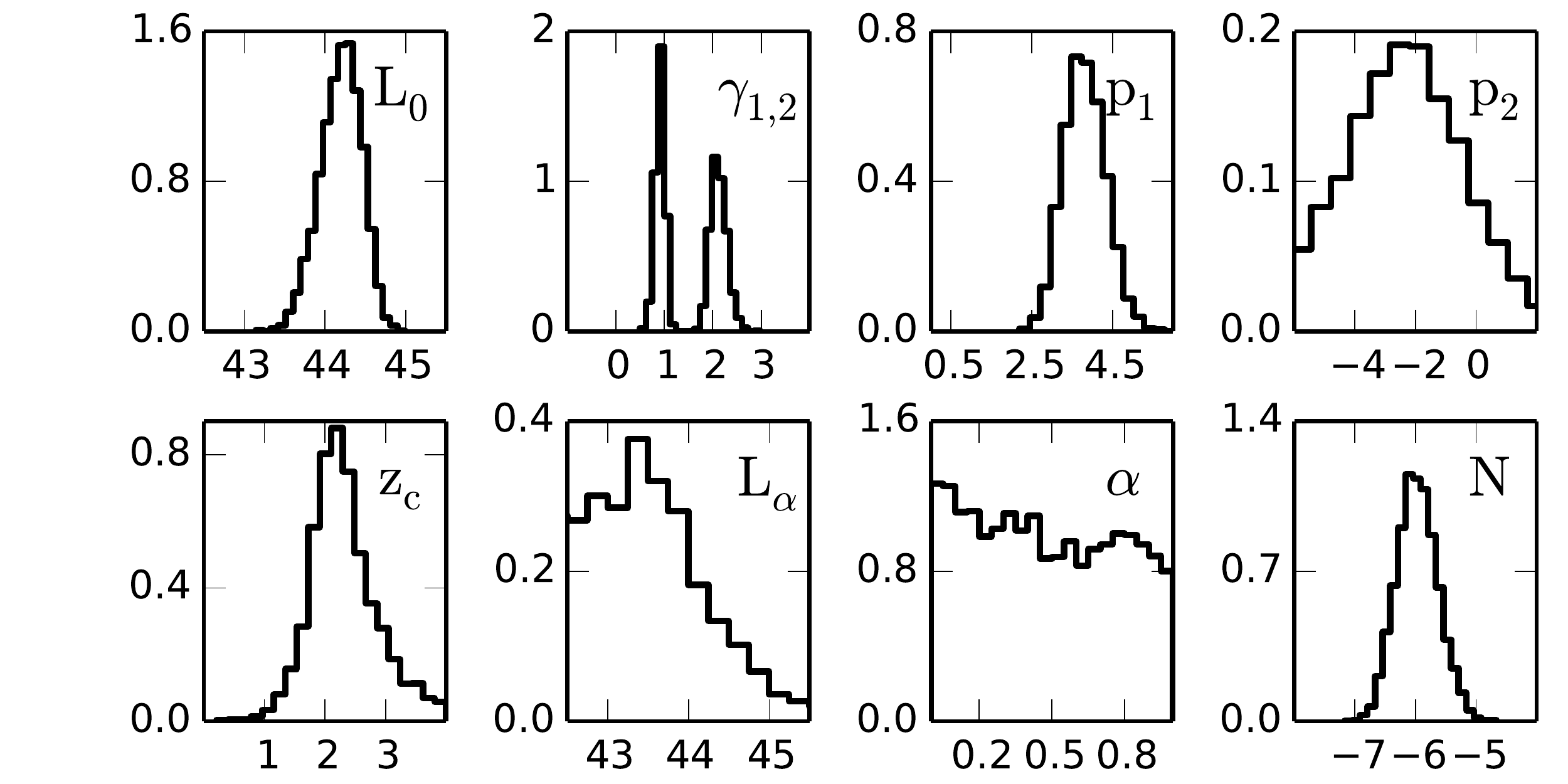} \\
    \includegraphics[width=0.8\textwidth]{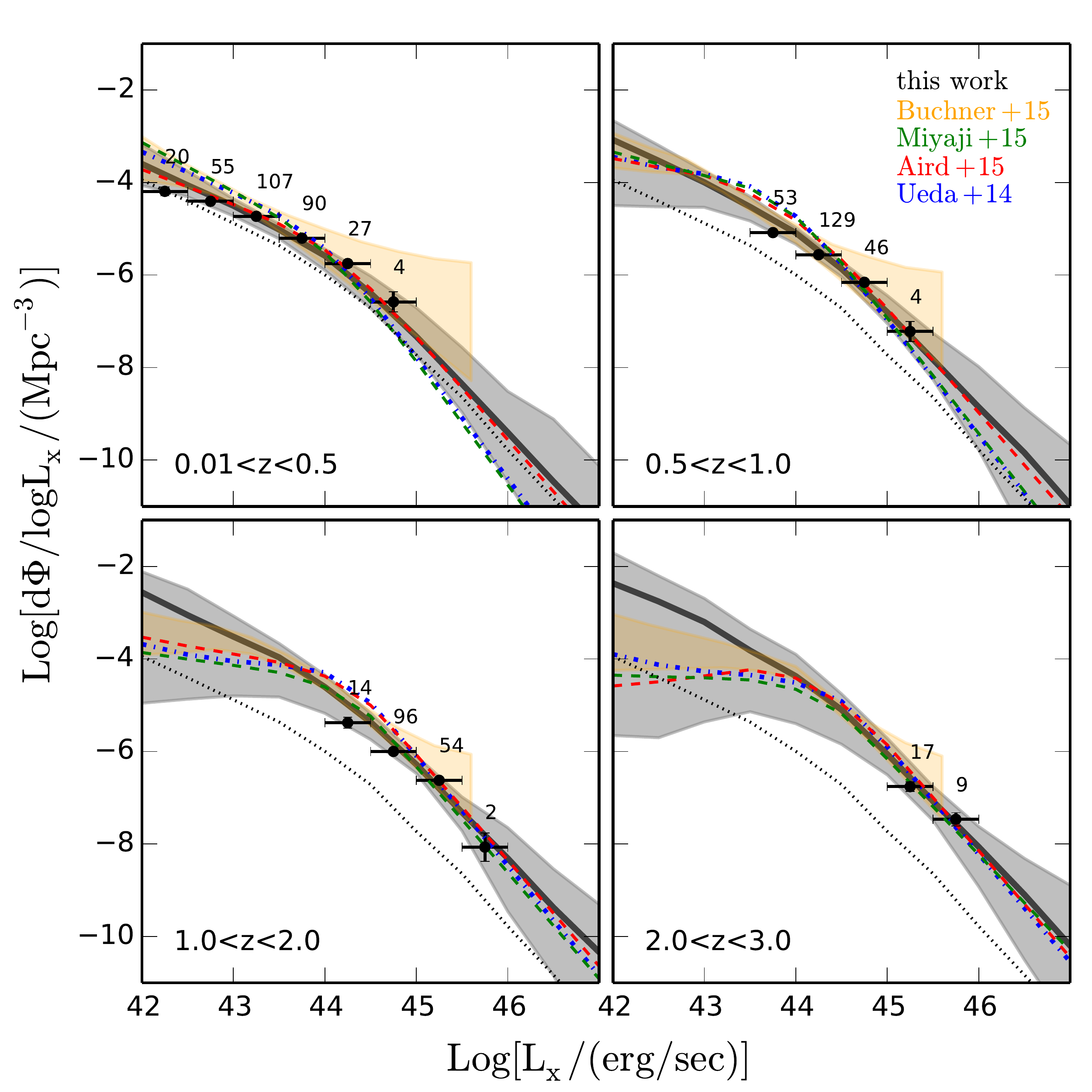}
    \end{tabular}
    \caption{X-ray luminosity function from the XXL-1000-AGN sample. The upper panels show the marginalised posterior distribution of the parameters in luminosity dependent density evolution (LDDE) model. The bottom panels show the estimated XLF as a function of luminosity in four redshift bins. The black solid line shows the mode of the distribution and the grey shaded area encloses the 90\% credible interval. The dotted black line is our estimated $\rm{z=0}$ XLF shown for reference in all plots. The black points are estimated using the $\rm{1/V_{max}}$ method in good agreement with our modelling. The numbers above the points show the number of objects per bin. They also show the parameter space occupied by the XXL-1000-AGN sample. The orange area and dashed lines show $\rm{2-10\,keV}$ XLF estimates from the literature (orange: \citet{Buchner2015}, green: \citet{Miyaji2015}, red:\citet{Aird2015}, blue: \citet{Ueda2014}). \label{fig:xlf}}
    \end{figure*}
    \end{center}

In Figure \ref{fig:xlf} we show the XLF estimated using the XXL-1000-AGN sample. The top panels show the posteriors of all model parameters. Parameters $\gamma_1$ and $\gamma_2$ are symmetric in equation \ref{eq:LF0}; therefore, they have the same double-peaked posterior distribution (shown here only once for simplicity). We note that when using a Bayesian approach, it is not necessary to fix any of the model parameters that are not constrained by our data. This is the case for parameter $\alpha$, which describes the decline of the faint end of the XLF at redshifts above $z_c$. The posterior distribution resembles the flat prior that we used during the parameter estimation.
Since we are using samples drawn from the multidimensional posterior to estimate the XLF, the uncertainty on any of the parameters is naturally incorporated in the uncertainty budget of the XLF (grey shaded area). In Table \ref{tab:xlf}, we give the mean of the posterior distribution and the standard deviation for each parameter.
The bottom panels of Fig. \ref{fig:xlf} show the resulting XLF using the LDDE model estimated with our sample as a function of luminosity and in four redshift bins. The black dashed line shows the estimated local luminosity function (eq. \ref{eq:LF0}) for reference. The red lines show the mode of the luminosity function distribution, while the grey shaded area shows the 90\% credible interval. As an independent method, we show the binned estimates calculated using the $\rm{1/V_{max}}$ method \citep{Schmidt1968,Page2000}. The numbers above the points show how many objects belong in each bin. We see that the two methods agree rather well within the uncertainties.

The XLF from this work is in good agreement with recent estimates of the AGN XLF \citep[][yellow area, green, blue, and red lines in Fig. \ref{fig:xlf}, respectively]{Buchner2015, Miyaji2015, Aird2015, Ueda2014}. We note that XLFs are in good agreement when compared within the parameter space that can be constrained by the data set used for the estimation, while large discrepancies, for example the faint end slope at $\rm{z>2}$, appear when extrapolating the models.

    \begin{figure}[h!]
    \includegraphics[width=\columnwidth]{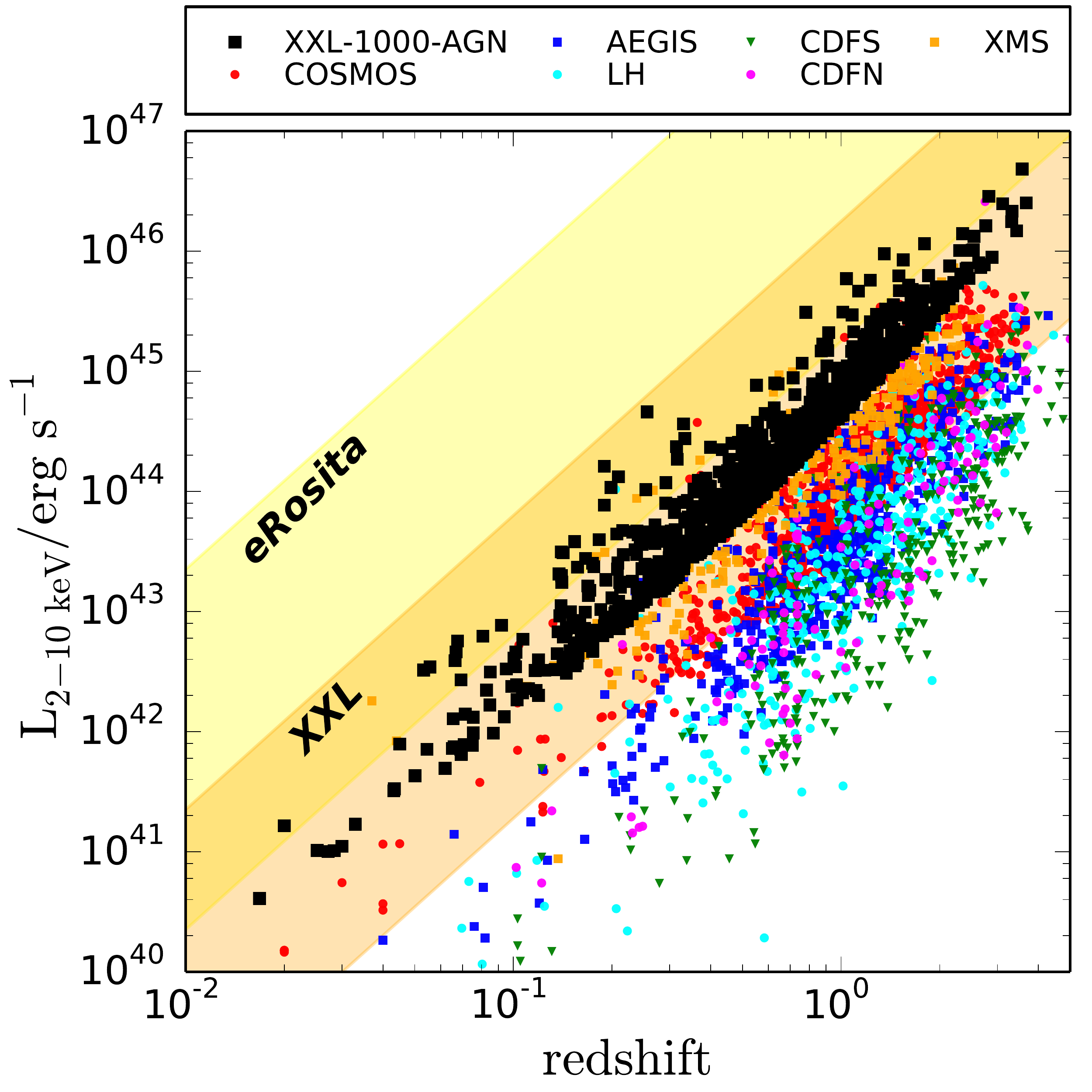} 
    \caption{Luminosity redshift plane for X-ray fields in the $\rm{2-10\,keV}$ energy band. The black points show the XXL-1000-AGN sample. The orange shaded area shows the parameter space covered by the full XXL catalogue. The yellow shaded area shows the expectation for the four-year all-sky coverage by eRosita.  The XXL survey is a stepping stone between current deep surveys (coloured points) and future all-sky surveys.
    \label{fig:Lxz}}
    \end{figure}

    \begin{table}
    \caption{Best fit values of the luminosity dependent density evolution model.}
    \label{tab:xlf}
    \centering
    \begin{tabular}{lr@{.}lr@{.}llr@{.}lr@{.}l}\hline
    \multicolumn{1}{c}{parameter}    & \multicolumn{2}{c}{mean} & \multicolumn{2}{c}{$\sigma$} &     \multicolumn{1}{c}{parameter}    & \multicolumn{2}{c}{mean} & \multicolumn{2}{c}{$\sigma$} \\ \hline
    $\rm{\log L_0}$      & 44&18    & 0&26 &     $\rm{z_c}$      & 2&30       & 0&5 \\
    $\rm{\gamma_1}$      & 0&91      & 0&10 &     $\rm{\log L_{\alpha}}$      & 43&00     & 1&0 \\
    $\rm{\gamma_2}$      & 2&13      & 0&17 &     $\rm{\alpha}$       & 0&48      & 0&29 \\
    $\rm{p_1}$      & 3&80      & 0&5 &     $\rm{\log N}$    & -6&00     & 0&3 \\
    $\rm{p_2}$      & -3&00     & 2&0 \\
    \hline
    \end{tabular}    
    \end{table}

\subsection{Large-scale structure}\label{sec:percolation}
Galaxy formation theories show that the
action of gravitational instability on a perturbed fluctuation
background gives rise to a wealth of large-scale structures, the
interconnections of which provide the so-called cosmic web or cosmic
foam \citep{Bond1996}. The complex and interconnected
structures of the cosmic web come in the variety of groups and
clusters of galaxies, filaments, walls, and voids, which are low-density
regions filling most of the volume of the Universe
\citep[e.g.][]{vandeWeygaert2008}. Observationally, it has been found
that when using galaxies as tracers of the large-scale structure, the wealth of structures in the Universe   can
clearly be identified  \citep[for
  recent studies, see][]{Parihar2014, James2007}. AGN have also been
used to trace the large-scale structure  by using a variety of
techniques, among which the two-point correlation function and BAOs 
\citep[e.g.][and references therein]{Cappelluti2012, Hutsi2014} or
their number counts \citep[e.g.][]{Dai2015}. 

In this section we present a preliminary analysis of the X-ray AGN
large-scale structures by applying the friends-of-friends algorithm in
order to identify compact or loose groupings of AGN and then we
assess their significance by using extended Monte Carlo simulations. 

\begin{figure*}
    \centering
    \includegraphics[width=0.85\linewidth]{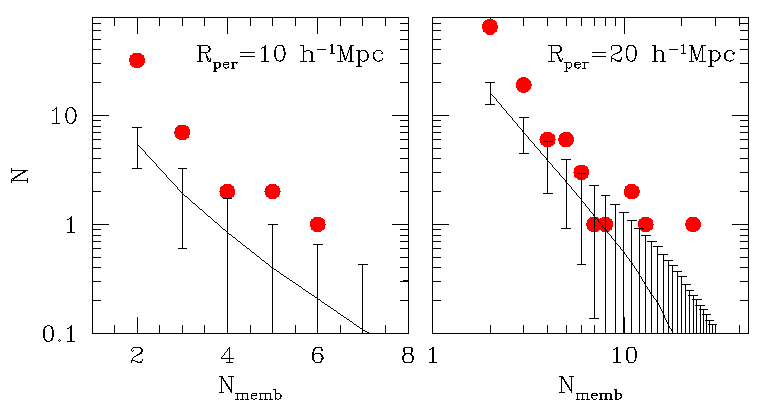} 
    \caption{Multiplicity function of AGN for two percolation radii,
      $R_{per}=10 h^{-1}$Mpc (a) and $R_{per}=20 h^{-1}$Mpc (b). The grey
      lines show the expected multiplicity of a random field with the
      same smoothed redshift distribution as the data. We see significant structures with 2-3
      members at both radii and one significant structure consisting
      of 23 members for $R_{per}=20 h^{-1}$Mpc.\label{fig:MultiAGN}}
    \end{figure*}

We use all the X-ray AGN for which we have either spectroscopic (589)
or photometric (383) redshifts. In Fig. \ref{fig:MultiAGN} we present as red points
the multiplicity function of AGN structures as a function of AGN members
and for two percolation radii ($R_{per}=10$ and 20 $h^{-1}$ Mpc). 

We also provide by the continuous line the mean random expectation from the 
application of our friends-of-friends algorithm on 10000 
random realisations having the same number of points, angular selection 
function, and redshift distribution as the real X-ray AGN. 
This is achieved by using the same celestial coordinates as the real data, but 
randomising the redshift of each AGN such that all observed redshifts are 
assigned to random points while no redshift is duplicated. 
Furthermore, we smooth the redshifts using a Gaussian kernel with zero 
mean and $\sigma=0.1$ (we  verified that varying $\sigma$ 
between 0.05 and 0.2 provides equivalent results).
It is evident that the most
significant X-ray AGN structures are those with two or three members
(for both percolation radii). 

\begin{figure}
\includegraphics[width=\columnwidth]{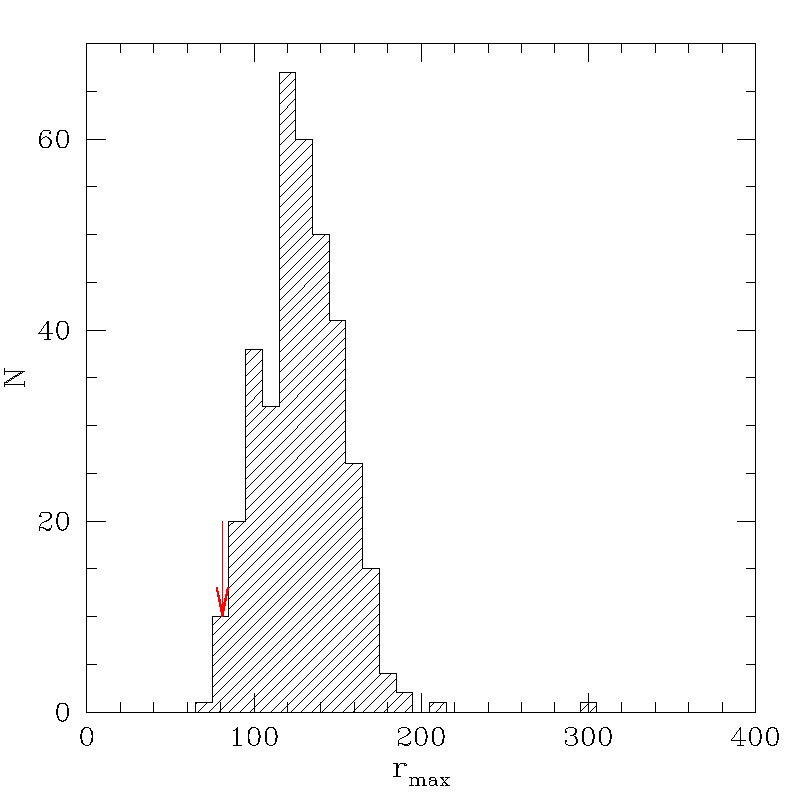} 
\caption{Radius distribution of overdensities with 23 members
  estimated in 10000 Monte Carlo simulations. The red arrow shows
  the radius of the overdensity with 23 members identified in the
  data. The compactness of the structure ($81h^{-1}Mpc$) makes it
  a candidate supercluster of AGN\label{fig:dmax20}}
\end{figure}

For the $R_{per}=20 \; h^{-1}$ Mpc case we also identify one
significant large structure containing 23 AGN, which has a
diameter of 81 $h^{-1}$ Mpc,  which leads us to consider it as a
candidate supercluster of X-ray AGN. 
The centre of mass of this structure has celestial coordinates
  $(\alpha, \delta)=(34.80^{\circ}, -5.17^{\circ})$ and a mean
  redshift $\langle z \rangle=0.14$, and it is related to
  the most abundant ({\em XLSSC-b}) supercluster of XXL bright clusters (see Paper II).
The significance of this AGN structure 
can be appreciated not only by the probability of observing such an
AGN membership, as provided by the comparison with the random
expectation in Fig. \ref{fig:MultiAGN}, but also by the compactness of this
structure. In Fig. \ref{fig:dmax20} we present the distribution of diameters of all
the structures with 23 members that have been found in 10000 Monte Carlo
simulations. Only in 3\% of the cases is the diameter of the random
structures  as small as that of the observed structure. 
We have also performed the same analysis using only sources with spectroscopic redshifts. The overdensity is also present in the restricted sample, which contains 16 members.

We conclude that the spatial distribution of XXL AGN shows a wealth of
significant large-scale structures, some of which are of  supercluster
size. The detected AGN supercluster, but also other compact structures
of various memberships, will be investigated in detail in a forthcoming paper.

\section{Released catalogue description}\label{sec:release}
With this paper we release the XXL-1000-AGN sample, the 1000 brightest X-ray point-like sources detected in the XMM-XXL field. The fluxes estimated by \textsc{Xamin 3.3} (columns 4-5) assume $\Gamma=1.7$, and $\rm{N_{\rm{H}}=2.6\times10^{20}cm^{-2}}$. The count-to-flux conversion factors are given in \S \ref{sec:det}. For the quantities estimated by the spectral fit (columns 6-17) we provide the mode of the marginalised posterior distribution, the 68\% credible interval around the mode, and the median of the distribution (see \S \ref{sec:Xspec_analysis}). All magnitudes and associated uncertainties are measured with Sextractor. They originate from SDSS/CFHTLS for XXL-N or BCS/DECam for XXL-S. They are given in AB, measured in a fixed circular aperture of 3'', and they are corrected to total and for Galactic extinction.
All fluxes are given in $\rm{erg\,s^{-1}cm^{-2}}$, magnitudes in AB. 
The contents of the catalogue are as follows:
\newline
\begin{itemize}
\item[] (1) Xcatname: unique source identification
\item[] (2) Xra: X-ray point source right ascension 
\item[] (3) Xdec: X-ray point source declination
\item[] (4) $\rm{Bflux}$: 0.5-2\,keV flux \textsc(Xamin 3.3)
\item[] (5) $\rm{CDFlux}$: 2-10\,keV flux \textsc(Xamin 3.3)
\item[] (6) $\rm{F_{mode}}$: $\rm{2-10\,keV}$ intrinsic (unabsorbed) flux, mode of PDF
\item[] (7) $\rm{F_{l}}$: $\rm{2-10\,keV}$ intrinsic (unabsorbed) flux, lower bound of 68\% credible interval
\item[] (8) $\rm{F_{h}}$: $\rm{2-10\,keV}$ intrinsic (unabsorbed) flux, upper bound of 68\% credible interval
\item[] (9) $\rm{F_{median}}$: $\rm{2-10\,keV}$ intrinsic (unabsorbed) flux, median of PDF
\item[] (10) $\rm{\Gamma_{mode}}$: photon index, mode of PDF
\item[] (11) $\rm{\Gamma_{l}}$: photon index,lower bound of 68\% credible interval
\item[] (12) $\rm{\Gamma_{h}}$: photon index, upper bound of 68\% credible interval
\item[] (13) $\rm{\Gamma_{median}}$: photon index, median of PDF
\item[] (14) $\rm{N_{H,mode}}$: hydrogen column density, mode of PDF
\item[] (15) $\rm{N_{H,l}}$: hydrogen column density, lower bound of 68\% credible interval
\item[] (16) $\rm{N_{H,h}}$: hydrogen column density, upper bound of 68\% credible interval
\item[] (17) $\rm{N_{H,median}}$: hydrogen column density, median of PDF
\item[] (18) Xflag: X-ray spectrum quality flag (see \S \ref{sec:Xspec_analysis}, Table \ref{tab:RatioF})
\item[] (19) CtpRa: counterpart right ascension
\item[] (20) CtpDec: counterpart declination
\item[] (21) g: g-band magnitude
\item[] (22) g$\rm{_{err}}$: g-band magnitude uncertainty
\item[] (23) r: r-band magnitude
\item[] (24) r$\rm{_{err}}$: r-band magnitude uncertainty
\item[] (25) i: i-band magnitude
\item[] (26) i$\rm{_{err}}$: i-band magnitude uncertainty
\item[] (27) z: z-band magnitude
\item[] (28) z$\rm{_{err}}$: z-band magnitude uncertainty
\item[] (29) $\rm{photo\_origin}$: parent photometric survey. A four-character code `griz', where each of the four characters can assume the value `-' in the case of missing data in the band,     S or C (for SDSS or CFHTL, XXL-N), or B or D (for BCS or DECam, XXL-S)
\item[] (30) $\rm{zspec}$: spectroscopic redshift, when available
\item[] (31) $\rm{zspec\_origin}$: spectroscopic redshift parent survey 
        \begin{itemize} 
        \item[-] AAT: \citet{Lidman2016} 
        \item[-] \citet{Akiyama2015}
        \item[-] GAMA: \citet{Hopkins2013}
        \item[-] LDSS03: \citet{Adami2011}
        \item[-] SDSSDR12 \citet{Menzel2015}
        \item[-] SDSS DR10
        \item[-] \citet{Simpson2006}
        \item[-] \citet{Simpson2012}
        \item[-] \citet{Stalin2010}
        \item[-] VIPERS: \citet{Garilli2014}
        \item[-] VVDS: \citet{LeFevre2013}
        \item[-] WHT: \citet{KoulouridisPXII}
        \end{itemize}
\item[] (32) $\rm{zphot}$: photometric redshift
\item[] (33) $\rm{zphot\_l}$: lower bound of $68\%$ credible interval
\item[] (34) $\rm{zphot\_h}$: upper bound of $68\%$ credible interval
\item[] (35) $\rm{zphot\_class}$: classification for best photo-z
        \begin{itemize}
                \item[-] passive = 1
                \item[-] star forming = 2
                \item[-] starburst = 3
                \item[-] AGN = 4
                \item[-] QSO = 5
        \end{itemize}
\item[] (36) Pstar: probability that the source is a star
\item[] (37) Poutlier: probability that a source  has\\$|z_{phot}-z_{spec}|>0.15\cdot(1+z_{spec})$
\end{itemize}
An example page of the final source catalogue is shown in Table \ref{tab:XXL1000}. A full printable version is available.  The full catalogue is available as a queryable database table XXL\_1000\_AGN via the XXL Master Catalogue browser\footnote{\url{http://cosmosdb.iasf-milano.inaf.it/XXL/}}. A copy is also  deposited at the  Centre de Donn\'ees astronomiques de Strasbourg (CDS)\footnote{\url{http://cdsweb.ustrasbg.fr}}

\section{Conclusions}\label{sec:conclusions}

    We presented the XXL-1000-AGN sample, the first release of the XXL point source catalogue, consisting of the 1000 brightest sources, selected in the $\rm{2-10\,keV}$ energy band. The flux limit of our catalogue is $\rm{F_{2-10\,keV}=4.8\times10^{-14}erg\,s^{-1}cm^{-2}}$.

    \begin{itemize}
            \item We built a multiwavelength catalogue creating SEDs from the far-ultraviolet to the mid-infrared and assigned the counterparts to the X-ray detections using the likelihood ratio technique. We retrieve counterparts for 97\% of our sources.

        \item Using machine learning classification, we assign the optimal class describing the broadband SEDs which improved the photometric redshift estimates.

        \item Modelling the X-ray spectra with an absorbed power-law distribution we find that the average photon index is $\rm{<\Gamma>=1.85\pm0.40}$. Our sample is dominated by unabsorbed sources (80\% with $\rm{N_{\rm{H}}}<10^{22}cm^{-2}$) with average hydrogen column density $\rm{\log<N_{\rm{H}}>=21.0\pm1.3}$.

        \item We present the median observed SED (AGN combined with the host galaxy emission) and show that the X-ray absorption, which was estimated from X-ray spectra,  shows the same general trend as the host galaxy absorption estimated by the SEDs. Low absorption in the X-rays in general corresponds to low absorption in the optical. This is consistent with the picture of the unified model of AGN where we observe the galaxy face-on.

        \item The XXL X-ray number counts are fully consistent with the Euclidean expectation and agree with previous deep (CDFS, COSMOS) and wide (H-ATLAS) XMM-Newton surveys.

        \item We present the best fit parameters for the LDDE XLF model up to $\rm{z=3}$. The XXL-1000-AGN sample poses constraints on the low luminosity -- low redshift, medium luminosity -- medium redshift X-ray luminosity function, in good agreement with recent estimates using deeper X-ray observations.

        \item An application of the friends-of-friends algorithm at $\rm{10h^{-1}Mpc}$ and $\rm{20h^{-1}Mpc}$ percolation radii shows significant structures with 2-3 members. Additionally, the analysis of $\rm{R_{per}=20h^{-1}Mpc}$ suggests the presence of a candidate supercluster of AGN with 23 members at redshift z=0.14.
The same result is retrieved using only sources with spec-z.

        \item We release the catalogue of the XXL-1000-AGN sample with positions, flux estimates from both  the pipeline and the X-ray spectra, optical magnitudes, and redshift information (spec-z 60\%  and photo-z 40\%).
        
    \end{itemize}

With the first data release, we provide a significant number of AGN detected with XMM-Newton, of comparable quality to modern deep X-ray surveys. In future publications we will expand the analysis presented in this work to the full XXL catalogue containing an unprecedented number of $\rm{\sim10^4}$ (X-ray) point-like $\rm{2-10\,keV}$ detected sources analysing the fully combined XMM pointings to reach maximum depth. A unique advantage of XXL collaboration is the combined
study of X-ray point-like sources and X-ray detected galaxy clusters in great numbers, which will
allow the  study of AGN with respect to their environment and,  vice-versa, the study of the impact of AGN
on clusters.  The full catalogue is planned to be released in incremental flux limits.

\begin{acknowledgements} 
We thank the anonymous referee for the useful comments on our work. FP and MR-C acknowledge support from DfG Transregio Programme TR33. PR acknowledges a grant from the Greek General Secretariat of Research and Technology in the framework of the programme Support of Postdoctoral Researchers. MB acknowledges support from the European Union’s FP7 under grant agreement 321913 (CIG, `SMBH evolution through cosmic time’). 
O.M. is grateful for the financial support provided by the European Union Seventh Framework Programme (FP7 2007-2013), grant agreement n\textsuperscript{o} 291823 Marie Curie FP7-PEOPLE-2011-COFUND (The new International Fellowship Mobility Programme for Experienced Researchers in Croatia - NEWFELPRO, project "AGN environs in XXL", Grant Agreement \#83). SA acknowledges a post-doctoral fellowship from TUBITAK-BIDEB through 2219 program.
XXL is an international project based around an XMM Very Large Programme surveying two 25 deg$^2$ extragalactic fields at a depth of ~5 x 10$^{-15}$ erg cm$^{-2}$ s$^{-1}$ in the [0.5-2] keV band for point-like sources. The XXL website is http://irfu.cea.fr/xxl. Multiband information and spectroscopic follow-up of the X-ray sources are obtained through a number of survey programmes, summarised at http://xxlmultiwave.pbworks.com/.
Based on observations obtained with MegaPrime/MegaCam, a joint project of CFHT and CEA/IRFU, at the Canada-France-Hawaii Telescope (CFHT) which is operated by the National Research Council (NRC) of Canada, the Institut National des Science de l'Univers of the Centre National de la Recherche Scientifique (CNRS) of France, and the University of Hawaii. 
This work is based in part on data products produced at Terapix available at the Canadian Astronomy Data Centre as part of the Canada-France-Hawaii Telescope Legacy Survey, a collaborative project of NRC and CNRS. 
Based on observations obtained with WIRCam, a joint project of CFHT, Taiwan, Korea, Canada, France, at the Canada-France-Hawaii Telescope (CFHT) which is operated by the National Research Council (NRC) of Canada, the Institut National des Sciences de l'Univers of the Centre National de la Recherche Scientifique of France, and the University of Hawaii.
This publication makes use of data products from the Wide-field Infrared Survey Explorer, which is a joint project of the University of California, Los Angeles, and the Jet Propulsion Laboratory/California Institute of Technology, funded by the National Aeronautics and Space Administration. 
This paper uses data from the VISTA Hemisphere Survey ESO programme ID: 179.A-2010 (PI. McMahon). 
This publication has made use of data from the VIKING survey from VISTA at the ESO Paranal Observatory, programme ID 179.A-2004. 
Based on data products from observations made with ESO Telescopes at the La Silla or Paranal Observatories under ESO programme ID 179.A-2006 (VIDEO survey). 
Data processing has been contributed by the VISTA Data Flow System at CASU, Cambridge and WFAU, Edinburgh. 
This work is based in part on data obtained as part of the UKIRT Infrared Deep Sky Survey. 
Funding for SDSS-III has been provided by the Alfred P. Sloan Foundation, the Participating Institutions, the National Science Foundation, and the U.S. Department of Energy Office of Science. The SDSS-III web site is http://www.sdss3.org/. SDSS-III is managed by the Astrophysical Research Consortium for the Participating Institutions of the SDSS-III Collaboration including the University of Arizona, the Brazilian Participation Group, Brookhaven National Laboratory, Carnegie Mellon University, University of Florida, the French Participation Group, the German Participation Group, Harvard University, the Instituto de Astrofisica de Canarias, the Michigan State/Notre Dame/JINA Participation Group, Johns Hopkins University, Lawrence Berkeley National Laboratory, Max Planck Institute for Astrophysics, Max Planck Institute for Extraterrestrial Physics, New Mexico State University, New York University, Ohio State University, Pennsylvania State University, University of Portsmouth, Princeton University, the Spanish Participation Group, University of Tokyo, University of Utah, Vanderbilt University, University of Virginia, University of Washington, and Yale University. 
This work is partly based on observations made with the Spitzer Space Telescope, which is operated by the Jet Propulsion Laboratory, California Institute of Technology under a contract with NASA. 
Some of the data presented in this paper were obtained from the Mikulski Archive for Space Telescopes (MAST). STScI is operated by the Association of Universities for Research in Astronomy, Inc., under NASA contract NAS5-26555. Support for MAST for non-HST data is provided by the NASA Office of Space Science via grant NNX09AF08G and by other grants and contracts.
Based in part on data acquired through the Australian Astronomical Observatory, under programs A/2013A/018 and A/2013B/001,and on observations at Cerro Tololo Inter-American Observatory, National Optical Astronomy Observatory (NOAO Prop. IDs 2013A-0618 and 2015A-0618), which is operated by the Association of Universities for Research in Astronomy (AURA) under a cooperative agreement with the National Science Foundation. This project used data obtained with the Dark Energy Camera (DECam), which was constructed by the Dark Energy Survey (DES) collaboration. Funding for the DES Projects has been provided by the U.S. Department of Energy, the U.S. National Science Foundation, the Ministry of Science and Education of Spain, the Science and Technology Facilities Council of the United Kingdom, the Higher Education Funding Council for England, the National Center for Supercomputing Applications at the University of Illinois at Urbana-Champaign, the Kavli Institute of Cosmological Physics at the University of Chicago, Center for Cosmology and Astro-Particle Physics at the Ohio State University, the Mitchell Institute for Fundamental Physics and Astronomy at Texas A\&M University, Financiadora de Estudos e Projetos, Fundação Carlos Chagas Filho de Amparo, Financiadora de Estudos e Projetos, Fundação Carlos Chagas Filho de Amparo à Pesquisa do Estado do Rio de Janeiro, Conselho Nacional de Desenvolvimento Científico e Tecnológico and the Ministério da Ciência, Tecnologia e Inovação, the Deutsche Forschungsgemeinschaft and the Collaborating Institutions in the Dark Energy Survey. The Collaborating Institutions are Argonne National Laboratory, the University of California at Santa Cruz, the University of Cambridge, Centro de Investigaciones Enérgeticas, Medioambientales y Tecnológicas–Madrid, the University of Chicago, University College London, the DES-Brazil Consortium, the University of Edinburgh, the Eidgen\"ossische Technische Hochschule (ETH) Z\"urich, Fermi National Accelerator Laboratory, the University of Illinois at Urbana-Champaign, the Institut de Ciències de l'Espai (IEEC/CSIC), the Institut de Física d'Altes Energies, Lawrence Berkeley National Laboratory, the Ludwig-Maximilians Universit\"at M\"unchen and the associated Excellence Cluster Universe, the University of Michigan, the National Optical Astronomy Observatory, the University of Nottingham, the Ohio State University, the University of Pennsylvania, the University of Portsmouth, SLAC National Accelerator Laboratory, Stanford University, the University of Sussex, and Texas A\&M University. 
\end{acknowledgements}



\begin{appendix}

\section{XXL-1000-AGN image cutouts}\label{cutouts}
In this appendix we present example X-ray spectra, multiwavelength SEDs, and cutout images for a total of 30 sources from the XXL-1000-AGN sample. We group them according to the Random Forest classification: QSO, AGN, starburst, star forming, and passive. We note   that the classification refers to the best class for the photo-z estimation (see also \S \ref{sec:photoz}). In each class we present three sources from the XXL-N field (top panels) and three sources from the XXL-S (bottom panels), selected randomly from our sample where the  only constraint is that it must  have a high S/N X-ray spectrum. From left to right the images are: 
\begin{enumerate}
\item X-ray spectrum, either PN or MOS as stated on the image.
\item SED from the far-ultraviolet to mid-infrared including the best fit model (red line) and the star model (grey line). The black dots show the available observations, while the white dots show the model magnitudes at the best fit redshift. The insert in the SED plot shows  99.9\% of the photo-z PDF. 
\item We also provide a single-filter cutout in the i-band (XXL-N:CFHTLS, XXL-S:BCS) unless otherwise stated. 
\end{enumerate}

The image cutouts are 30''x30'' and the orientation is north-up, east-left. The dashed green circle in centred on the X-ray position and has a radius of 5''. The circle  shows the chosen counterpart and has a radius of 1.5''.

\begin{center}
\begin{figure*}
\begin{tabular}{cc}
\centering
\begin{adjustbox}{valign=t}
\begin{tabular}{cccc}
\multicolumn{4}{l}{3XLSS J023322.1-045505 } \\
  \includegraphics[width=0.24\hsize]{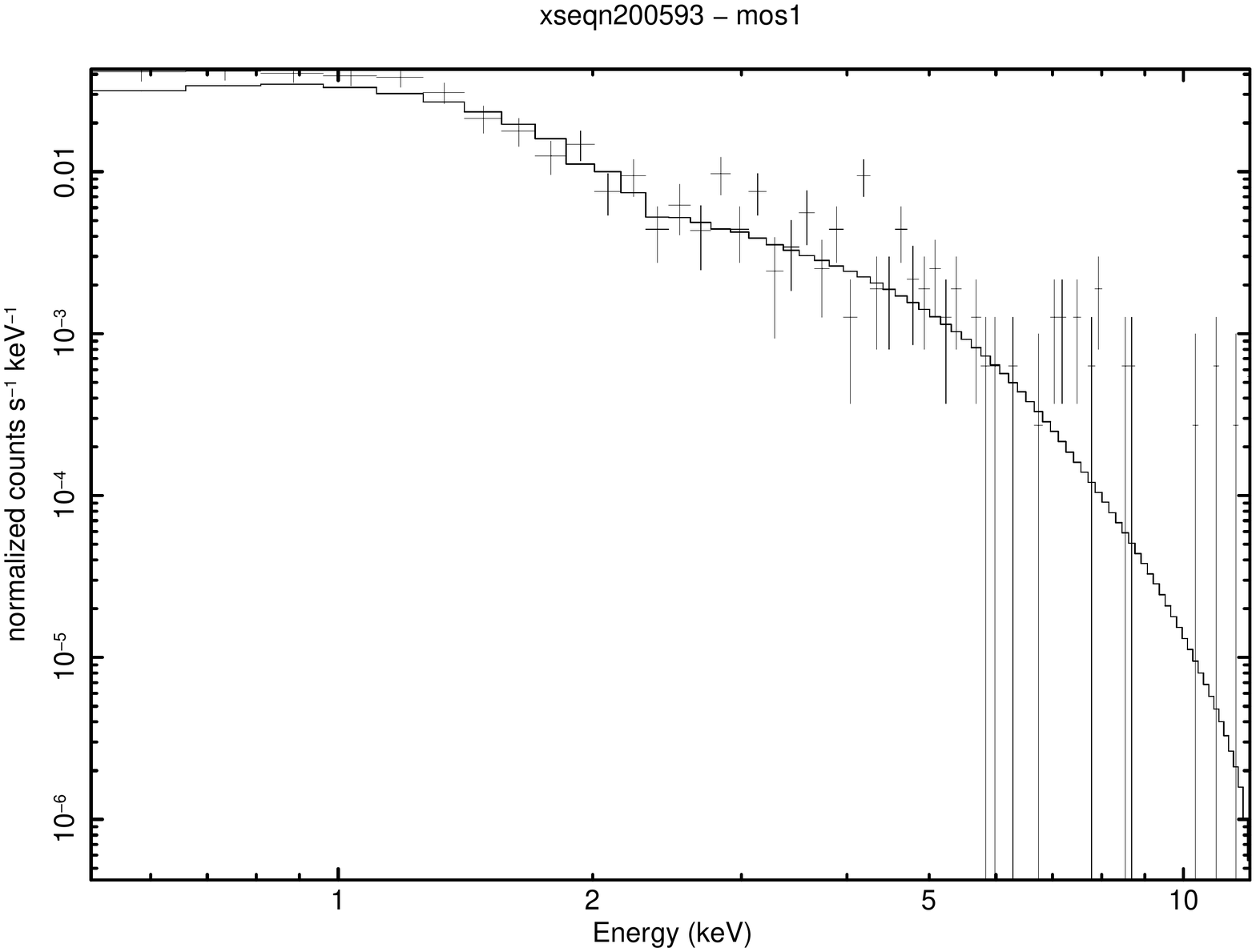} & 
  \includegraphics[width=0.25\hsize]{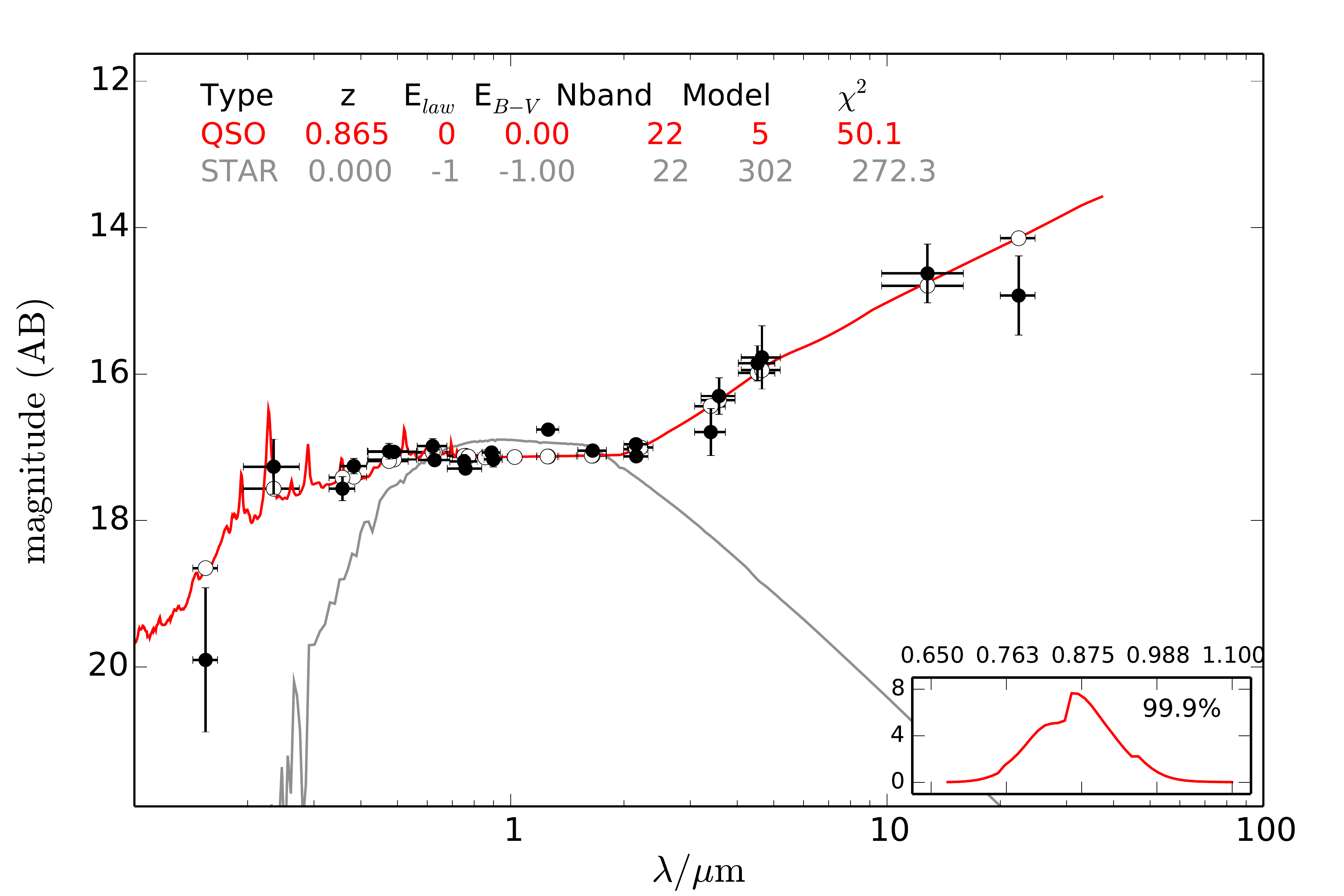} &
  \includegraphics[width=0.17\hsize]{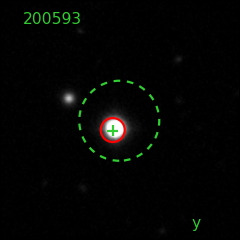} &
  \includegraphics[width=0.17\hsize]{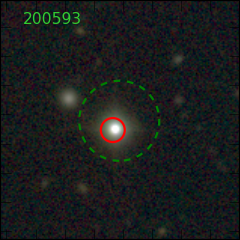} \\

\multicolumn{4}{l}{3XLSS J022935.1-055208 } \\
  \includegraphics[width=0.24\hsize]{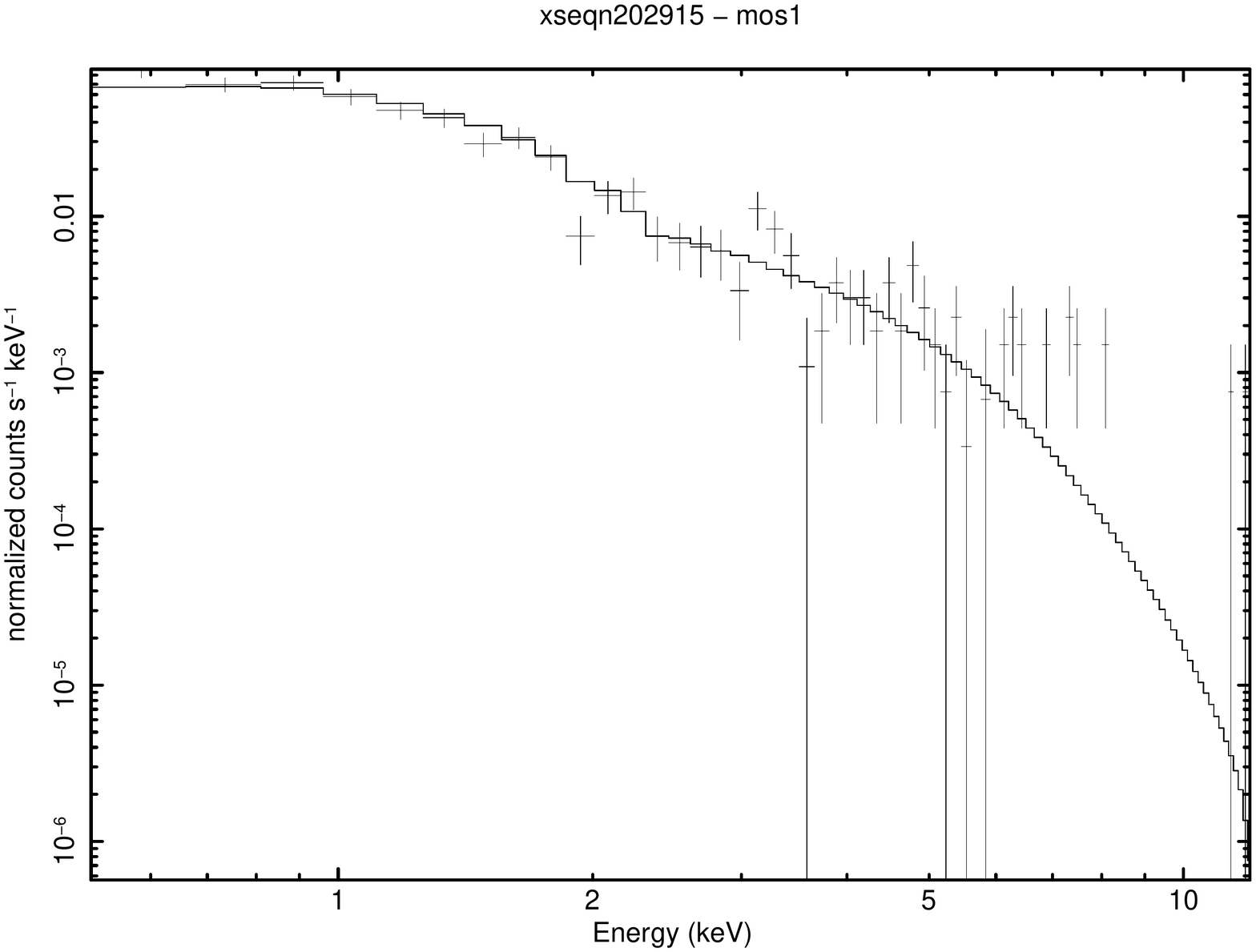} & 
  \includegraphics[width=0.25\hsize]{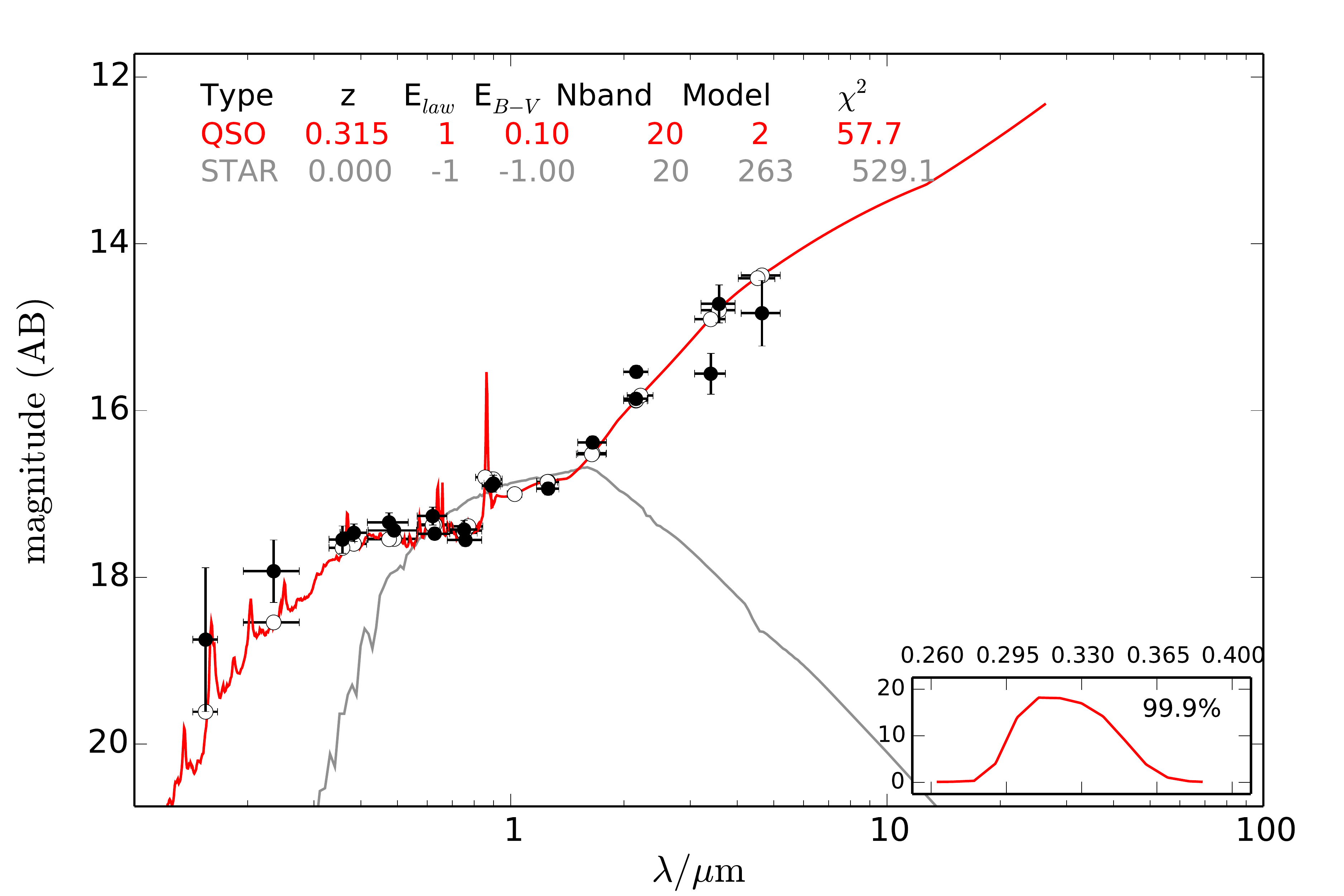} &
  \includegraphics[width=0.17\hsize]{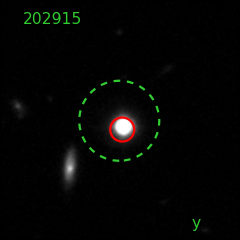} &
  \includegraphics[width=0.17\hsize]{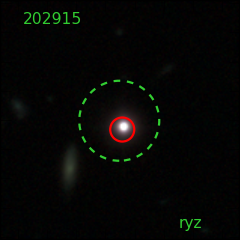} \\

\multicolumn{4}{l}{3XLSS J022711.8-045036  } \\
  \includegraphics[width=0.24\hsize]{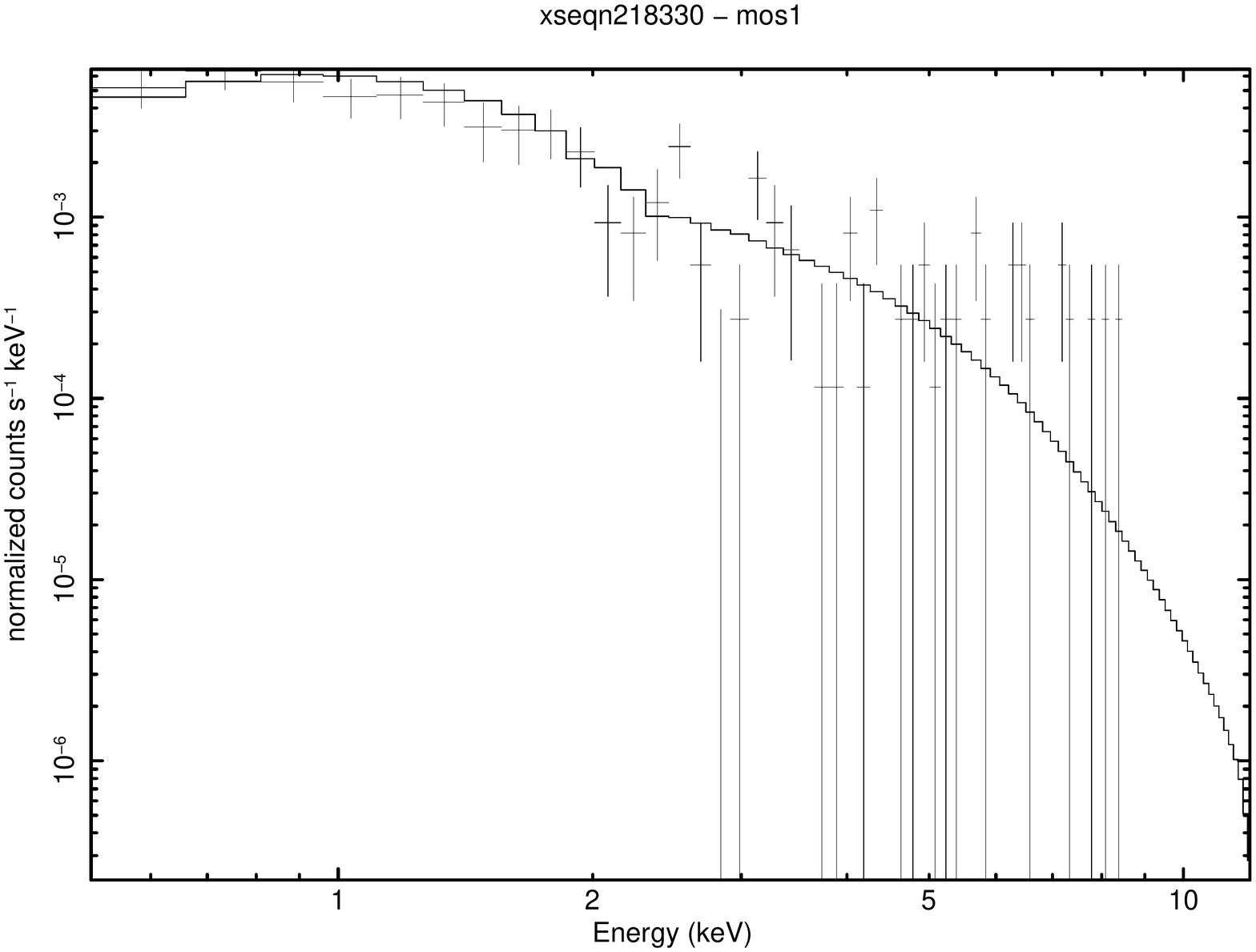} & 
  \includegraphics[width=0.25\hsize]{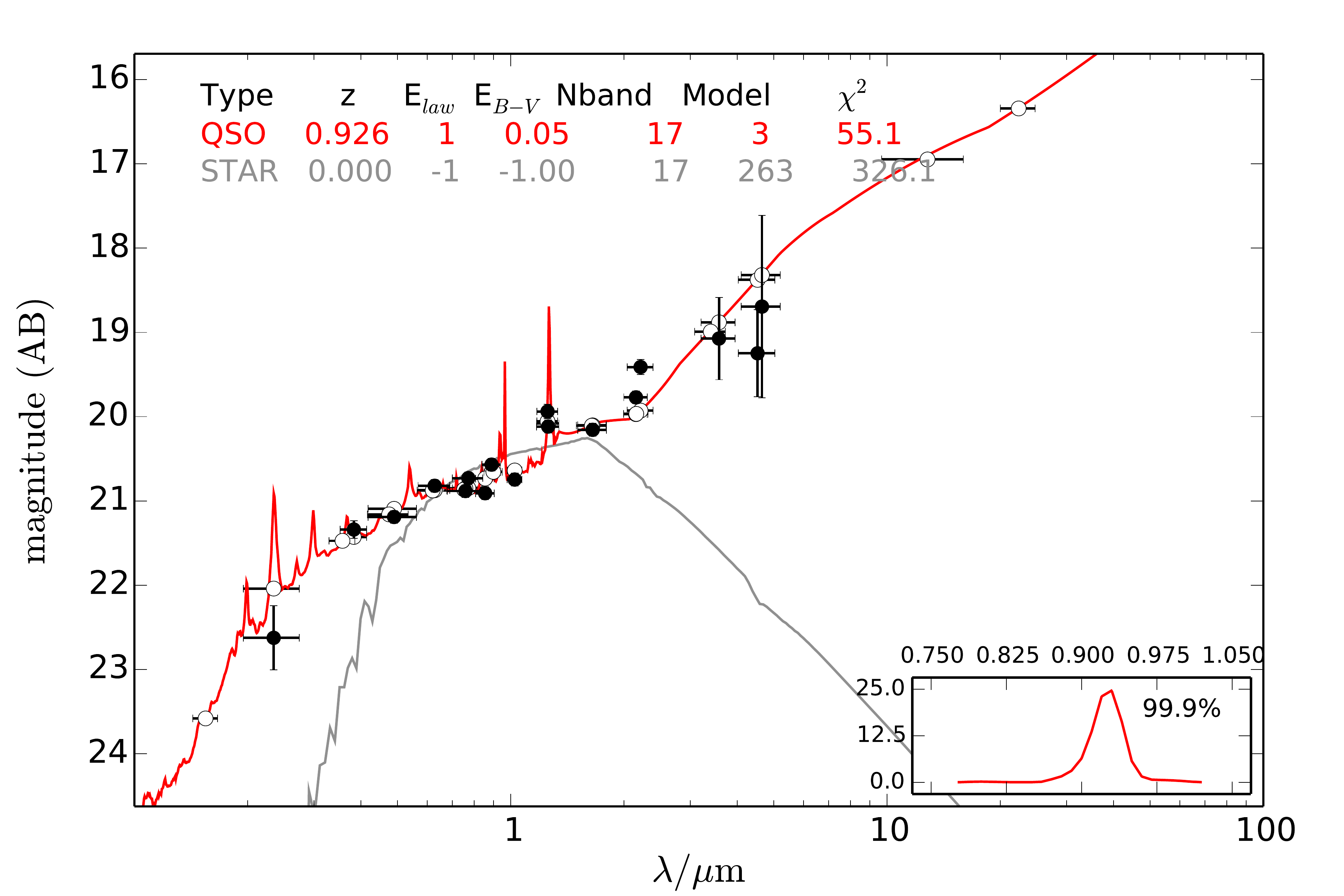} &
  \includegraphics[width=0.17\hsize]{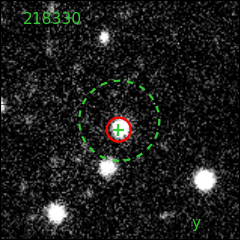} &
  \includegraphics[width=0.17\hsize]{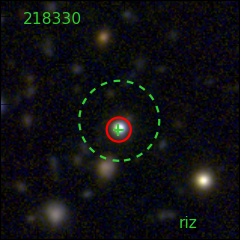} \\
  
\multicolumn{4}{l}{3XLSS J022105.5-044101 } \\
  \includegraphics[width=0.24\hsize]{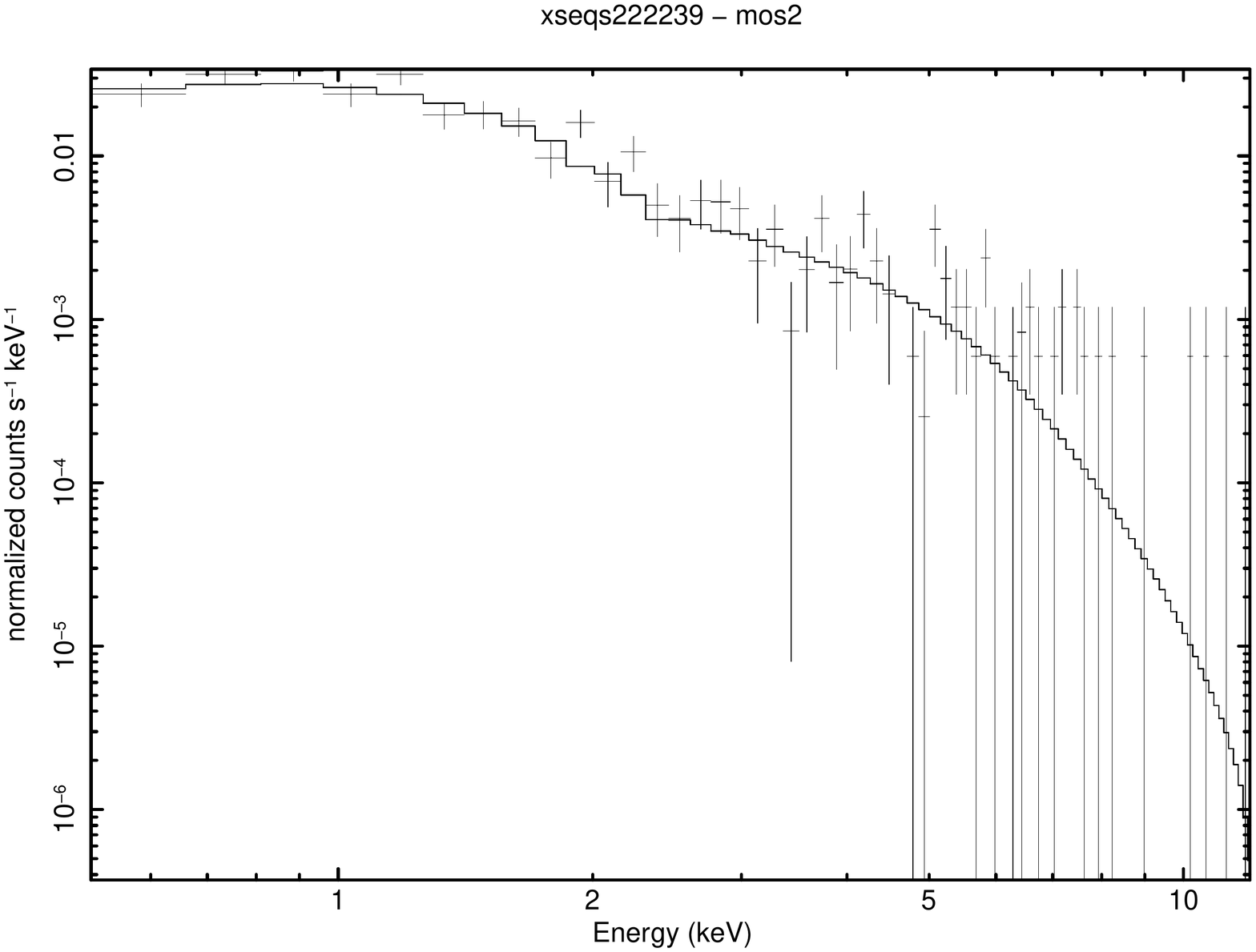} &
  \includegraphics[width=0.25\hsize]{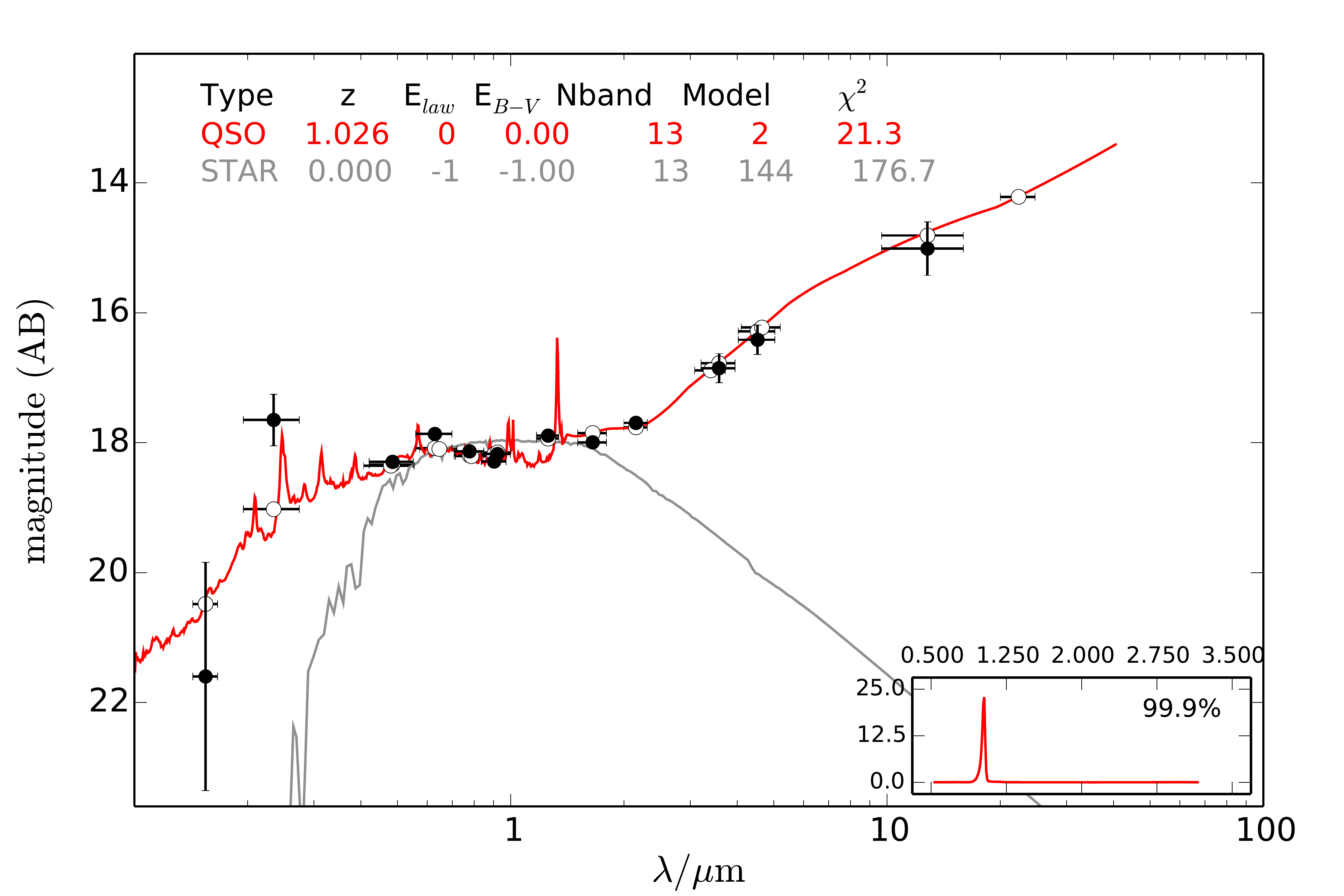} &
  \includegraphics[width=0.17\hsize]{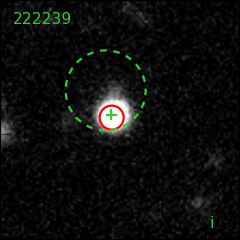}&
  \includegraphics[width=0.17\hsize]{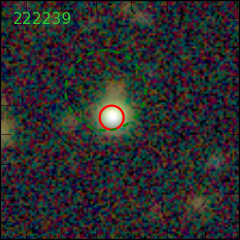}\\

\multicolumn{4}{l}{3XLSS J020423.1-040851 } \\
  \includegraphics[width=0.24\hsize]{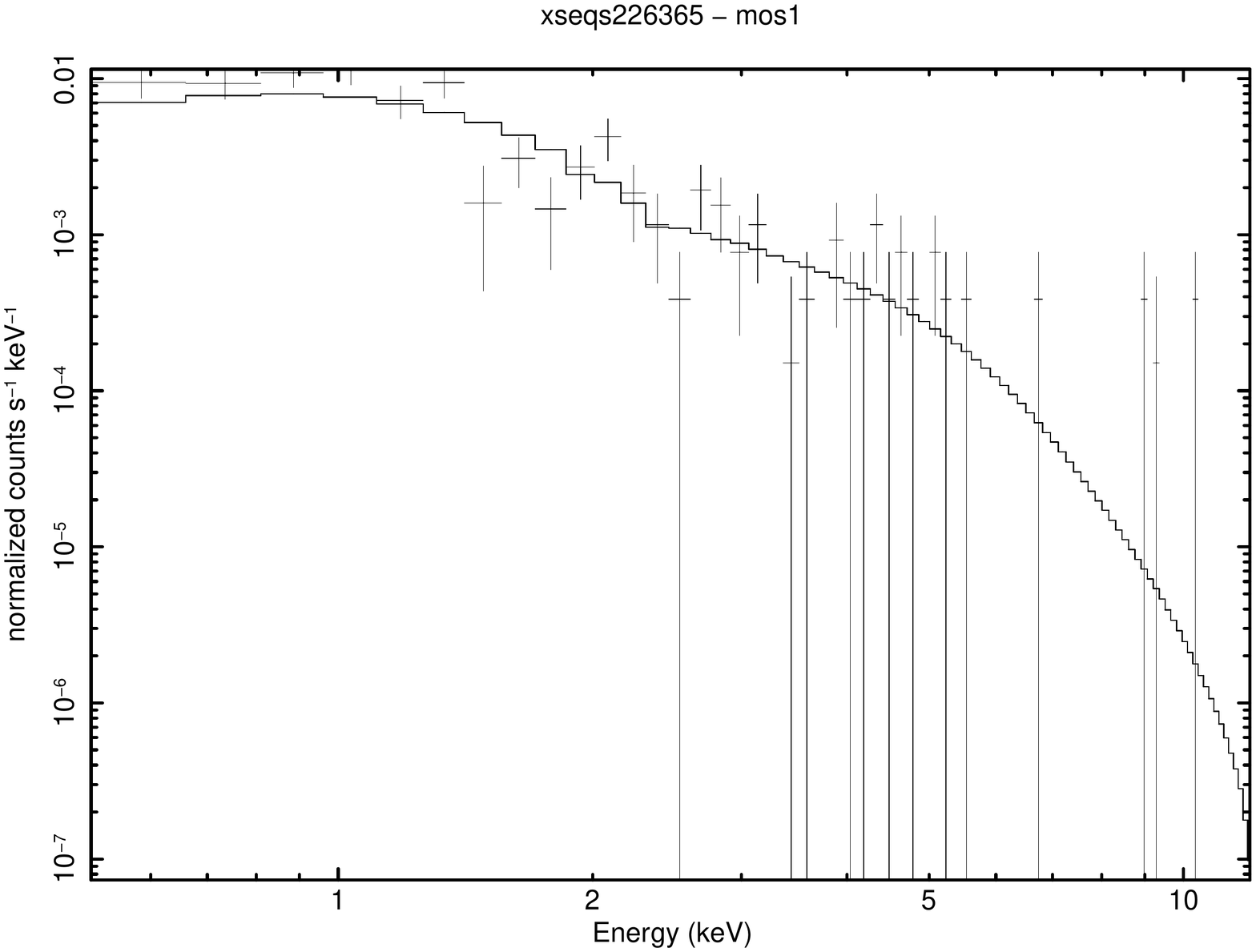} &
  \includegraphics[width=0.25\hsize]{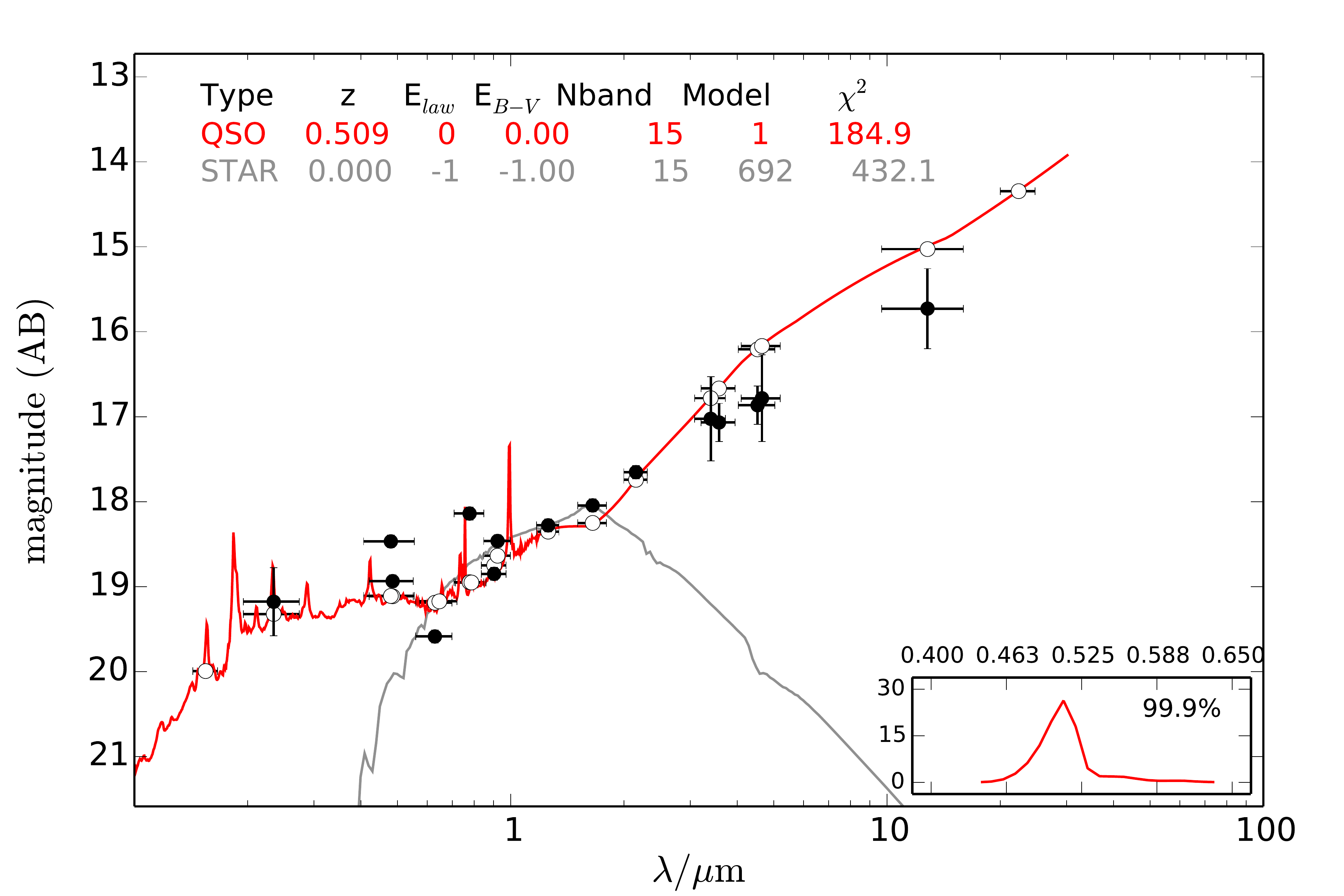} &
  \includegraphics[width=0.17\hsize]{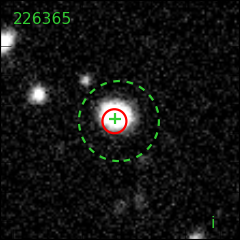} &
  \includegraphics[width=0.17\hsize]{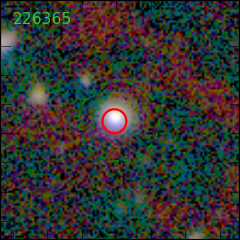}\\

\multicolumn{4}{l}{3XLSS J023304.3-053946 1} \\
  \includegraphics[width=0.24\hsize]{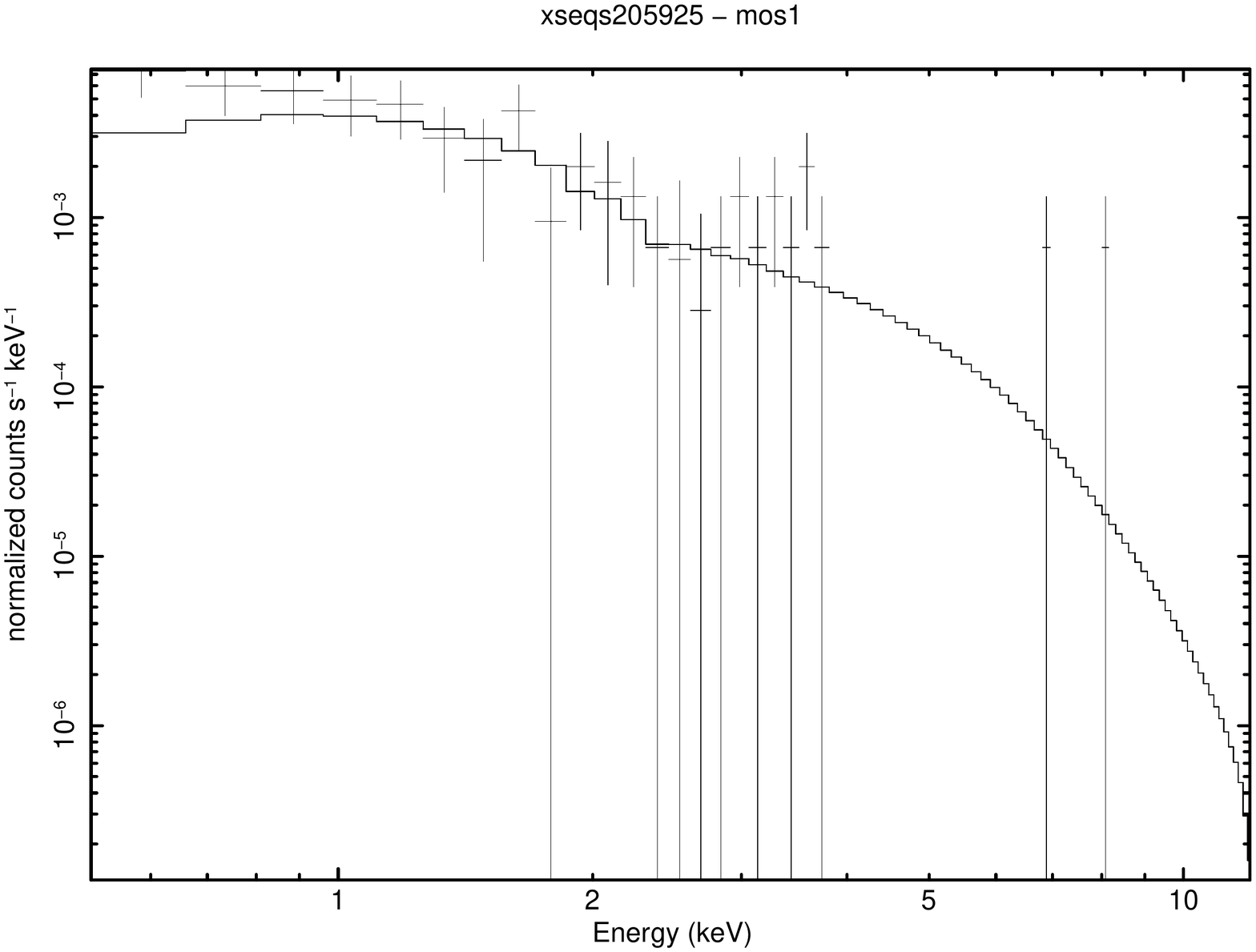} &
  \includegraphics[width=0.25\hsize]{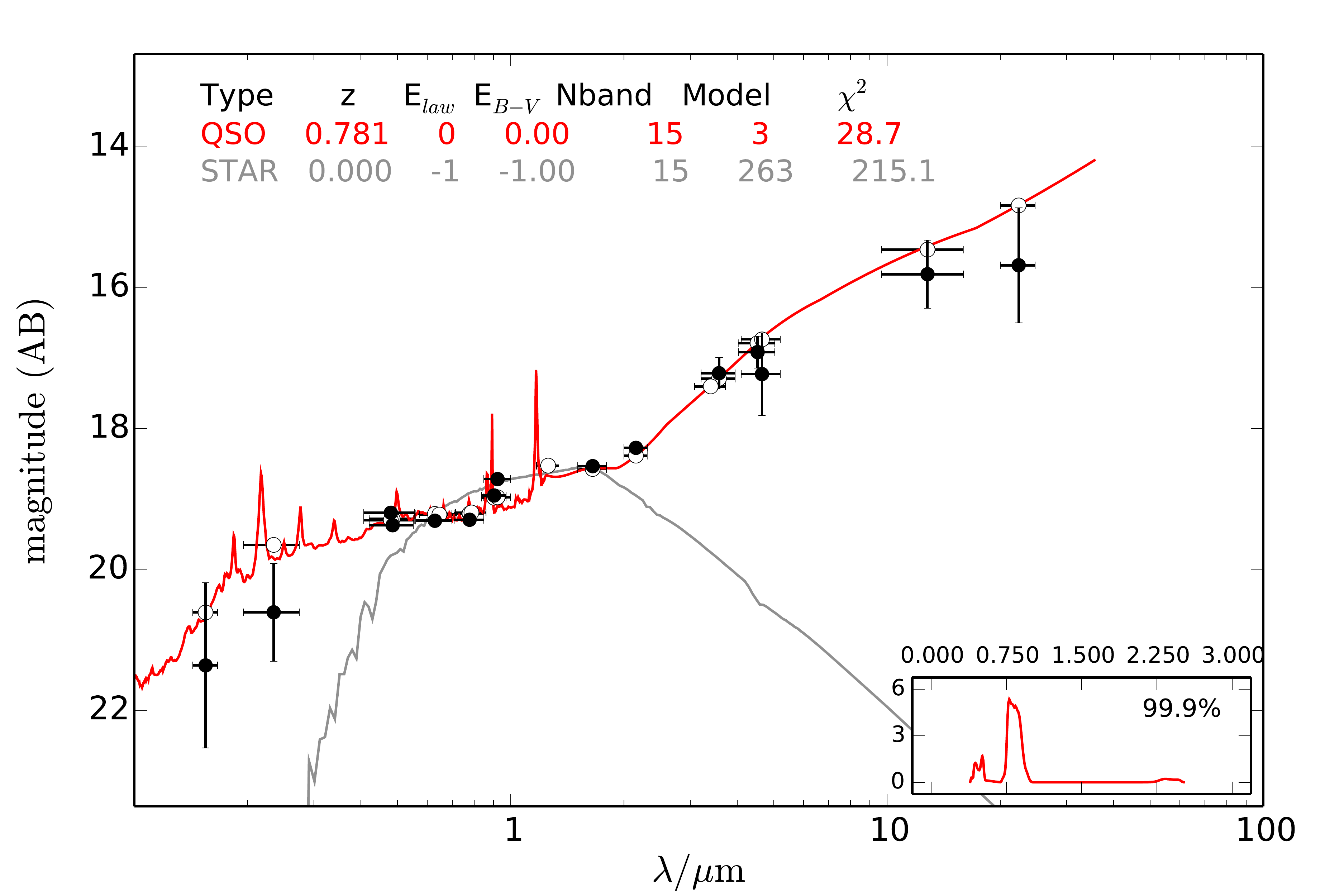} &
  \includegraphics[width=0.17\hsize]{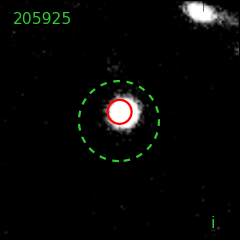}&
  \includegraphics[width=0.17\hsize]{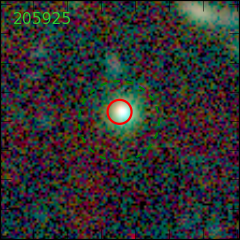}\\

\end{tabular}
\end{adjustbox}
\end{tabular}
\caption{Sources classified as QSO. The panels from left to right are: X-ray spectrum, SED from far-ultraviolet to mid-infrared with best fit photo-z solution (red line) and star solution (grey line). The inset is the 99.9\% PDF of the photo-z solution. Single-filter image, three-colour composite.\label{app:QSO}}
\end{figure*}
\end{center}

\begin{center}
\begin{figure*}
\centering
\begin{tabular}{cc}
\begin{adjustbox}{valign=t}
\begin{tabular}{cccc}
\multicolumn{4}{l}{3XLSS J021835.7-053758 } \\
  \includegraphics[width=0.24\hsize]{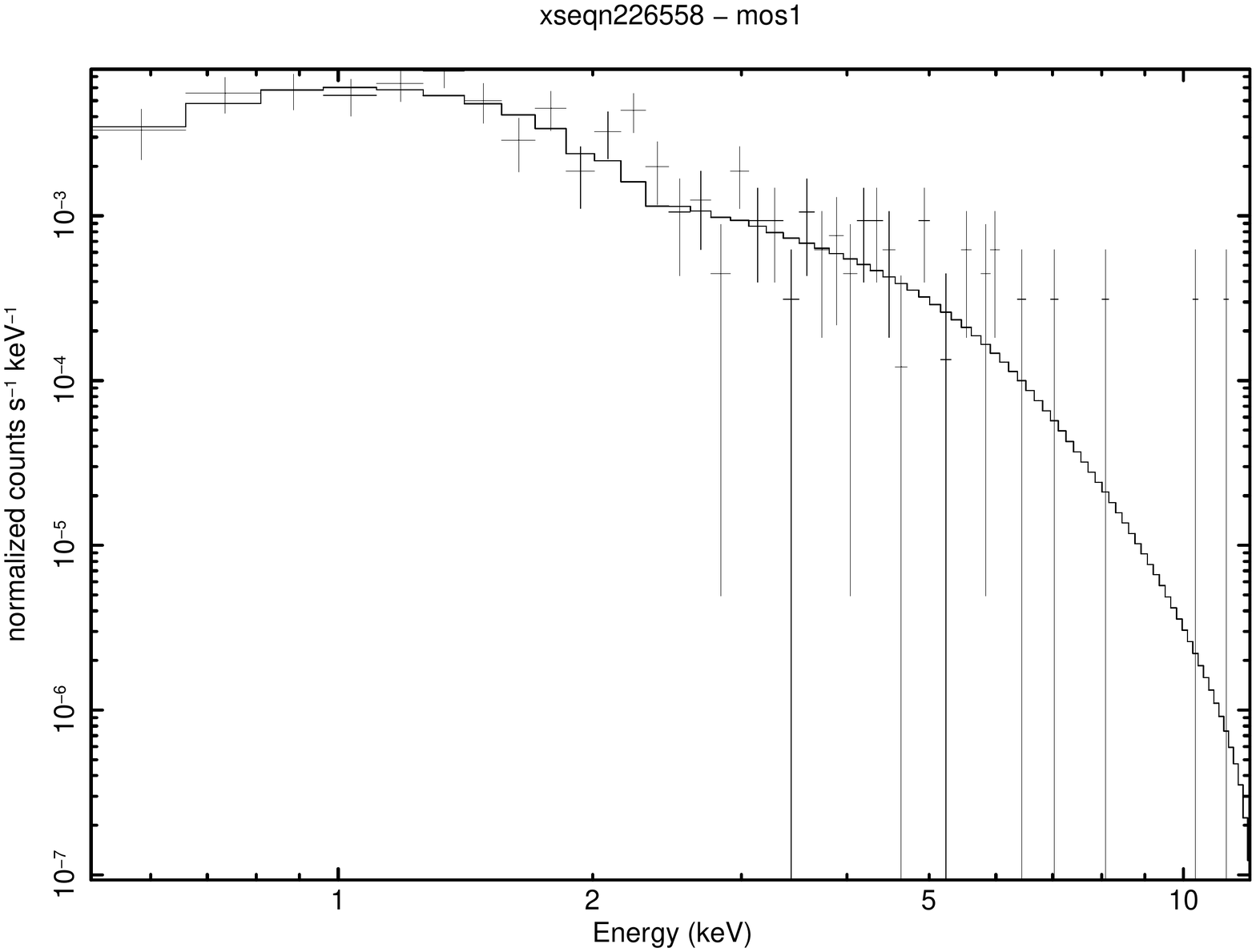} & 
  \includegraphics[width=0.25\hsize]{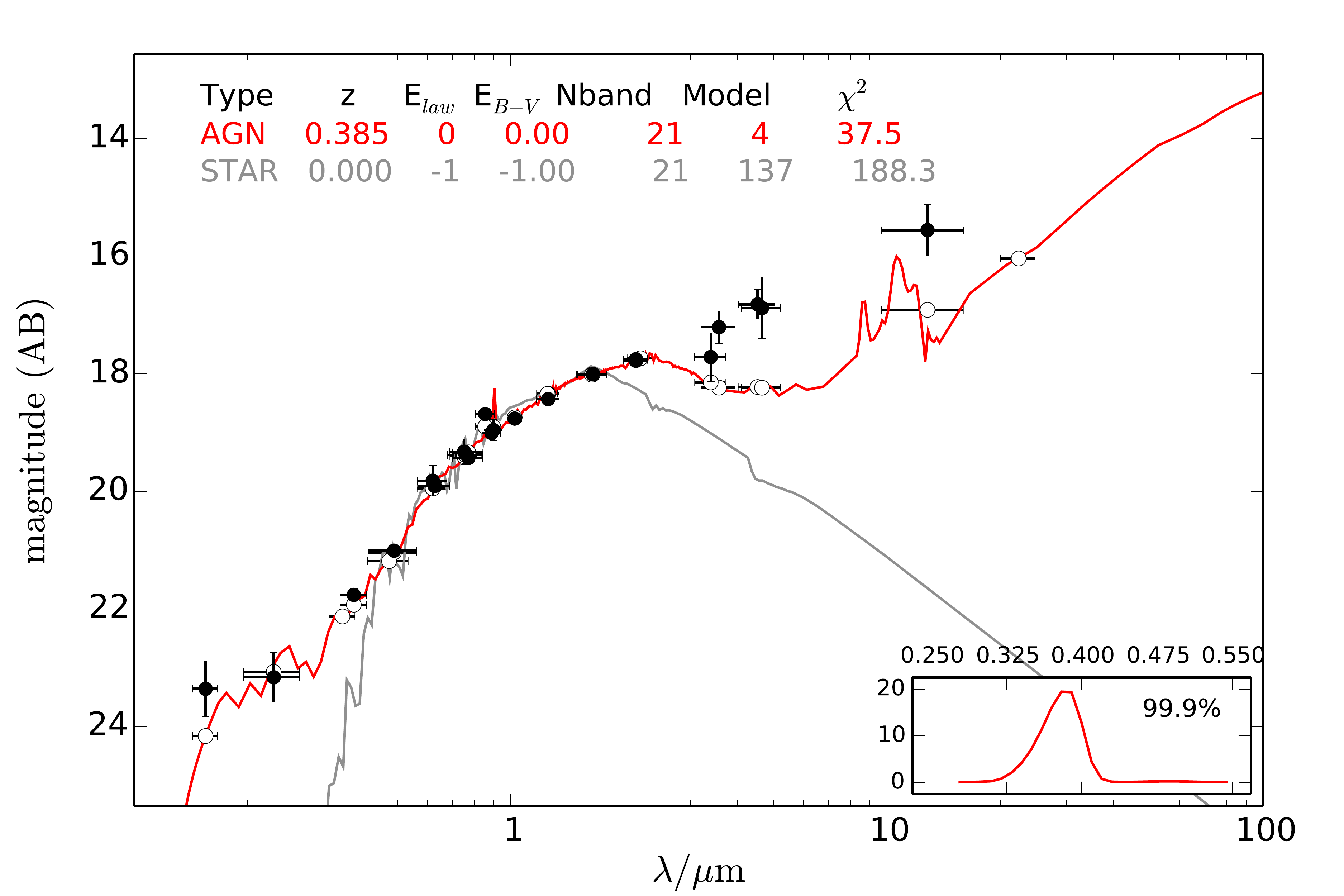} &
  \includegraphics[width=0.17\hsize]{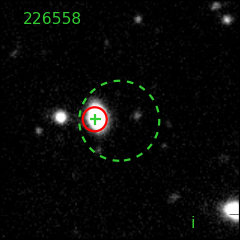} &
  \includegraphics[width=0.17\hsize]{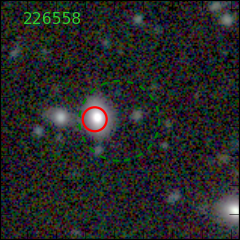} \\

\multicolumn{4}{l}{3XLSS J022845.3-050235 } \\
  \includegraphics[width=0.24\hsize]{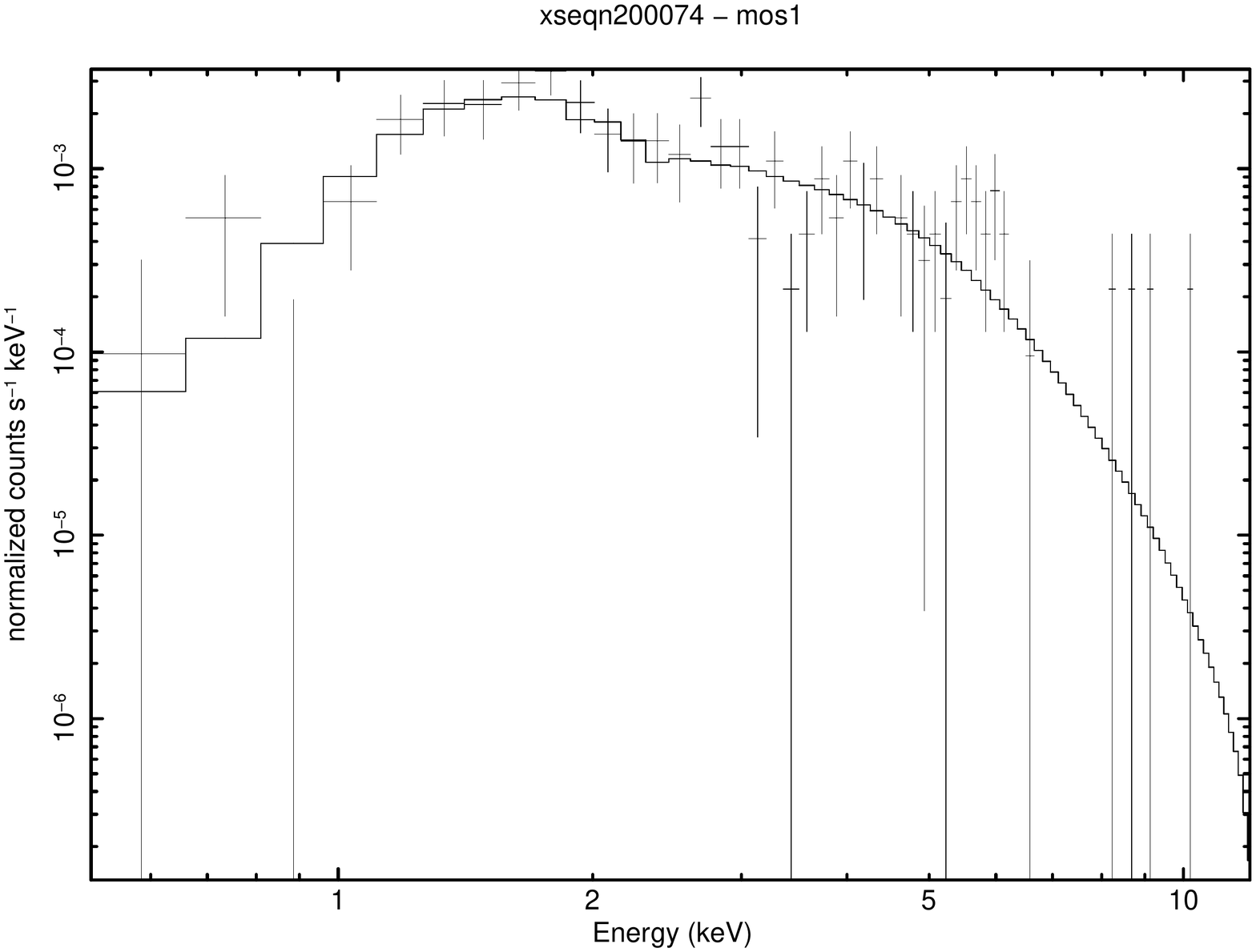} & 
  \includegraphics[width=0.25\hsize]{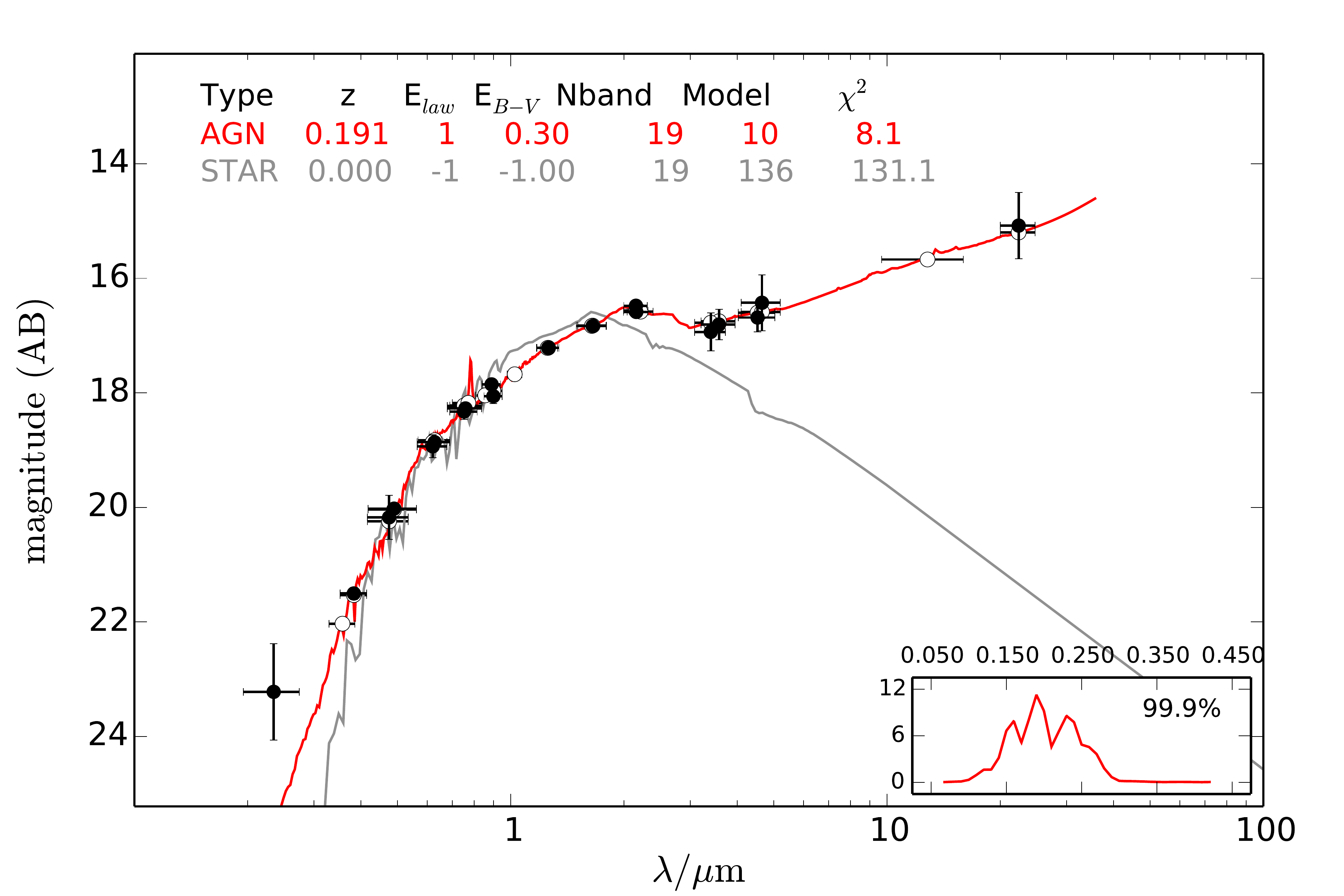} &
  \includegraphics[width=0.17\hsize]{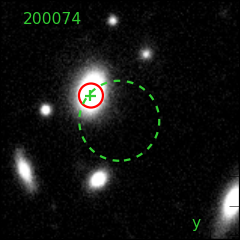} &
  \includegraphics[width=0.17\hsize]{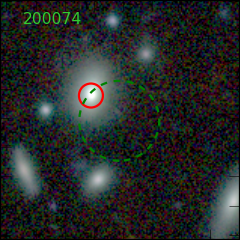} \\

\multicolumn{4}{l}{3XLSS J023040.6-040752 } \\
  \includegraphics[width=0.24\hsize]{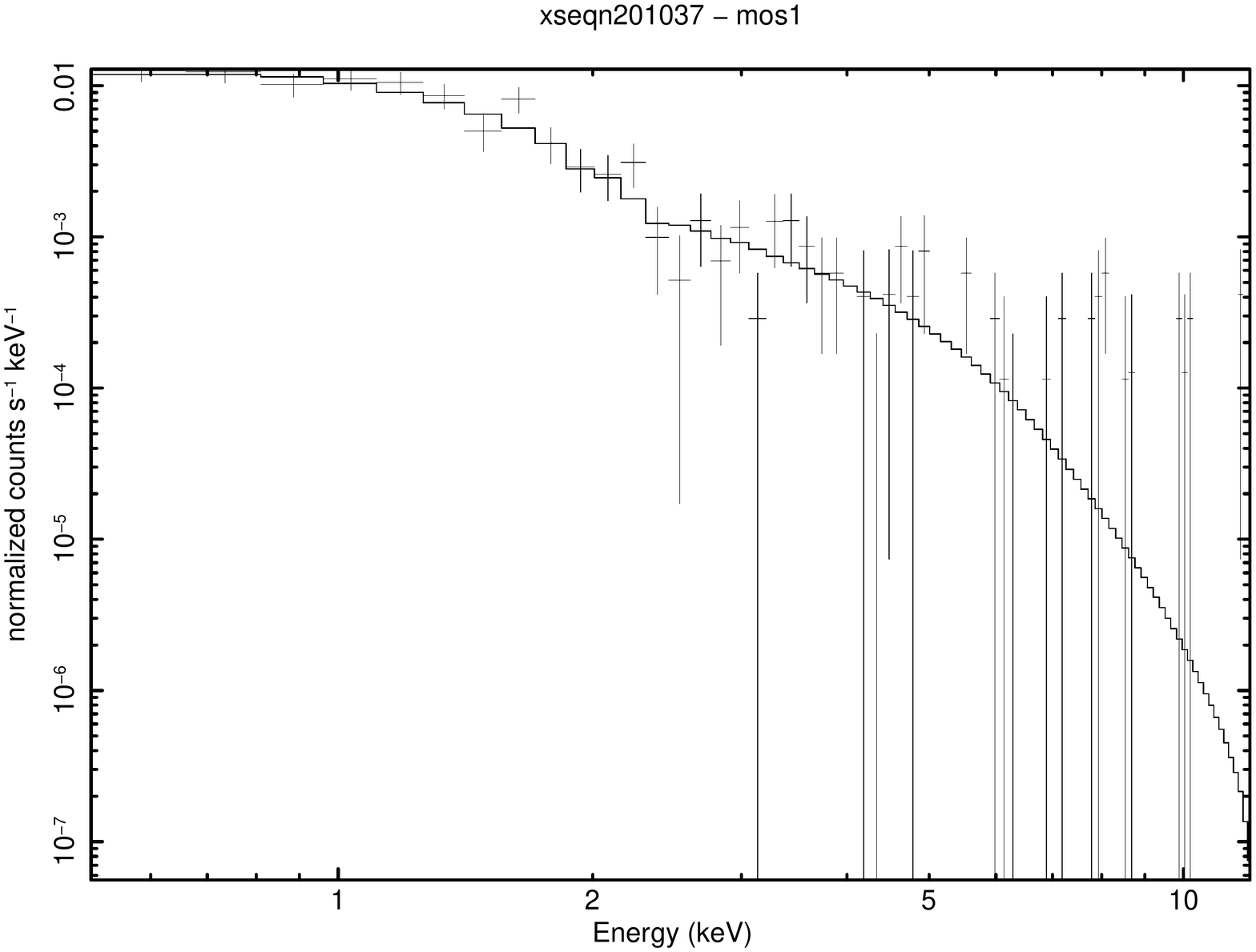} & 
  \includegraphics[width=0.25\hsize]{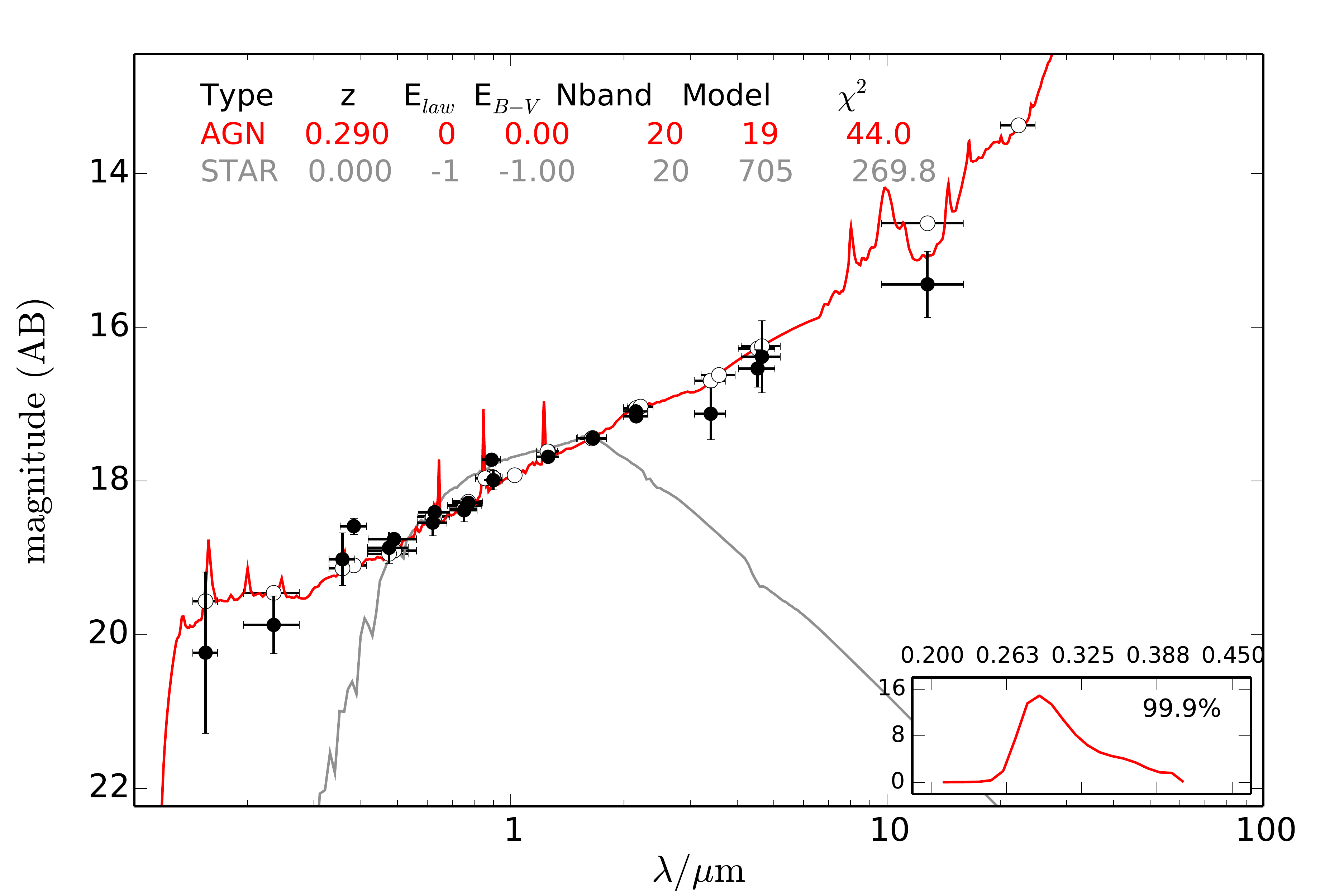} &
  \includegraphics[width=0.17\hsize]{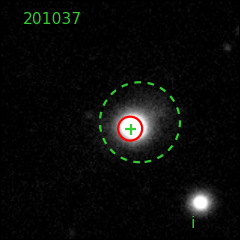} &
  \includegraphics[width=0.17\hsize]{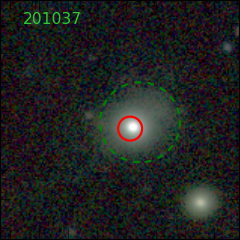} \\
  
\multicolumn{4}{l}{3XLSS J020106.4-064859 } \\
  \includegraphics[width=0.24\hsize]{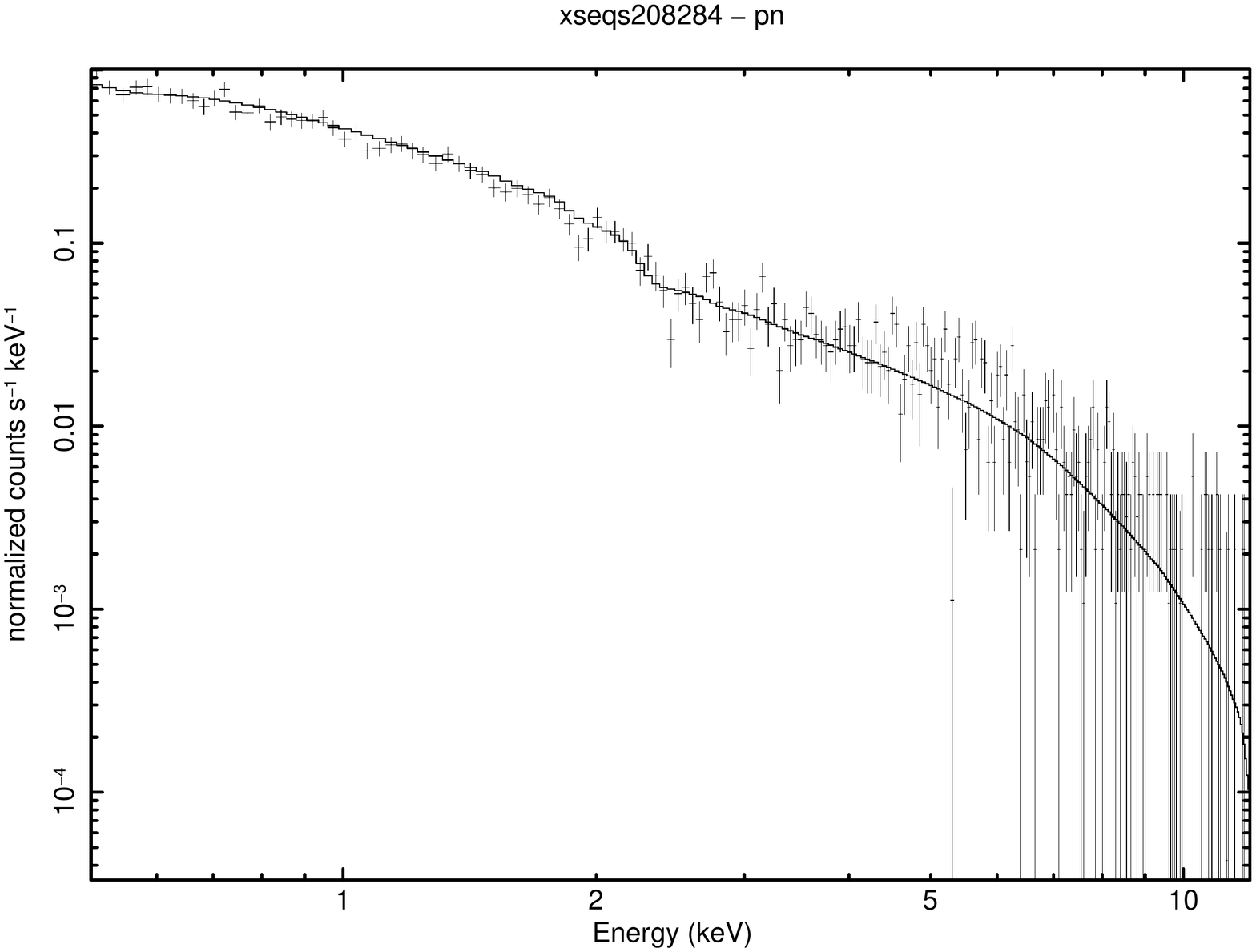} &
  \includegraphics[width=0.25\hsize]{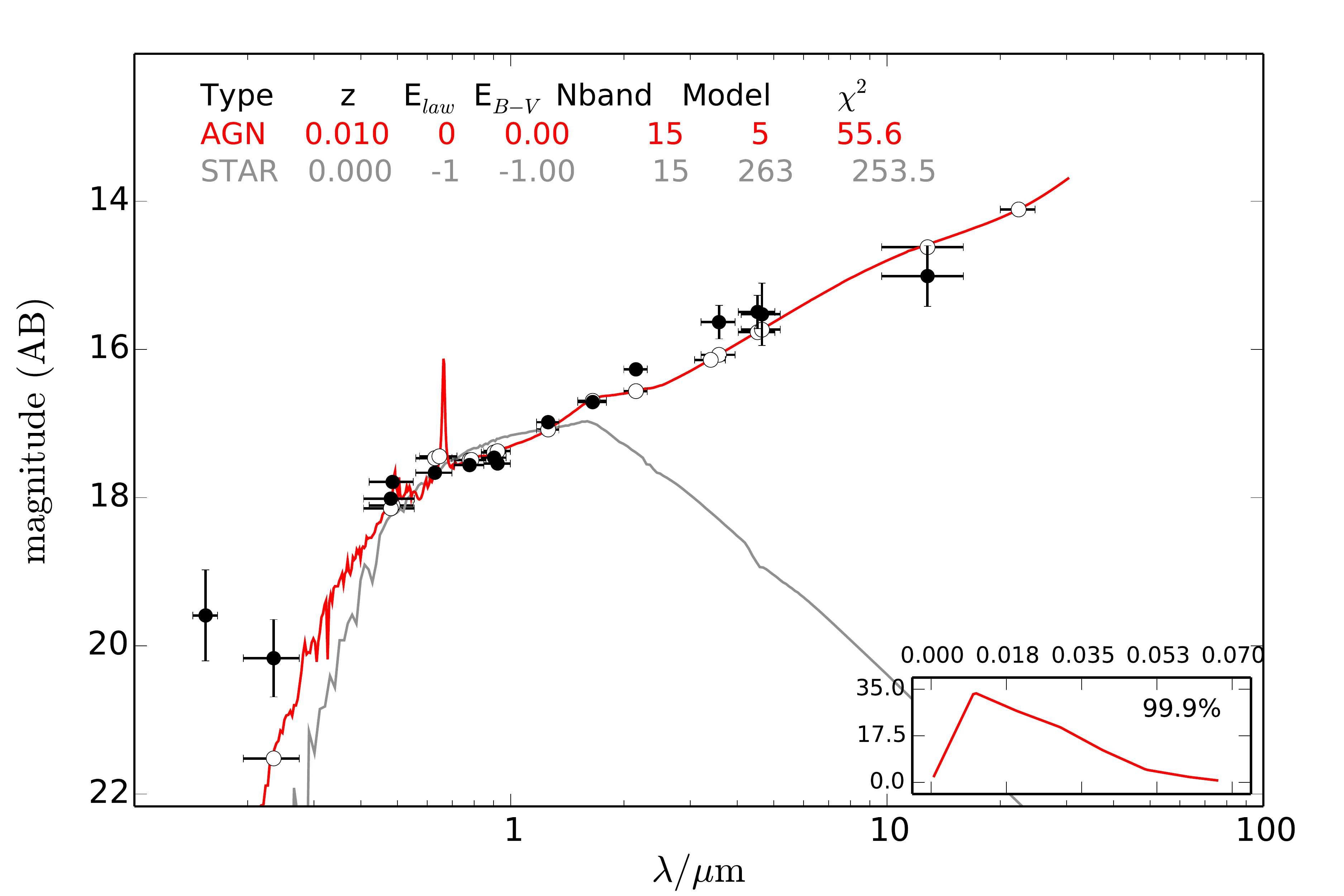} &
  \includegraphics[width=0.17\hsize]{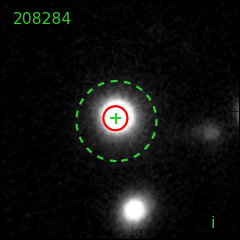}&
  \includegraphics[width=0.17\hsize]{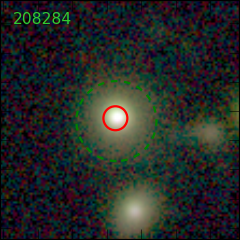}\\

\multicolumn{4}{l}{3XLSS J021657.6-032500 } \\
  \includegraphics[width=0.24\hsize]{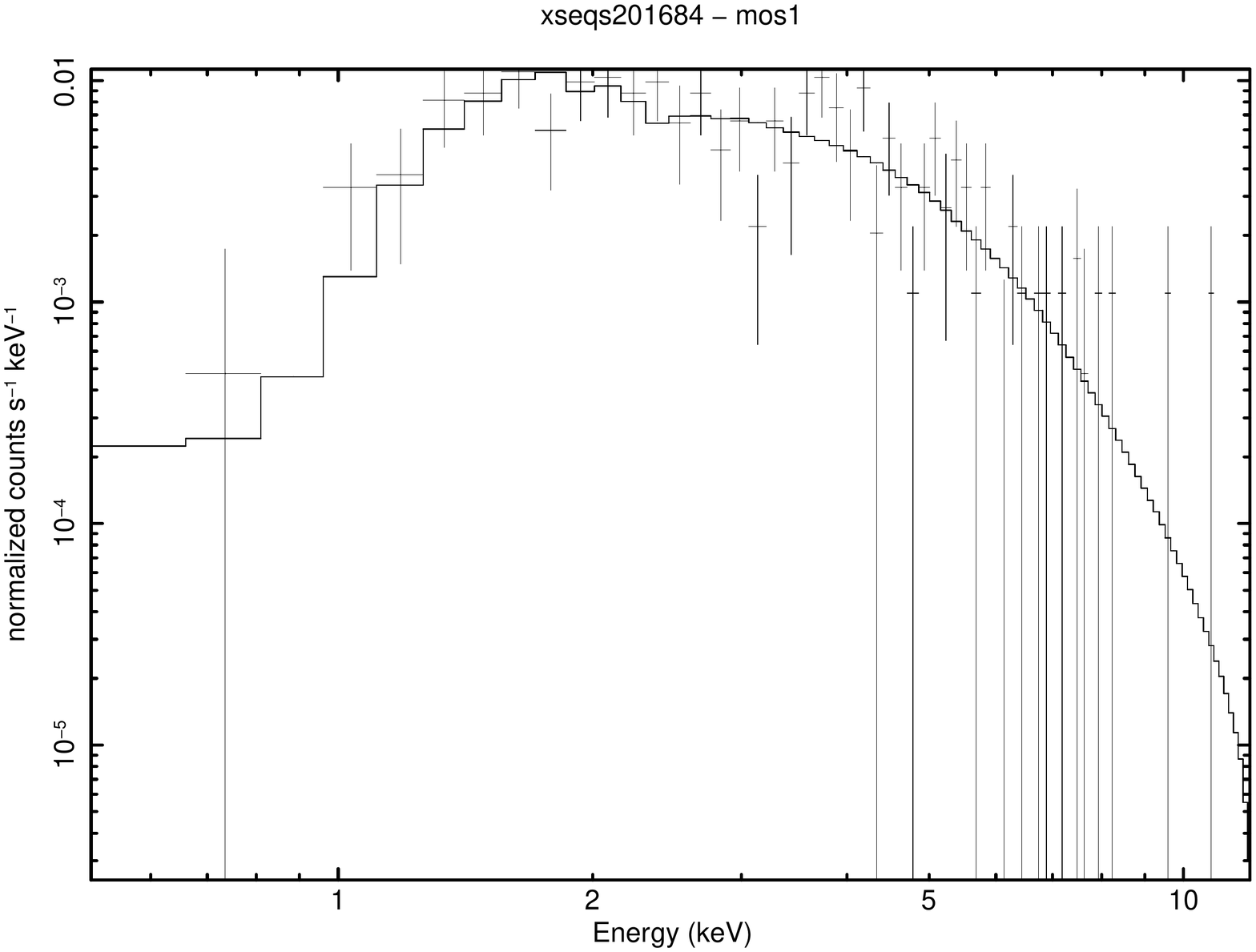} &
  \includegraphics[width=0.25\hsize]{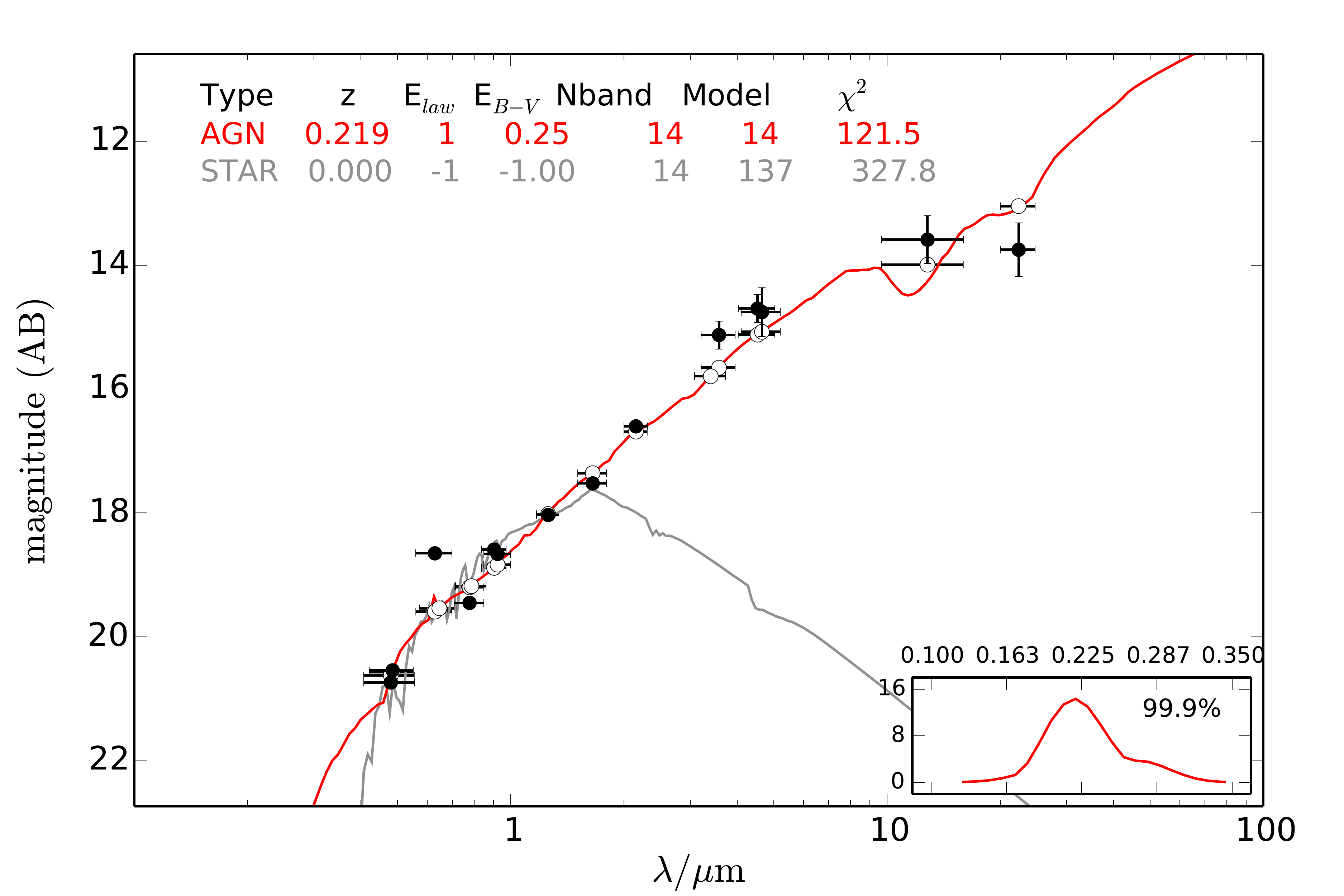} &
  \includegraphics[width=0.17\hsize]{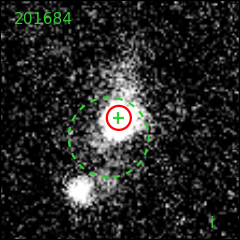}&
  \includegraphics[width=0.17\hsize]{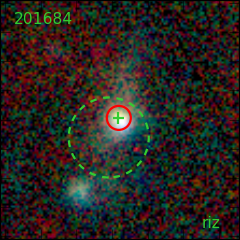}\\

\multicolumn{4}{l}{3XLSS J022935.1-055208 } \\
  \includegraphics[width=0.24\hsize]{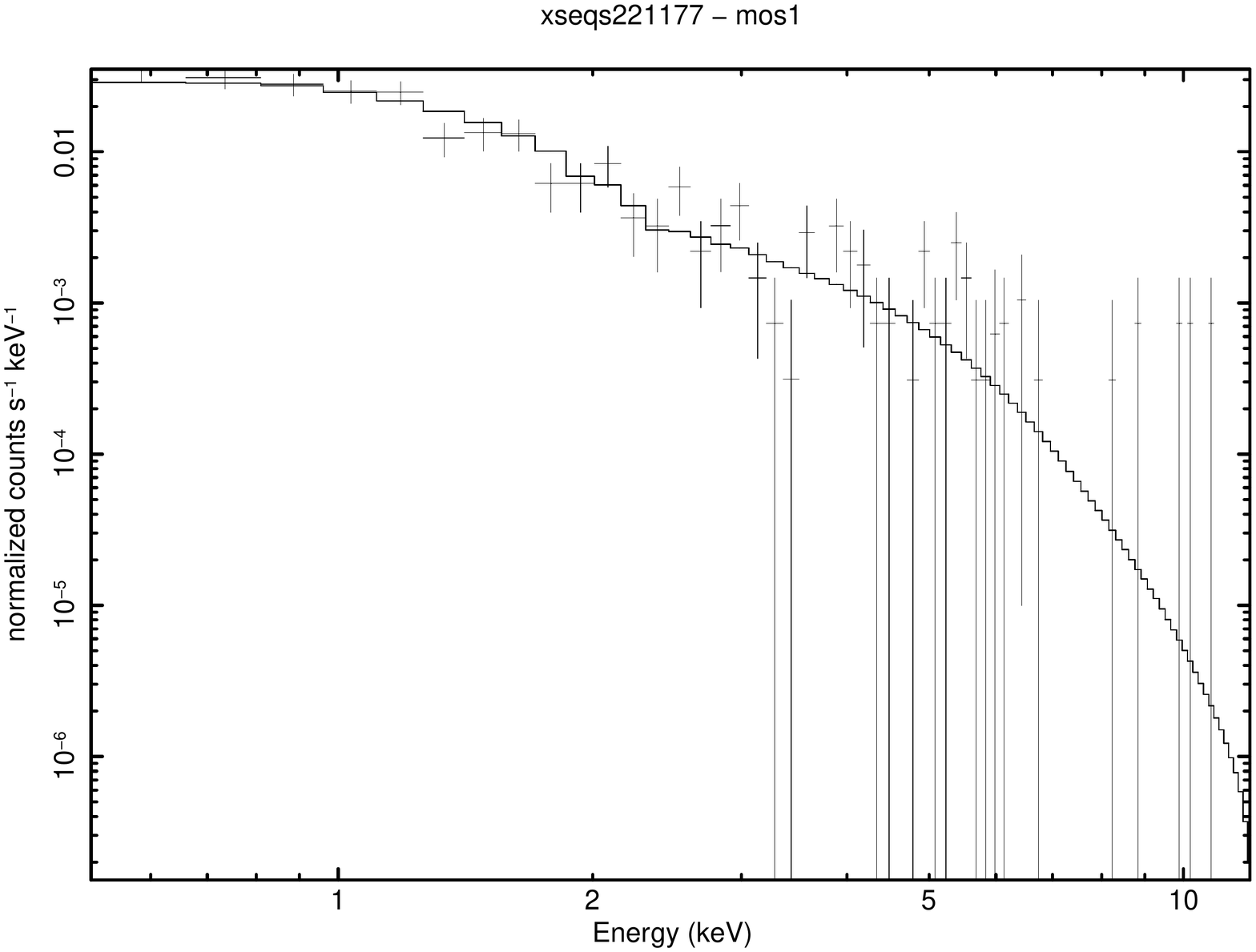} &
  \includegraphics[width=0.25\hsize]{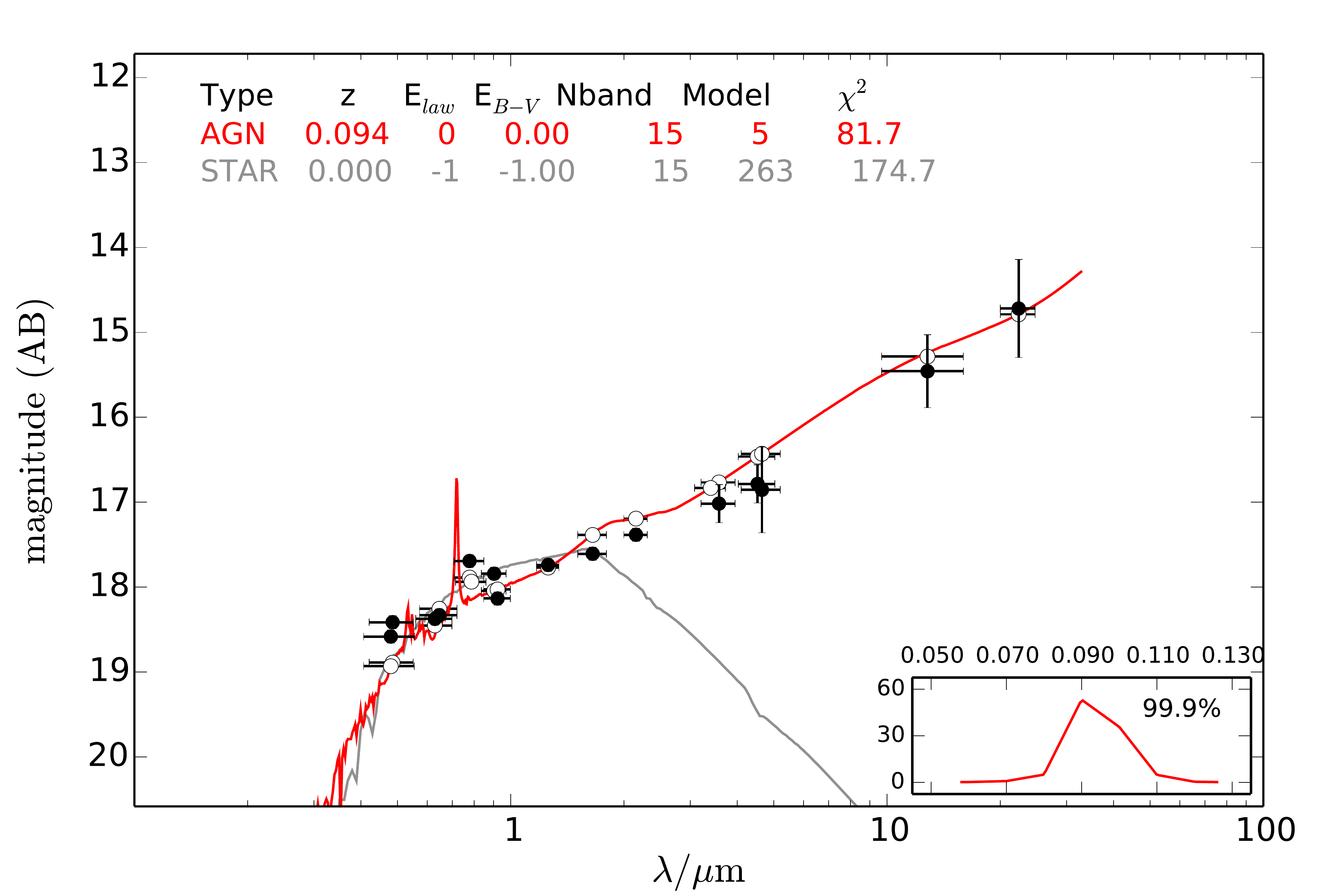} &
  \includegraphics[width=0.17\hsize]{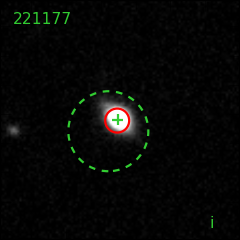}&
  \includegraphics[width=0.17\hsize]{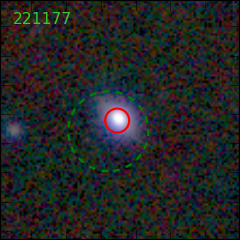}\\

\end{tabular}
\end{adjustbox}
\end{tabular}
\caption{Sources classified as AGN. Panels as in Fig. \ref{app:QSO}}
\end{figure*}
\end{center}

\begin{center}
\begin{figure*}
\centering
\begin{tabular}{cc}
\begin{adjustbox}{valign=t}
\begin{tabular}{cccc}
\multicolumn{4}{l}{3XLSS J020207.0-055858 2} \\
  \includegraphics[width=0.24\hsize]{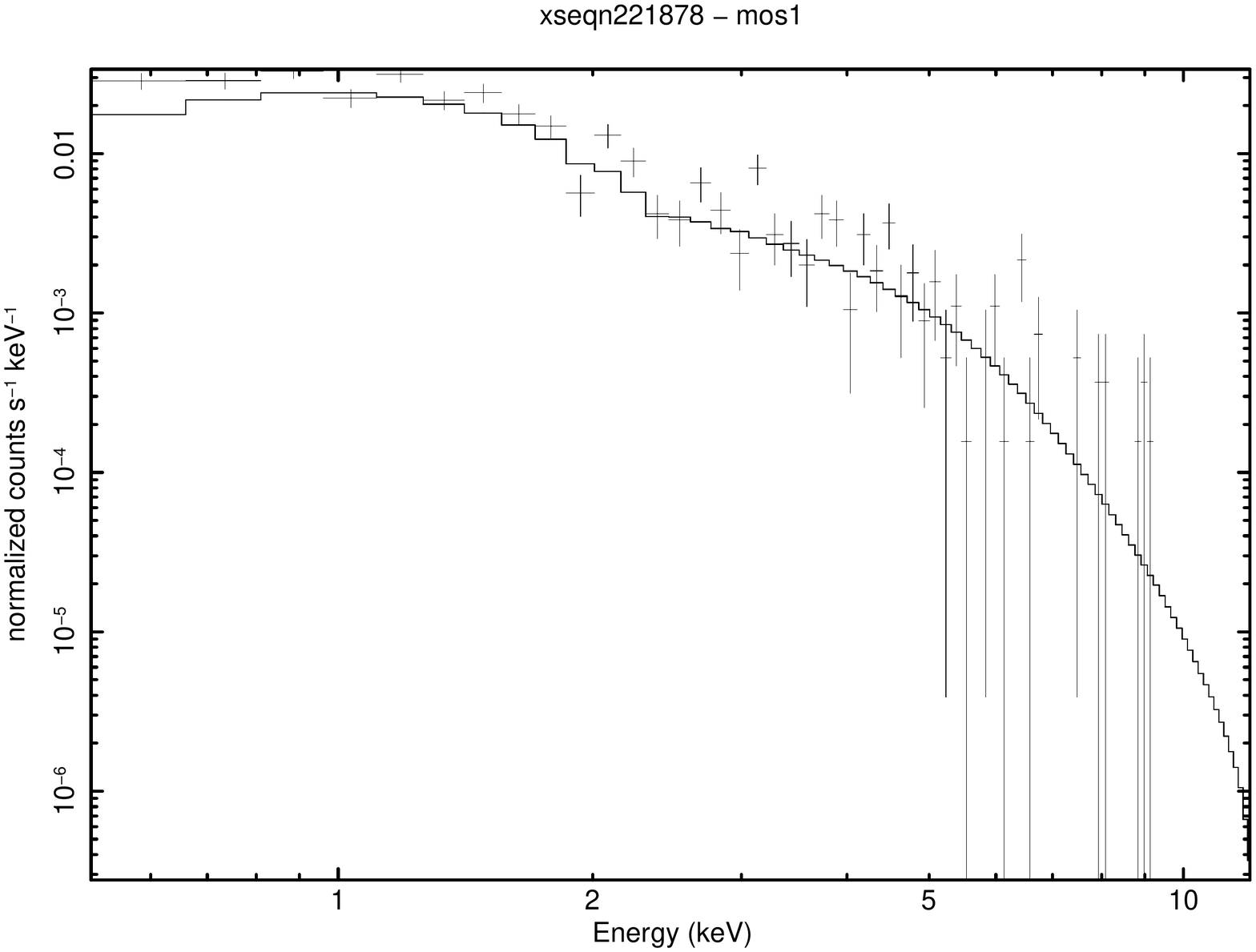} & 
  \includegraphics[width=0.25\hsize]{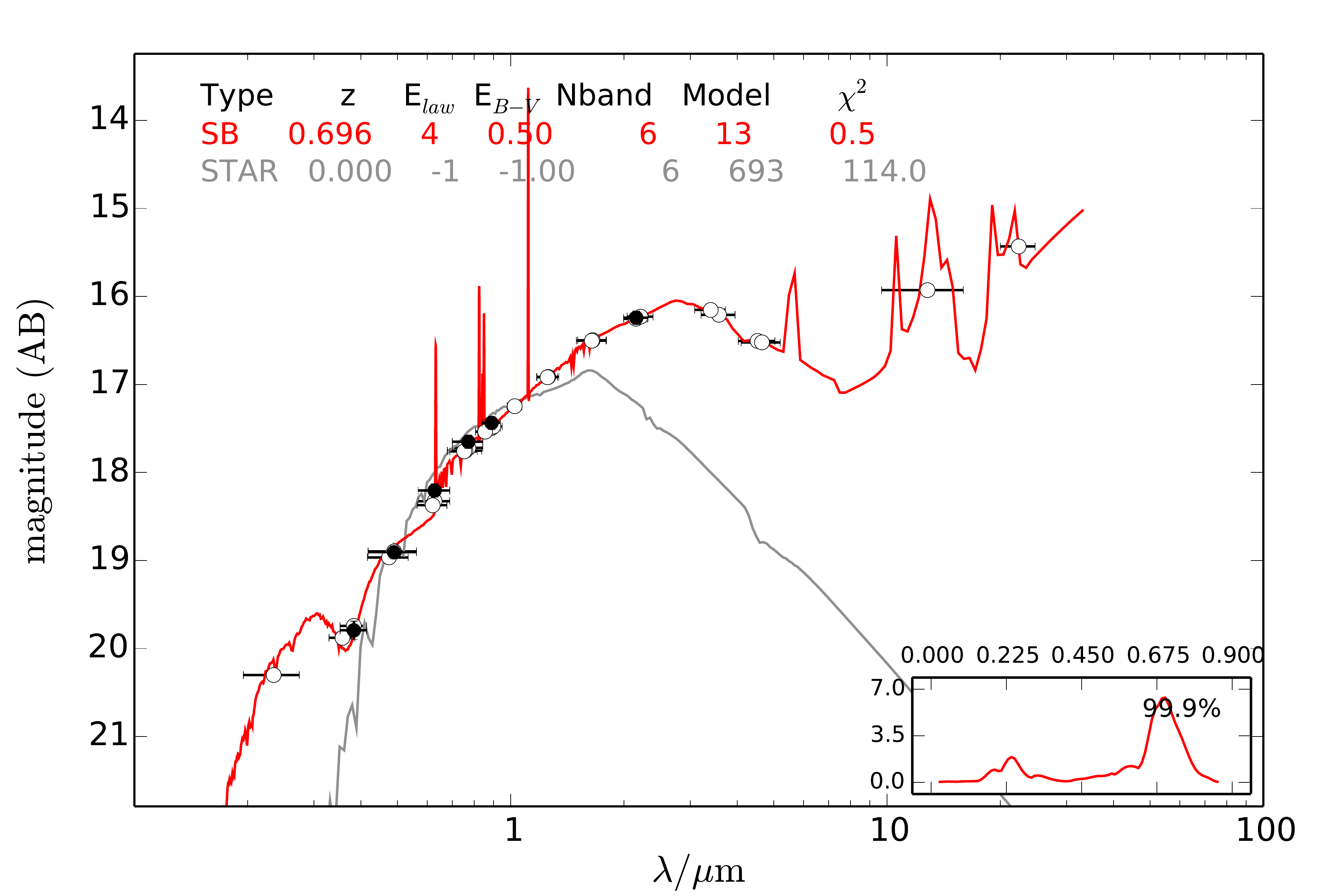} &
  \includegraphics[width=0.17\hsize]{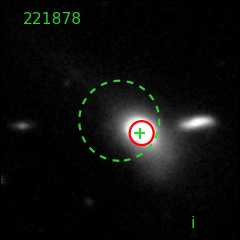} &
  \includegraphics[width=0.17\hsize]{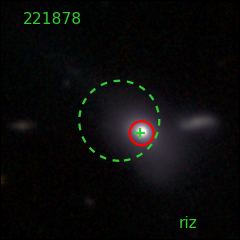} \\

\multicolumn{4}{l}{3XLSS J021606.0-051720 } \\
  \includegraphics[width=0.24\hsize]{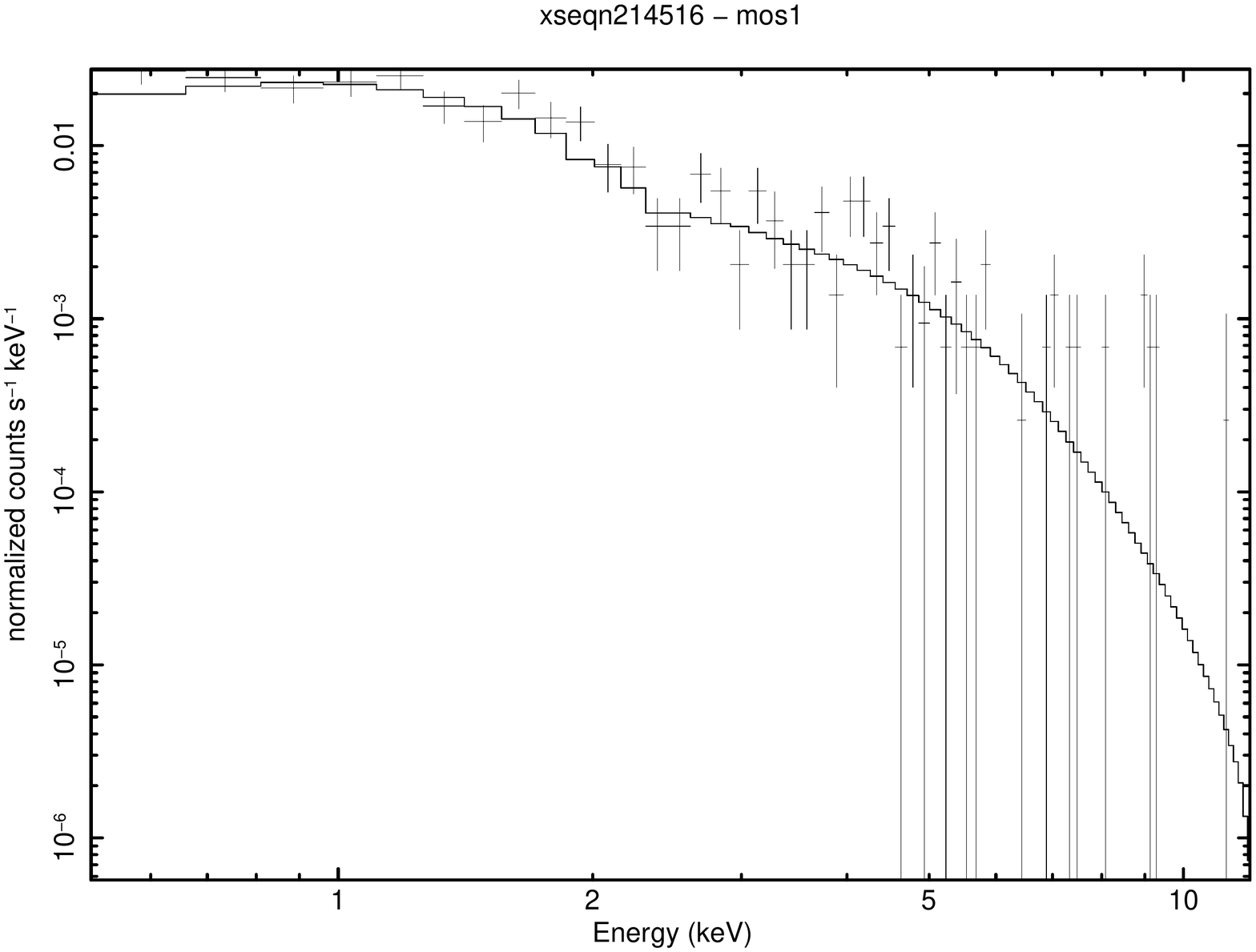} & 
  \includegraphics[width=0.25\hsize]{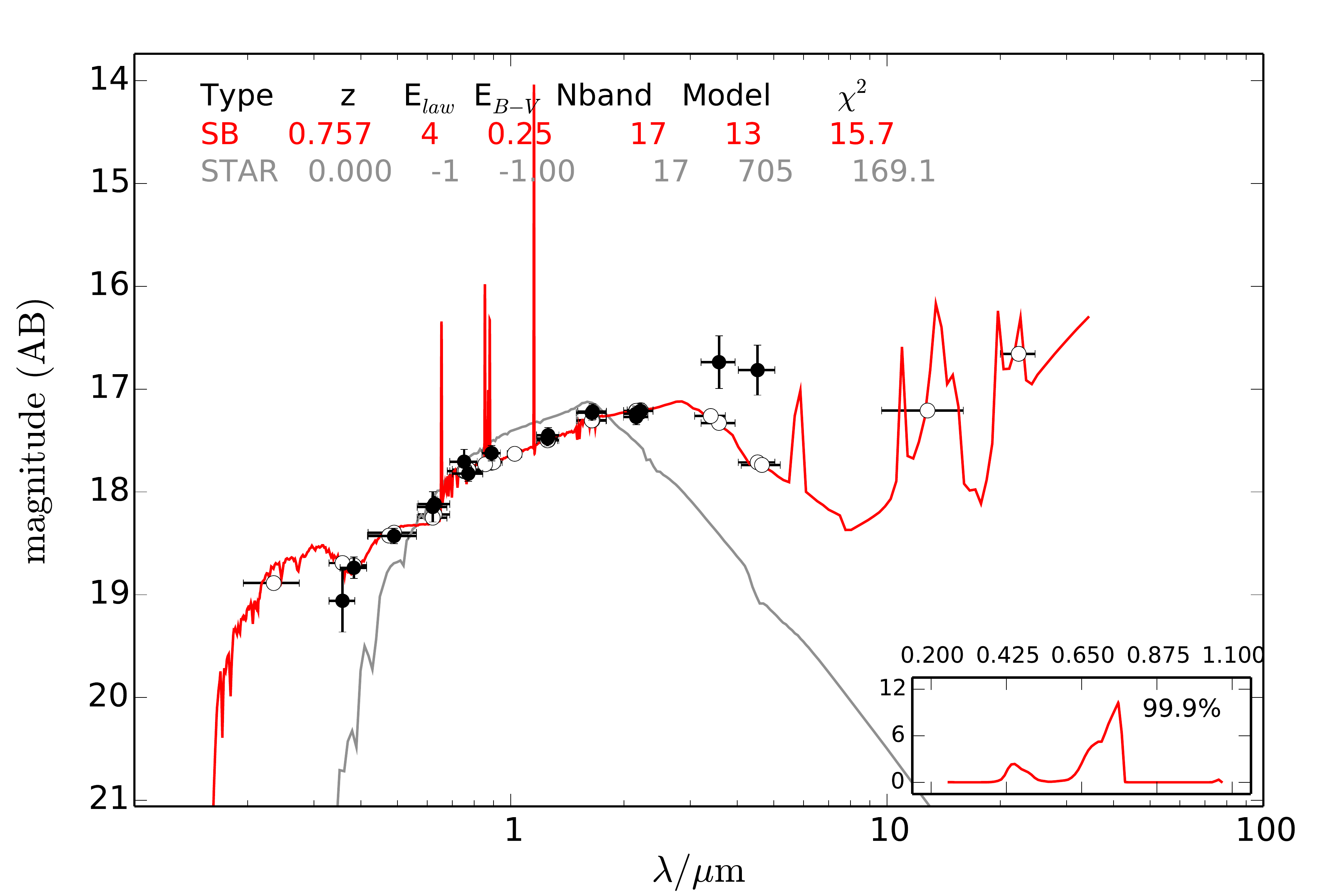} &
  \includegraphics[width=0.17\hsize]{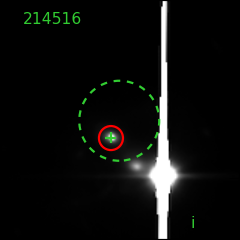} &
  \includegraphics[width=0.17\hsize]{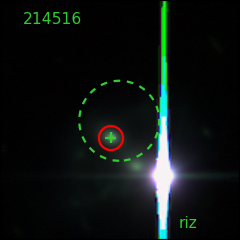} \\

\multicolumn{4}{l}{3XLSS J020534.0-073705 } \\
  \includegraphics[width=0.24\hsize]{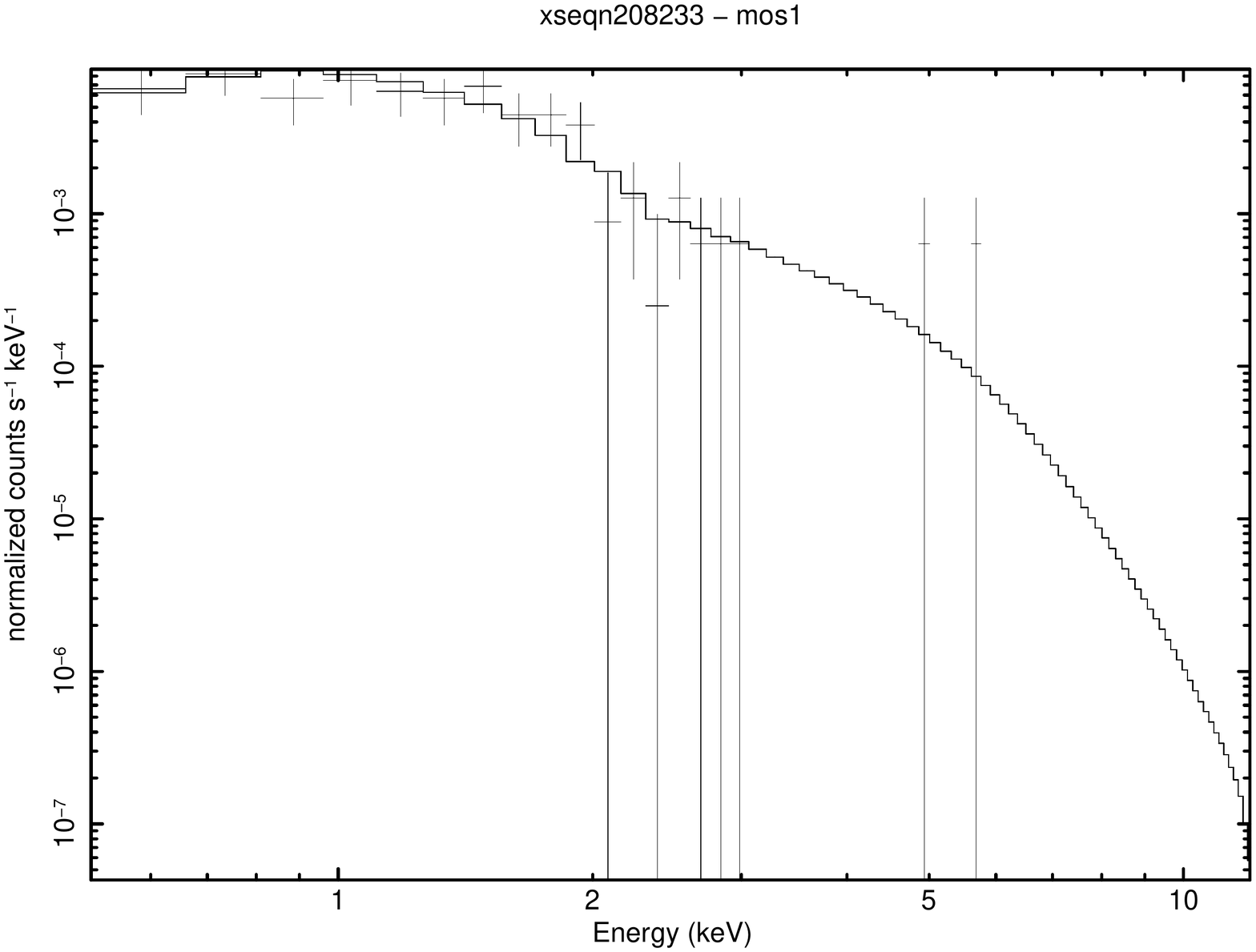} & 
  \includegraphics[width=0.25\hsize]{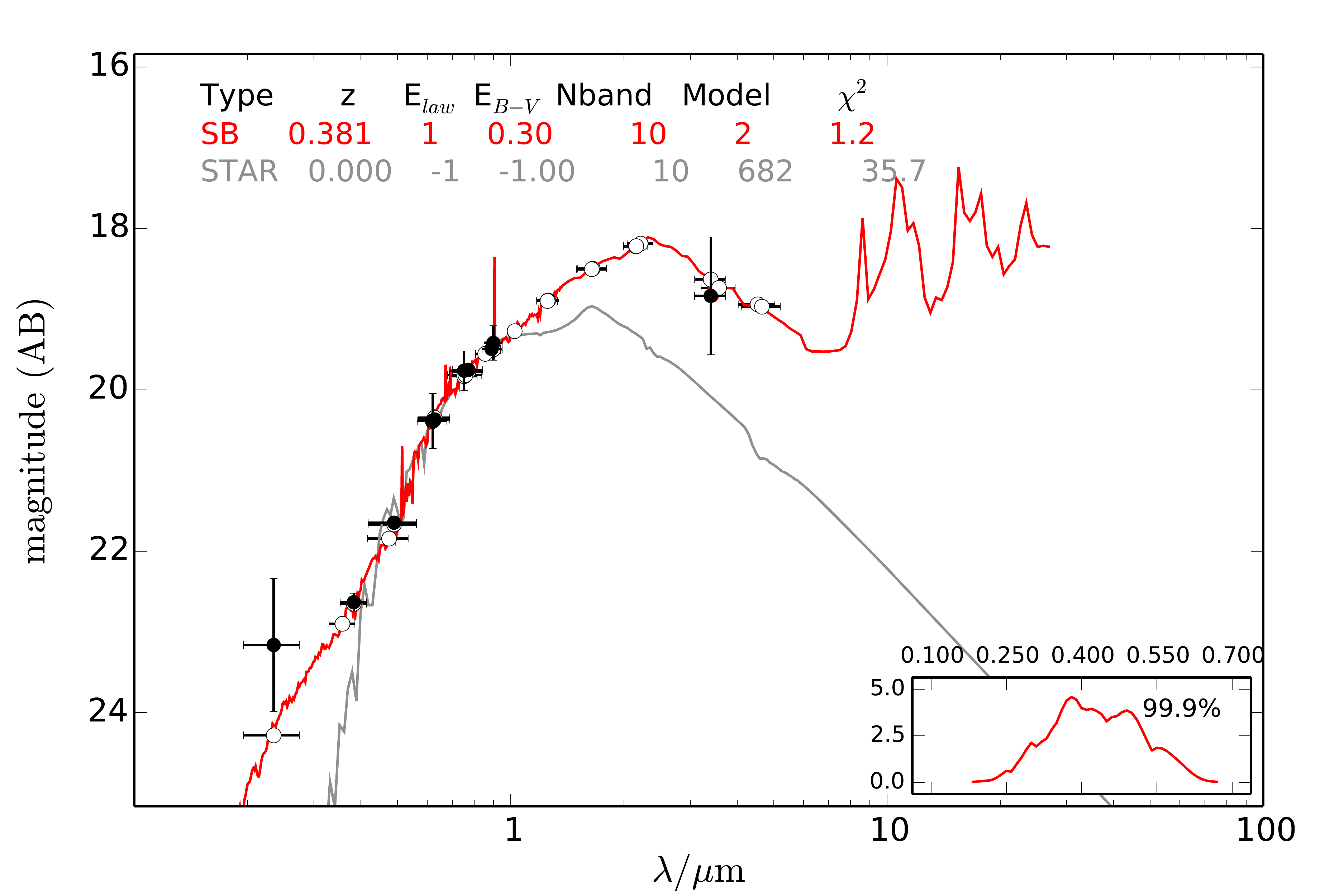} &
  \includegraphics[width=0.17\hsize]{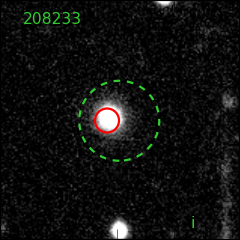} &
  \includegraphics[width=0.17\hsize]{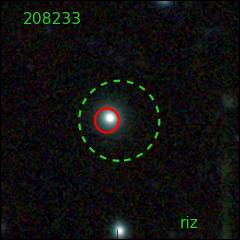} \\
  
\multicolumn{4}{l}{3XLSS J020559.6-063736  } \\
  \includegraphics[width=0.24\hsize]{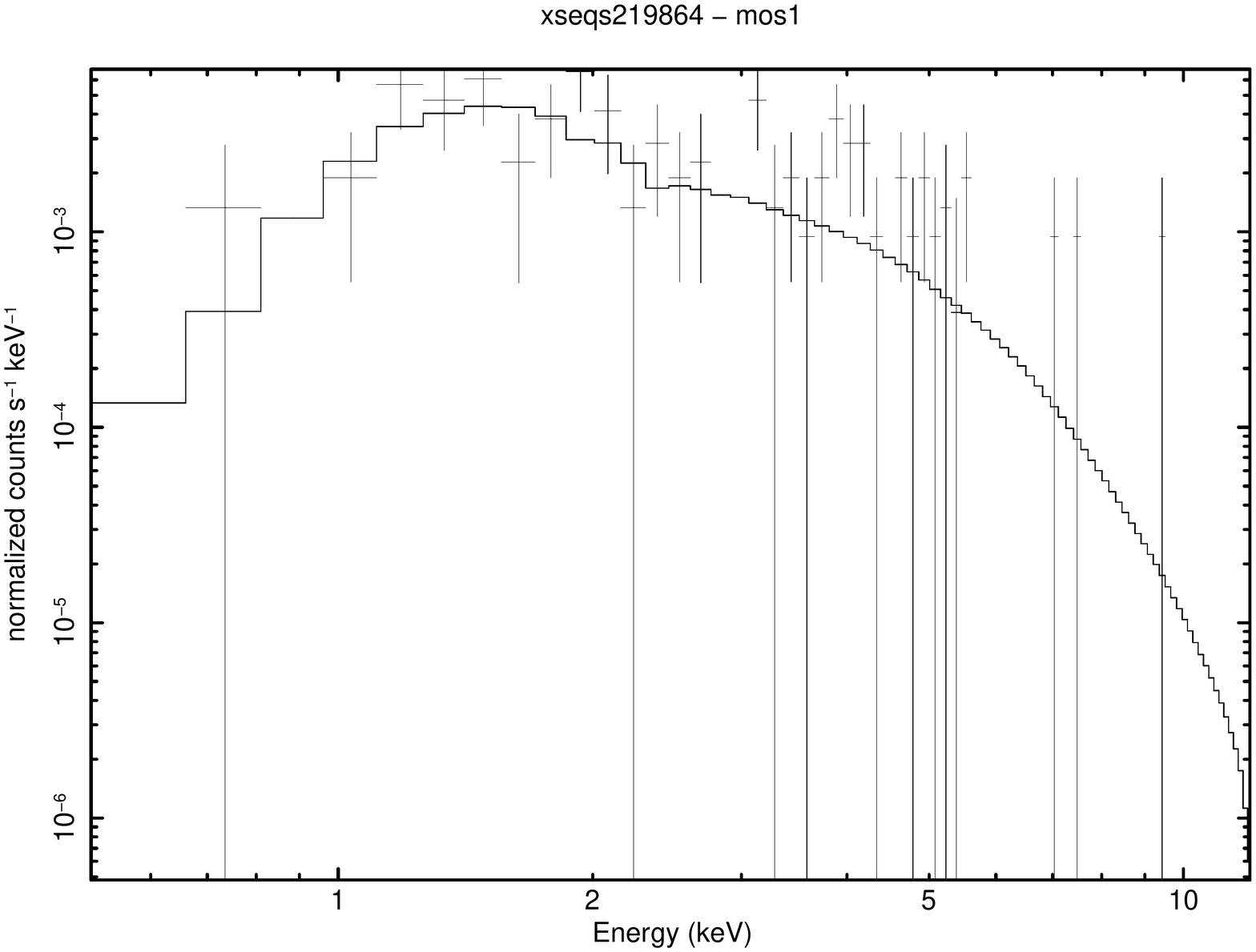} &
  \includegraphics[width=0.25\hsize]{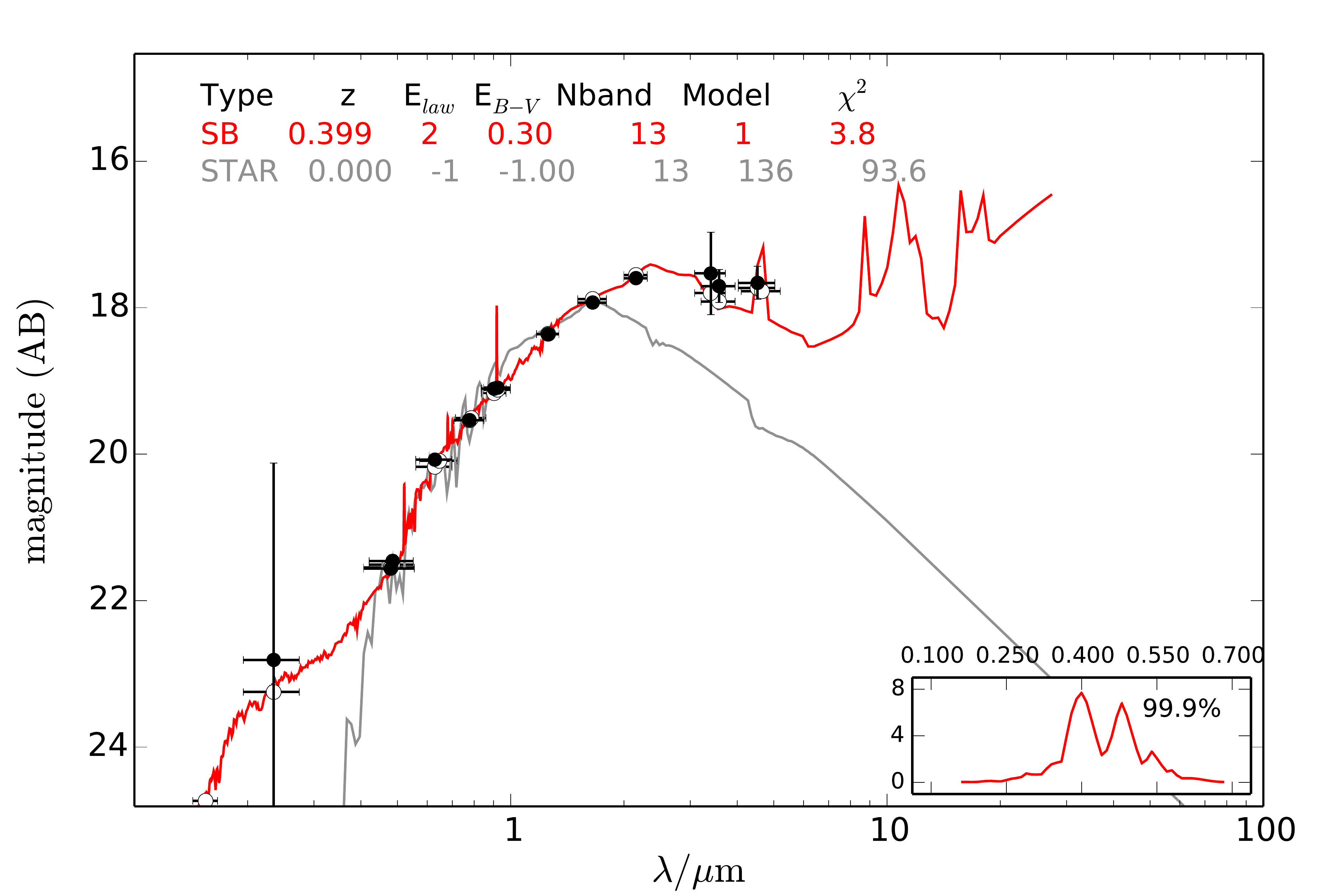} &
  \includegraphics[width=0.17\hsize]{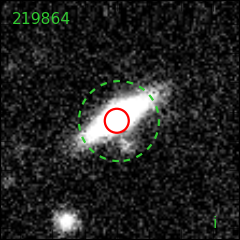}&
  \includegraphics[width=0.17\hsize]{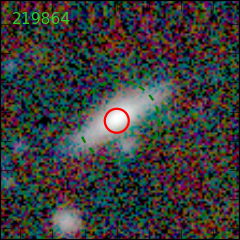}\\

\multicolumn{4}{l}{3XLSS J022541.3-032624 } \\
  \includegraphics[width=0.24\hsize]{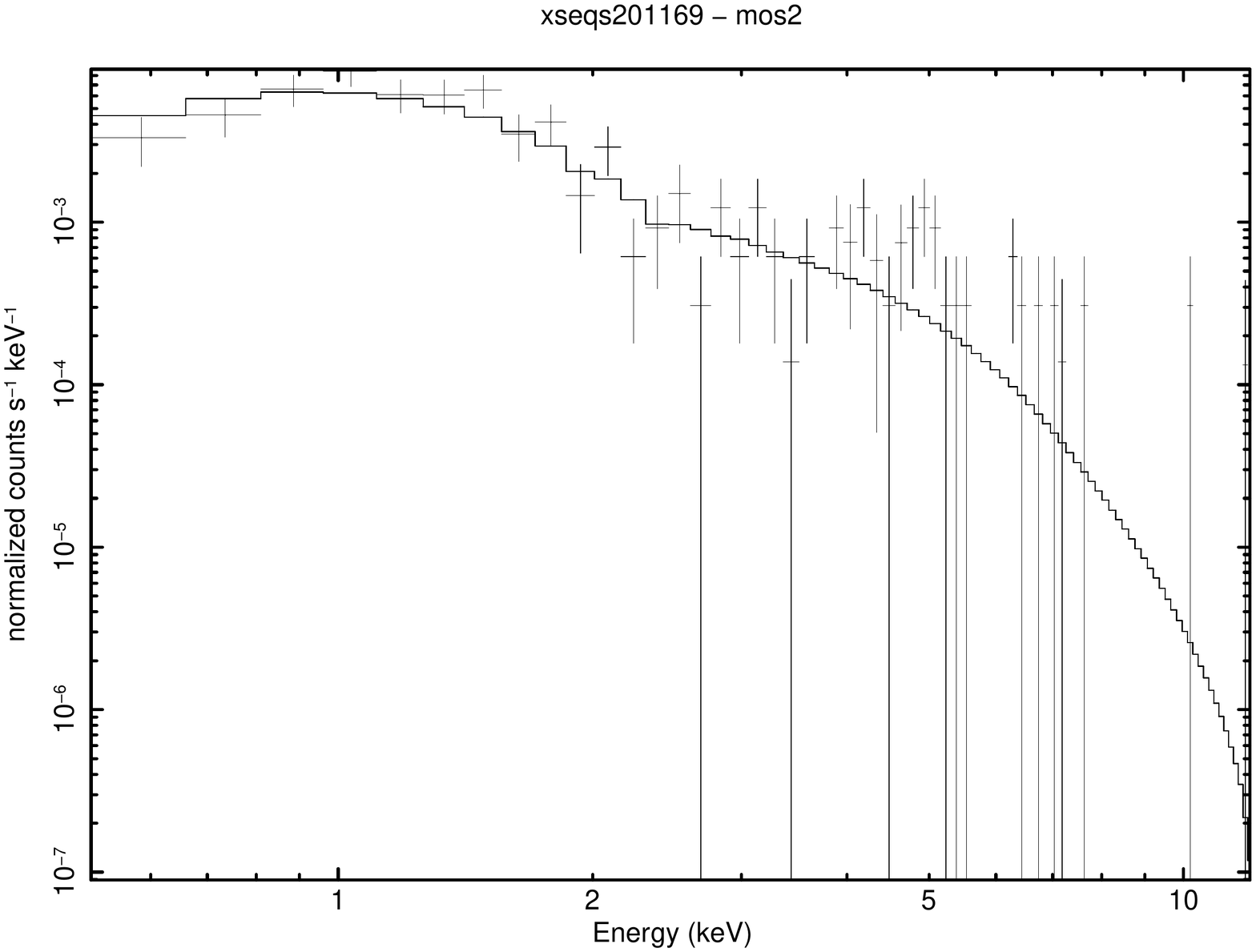} &
  \includegraphics[width=0.25\hsize]{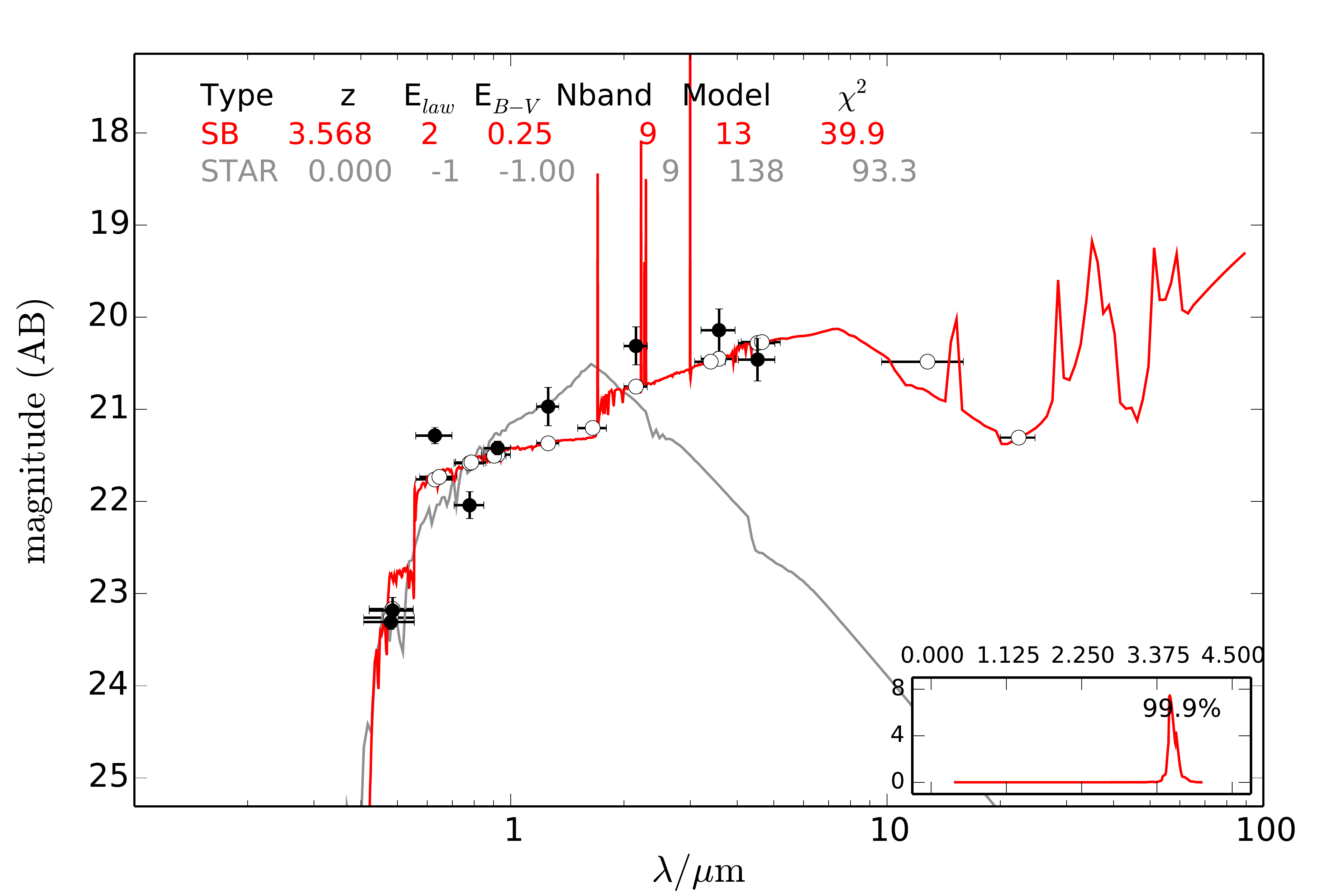} &
  \includegraphics[width=0.17\hsize]{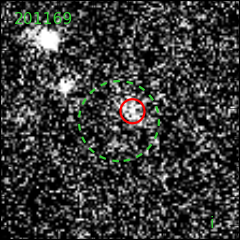}&
  \includegraphics[width=0.17\hsize]{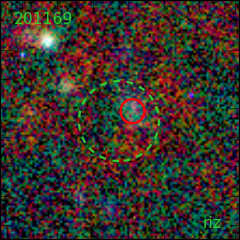}\\

\multicolumn{4}{l}{ 3XLSS J020326.7-060131} \\
  \includegraphics[width=0.24\hsize]{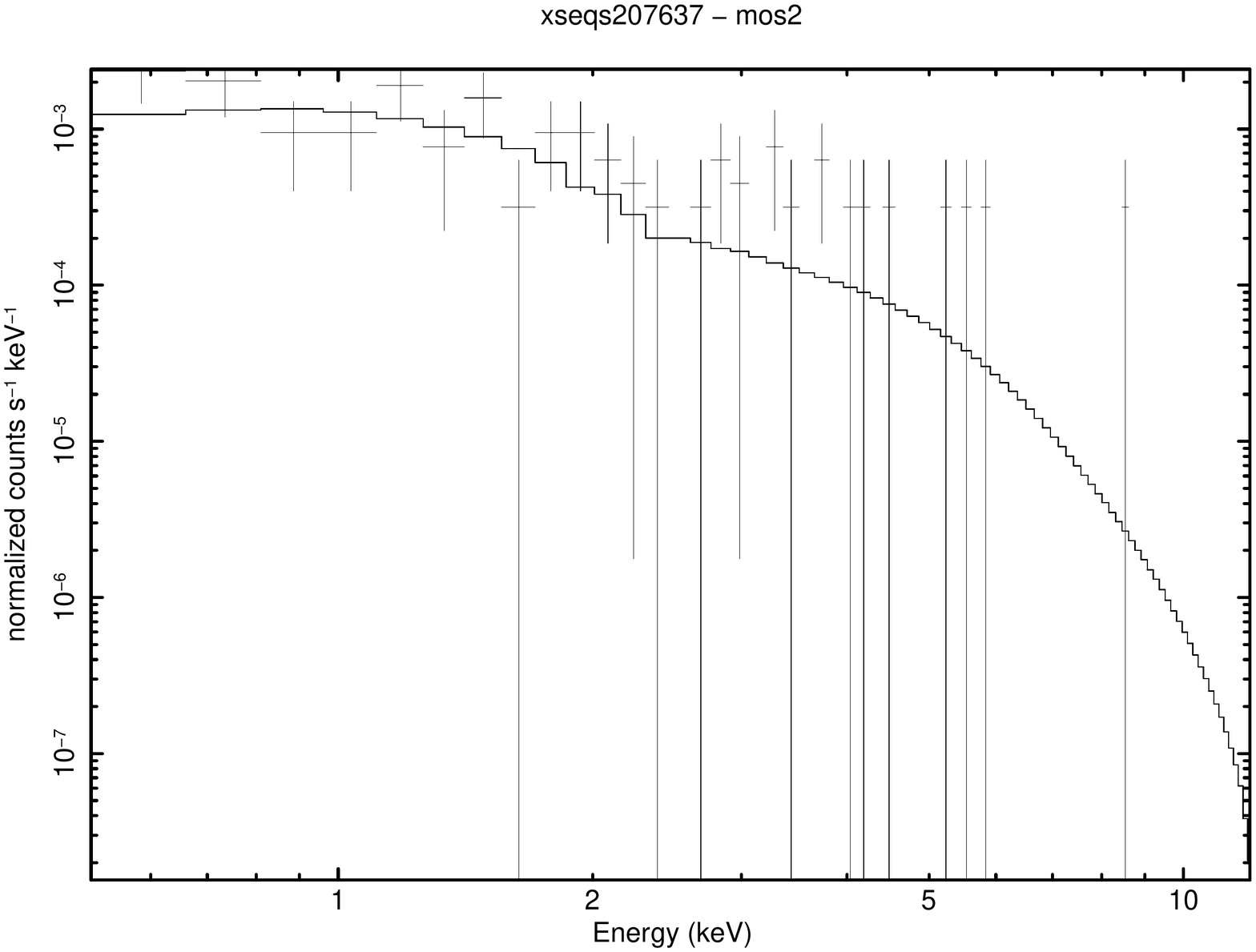} &
  \includegraphics[width=0.25\hsize]{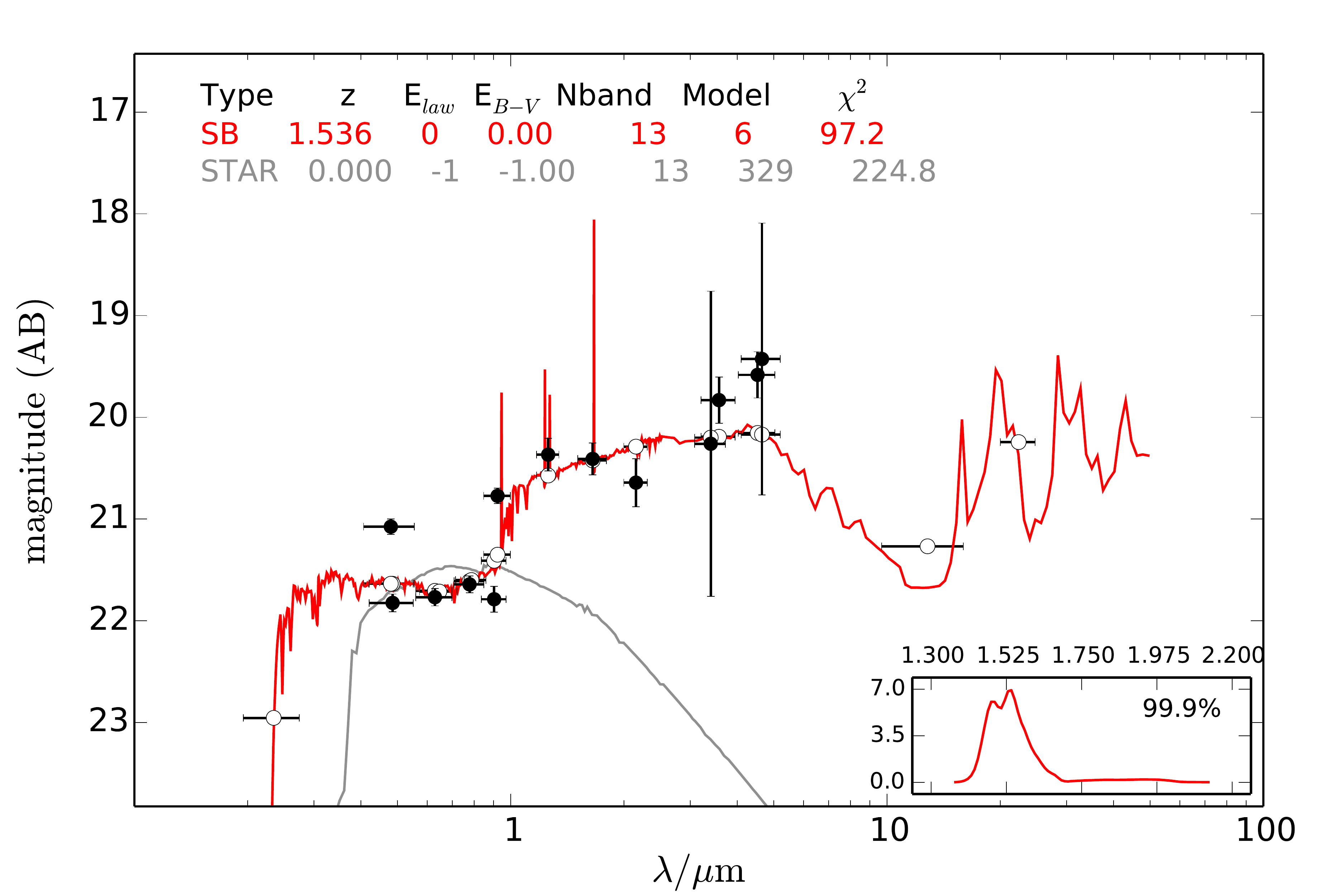} &
  \includegraphics[width=0.17\hsize]{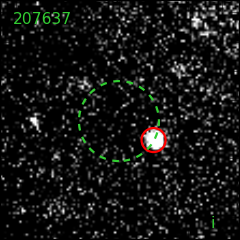}&
  \includegraphics[width=0.17\hsize]{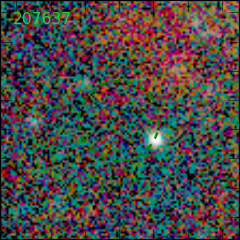}\\

\end{tabular}
\end{adjustbox}
\end{tabular}
\caption{Sources classified as starburst. Panels as in Fig. \ref{app:QSO}}
\end{figure*}
\end{center}

\begin{center}
\begin{figure*}
\centering
\begin{tabular}{cc}
\begin{adjustbox}{valign=t}
\begin{tabular}{cccc}
\multicolumn{4}{l}{3XLSS J021355.1-055120 } \\
  \includegraphics[width=0.24\hsize]{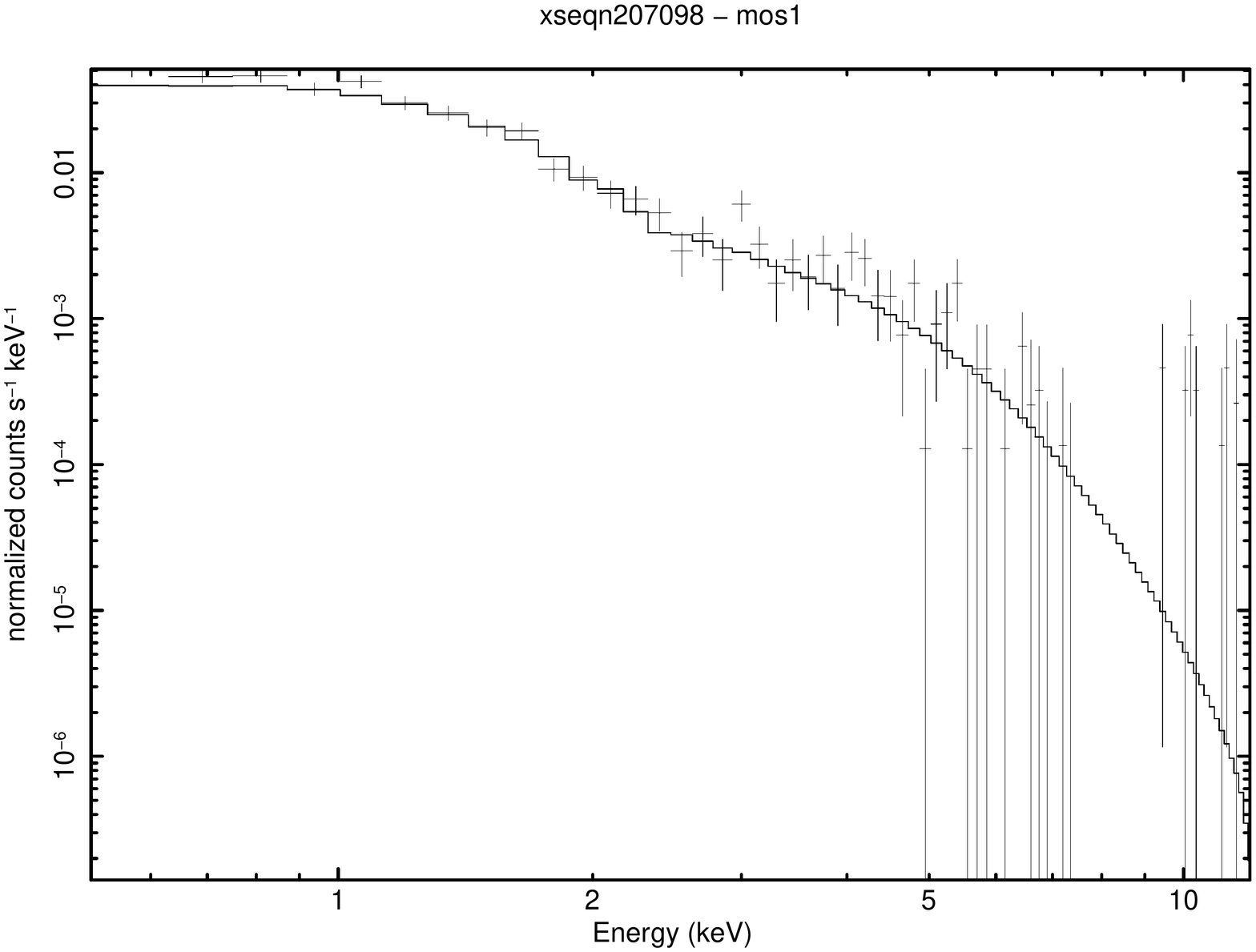} & 
  \includegraphics[width=0.25\hsize]{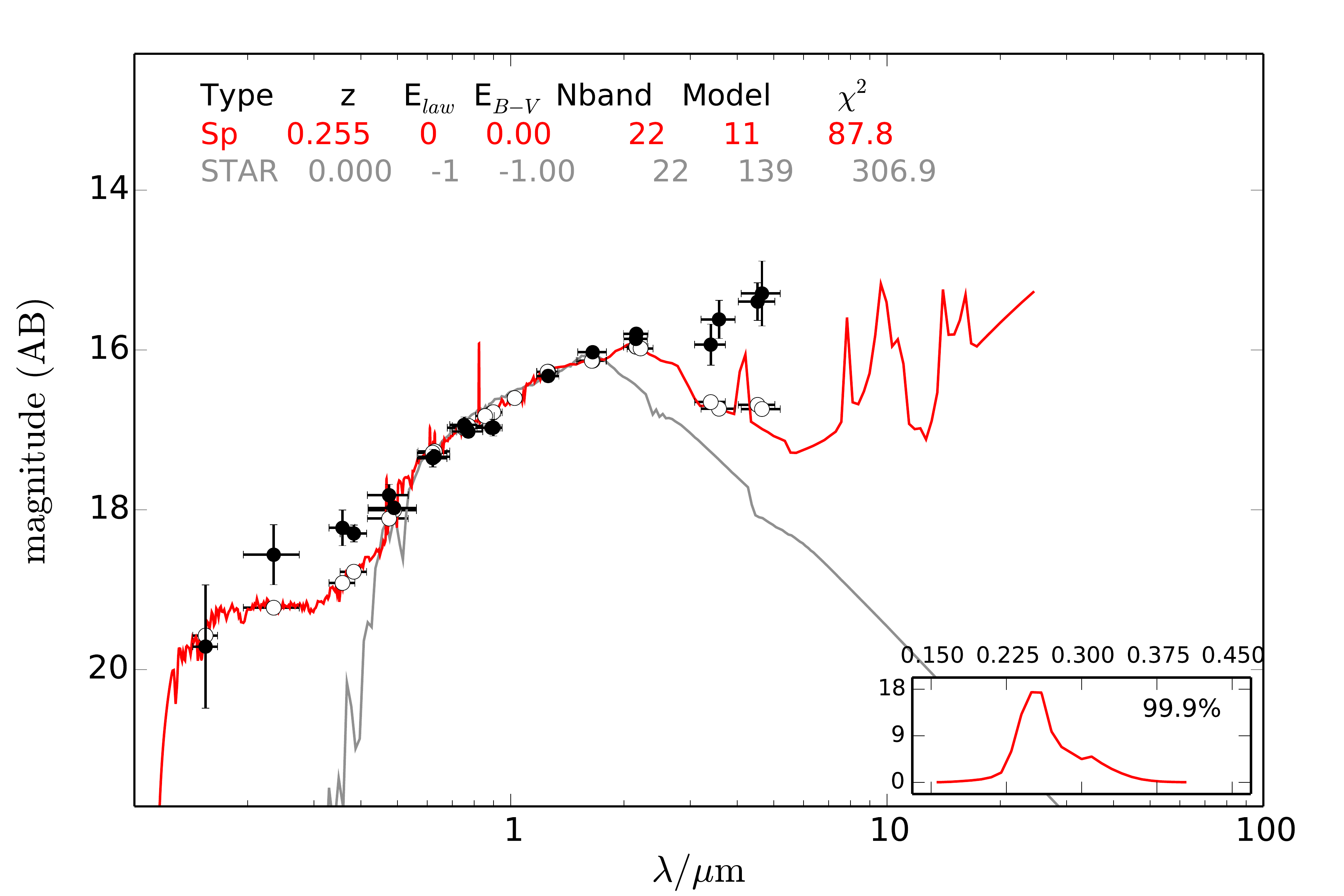} &
  \includegraphics[width=0.17\hsize]{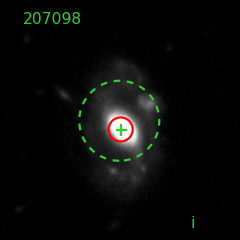} &
  \includegraphics[width=0.17\hsize]{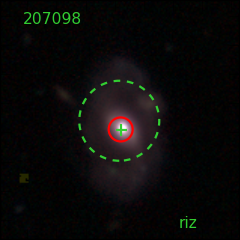} \\

\multicolumn{4}{l}{3XLSS J022700.8-042020 } \\
  \includegraphics[width=0.24\hsize]{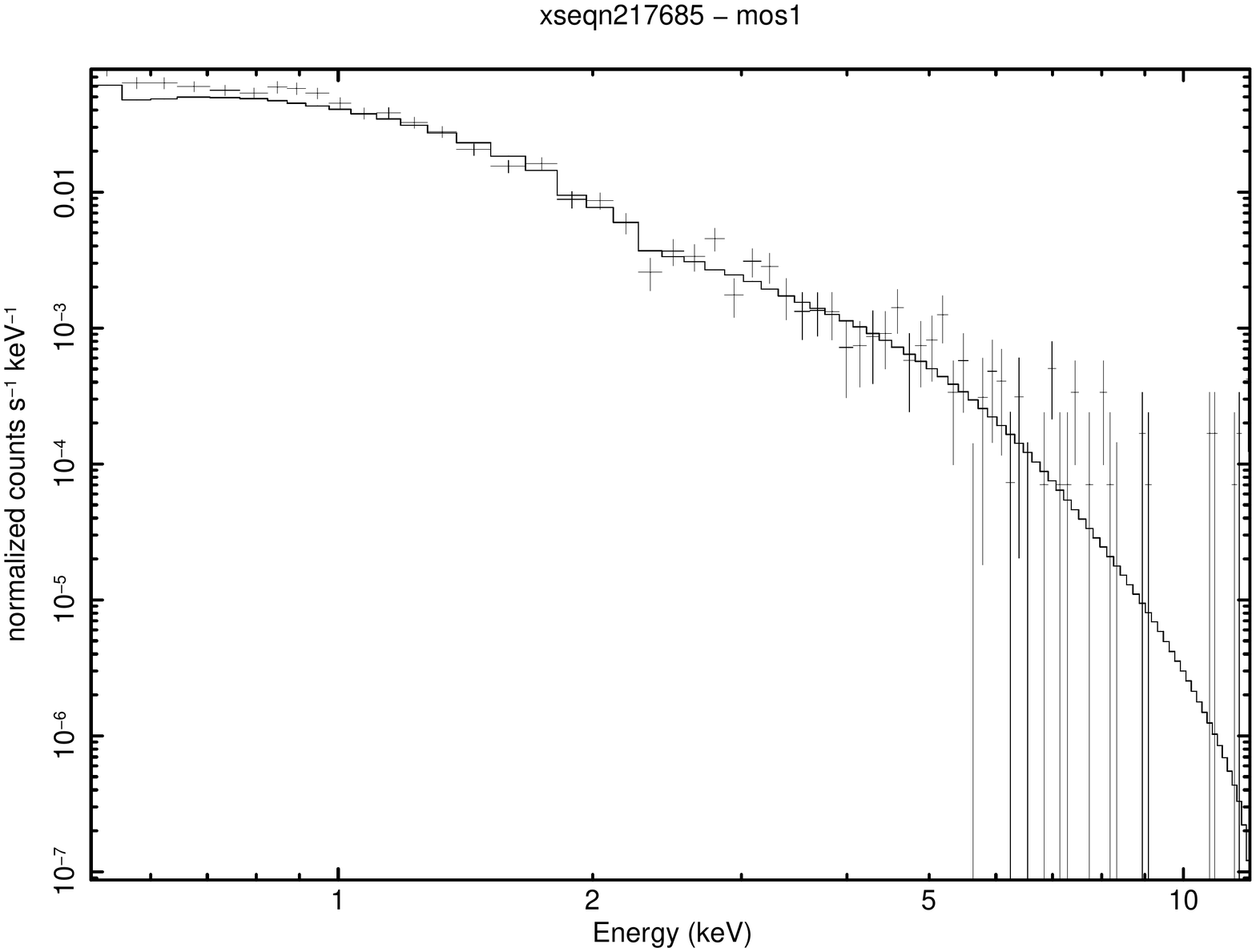} & 
  \includegraphics[width=0.25\hsize]{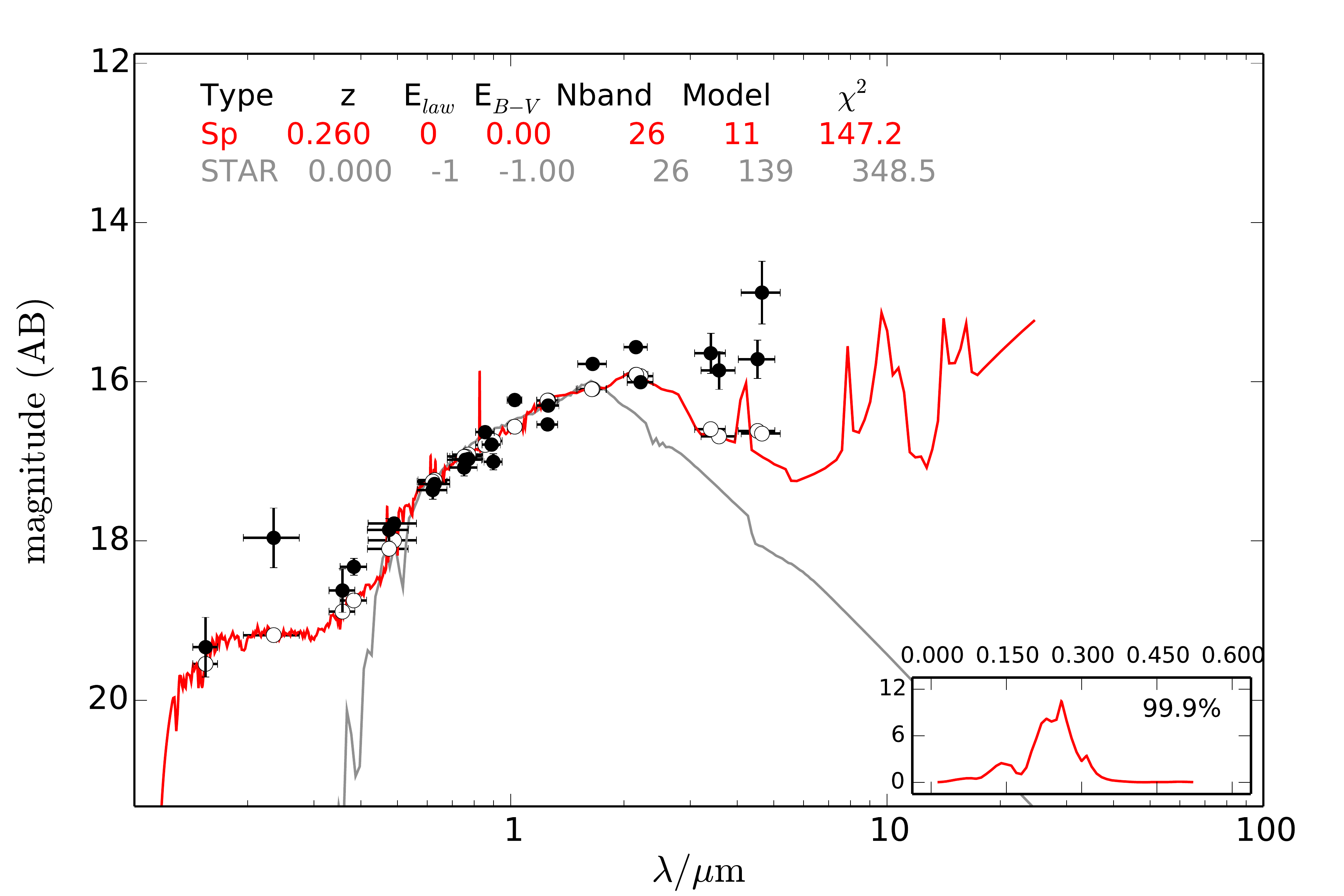} &
  \includegraphics[width=0.17\hsize]{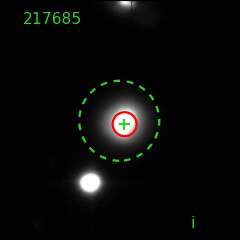} &
  \includegraphics[width=0.17\hsize]{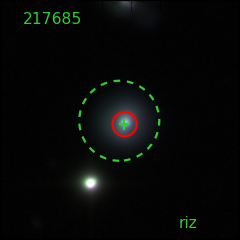} \\

\multicolumn{4}{l}{3XLSS J020840.6-062715 } \\
  \includegraphics[width=0.24\hsize]{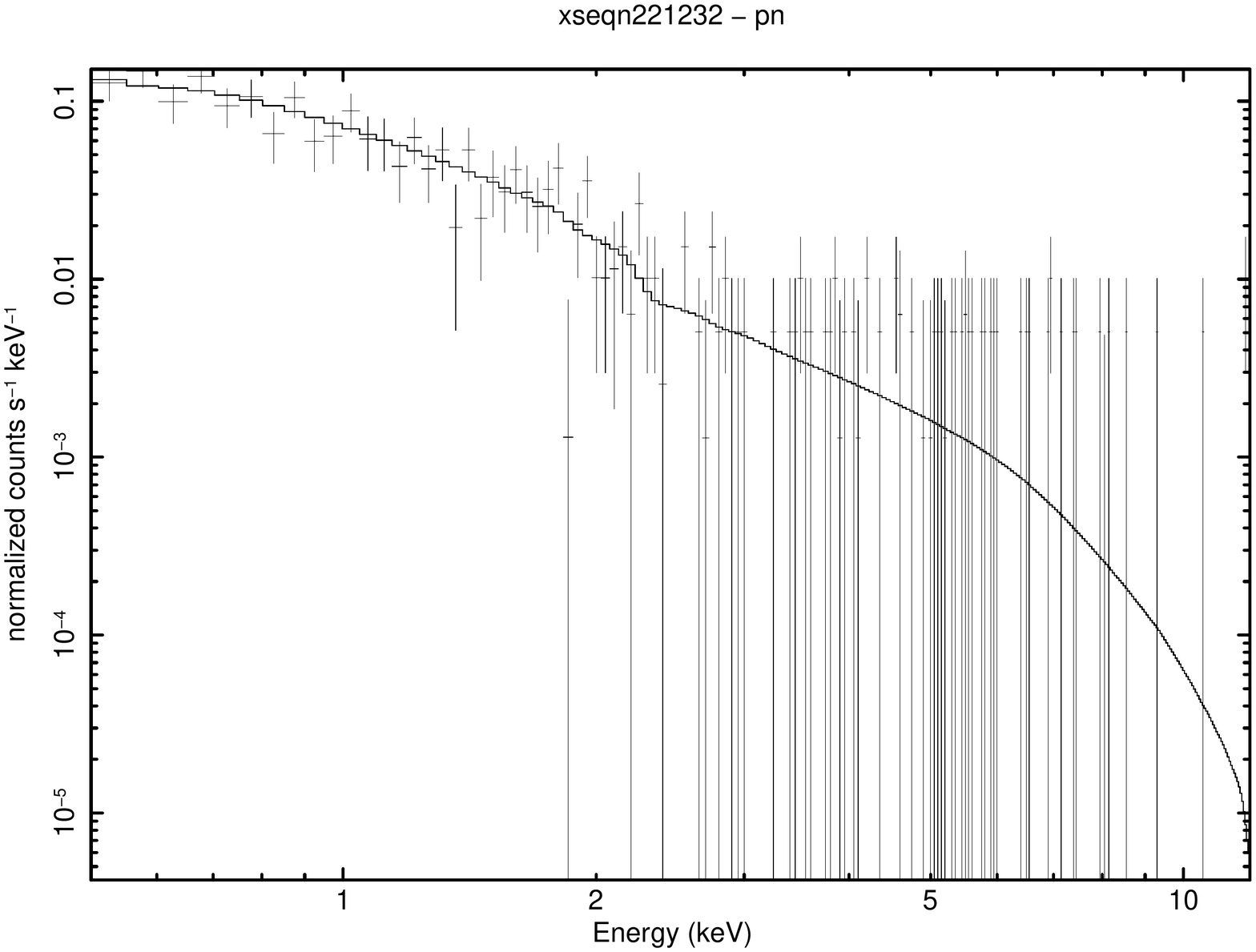} & 
  \includegraphics[width=0.25\hsize]{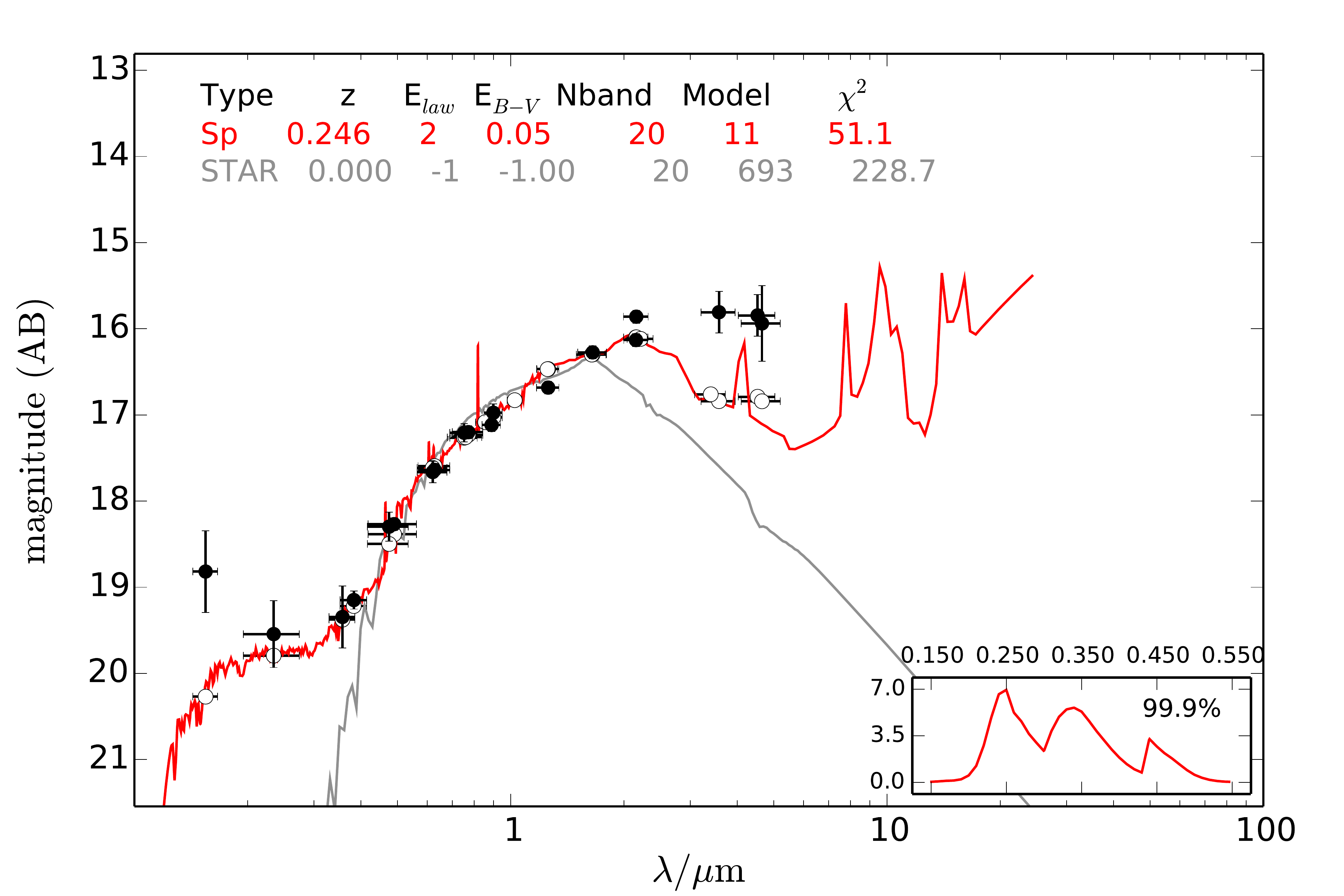} &
  \includegraphics[width=0.17\hsize]{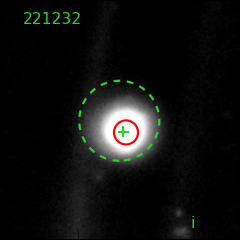} &
  \includegraphics[width=0.17\hsize]{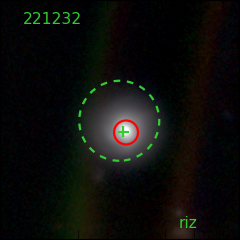} \\
  
\multicolumn{4}{l}{3XLSS J023322.1-045505 } \\
  \includegraphics[width=0.24\hsize]{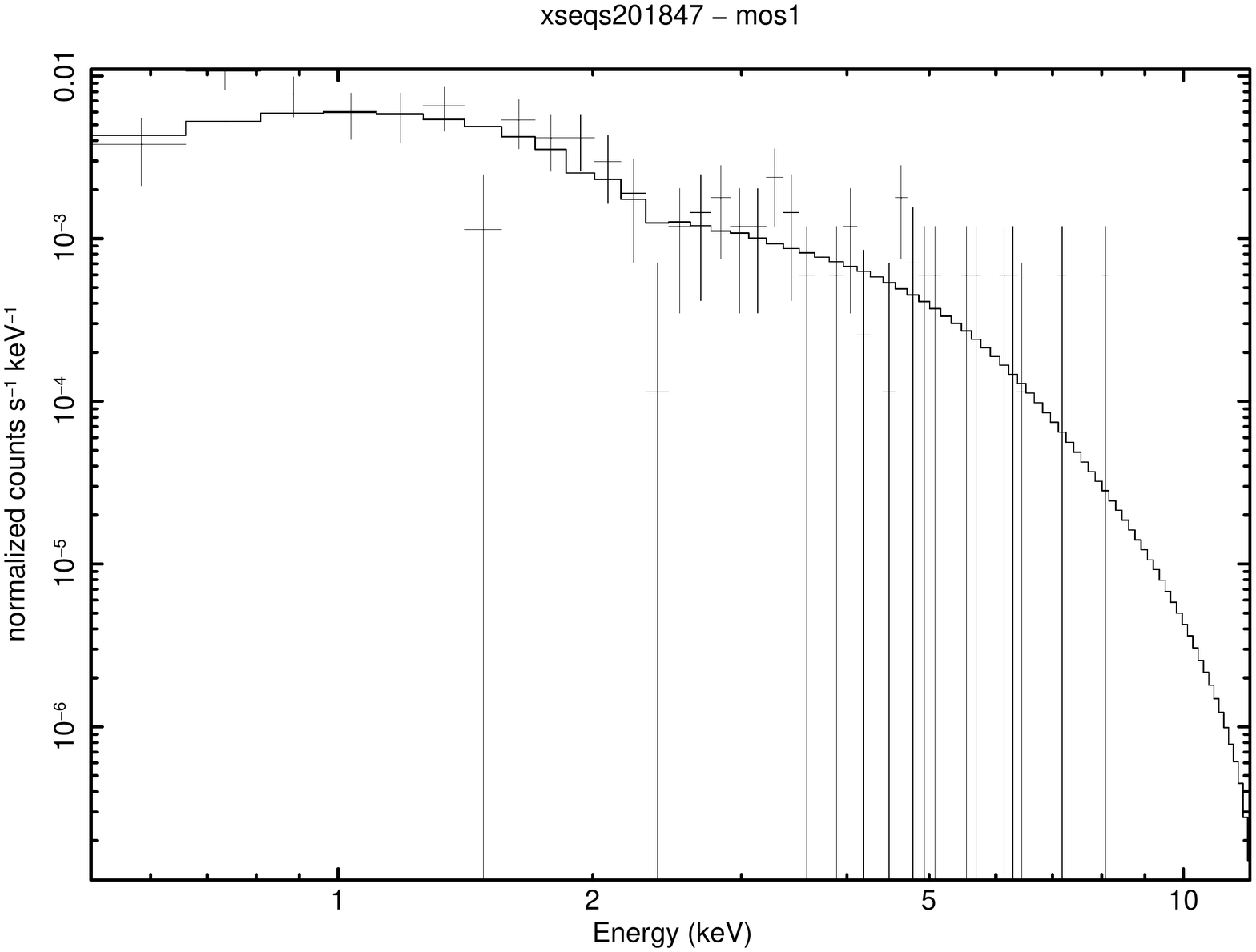} &
  \includegraphics[width=0.25\hsize]{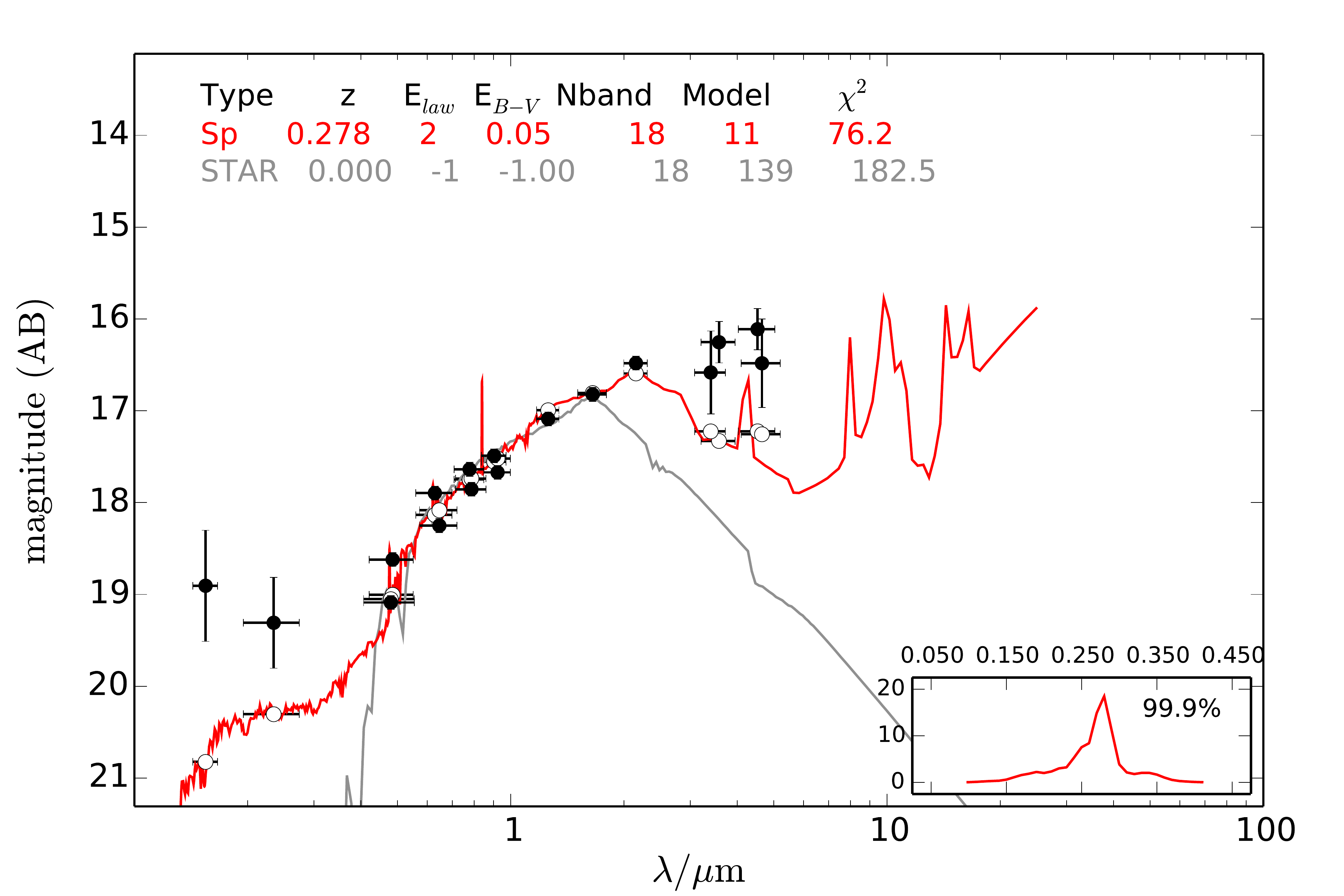} &
  \includegraphics[width=0.17\hsize]{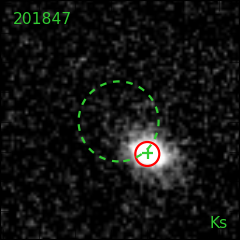}&
  \includegraphics[width=0.17\hsize]{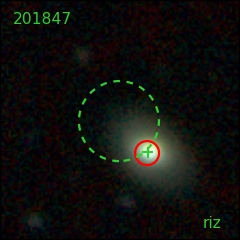}\\

\multicolumn{4}{l}{3XLSS J020207.0-055858 } \\
  \includegraphics[width=0.24\hsize]{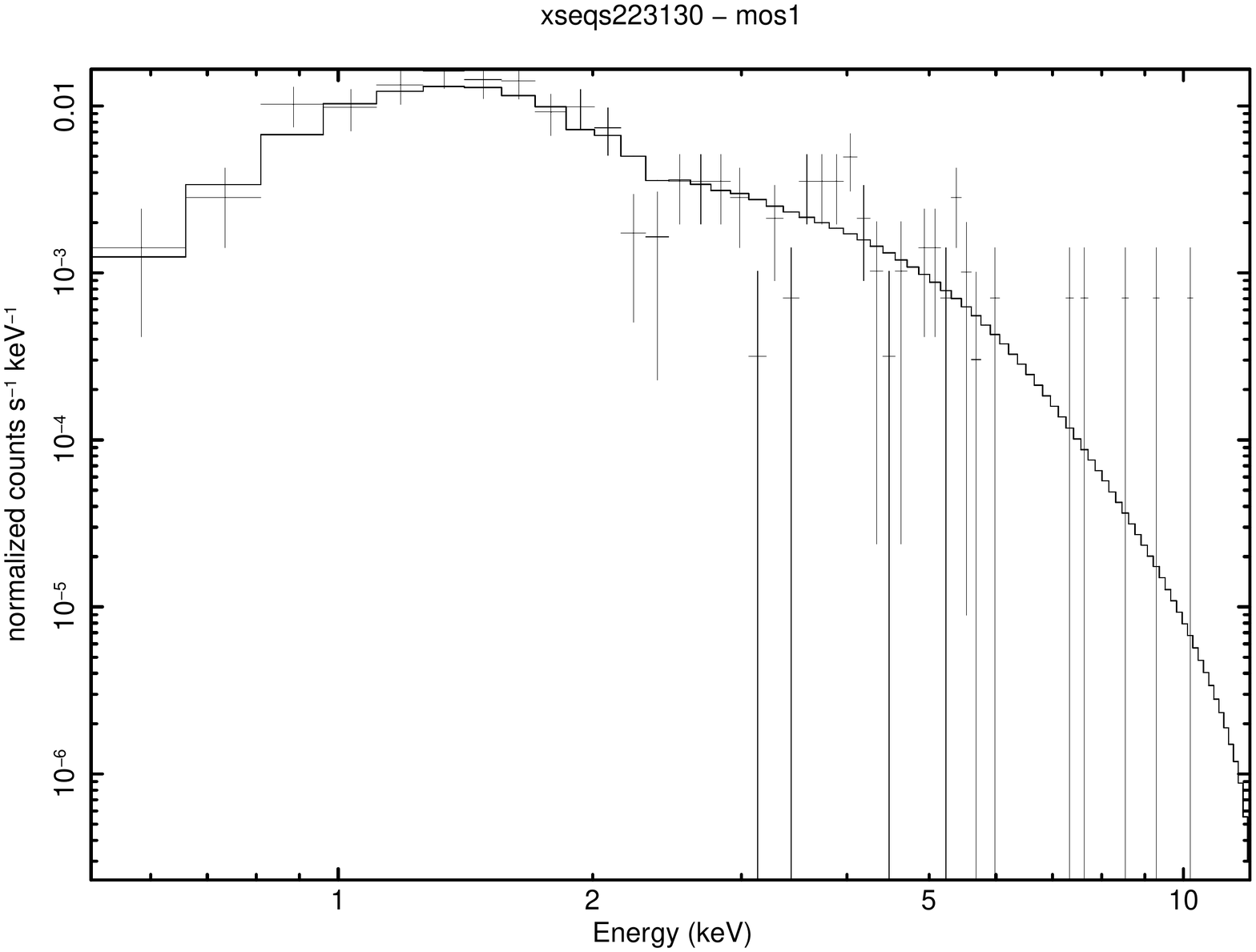} &
  \includegraphics[width=0.25\hsize]{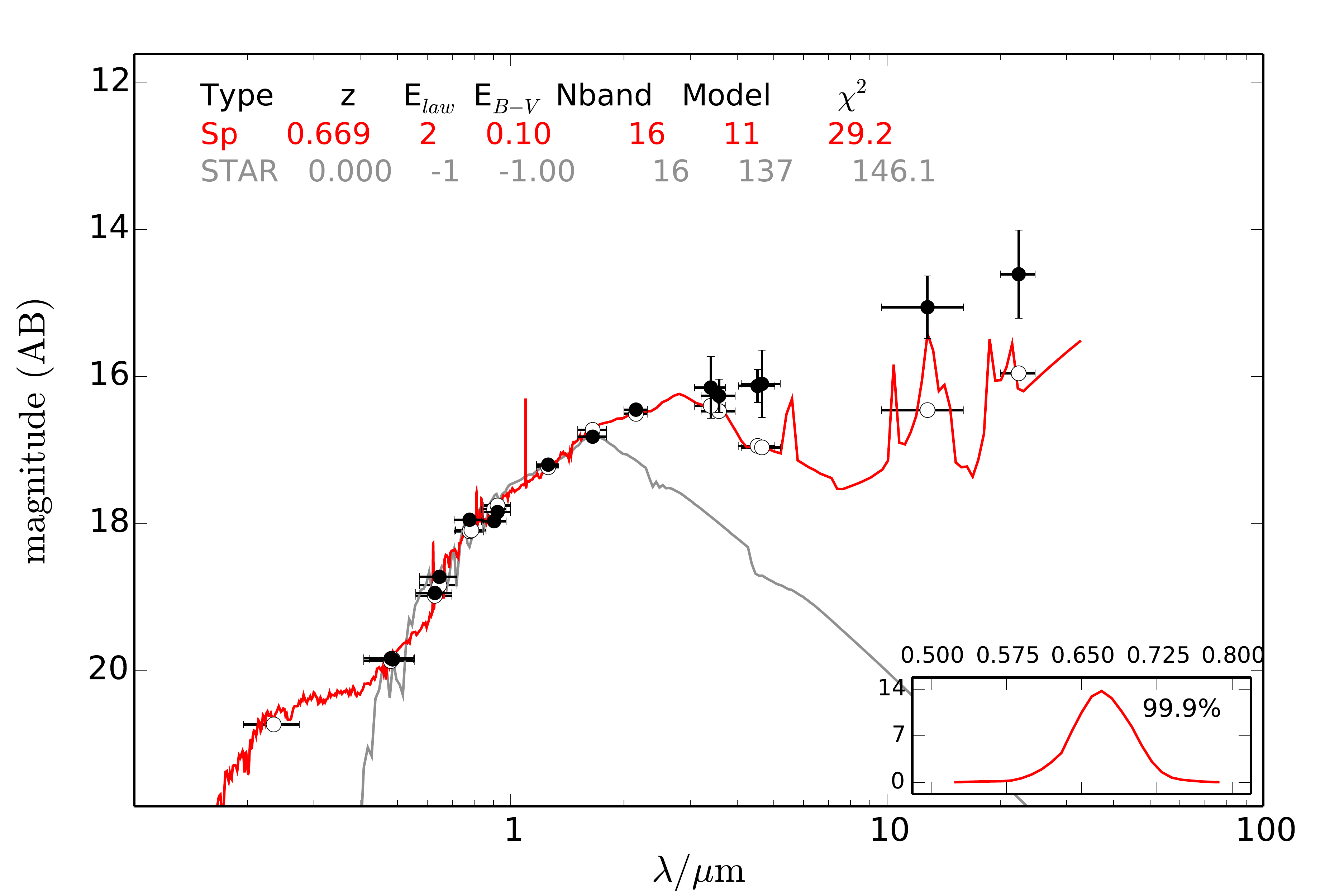} &
  \includegraphics[width=0.17\hsize]{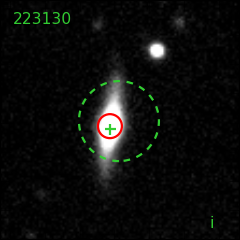}&
  \includegraphics[width=0.17\hsize]{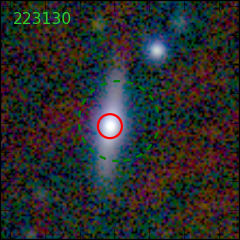}\\

\multicolumn{4}{l}{3XLSS J022707.7-050816 } \\
  \includegraphics[width=0.24\hsize]{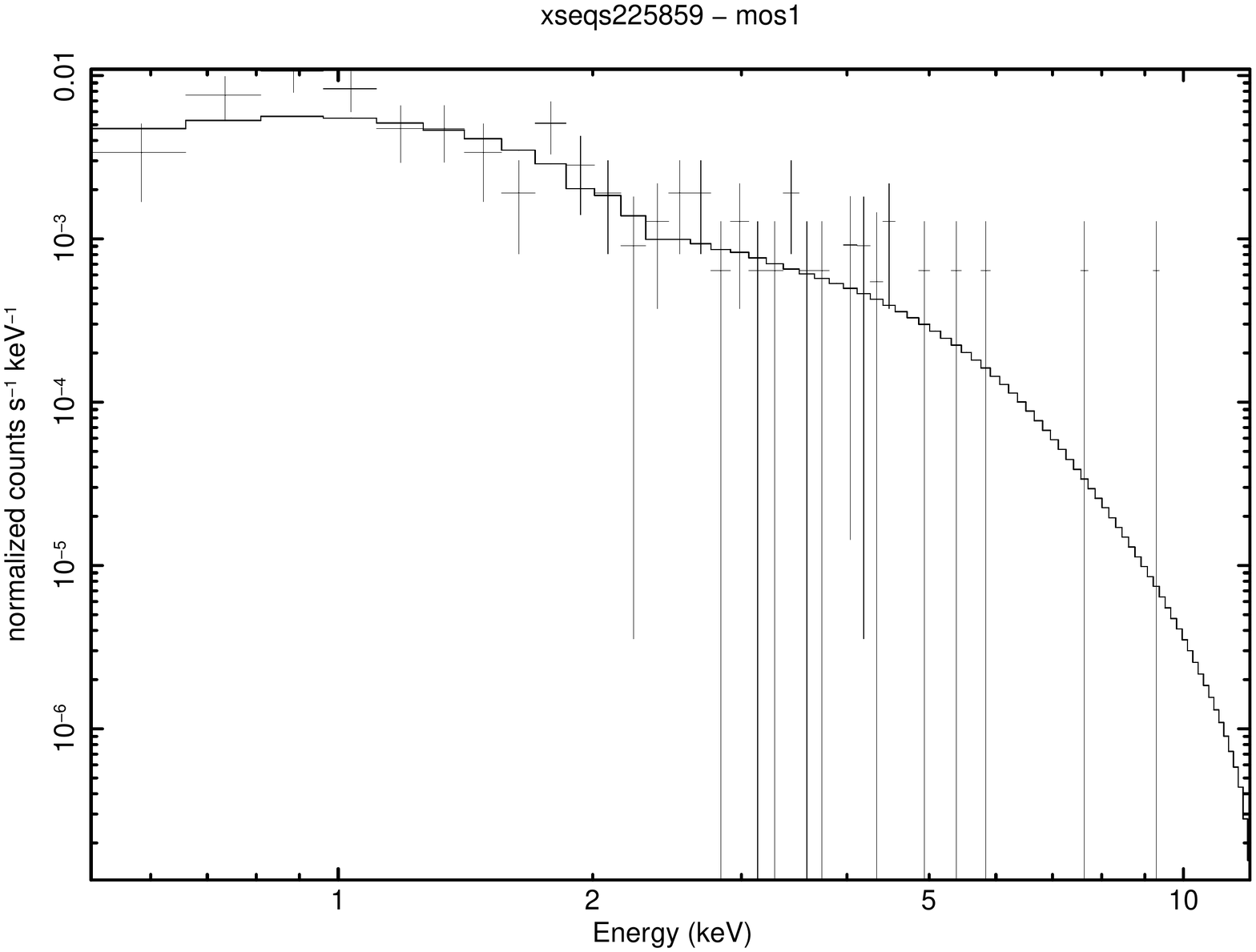} &
  \includegraphics[width=0.25\hsize]{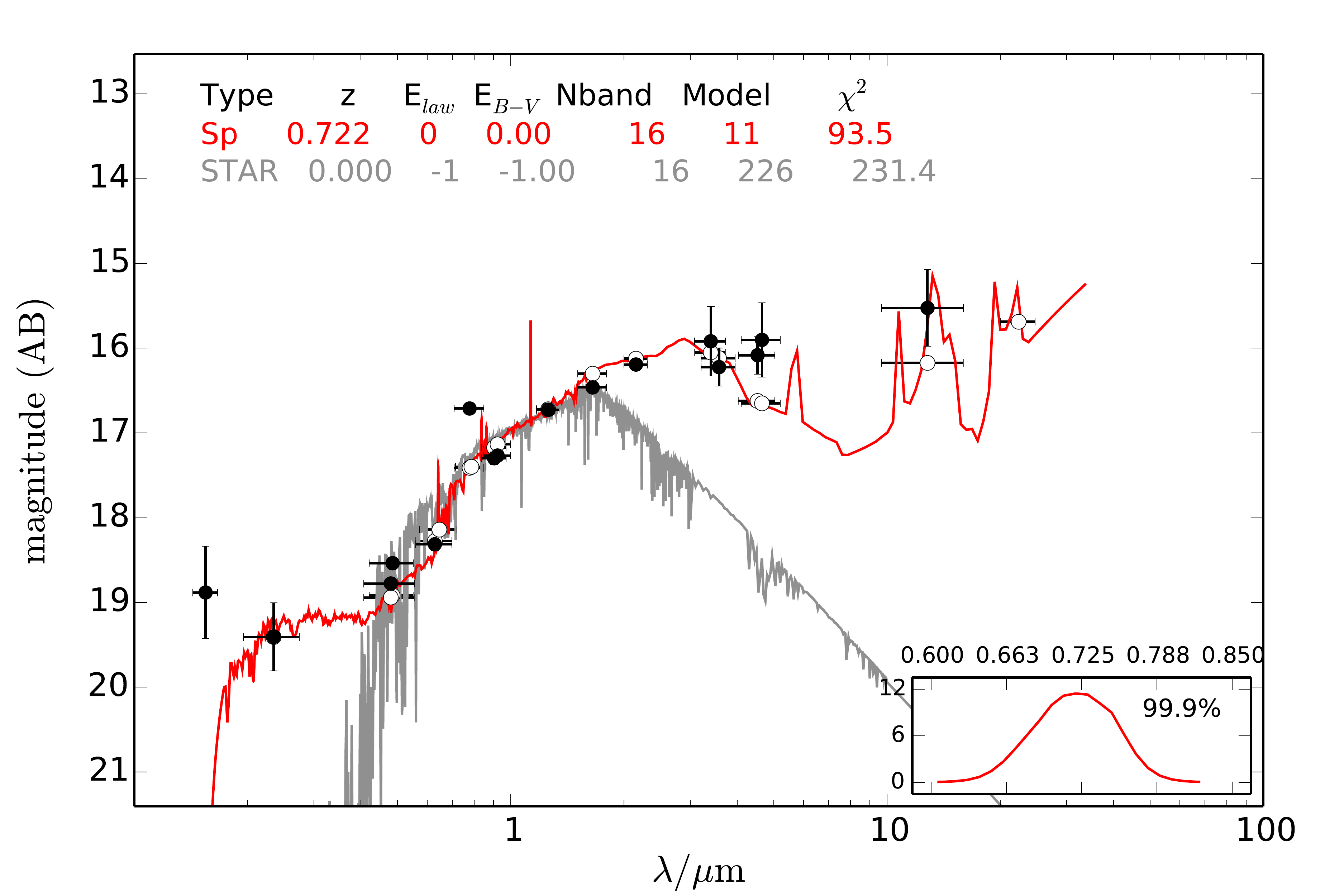} &
  \includegraphics[width=0.17\hsize]{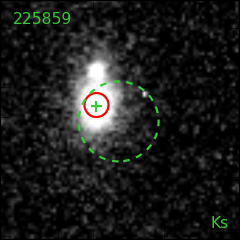}&
  \includegraphics[width=0.17\hsize]{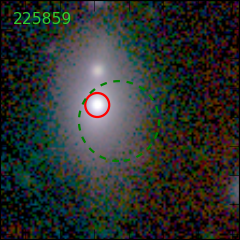}\\

\end{tabular}
\end{adjustbox}
\end{tabular}
\caption{Sources classified as star forming. Panels as in Fig. \ref{app:QSO}}
\end{figure*}
\end{center}

\begin{center}
\begin{figure*}
\centering
\begin{tabular}{cc}
\begin{adjustbox}{valign=t}
\begin{tabular}{cccc}
\multicolumn{4}{l}{3XLSS J023209.7-051942} \\
  \includegraphics[width=0.24\hsize]{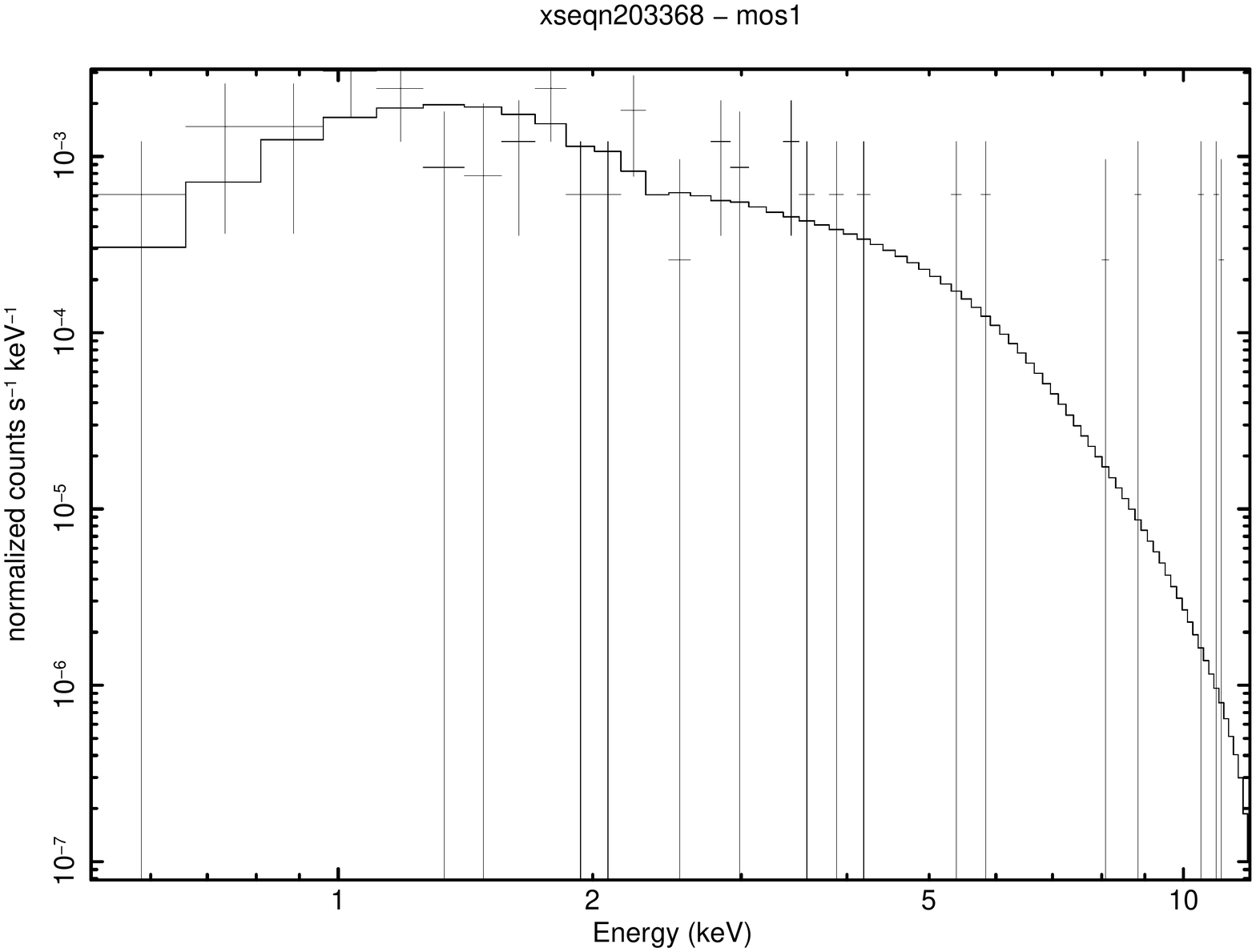} & 
  \includegraphics[width=0.25\hsize]{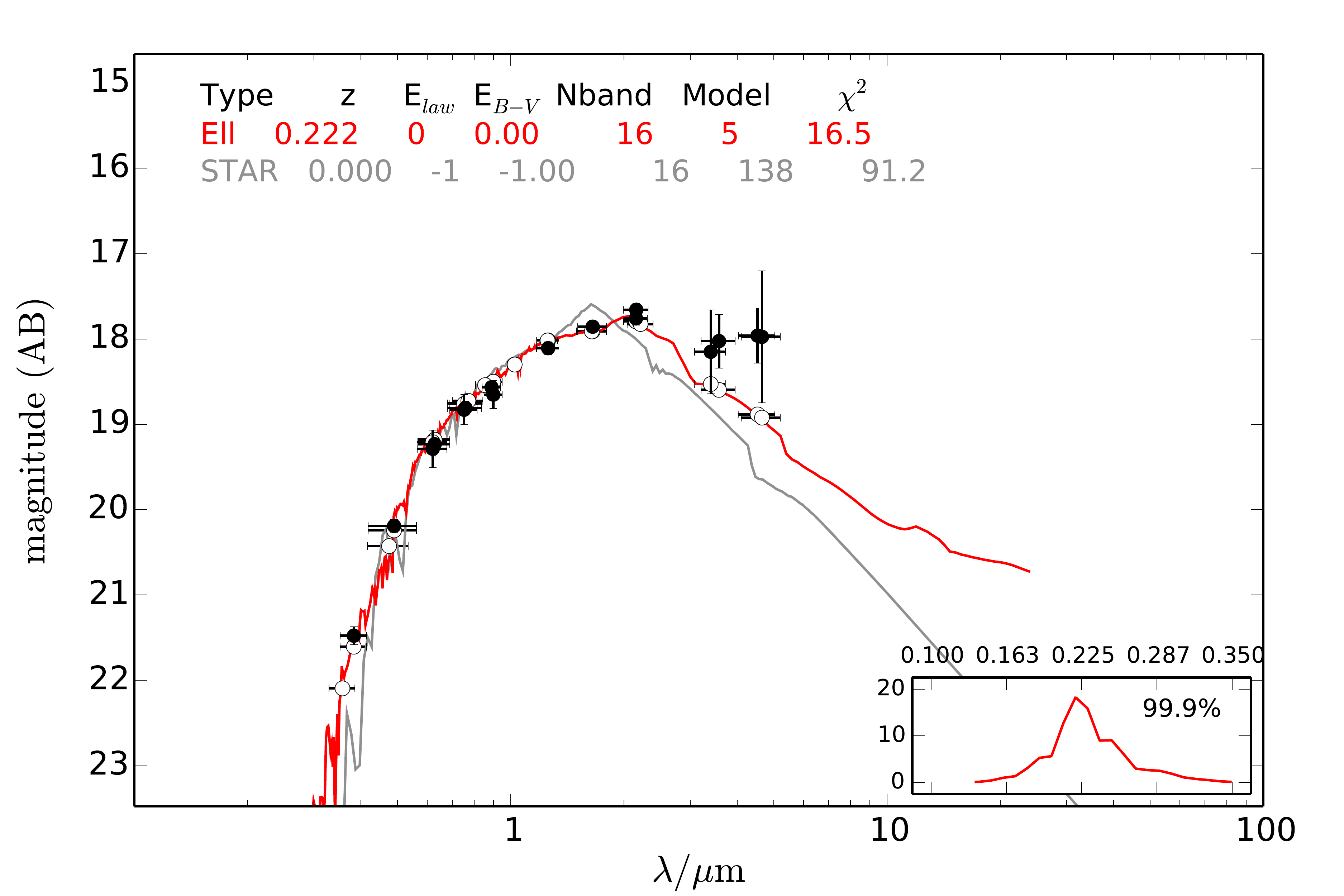} &
  \includegraphics[width=0.17\hsize]{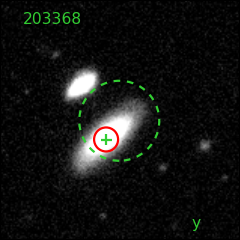} &
  \includegraphics[width=0.17\hsize]{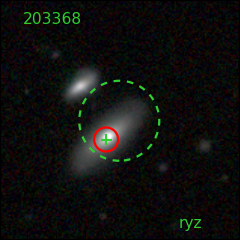} \\

\multicolumn{4}{l}{3XLSS J021329.1-053909} \\
  \includegraphics[width=0.24\hsize]{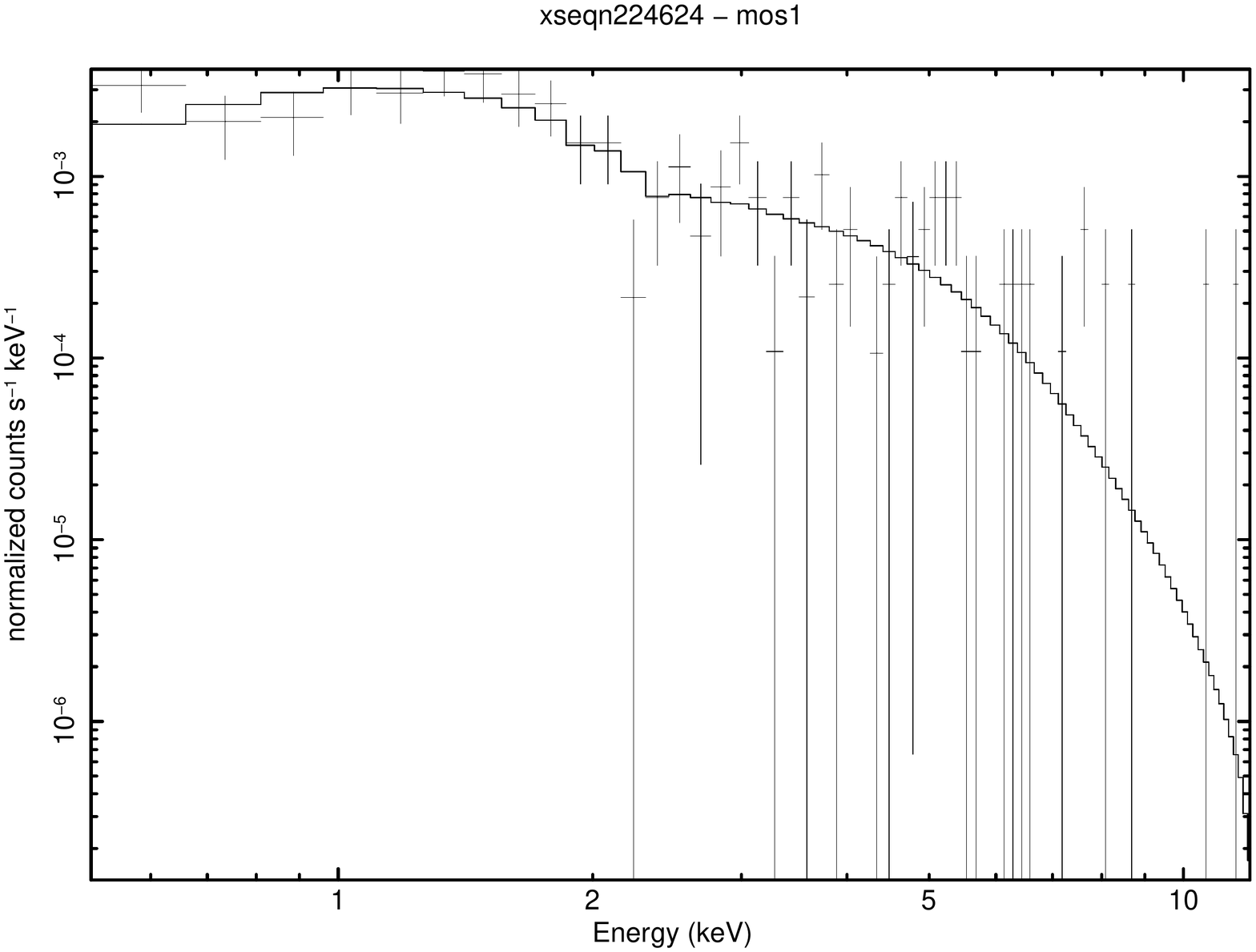} & 
  \includegraphics[width=0.25\hsize]{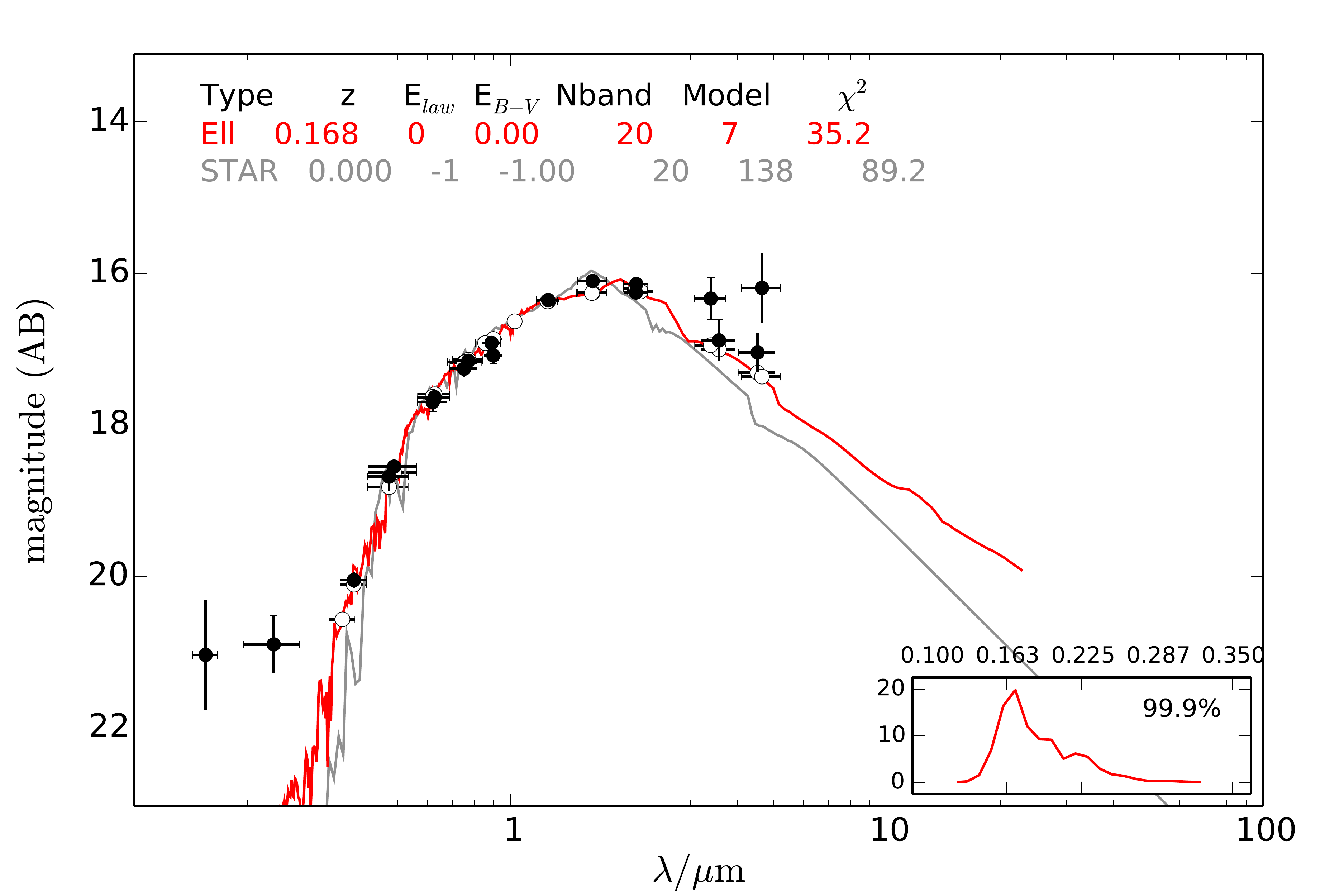} &
  \includegraphics[width=0.17\hsize]{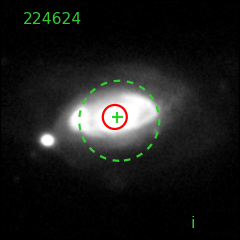} &
  \includegraphics[width=0.17\hsize]{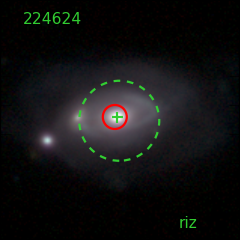} \\

\multicolumn{4}{l}{3XLSS J022258.0-041840} \\
  \includegraphics[width=0.24\hsize]{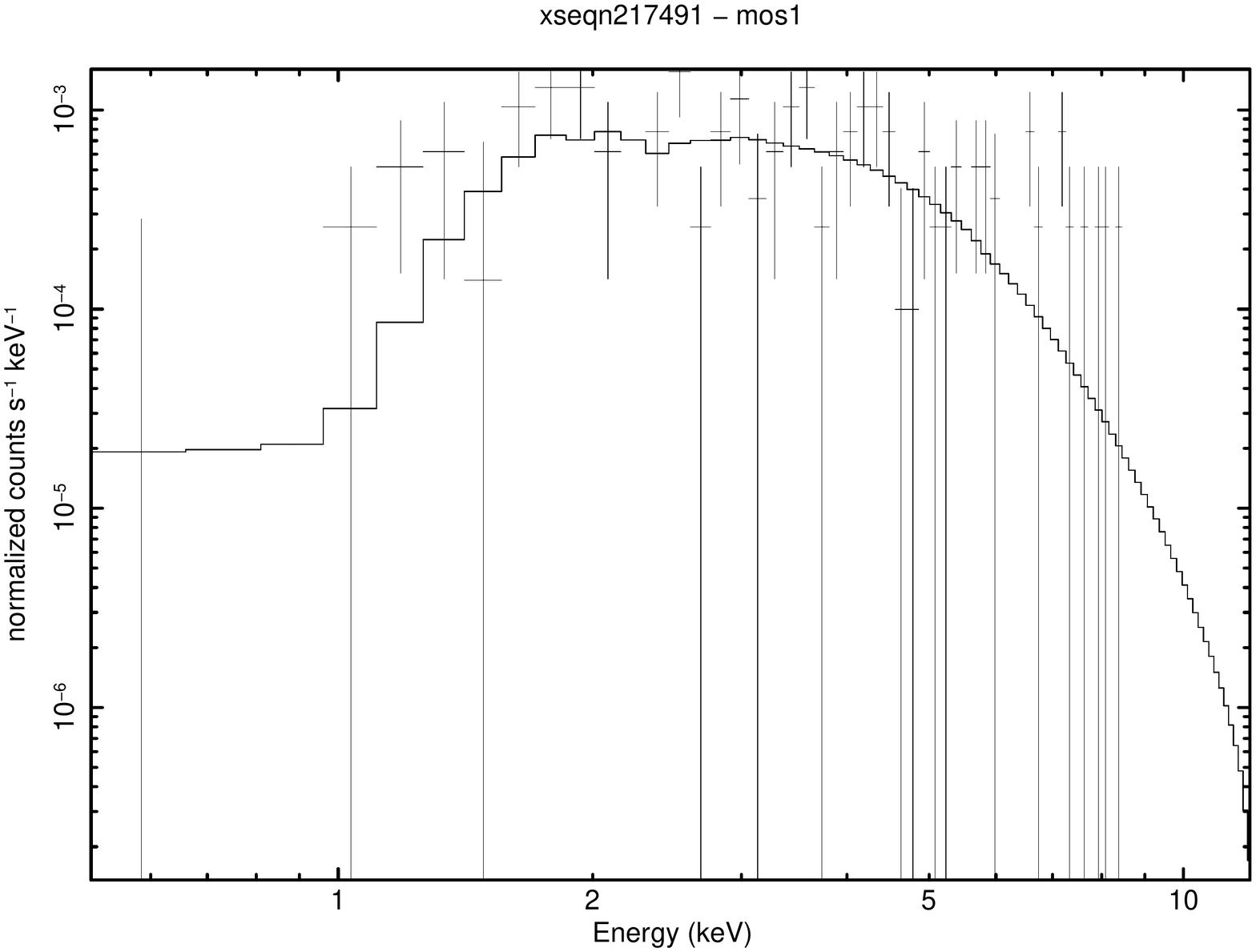} & 
  \includegraphics[width=0.25\hsize]{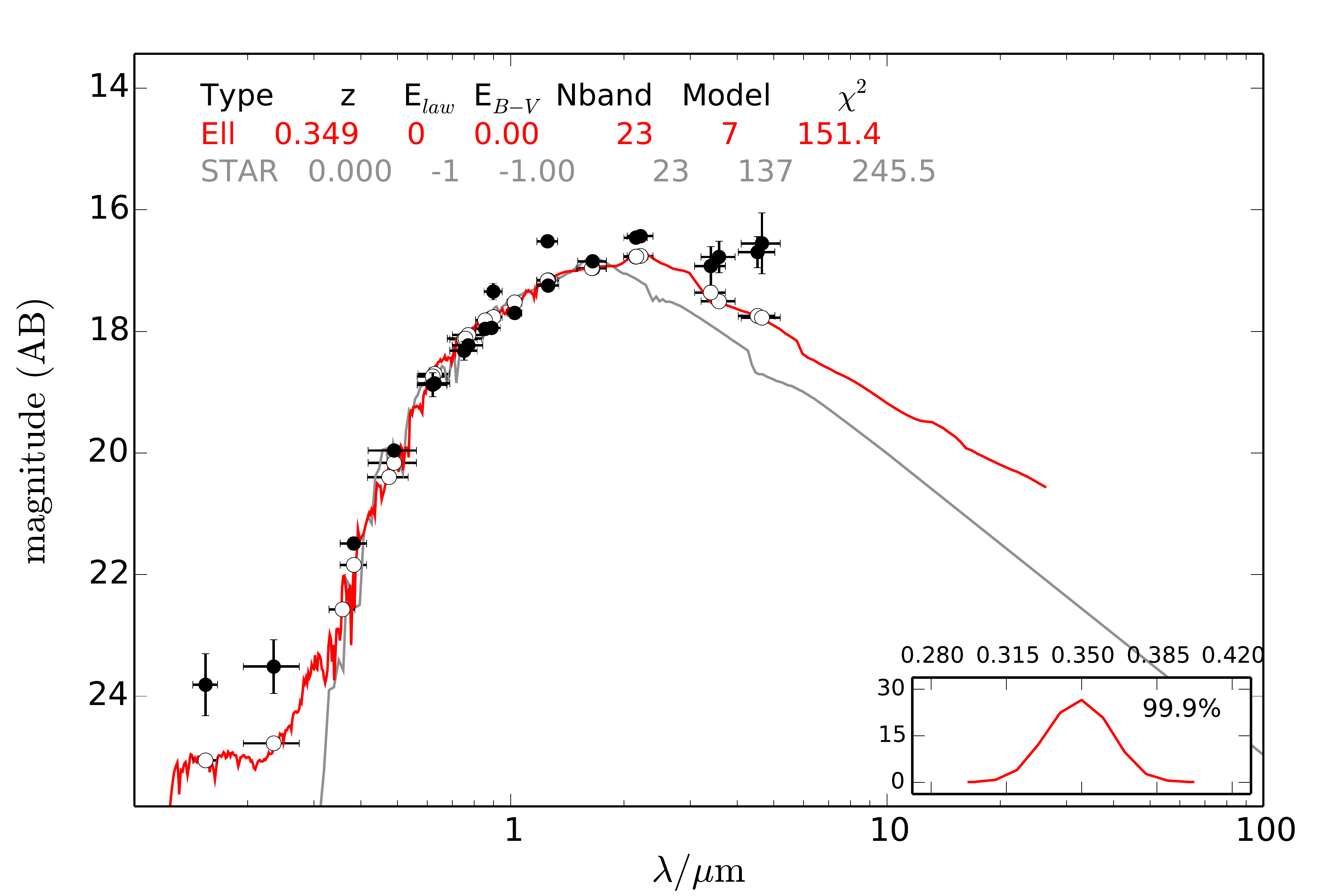} &
  \includegraphics[width=0.17\hsize]{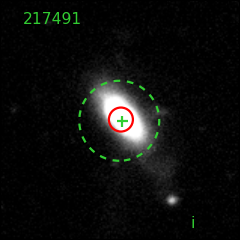} &
  \includegraphics[width=0.17\hsize]{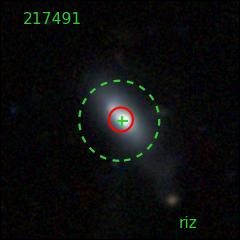} \\
  
\multicolumn{4}{l}{3XLSS J022143.9-053146 } \\
  \includegraphics[width=0.24\hsize]{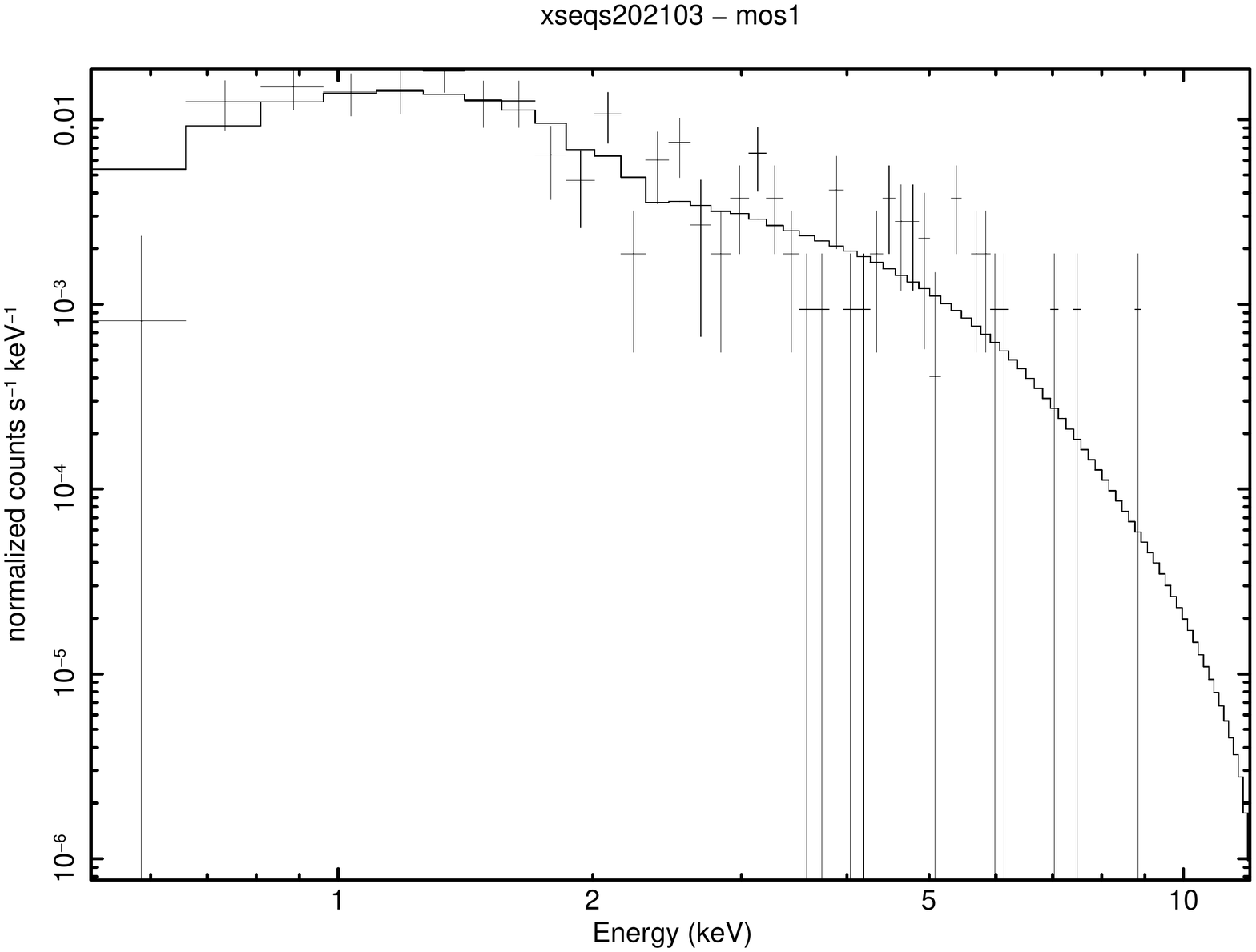} &
  \includegraphics[width=0.25\hsize]{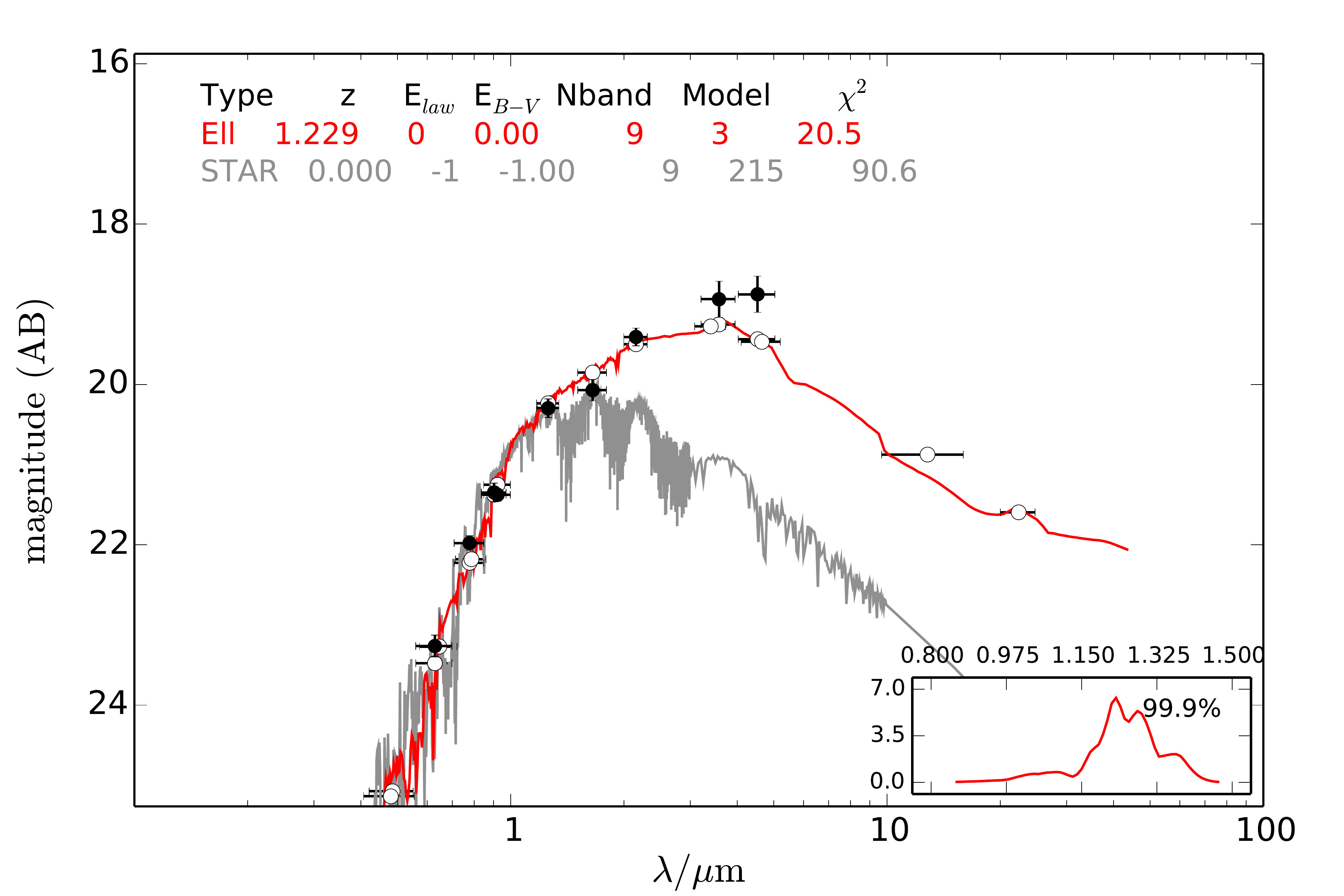} &
  \includegraphics[width=0.17\hsize]{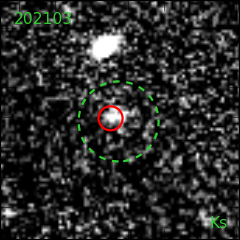}&
  \includegraphics[width=0.17\hsize]{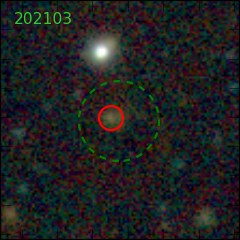}\\

\multicolumn{4}{l}{3XLSS J023440.2-050231 } \\
  \includegraphics[width=0.24\hsize]{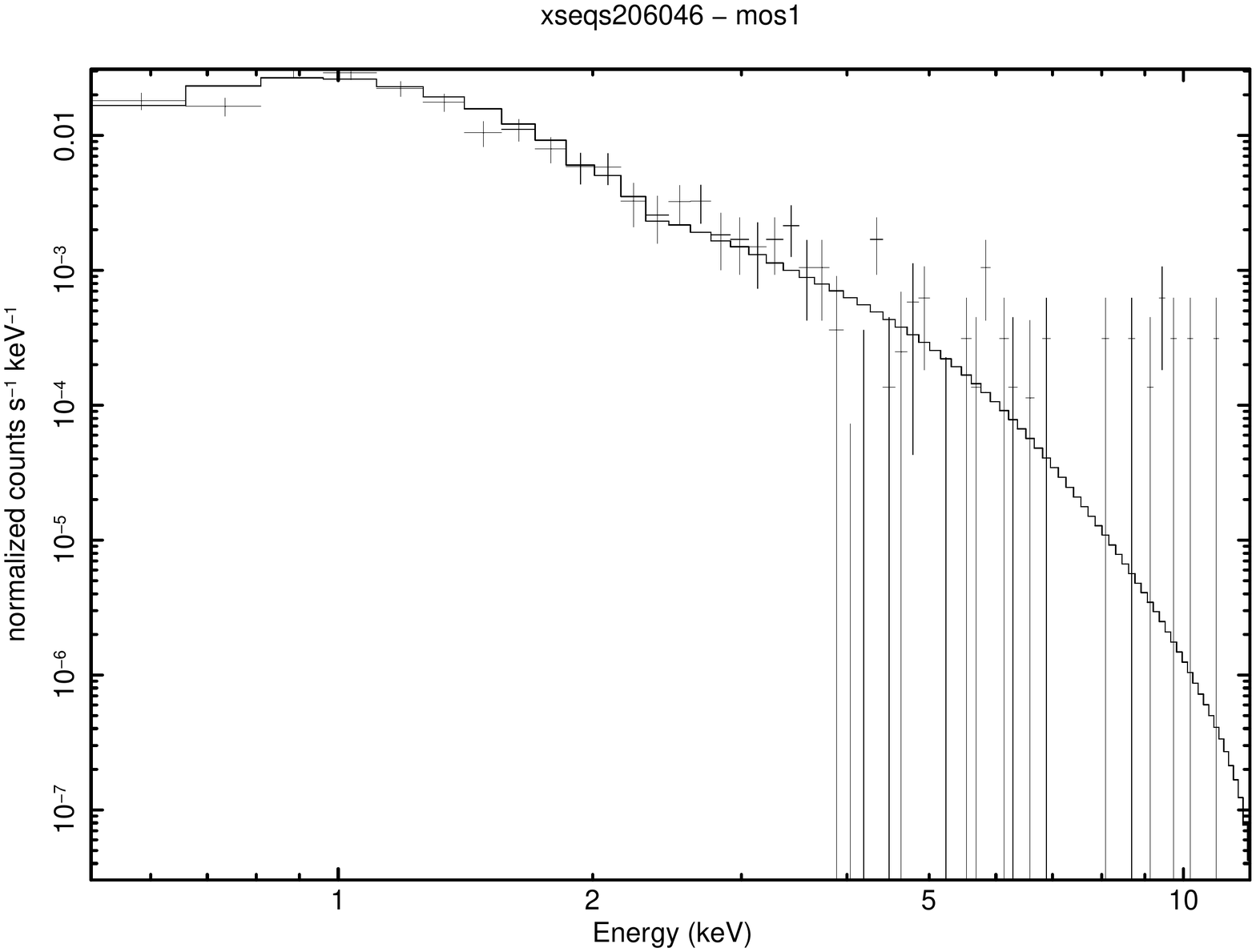} &
  \includegraphics[width=0.25\hsize]{Id001279032.pdf} &
  \includegraphics[width=0.17\hsize]{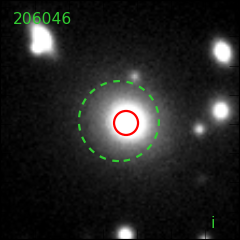} &
  \includegraphics[width=0.17\hsize]{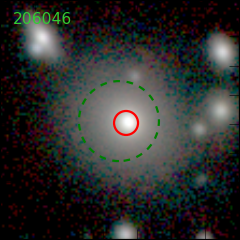}\\

\multicolumn{4}{l}{3XLSS J022012.0-034111} \\
  \includegraphics[width=0.24\hsize]{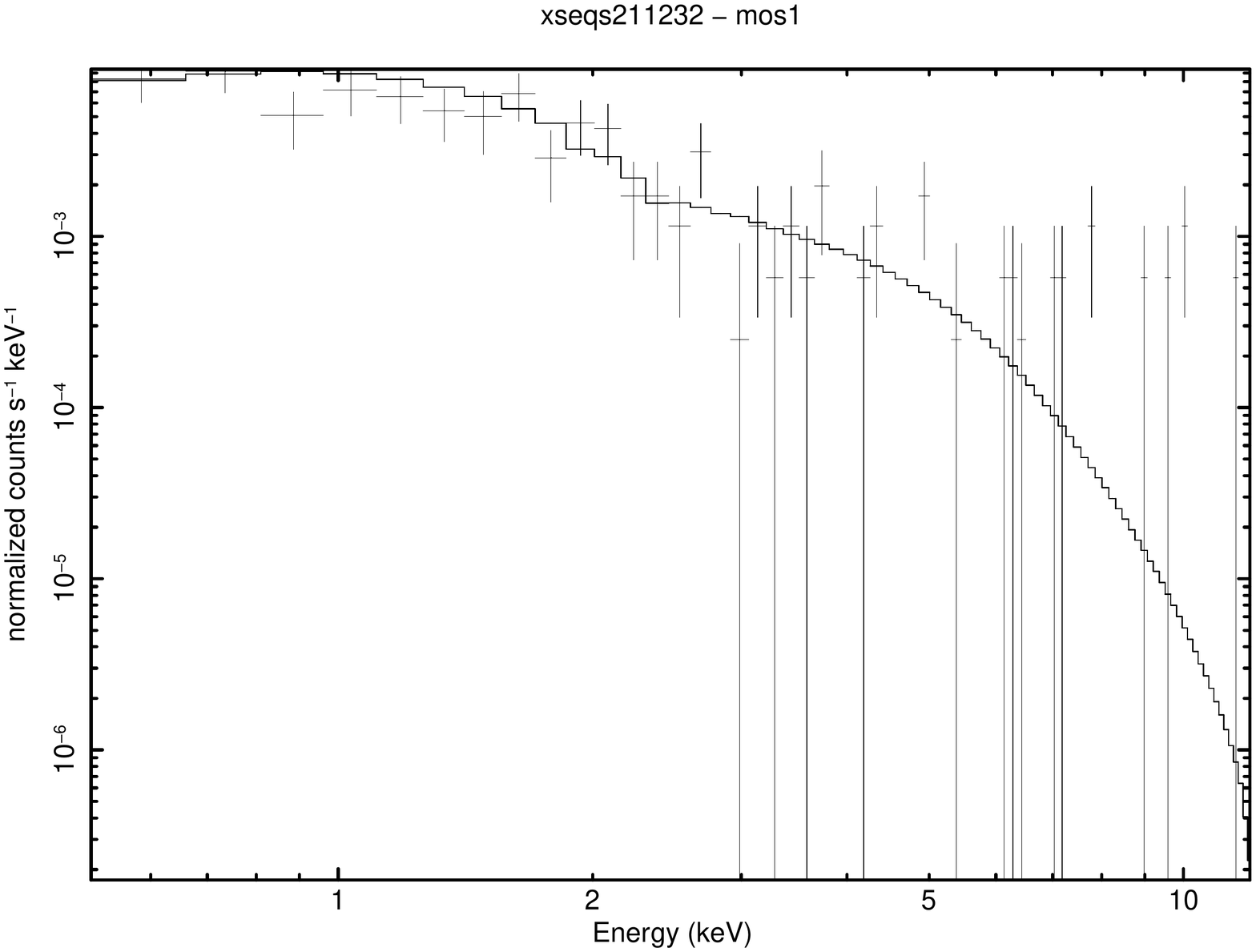} &
  \includegraphics[width=0.25\hsize]{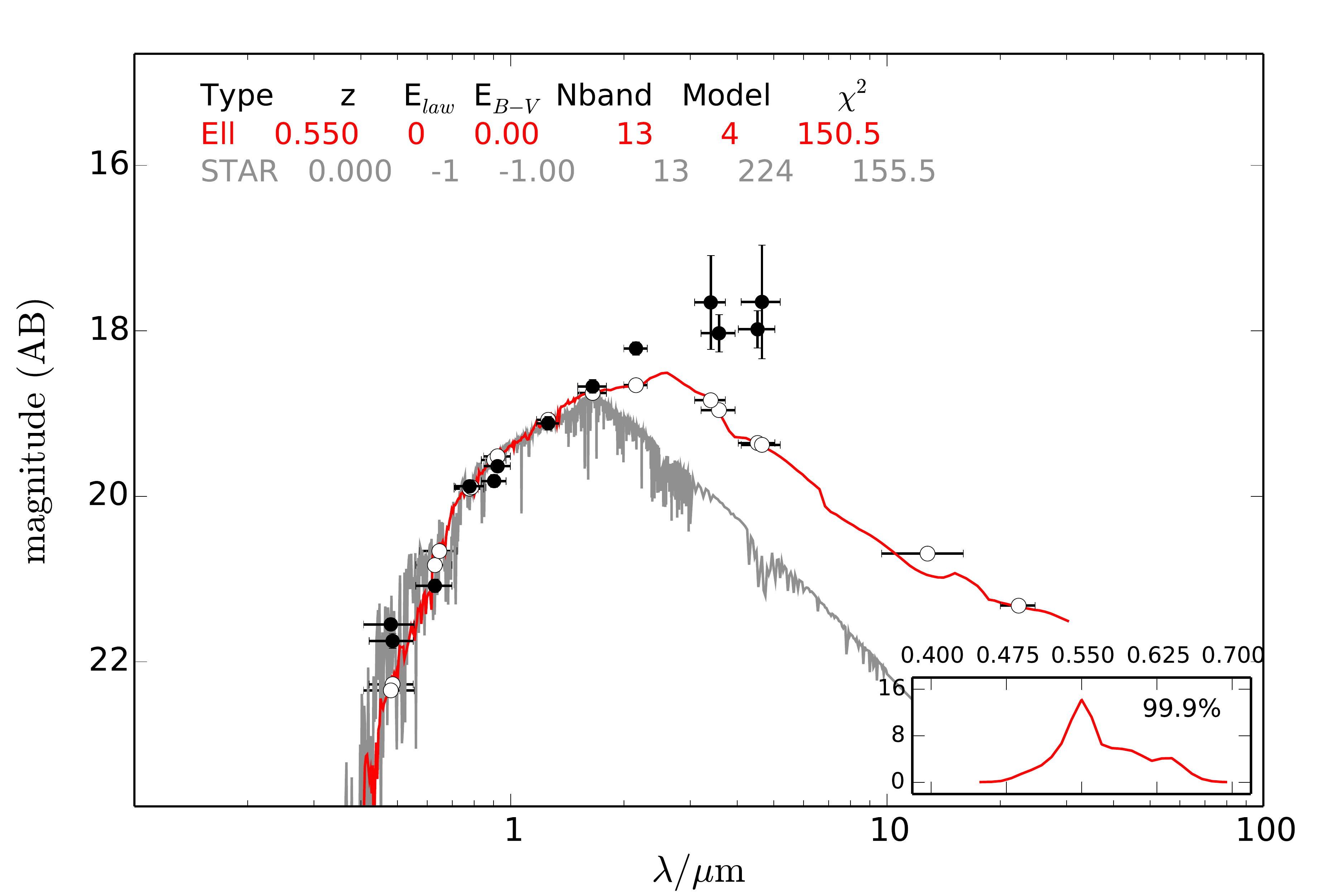} &
  \includegraphics[width=0.17\hsize]{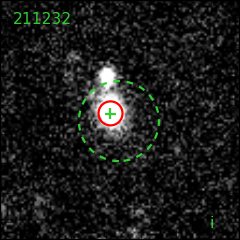}&
  \includegraphics[width=0.17\hsize]{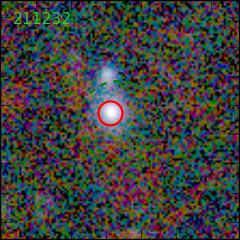}\\

\end{tabular}
\end{adjustbox}
\end{tabular}
\caption{Sources classified as passive. Panels as in Fig. \ref{app:QSO}}
\end{figure*}
\end{center}

\pagebreak
\onecolumn

\section{Probability density functions for $\rm{\log{N_{\rm{H}}}}$}\label{NH_PDF}
We provide the probability density functions (PDFs) of the $\rm{\log{N_{\rm{H}}}}$ in four hardness ratio (HR) bins. These curves can be used to draw random values of the $\rm{\log{N_{\rm{H}}}}$ when only HR is available for an X-ray source.

\begin{table}[h!]
\label{tab:NH_PDF}
\centering
\begin{tabular}{ccccc}
\hline
    \multirow{2}{*}{$\rm{\log{N_{\rm{H}}}}$} & \multicolumn{4}{c}{hardness ratio bin}\\ \cline{2-5}
                     & (-1.0, -0.25) & (-0.25, 0.25) & (0.25, 0.75) & (0.75, 1.00)\\ \hline
  19.08 & 0.271  &   0.060  &  0.013  &  0.006 \\
  19.26 & 0.411  &   0.099  &  0.021  &  0.014 \\
  19.43 & 0.442  &   0.110  &  0.023  &  0.017 \\
  19.61 & 0.455  &   0.115  &  0.025  &  0.020 \\
  19.78 & 0.456  &   0.118  &  0.027  &  0.023 \\
  19.96 & 0.445  &   0.118  &  0.028  &  0.025 \\
  20.13 & 0.428  &   0.116  &  0.031  &  0.025 \\
  20.31 & 0.402  &   0.113  &  0.033  &  0.023 \\
  20.48 & 0.375  &   0.109  &  0.035  &  0.021 \\
  20.66 & 0.355  &   0.107  &  0.038  &  0.018 \\
  20.83 & 0.337  &   0.121  &  0.040  &  0.015 \\
  21.01 & 0.317  &   0.137  &  0.037  &  0.013 \\
  21.18 & 0.302  &   0.151  &  0.035  &  0.012 \\
  21.36 & 0.249  &   0.224  &  0.037  &  0.013 \\
  21.53 & 0.202  &   0.351  &  0.050  &  0.013 \\
  21.71 & 0.130  &   0.531  &  0.077  &  0.014 \\
  21.88 & 0.087  &   0.633  &  0.145  &  0.015 \\
  22.06 & 0.047  &   0.581  &  0.336  &  0.019 \\
  22.23 & 0.039  &   0.530  &  0.687  &  0.033 \\
  22.41 & 0.039  &   0.401  &  0.862  &  0.077 \\
  22.58 & 0.028  &   0.360  &  0.994  &  0.222 \\
  22.76 & 0.008  &   0.246  &  0.825  &  0.632 \\
  22.93 & 0.004  &   0.141  &  0.490  &  1.004 \\
  23.11 & 0.002  &   0.111  &  0.365  &  0.865 \\
  23.28 & 0.001  &   0.062  &  0.208  &  0.837 \\
  23.46 & 0.001  &   0.024  &  0.108  &  0.707 \\
  23.63 & 0.001  &   0.018  &  0.085  &  0.444 \\
  23.81 & 0.001  &   0.017  &  0.039  &  0.294 \\
  23.98 & 0.001  &   0.015  &  0.006  &  0.197 \\
  24.16 & 0.001  &   0.009  &  0.002  &  0.072 \\
  24.33 & 0.001  &   0.002  &  0.002  &  0.009 \\
  24.51 & 4.3E-4 &   5.5E-4 &  0.001  &  0.00  \\
  24.68 & 0.00   &   0.00   &  0.00   &  0.00  \\ \hline
\end{tabular}
\end{table}

\onecolumn
 \begin{landscape}

\section{Excerpt from the XXL-1000-AGN catalogue}\label{cat:XXL1000AGN}
\begin{longtab}
\fontsize{6.5}{7.5}\selectfont
  \setlength{\tabcolsep}{1pt}
  \begin{longtable}{lrrrrrrrrrrrrrrrrrrrrrrrrrrrrrrrrrrrr }  
   \caption{\label{tab:XXL1000} Excerpt from the XXL-1000-AGN catalogue shown here for guidance. All fluxes are given in $\rm{erg~s^{-1}cm^{-2}}$, magnitudes in AB. See \S \ref{sec:release} for a detailed description of the catalogue content. The columns are as follows:
 (1) Xcatname, 
 (2) X-ray Right Ascension (J2000),
 (3) X-ray Declination (J2000),
 (4) $\rm{0.5-2\,keV}$ flux determined by \textsc{Xamin 3.3}, 
 (5) $\rm{2-10\,keV}$ flux determined by \textsc{Xamin 3.3}, 
 (6) spectral fit $\rm{F_{2-10\,keV}}$ mode of posterior distribution,
 (7)  68\% confidence lower limit for fit $\rm{F_{2-10\,keV}}$,
 (8)  68\% confidence upper limit for fit $\rm{F_{2-10\,keV}}$,
 (9)  spectral fit $\rm{F_{2-10\,keV}}$ median of posterior distribution,
 (10) spectral fit photon index, $\rm{\Gamma}$, mode of posterior distribution,
 (11) 68\% confidence lower limit for $\rm{\Gamma}$,
 (12) 68\% confidence upper limit for $\rm{\Gamma}$,
 (13) spectral fit $\rm{\Gamma}$ median of posterior distribution,
 (14) spectral fit hydrogen column density, $\rm{N_{\rm{H}}}$ mode of posterior distribution,
 (15) 68\% confidence lower limit for $\rm{N_{\rm{H}}}$,
 (16) 68\% confidence upper limit for $\rm{N_{\rm{H}}}$,
 (17) spectral fit $\rm{N_{\rm{H}}}$ median of posterior distribution,
 (18) X-ray spectrum quality flag,
 (19) Counterpart Right Ascension (J2000),
 (20) Counterpart Declination (J2000),
 (21) g-band magnitude,
 (22) error on g-band magnitude,
 (23) r-band magnitude,
 (24) error on r-band magnitude,
 (25) i-band magnitude,
 (26) error on i-band magnitude,
 (27) z-band magnitude,
 (28) error on z-band magnitude,
 (29) four-character code for parent survey(s) (see \S\ref{sec:release}),
 (30) spectroscopic redshift,
 (31) spectroscopy parent survey,
 (32) photometric redshift,
 (33) 68\% confidence photometric redshift lower limit,
 (34) 68\% confidence photometric redshift upper limit,
 (35) classification for best zphot,
 (36) Probability of being a star,
 (37) Probability of being an outlier.}\\
    \hline
3XLSS&2&3&4&5&6&7&8&9&10&11&12&13&14&15&16&17&18&19&20&21&22&23&24&25&26&27&28&29&30&31&32&33&34&35&36&37\\ \hline
J020019.6-06475& 30.0819& -6.7976&3.92e-14&8.77e-14&9.20e-14&7.83e-14&1.07e-13&9.20e-14&1.57&1.47&1.69&1.59&1.12e+19&1.08e+19&1.80e+20&7.38e+19&2& 30.0821& -6.7978&20.09&0.22&19.17&0.13&18.77&0.10&18.38&0.08&{\tt SSSS}&     &  &0.30&0.24&0.65&3&0.10&0.11\\
J020044.1-07195& 30.1841& -7.3310&        &2.18e-13&4.31e-12&2.06e-12&7.32e-12&3.82e-12&1.42&1.03&2.53&1.92&1.19e+23&6.96e+22&1.62e+23&1.04e+23&4& 30.1823& -7.3295&19.33&0.16&18.81&0.11&18.64&0.08&17.91&0.07&{\tt SSSS}&     &  &0.43&0.43&0.44&5&0.22&0.63\\
J020056.4-06310& 30.2351& -6.5191&1.56e-14&5.17e-14&2.53e-14&9.45e-15&4.93e-14&2.05e-14&1.55&1.03&2.13&1.82&3.20e+22&8.93e+20&1.51e+23&8.21e+21&4& 30.2359& -6.5194&22.13&0.01&21.40&0.01&21.58&0.01&21.56&0.03&{\tt CCCC}&     &  &1.35&1.30&1.42&5&0.02&0.21\\
J020057.9-06183& 30.2415& -6.3089&        &1.06e-13&9.41e-16&1.17e-16&7.54e-15&2.13e-15&1.04&1.03&2.20&1.83&5.42e+23&1.39e+20&1.41e+24&1.31e+22&4&        &        &     &    &     &    &     &    &     &    &{\tt -{}-{}-{}-{}}&     &  &    &    &    & &    &    \\
J020058.8-06582& 30.2453& -6.9729&1.81e-14&5.14e-14&3.83e-14&2.72e-14&4.80e-14&3.66e-14&2.11&1.77&2.51&2.18&1.30e+22&8.36e+21&1.69e+22&1.17e+22&2& 30.2453& -6.9729&23.06&0.02&22.36&0.02&21.53&0.01&21.05&0.02&{\tt CCCC}&     &  &0.83&0.71&0.90&4&0.01&0.10\\
J020100.3-07085& 30.2515& -7.1479&2.31e-14&5.44e-14&1.25e-13&9.08e-14&1.53e-13&1.16e-13&1.03&1.02&2.38&1.17&2.54e+20&5.61e+19&6.16e+21&6.19e+20&4& 30.2511& -7.1493&20.08&0.20&20.10&0.19&19.73&0.15&19.45&0.13&{\tt SSSS}&     &  &1.97&1.94&2.07&5&0.01&0.18\\
J020106.4-06485& 30.2767& -6.8166&1.45e-13&1.41e-12&1.29e-12&1.27e-12&1.31e-12&1.29e-12&1.00&1.00&1.01&1.00&1.03e+19&1.02e+19&4.83e+19&1.32e+19&1& 30.2770& -6.8158&     &    &     &    &     &    &     &    &{\tt -{}-{}-{}-{}}&     &  &0.19&0.13&0.23&1&0.31&0.33\\
J020109.2-06432& 30.2886& -6.7232&3.84e-14&1.21e-13&1.33e-13&1.20e-13&1.49e-13&1.34e-13&1.76&1.60&1.91&1.76&5.61e+21&4.50e+21&6.65e+21&5.46e+21&1& 30.2890& -6.7231&20.57&0.00&19.57&0.15&19.09&0.12&18.62&0.10&{\tt CSSS}&     &  &0.38&0.35&0.42&2&0.00&0.16\\
J020113.6-07011& 30.3067& -7.0201&1.28e-14&5.37e-14&2.10e-14&9.94e-15&3.71e-14&1.80e-14&1.69&1.39&2.33&1.89&1.19e+19&1.14e+19&1.36e+21&2.52e+20&4& 30.3076& -7.0204&23.29&0.03&21.94&0.01&20.91&0.01&20.59&0.02&{\tt CCCC}&     &  &0.59&0.53&0.62&3&0.00&0.07\\
J020115.9-05152& 30.3163& -5.2573&1.17e-14&6.39e-14&2.16e-15&1.15e-16&1.71e-14&3.08e-15&1.51&1.03&2.29&1.89&1.85e+23&6.45e+19&5.51e+23&5.35e+21&2& 30.3161& -5.2576&22.83&0.02&21.99&0.02&21.27&0.01&20.20&0.21&{\tt CCCS}&1.641&12&1.51&1.48&1.57&5&0.05&0.21\\
J020119.2-06172& 30.3302& -6.2905&8.73e-14&1.48e-13&2.17e-13&1.94e-13&2.44e-13&2.17e-13&1.85&1.76&1.93&1.85&1.06e+21&2.25e+19&1.77e+21&2.54e+20&1& 30.3301& -6.2903&19.04&0.12&18.58&0.09&18.47&0.09&18.37&0.09&{\tt SSSS}&1.443& 7&1.46&1.35&1.50&5&0.03&0.19\\
J020121.3-06172& 30.3389& -6.2903&        &1.06e-13&1.29e-16&1.21e-16&3.59e-14&6.09e-15&1.93&1.03&2.33&1.92&2.45e+23&4.25e+20&1.04e+24&1.06e+22&4&        &        &     &    &     &    &     &    &     &    &{\tt -{}-{}-{}-{}}&     &  &    &    &    & &    &    \\
J020124.5-06192& 30.3524& -6.3234&        &8.65e-14&3.07e-13&1.27e-13&5.45e-13&2.45e-13&1.60&1.03&2.30&1.88&5.93e+23&1.88e+23&9.61e+23&3.99e+23&4& 30.3527& -6.3226&24.89&0.15&25.13&0.29&24.51&0.23&24.12&0.33&{\tt CCCC}&     &  &1.98&0.69&2.27&3&0.12&0.34\\
J020125.6-05380& 30.3569& -5.6360&        &1.19e-13&1.52e-13&4.71e-14&3.55e-13&1.09e-13&1.52&1.03&2.22&1.88&1.02e+23&1.37e+22&3.23e+23&5.11e+22&4& 30.3577& -5.6379&     &    &     &    &     &    &     &    &{\tt -{}-{}-{}-{}}&     &  &0.86&0.84&0.88&1&0.00&0.00\\
J020127.4-07170& 30.3642& -7.2841&        &1.36e-13&1.24e-13&5.00e-14&2.00e-13&9.77e-14&1.55&1.03&2.04&1.73&5.90e+23&6.81e+22&1.35e+24&1.93e+23&4&        &        &     &    &     &    &     &    &     &    &{\tt -{}-{}-{}-{}}&     &  &    &    &    & &    &    \\
J020135.2-07172& 30.3967& -7.2915&        &1.74e-13&6.09e-14&1.22e-16&1.73e-13&1.28e-14&1.91&1.03&2.25&1.87&3.65e+23&9.71e+20&1.23e+24&2.52e+22&4& 30.3964& -7.2929&24.85&0.13&24.27&0.13&23.96&0.12&     &    &{\tt CCC-{}}&     &  &0.22&0.01&3.51&4&0.05&0.39\\
J020135.4-05084& 30.3976& -5.1465&        &5.04e-14&5.48e-15&1.81e-15&1.34e-14&4.61e-15&1.89&1.31&2.41&1.95&2.44e+21&6.62e+19&1.25e+22&1.31e+21&4& 30.3976& -5.1464&23.77&0.04&22.79&0.03&21.66&0.02&20.57&0.23&{\tt CCCS}&     &  &0.72&0.70&0.80&2&0.02&0.02\\
J020138.7-06223& 30.4114& -6.3773&        &1.66e-13&1.31e-12&5.62e-13&3.38e-12&1.20e-12&1.04&1.03&2.98&1.75&5.73e+23&2.63e+23&9.49e+23&4.71e+23&2&        &        &     &    &     &    &     &    &     &    &{\tt -{}-{}-{}-{}}&     &  &    &    &    & &    &    \\
J020141.7-05144& 30.4241& -5.2452&3.37e-14&8.30e-14&7.64e-14&5.82e-14&9.42e-14&7.50e-14&1.83&1.46&2.07&1.78&2.26e+21&1.10e+21&3.57e+21&1.85e+21&2& 30.4243& -5.2454&20.40&0.00&19.46&0.15&18.92&0.10&18.57&0.11&{\tt CSSS}&0.210& 7&0.17&0.16&0.22&2&0.00&0.02\\
J020145.6-06093& 30.4404& -6.1592&3.97e-14&5.41e-14&5.02e-14&4.28e-14&5.60e-14&4.92e-14&1.99&1.87&2.14&2.01&3.60e+21&2.05e+21&5.99e+21&3.11e+21&1& 30.4402& -6.1595&21.10&0.00&20.83&0.01&20.47&0.22&20.30&0.25&{\tt CCSS}&1.640& 7&1.56&1.50&1.61&5&0.00&0.38\\
J020147.4-04210& 30.4479& -4.3515&        &7.99e-14&6.31e-14&2.62e-14&2.13e-13&7.17e-14&1.04&1.03&1.89&1.62&8.41e+23&1.97e+23&1.69e+24&3.97e+23&4&        &        &     &    &     &    &     &    &     &    &{\tt -{}-{}-{}-{}}&     &  &    &    &    & &    &    \\
J020153.6-05195& 30.4737& -5.3325&        &6.95e-14&1.73e-13&5.04e-14&5.95e-13&1.53e-13&1.04&1.03&2.42&1.92&1.99e+23&7.57e+22&5.23e+23&1.84e+23&3& 30.4739& -5.3324&21.60&0.01&20.36&0.00&19.98&0.17&19.49&0.15&{\tt CCSS}&0.348& 7&0.36&0.34&0.38&3&0.01&0.13\\
J020157.2-06380& 30.4887& -6.6358&2.15e-14&5.79e-14&4.28e-14&2.97e-14&5.37e-14&4.06e-14&1.65&1.46&1.84&1.67&1.15e+19&1.11e+19&3.08e+20&1.16e+20&2& 30.4885& -6.6350&22.22&0.01&21.63&0.01&21.45&0.01&20.97&0.02&{\tt CCCC}&     &  &1.09&1.05&1.16&4&0.02&0.18\\
J020159.7-04115& 30.4988& -4.1983&1.97e-14&6.80e-14&6.22e-14&4.07e-14&1.06e-13&6.12e-14&1.56&1.22&1.83&1.58&9.77e+19&1.12e+19&7.78e+20&1.91e+20&4& 30.4994& -4.1979&20.92&0.00&20.93&0.01&20.48&0.22&19.91&0.18&{\tt CCSS}&     &  &1.66&1.60&1.90&5&0.00&0.25\\
J020207.0-05585& 30.5293& -5.9830&3.24e-13&6.68e-13&5.66e-13&5.38e-13&5.92e-13&5.66e-13&1.95&1.91&2.01&1.96&8.36e+20&6.78e+20&9.27e+20&7.93e+20&1& 30.5286& -5.9834&18.90&0.00&18.19&0.00&17.61&0.00&17.43&0.00&{\tt CCCC}&0.189& 7&0.70&0.65&0.74&3&0.30&0.24\\
J020209.1-04103& 30.5383& -4.1770&        &5.47e-14&1.19e-16&1.15e-16&2.14e-15&9.21e-16&1.69&1.23&2.33&1.89&1.19e+23&1.72e+21&1.22e+24&1.89e+22&4& 30.5392& -4.1767&24.67&0.10&24.82&0.16&24.63&0.22&     &    &{\tt CCC-{}}&     &  &1.82&0.13&2.66&3&0.07&0.39\\
J020215.1-07414& 30.5633& -7.6945&5.37e-14&9.16e-14&9.78e-14&8.46e-14&1.21e-13&1.01e-13&1.74&1.60&1.84&1.74&1.14e+19&1.10e+19&2.18e+20&9.26e+19&2& 30.5634& -7.6949&20.34&0.26&20.04&0.19&20.04&0.18&20.16&0.21&{\tt SSSS}&     &  &1.45&1.40&1.48&5&0.08&0.51\\
J020221.5-04090& 30.5897& -4.1518&        &6.13e-14&2.46e-13&2.74e-14&8.55e-13&1.32e-13&1.04&1.03&2.01&1.72&7.51e+23&9.52e+22&1.64e+24&2.64e+23&4&        &        &     &    &     &    &     &    &     &    &{\tt -{}-{}-{}-{}}&     &  &    &    &    & &    &    \\
J020225.9-06115& 30.6080& -6.1985&1.15e-14&6.67e-14&7.51e-14&5.96e-14&8.49e-14&7.19e-14&1.52&1.25&1.80&1.56&1.44e+22&9.69e+21&1.92e+22&1.33e+22&1& 30.6083& -6.1990&19.78&0.17&19.89&0.17&19.69&0.15&19.52&0.16&{\tt SSSS}&0.561& 8&0.76&0.72&0.77&5&0.03&0.25\\
J020228.5-04504& 30.6192& -4.8462&6.27e-14&1.30e-13&1.29e-13&1.05e-13&1.46e-13&1.26e-13&1.80&1.63&1.99&1.81&5.85e+22&2.53e+22&1.11e+23&4.44e+22&1& 30.6197& -4.8461&20.35&0.00&19.78&0.17&19.42&0.14&18.76&0.10&{\tt CSSS}&0.293& 7&0.31&0.27&0.34&4&0.01&0.12\\
J020230.1-04424& 30.6256& -4.7123&        &9.83e-14&1.24e-16&1.18e-16&7.16e-15&2.03e-15&1.75&1.03&2.22&1.85&4.98e+23&1.34e+21&1.22e+24&1.76e+22&4& 30.6265& -4.7138&     &    &     &    &     &    &     &    &{\tt -{}-{}-{}-{}}&     &  &2.84&0.10&3.39&3&0.07&0.29\\
J020231.3-04224& 30.6307& -4.3786&2.14e-14&6.98e-14&2.18e-13&1.45e-13&3.86e-13&2.44e-13&1.04&1.03&2.99&1.41&5.92e+23&1.33e+23&1.23e+24&3.15e+23&4& 30.6298& -4.3797&     &    &     &    &     &    &     &    &{\tt -{}-{}-{}-{}}&4.270& 7&1.73&1.72&1.79&3&0.02&0.11\\
J020232.3-06422& 30.6348& -6.7077&2.34e-14&5.29e-14&5.35e-14&4.51e-14&6.10e-14&5.26e-14&2.09&1.99&2.20&2.11&2.92e+19&1.07e+19&1.23e+20&6.00e+19&2& 30.6349& -6.7079&19.09&0.14&18.33&0.09&17.98&0.07&17.69&0.06&{\tt SSSS}&0.075& 7&0.08&0.05&0.13&2&0.02&0.02\\
J020237.6-06143& 30.6569& -6.2433&7.99e-14&1.60e-13&1.85e-13&1.50e-13&2.19e-13&1.80e-13&2.01&1.84&2.27&2.06&1.39e+21&8.58e+20&1.73e+21&1.19e+21&1& 30.6567& -6.2430&19.89&0.22&20.01&0.22&20.03&0.22&19.96&0.01&{\tt SSSC}&0.005& 5&0.02&0.02&0.02&1&0.33&0.50\\
J020238.5-06142& 30.6604& -6.2398&6.30e-14&1.79e-13&1.52e-12&1.20e-12&1.74e-12&1.44e-12&1.30&1.17&1.44&1.32&1.17e+19&1.12e+19&7.01e+20&1.82e+20&2& 30.6581& -6.2403&19.39&0.17&19.37&0.16&19.53&0.00&19.47&0.00&{\tt SSCC}&     &  &0.02&0.02&0.02&1&0.30&0.50\\
J020238.7-04121& 30.6615& -4.2029&        &9.46e-14&5.63e-15&1.83e-15&1.85e-14&5.20e-15&1.69&1.47&2.60&2.02&1.36e+21&6.61e+19&6.03e+22&3.28e+21&4&        &        &     &    &     &    &     &    &     &    &{\tt -{}-{}-{}-{}}&     &  &    &    &    & &    &    \\
J020239.4-04085& 30.6644& -4.1500&1.55e-14&7.16e-14&1.44e-13&1.15e-13&1.67e-13&1.38e-13&1.02&1.01&1.19&1.14&1.05e+20&1.10e+19&2.24e+20&8.75e+19&2& 30.6646& -4.1503&     &    &     &    &     &    &     &    &{\tt -{}-{}-{}-{}}&0.043& 5&0.03&0.01&0.05&3&0.10&0.09\\
J020245.1-06081& 30.6882& -6.1383&5.38e-15&5.24e-14&6.86e-14&5.34e-14&7.92e-14&6.45e-14&1.03&1.03&2.71&1.19&2.21e+22&1.38e+22&3.65e+22&2.22e+22&2& 30.6884& -6.1390&24.56&0.09&23.45&0.05&22.68&0.03&21.92&0.04&{\tt CCCC}&     &  &1.08&1.01&1.13&3&0.04&0.07\\
J020246.5-05314& 30.6939& -5.5299&        &7.46e-14&3.78e-14&1.26e-14&1.29e-13&3.71e-14&1.29&1.03&2.57&1.90&1.49e+23&3.32e+22&4.44e+23&1.00e+23&4& 30.6938& -5.5299&24.55&0.07&23.31&0.05&22.26&0.02&21.67&0.03&{\tt CCCC}&     &  &0.92&0.87&1.05&3&0.05&0.07\\
J020259.6-05125& 30.7485& -5.2147&        &7.05e-14&1.22e-16&1.16e-16&2.42e-15&1.12e-15&1.94&1.03&2.34&1.93&6.68e+23&2.60e+20&1.43e+24&1.45e+22&4& 30.7483& -5.2152&24.32&0.06&24.05&0.10&23.80&0.10&23.41&0.16&{\tt CCCC}&     &  &1.10&0.26&1.33&3&0.12&0.30\\
J020302.8-07414& 30.7617& -7.6965&2.52e-13&4.07e-13&3.09e-13&2.94e-13&3.23e-13&3.08e-13&2.16&2.13&2.19&2.16&1.09e+19&1.07e+19&3.63e+19&2.52e+19&1& 30.7631& -7.6978&     &    &     &    &     &    &     &    &{\tt -{}-{}-{}-{}}&0.067& 5&0.01&0.01&0.01&1&0.22&0.44\\
J020303.1-07143& 30.7630& -7.2442&4.72e-14&1.09e-13&7.76e-14&6.33e-14&8.93e-14&7.53e-14&1.82&1.72&1.94&1.84&1.12e+19&1.09e+19&1.43e+20&7.11e+19&2& 30.7630& -7.2440&21.13&0.01&20.35&0.22&20.32&0.19&18.84&0.17&{\tt CSSS}&     &  &0.98&0.97&0.98&4&0.10&0.19\\
J020306.4-04593& 30.7770& -4.9933&4.33e-14&1.02e-13&7.10e-14&6.08e-14&8.23e-14&7.12e-14&1.88&1.80&2.00&1.90&1.12e+19&1.09e+19&1.52e+20&6.12e+19&2& 30.7769& -4.9933&19.26&0.14&19.10&0.12&19.35&0.11&19.28&0.12&{\tt SSSS}&1.004&12&0.79&0.64&0.82&5&0.03&0.22\\
J020307.3-04262& 30.7805& -4.4396&        &7.95e-14&7.15e-12&1.28e-16&2.07e-11&8.58e-14&1.62&1.03&2.38&1.92&1.07e+23&8.66e+20&9.07e+23&2.52e+22&4& 30.7816& -4.4402&22.04&0.01&21.85&0.01&21.86&0.02&22.00&0.06&{\tt CCCC}&     &  &0.05&0.04&0.25&3&0.16&0.30\\
J020307.4-04321& 30.7811& -4.5380&2.40e-14&5.20e-14&6.47e-14&5.35e-14&7.74e-14&6.43e-14&1.42&1.30&1.59&1.46&4.01e+21&1.13e+19&8.63e+21&7.35e+20&2& 30.7812& -4.5379&21.81&0.01&22.05&0.01&21.37&0.01&21.21&0.02&{\tt CCCC}&     &  &2.19&2.14&2.24&5&0.15&0.27\\
J020310.1-05360& 30.7925& -5.6021&3.06e-14&6.21e-14&5.66e-14&3.37e-14&9.28e-14&5.59e-14&1.59&1.18&2.27&1.78&3.27e+21&2.69e+20&7.76e+21&1.35e+21&4& 30.7931& -5.6022&21.74&0.01&20.85&0.01&20.34&0.00&20.19&0.21&{\tt CCCS}&0.308& 5&0.36&0.33&0.44&2&0.00&0.12\\
J020312.7-05323& 30.8030& -5.5427&2.26e-14&5.47e-14&6.98e-14&4.99e-14&9.30e-14&6.79e-14&1.51&1.27&1.72&1.52&2.41e+20&1.10e+19&7.28e+20&1.66e+20&4& 30.8030& -5.5439&20.41&0.24&20.57&0.23&20.49&0.22&20.05&0.19&{\tt SSSS}&0.812& 8&0.64&0.62&0.66&5&0.04&0.40\\
J020319.3-07184& 30.8307& -7.3134&2.65e-14&5.22e-14&2.58e-14&2.05e-14&3.72e-14&2.74e-14&2.14&1.89&2.29&2.12&7.48e+20&3.78e+19&1.63e+21&2.43e+20&4& 30.8308& -7.3137&18.23&0.09&18.04&0.07&17.96&0.06&18.12&0.07&{\tt SSSS}&     &  &1.16&1.06&1.26&5&0.35&0.66\\
J020320.6-07375& 30.8360& -7.6314&2.40e-14&4.96e-14&6.22e-14&4.68e-14&7.36e-14&5.93e-14&2.17&1.85&2.47&2.18&5.16e+21&4.09e+21&7.29e+21&5.31e+21&2& 30.8370& -7.6322&20.30&0.00&19.37&0.15&18.84&0.10&18.50&0.10&{\tt CSSS}&     &  &0.22&0.17&0.26&3&0.14&0.57\\
J020326.1-05553& 30.8591& -5.9251&3.32e-14&6.31e-14&6.57e-14&5.41e-14&7.89e-14&6.61e-14&1.69&1.57&1.82&1.71&1.11e+19&1.09e+19&1.12e+20&5.52e+19&2& 30.8597& -5.9254&18.95&0.13&18.88&0.12&18.60&0.10&18.25&0.00&{\tt SSSC}&0.429& 7&0.45&0.44&0.45&4&0.17&0.44\\\hline
  \end{longtable}
\end{longtab}
 \end{landscape}
\end{appendix}

\end{document}